\pdfoutput=1
\documentclass{pasa}
\usepackage{lscape}
\usepackage{longtable}

\title[The Ursa Major cluster redefined as a `supergroup']{The Ursa Major cluster redefined as a `supergroup'}
\author[K. Wolfinger et al.]{K.~Wolfinger$^1$, V.~A.~Kilborn$^1$, E.~V.~Ryan-Weber$^1$, B.~S.~Koribalski$^2$\\
$^1$ Centre for Astrophysics \& Supercomputing, Swinburne University of Technology, Mail H39, PO Box 218, Hawthorn, VIC 3122, Australia\\
$^2$ CSIRO Astronomy and Space Science, PO Box 76, Epping, NSW 1710, Australia}

\usepackage[authoryear]{natbib}
\bibpunct{(}{)}{;}{a}{}{,}
\usepackage{amsmath}
\usepackage{amssymb}

\begin{document}

\begin{abstract}
We identify gravitationally bound structures in the Ursa Major region using positions, velocities and photometry from the Sloan Digital Sky Survey (SDSS DR7) and the Third Reference Catalogue of Bright Galaxies (RC3). A friends-of-friends algorithm is extensively tested on mock galaxy lightcones and then implemented on the real data to determine galaxy groups whose members are likely to be physically and dynamically associated with one another. We find several galaxy groups within the region that are likely bound to one another and in the process of merging. We classify 6 galaxy groups as the Ursa Major `supergroup', which are likely to merge and form a poor cluster with a mass of $\sim 8 \times 10^{13}$~M$_{\odot}$. Furthermore, the Ursa Major supergroup as a whole is likely bound to the Virgo cluster, which will eventually form an even larger system in the context of hierarchical structure formation.

To investigate the evolutionary state of the galaxy groups in the Ursa Major region, we examine the location of the brightest group galaxy (BGG) as well as galaxy properties such as colour, morphological type, luminosity and HI content. The galaxy properties are examined with respect to group/non-group membership as well as a function of their local projected surface density. We find that (\textit{i}) the BGGs often show a relatively large spatial and kinematic offset from the group centres, (\textit{ii}) the increase in the fraction of early-type galaxies with local projected surface density is shallow, (\textit{iii}) galaxy colours tend to be redder with increasing density, (\textit{iv}) there is no trend towards brighter galaxies with increasing density and (\textit{v}) several galaxies with HI excess are found in low density environments, whereas HI deficient galaxies tend to reside in the galaxy groups. We note that several galaxy groups contain members with disturbed HI content or HI tails. Given the low velocity dispersion of the groups and the lack of X-ray emission, these disturbances are likely due to galaxy-galaxy interactions. We conclude that the galaxy groups in the Ursa Major region are in an early evolutionary state and the properties of their member galaxies are similar to those in the field. 
\end{abstract}


\maketitle

\section{Introduction}

Conglomerations of galaxy groups that will merge to form a cluster are called `supergroups' (\citealt{gonzalez2005}; \citealt{brough2006GEMS}; \citealt{brough2006}). For example, SG1120-12 (at $z = 0.37$) is comprised of 4 distinct galaxy groups that are gravitationally bound to one another and will merge into a cluster comparable in mass to Coma by $z \sim 0$ (\citealt{gonzalez2005}; \citealt{tran2009}). Supergroups provide an ideal opportunity to study the effect of `pre-processing' in the group environment. This may be responsible for the observed differences between the properties of galaxies residing in dense regions and those in the field.

Observations have shown that massive, bright, red and passive early-type galaxies are commonly found in dense environments, whereas low mass, faint, blue and star forming galaxies dominate the less dense environments (e.g. \citealt{dressler1980}; \citealt{butcher1984}; \citealt{kauffmann2004}; \citealt{blanton2005}; \citealt{girardi2005}; \citealt{ellison2009}). Galaxy properties begin to change from the ones in the field to those of the dense environment at $\sim$2-3 virial cluster radii (e.g. \citealt{lewis2002}). The projected surface density at which these changes occur corresponds to $\sim 1$~Mpc$^{-2}$, which is similar to the density of poor groups \citep{gomez2003}. 

The distinction between the cluster and group environment is not unambiguous. In dynamically evolved groups, hot intra-group X-ray gas has been observed, which used to be a distinctive feature of the cluster environment (e.g. \citealt{mulchaey1998}; \citealt{rasmussen2006}). Furthermore, these dynamically evolved groups have bright ellipticals in the centre and a high fraction of early-type galaxies which used to be typical for the cluster environment. Whilst galaxy groups and clusters can be comparable in size (1-5 Mpc across), the term `cluster' usually refers to a numerous overdensity of galaxies (hundreds or even more than a thousand galaxies), whereas the term `group' is used for a smaller accumulations (ranging from two to tens, but also up to a few hundred galaxies - see Table 1 in \citealt{kilborn2009} for examples of detailing galaxy groups). Therefore, environment can be measured by the density of galaxies per Mpc$^3$. In general, clusters are more massive ($\geq 10^{14}$~M$_{\odot}$; e.g. \citealt{gavazzi2009}) than groups ($10^{12}-10^{13}$~M$_{\odot}$, \citealt{brough2006}). 

This is the second paper in a series analyzing the nearby Ursa Major region. The Ursa Major `cluster' has been previously defined in the literature \citep{tully1996} and we summarize this in the first paper \citep{wolfinger2013}. Clusters usually have a broader velocity distribution, a hot intracluster gas which can strip the galaxies of their HI gas and an enhanced population of both the early-type (S0 and E) galaxies and the HI depleted spiral galaxies. For example, (\textit{i}) the velocity dispersion of Virgo is 715~km~s$^{-1}$ whereas the velocity dispersion of Ursa Major is only 148~km~s$^{-1}$, (\textit{ii}) the number of early-type galaxies is 66 for Virgo as opposed to 9 for Ursa Major and (\textit{iii}) the number of late-type galaxies is 91 for Virgo and 53 for Ursa Major (see \citealt{tully1996} for quantitative differences between Ursa Major and Virgo/Fornax). Given that the majority of the Ursa Major members are HI-rich spiral galaxies with a small number of dwarf galaxies and a low abundance of elliptical galaxies as well as a total mass $< 10^{14}$~M$_{\odot}$, a low velocity dispersion and the non-detection of X-ray emitting intra-cluster gas, make it apparent that Ursa Major does not meet the criteria for a cluster. However, it is also clear that its structure is more complex than a simple group. Our hypothesis is that Ursa Major is a supergroup, i.e. Ursa Major is comprised of distinct galaxy groups that are gravitationally bound to one another and will eventually merge to form a cluster.

The aim of this paper is to identify gravitationally bound structures in the Ursa Major region and to investigate the evolutionary state of these structures. This paper is organized as follows: in Section~\ref{sec:sample} the optical and HI data of the galaxy sample is presented. A friends-of-friends algorithm \citep{huchra1982} is tested on mock galaxy catalogues that are constrained by the real data to optimize the choice of input parameters (Section~\ref{sec:structures}). In Section~\ref{sec:results} an overview of the structures found in the Ursa Major region, their properties and their dynamics are presented. The evolutionary state of the Ursa Major galaxy groups are examined by investigating the location of the brightest group galaxy (BGG; Section~\ref{sec:BGG}) and studying galaxy properties as a function of environment (Section~\ref{sec:GalProp}). Trends in colour, morphological type, luminosity and HI content as a function of environment could indicate a transformation occurring in the galaxies. In Section~\ref{sec:discussion} we compare our picture of the Ursa Major supergroup to previous studies (e.g. \citealt{tully1996}; \citealt{karachentsev2013}) and ask the question: `Will Ursa Major and Virgo merge to form an even larger system?'. Throughout this paper we assume $H_0 = 73$~km~s$^{-1}$~Mpc$^{-1}$.

\section{The galaxy sample}
\label{sec:sample}

\subsection{Optical data}
\label{subsec:opt}

In order to analyse the region a velocity-limited ($300 \leq v_{\mathrm{LG}} \leq 3000$~km~s$^{-1}$), absolute-magnitude-limited ($\mathrm{M}_r \leq -15.3$~mag) and stellar-mass-limited ($\mathrm{M_{\ast}} \geq 10^8~\mathrm{M}_{\odot}$) sample of 1209 galaxies with spectroscopic redshifts residing within $9^{h}30^{m}\leq\alpha\leq+14^{h}30^{m}$ and $20^{\circ}\leq\delta\leq+65^{\circ}$ was obtained from the Sloan Digital Sky Survey (SDSS DR7; \citealt{abazajian2009}) and the NASA/IPAC Extragalactic Database (NED). The region of interest includes the Ursa Major `cluster' and its surroundings spanning approximately 400~Mpc$^2$ at the distance of the `cluster' (17.1~Mpc; \citealt{tully2008}).

\begin{figure*}
\begin{center}
\begin{tabular}{p{7cm} p{0.5cm} p{7cm}} 
\includegraphics[trim=0.1cm 0.5cm 6.5cm 0.5cm, clip=true,width=7.5cm]{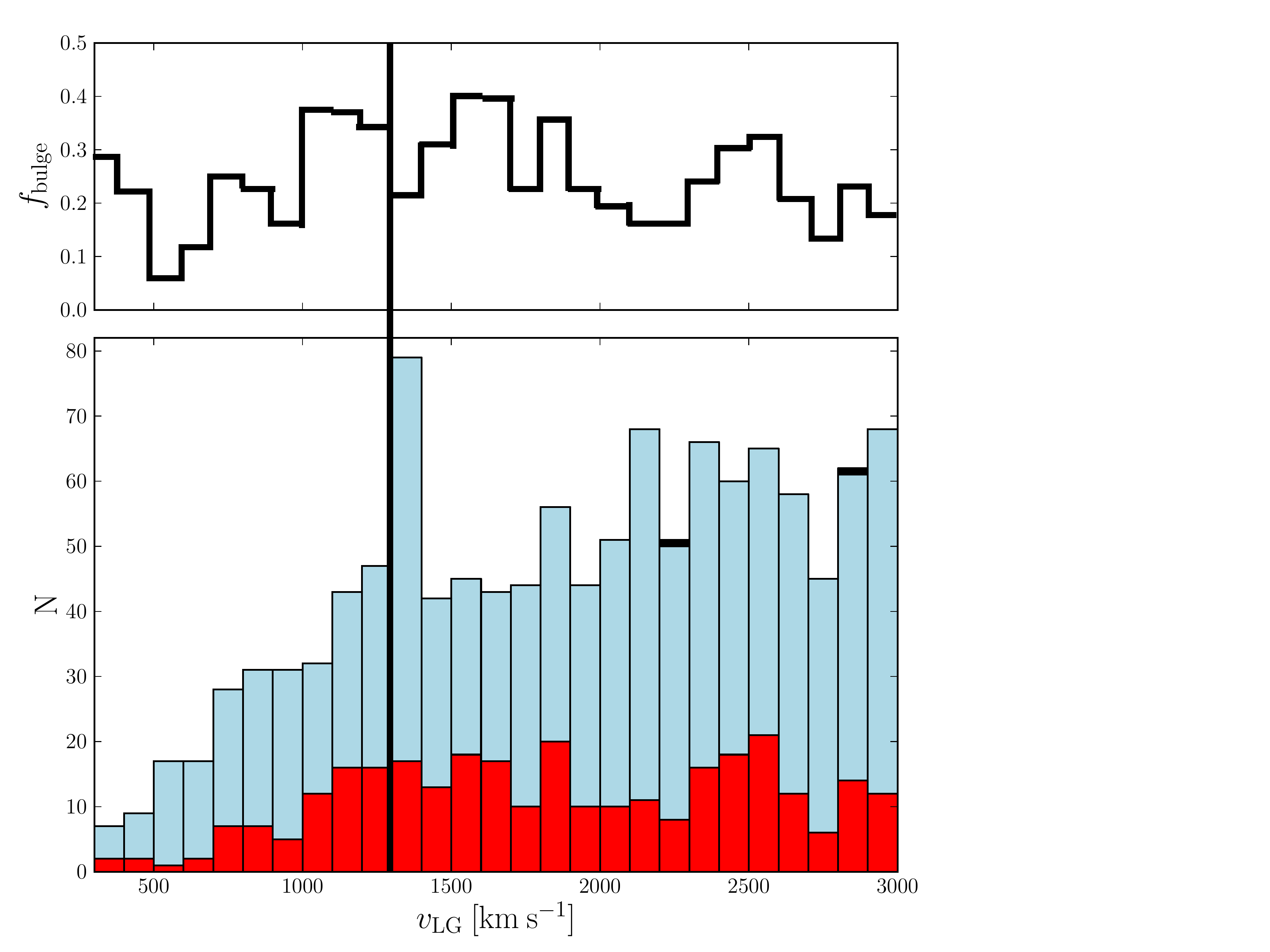}& &
\includegraphics[trim=0.1cm 0.5cm 6.5cm 0.5cm, clip=true,width=7.5cm]{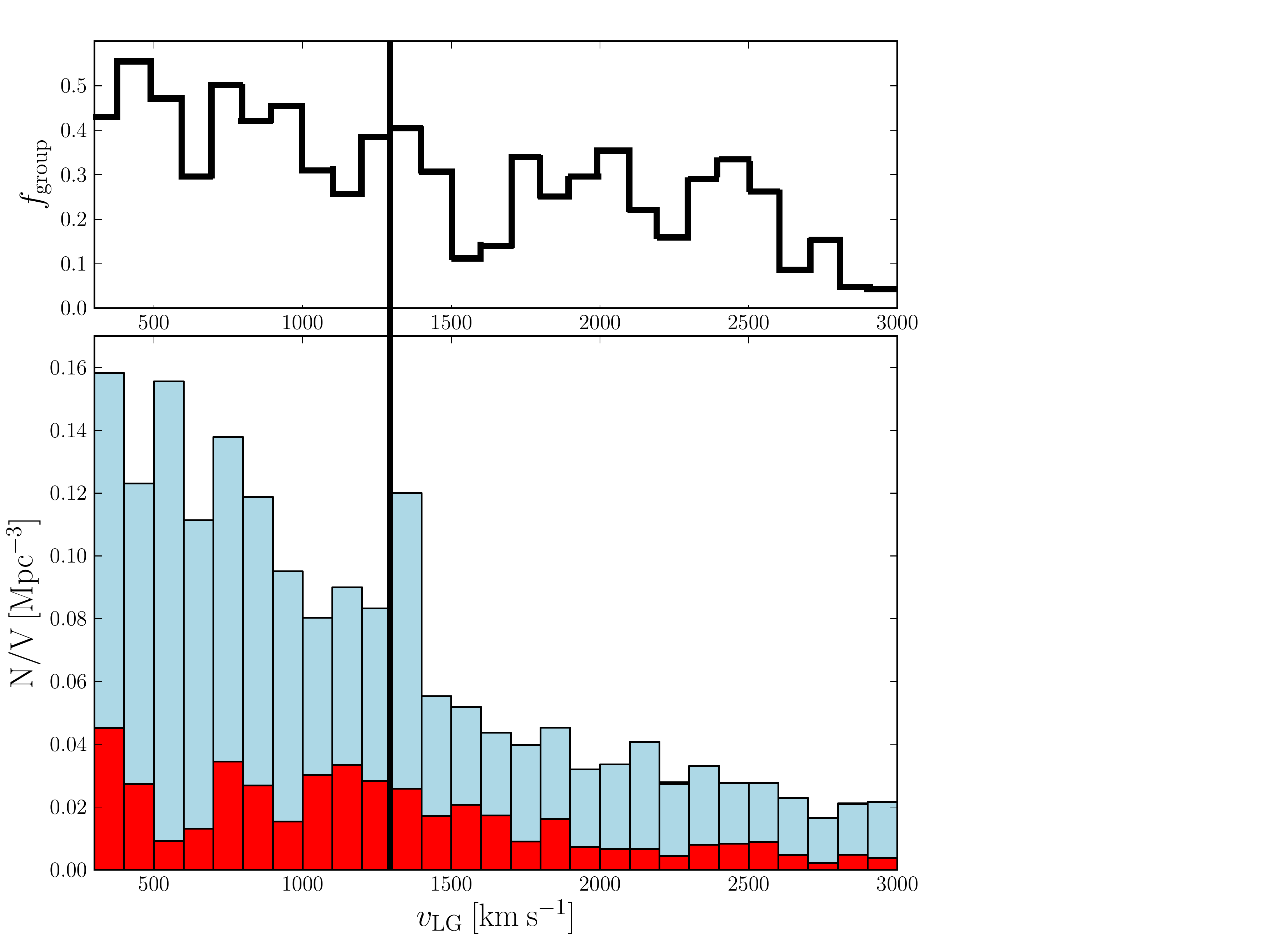}\\
\end{tabular}
\caption{(\textit{top left}) The fraction of bulge dominated galaxies in the complete sample per velocity bin. (\textit{bottom left}) The absolute velocity distribution of the complete sample wherein bulge dominated galaxies are shown in red, disk dominated galaxies are shown in blue and galaxies with RC3 data without morphological T-type available are shown in black (stacked histogram). The majority of galaxies in the Ursa Major region are disk dominated galaxies. The overdensity appearing at $\sim 1300$~km~s$^{-1}$ (marked with a vertical line) indicates the presence of the Ursa Major `cluster' without an apparent overdensity in early-type galaxies as one might expect for a dense environment. Note that the volume corresponding to a velocity bin increases from low to high velocities, nevertheless, the overdensity is apparent. (\textit{top right}) The fraction of galaxies in the complete sample that are residing in groups (as determined in Section~\ref{sec:est_prop}) per velocity bin. (\textit{bottom right}) The velocity distribution per unit volume of the complete sample (stacked histogram). At low recession velocities ($v_{\mathrm{LG}} \leq 1500$~km~s$^{-1}$) 30 to 60~per~cent of the galaxies reside in groups (corresponding to a high number density of galaxies or clustered galaxies), whereas at higher velocities only 10 to 30~per~cent of galaxies are found in dense regions (corresponding to a low number density of galaxies or a less clustered galaxy distribution).}
\label{fig:hist_vlg}
\end{center}
\end{figure*}

The main SDSS galaxy sample (photometry and spectroscopy) is highly complete to an approximate apparent $r$-band Petrosian magnitude of 17.77 \citep{strauss2002}, which corresponds to an absolute $r$-band magnitude of $-15.3$ at the upper velocity limit of 3000~km~s$^{-1}$. However, incompleteness in the spectroscopic sample occurs in high density regions due to fiber constraints (fibers can not be assigned to targets with separations $< 55''$) as well as for the brightest galaxies due to fiber magnitude limits (to avoid saturation and crosstalk).  Therefore, the SDSS catalogue is supplemented with galaxies listed in NED (galaxy positions and velocities) for which SDSS $ugriz$ photometry are available -- the NED galaxies are matched to the closest SDSS galaxy within 10~arcsec with photometric measurements only. Note that for 17 galaxies SDSS $ugriz$ photometry are unavailable due to nearby bright stars or artifacts (e.g. satellite or meteor trails). The RC3 catalogue (Third Reference Catalogue of Bright Galaxies; \citealt{devaucouleurs1991}) is complete for galaxies within the studied velocity range with apparent diameters larger than 1~arcmin at the D25 isophotal level and a total $B$-band magnitude limit of 15.5, which is used to complement the data (diameters, magnitudes, colours and morphological types). 

The full sample of galaxies listed in SDSS and NED contains $\sim 2100$ galaxies. Galaxies with angular separations $\theta_{1,2} \leq 10$~arcsec are assumed to be the same galaxy. These doubles are automatically removed from the sample. Galaxies with angular separations 10~arcsec$ < \theta_{1,2} \leq 5$~arcmin are inspected by eye to avoid fragmentation. In cases where the SDSS detections are part of a galaxy and not individual galaxies, the SDSS objects are removed from the sample and the RC3 data for the galaxy is utilized (where available). 

The 1209 galaxies that are in our complete sample fulfill the three selection criteria ($\mathrm{M}_r \leq -15.3$~mag, $300 \leq v_{\mathrm{LG}} \leq 3000$~km~s$^{-1}$ and $\mathrm{M_{\ast}} \geq 10^8~\mathrm{M}_{\odot}$). The absolute magnitudes are corrected for Galactic extinction \citep{schlegel1998} and inclination \citep{driver2008}. Heliocentric velocities are transformed into the Local Group rest frame using the expression in \citet{yahil1977}:
\begin{equation} \label{eq:v_LG}
v_{\mathrm{LG}} = cz + 295 \:\mathrm{sin}\,l \:\mathrm{cos}\,b - 79.1 \:\mathrm{cos}\,l \:\mathrm{cos}\,b -37.6 \:\mathrm{sin}\,b
\end{equation}
with $l$ and $b$ being the galactic coordinates of the galaxies. The velocity distribution of the complete sample of galaxies in the Ursa Major region is shown in Figure~\ref{fig:hist_vlg} wherein the bulge dominated galaxies are shown in red and disk dominated galaxies are shown in blue. The bulge-to-disc ratios are measured by the concentration index of the r-band light $cr = r90/r50$, where $r90$ and $r50$ are the radii within which 90~per~cent and 50~per~cent of the Petrosian flux. Early-type galaxies have higher $cr$ than later types. A separator of 2.6 is adopted to define bulge dominated and disk dominated galaxies as previously used within the SDSS collaboration (\citealt{strateva2001}; \citealt{shimasaku2001}). For galaxies with morphological T-types obtained from RC3, T-types ranging from -5 up to 0 correspond to E to S0a galaxies (bulge dominated), whereas $ 0< \mathrm{T-type} \leq 10$ represents Sa to Irr galaxies (disk dominated). Galaxies with RC3 data without morphological T-type available are shown in black.

The \textit{top left} panel in Figure~\ref{fig:hist_vlg} shows the fraction of bulge dominated galaxies in the complete sample per velocity bin. The majority of galaxies in the Ursa Major region are late-type galaxies. The \textit{bottom left} panel shows the velocity distribution of the complete sample. The volumes corresponding to the velocity bins grow non-linearly from the low to high redshift Universe and one might expect a smooth, steady increase in the number of galaxies for an absolute magnitude limited sample. The peak at $\sim 1300$~km~s$^{-1}$ indicates the presence of the Ursa Major `cluster' without an apparent overdensity in bulge dominated galaxies as one might expect for the denser environment (further discussed in Section~\ref{subsubsec:morph}). The velocity bins $v_{\mathrm{LG}} \geq 1500$~km~s$^{-1}$ show no significant increase in the absolute number of galaxies in the growing volumes.

The \textit{top right} panel in Figure~\ref{fig:hist_vlg} shows the fraction of galaxies in the complete sample that are residing in groups (with number of group members N$_{\mathrm{m}} \geq 4$ as identified in Section~\ref{sec:groups}) per velocity bin. The \textit{bottom right} panel shows the velocity distribution of the complete sample per unit volume. At low recession velocities ($v_{\mathrm{LG}} \leq 1500$~km~s$^{-1}$) 30 to 60~per~cent of the galaxies reside in groups (corresponding to a high number of galaxies per Mpc$^3$ or clustered galaxies), whereas at higher velocities only 10 to 30~per~cent of galaxies are found in dense regions (corresponding to a low number density of galaxies or a less clustered galaxy distribution) indicating a high fraction of non-group galaxies, e.g. isolated galaxies or galaxies residing in small groups (with N$_{\mathrm{m}} < 4$). The drop in $f_{group}$ just beyond 1300~km~s$^{-1}$ suggests a picture of the Ursa Major cluster/groups as having low density regions behind it.

\begin{figure}
\begin{center}
\includegraphics[trim=0cm 0cm 1cm 1cm, clip=true, width=7.5cm]{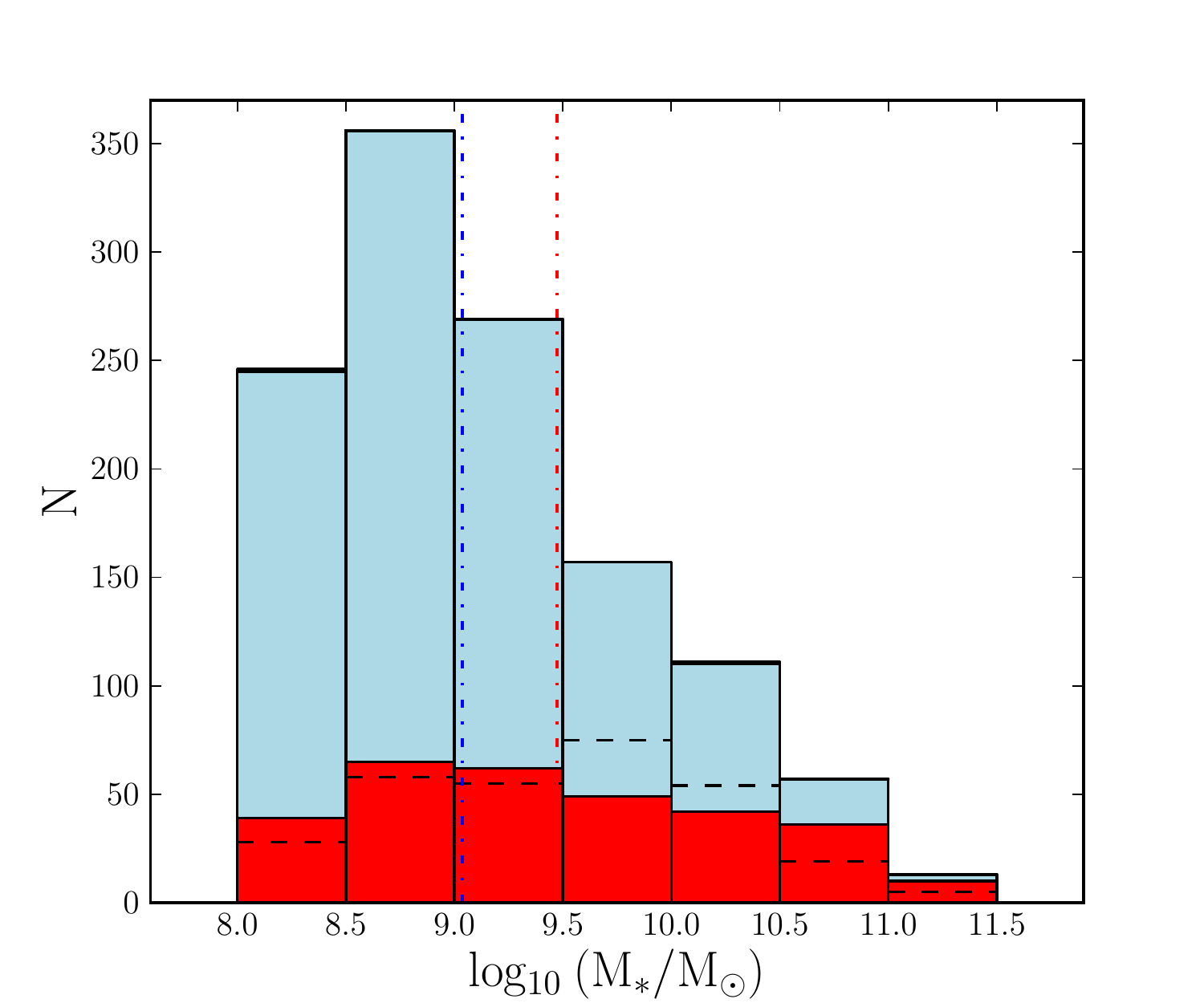} 
\includegraphics[trim=0cm 0cm 1cm 1cm, clip=true, width=7.5cm]{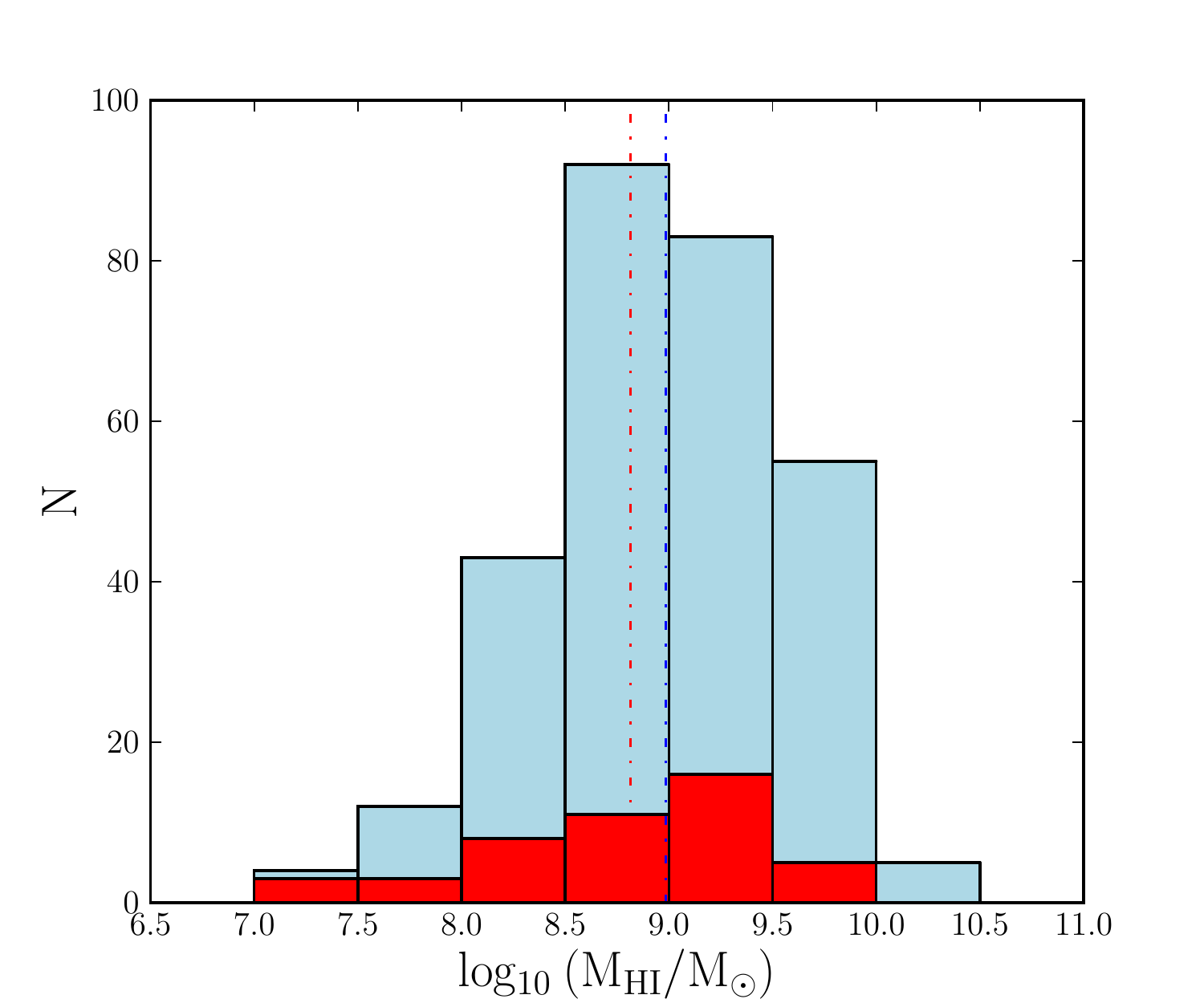}
\caption{Stellar and HI mass histograms in the Ursa Major region. (\textit{top}) Stellar mass histogram of the complete sample wherein bulge dominated galaxies are shown in red and disk dominated galaxies are shown in blue. Stellar masses are estimated using Bell et al. (2003). The average mass of bulge dominated galaxies is larger than the average mass of disk dominated galaxies (dashed vertical lines in red and blue, respectively). The dashed horizontal lines (black) indicate the number of galaxies for which HI masses are available. (\textit{bottom}) The HI mass histogram similar to the stellar mass histogram. There are very few gas-rich early-type galaxies (bulge dominated) in the Ursa Major region ($\mathrm{M_{HI}} \geq 10^{9.5}~\mathrm{M}_{\odot}$). On average disk dominated galaxies tend to have larger HI masses than bulge dominated galaxies (blue dashed as opposed to the red dashed line).}
\label{fig:hist_mass}
\end{center}
\end{figure}

The stellar $M/L$ ratio in Bell et al. (2003) is given as a function of SDSS colour: 
\begin{equation} \label{eq:stellarM_L}
\mathrm{log}_{10} (\mathrm{M}_{\ast}/\mathrm{L}_g) = -0.499 + [1.519 \times (g-r)] 
\end{equation}
The galaxy luminosity in a certain band is given by: 
\begin{equation} \label{eq:L_g}
\frac{\mathrm{L}_g}{\mathrm{L}_{\odot}} = 10^{-0.4 (\mathrm{M}_g-\mathrm{M}_{\odot,g})} 
\end{equation}
Therefore the stellar mass is calculated using 
\begin{equation} \label{eq:stellarM}
\mathrm{log}_{10} \mathrm{M}_{\ast} = -0.499 + [1.519 \times (g-r)] - 0.4 \times (\mathrm{M}_g - \mathrm{M}_{\odot,g})
\end{equation}
in which the absolute magnitude of a galaxy as a function of apparent magnitude and distance (in parsec) is given by: 
\begin{equation} \label{eq:m_abs}
\mathrm{M}_g = m_g - 2.5 \times \mathrm{log}_{10} (d/10)^2 
\end{equation}
An absolute $g$-band magnitude of the sun $\mathrm{M}_{\odot,g} = 5.45$ \citep{blanton2003} is adopted. Similarly the stellar mass can be calculated for the RC3 data (using the absolute B-band magnitude $\mathrm{M}_B$ and $B-V$ colour) with the stellar $M/L$ ratio given in \citet{bell2003} and adopting an absolute $B$-band magnitude of the sun $\mathrm{M}_{\odot,B} = 5.48$ \citep{binney1998}. In cases where the stellar mass estimates from SDSS and RC3 disagree, the galaxy sizes are compared to investigate if the discrepancy is due to fiber magnitude limits. If the size measured in SDSS ($r90$) is much smaller than the diameter in RC3 (D25), the stellar mass estimate from RC3 is preferred. For 134 galaxies RC3 data is utilized (magnitude, stellar mass, colour and morphological type). The stellar mass distribution of the complete sample is shown in Figure~\ref{fig:hist_mass} (\textit{top}) wherein bulge dominated galaxies or RC3 galaxies with $-5 \leq \mathrm{T-types} \leq 0$ are shown in red and disk dominated galaxies or RC3 galaxies with $0< \mathrm{T-types} \leq 10$ are shown in blue. On average bulge dominated galaxies tend to have larger stellar masses than disk dominated galaxies (red dashed line at $\mathrm{M_{\ast}} = 10^{9.5}~\mathrm{M}_{\odot}$ as opposed to the blue dashed line at $\mathrm{M_{\ast}} = 10^9~\mathrm{M}_{\odot}$). However, given that the average stellar mass for early/late type galaxies lie in the same bin, it is unlikely that the offsets are statistically significant. The dashed horizontal lines (black) indicate the number of galaxies for which HI masses are available (as discussed in the following section).

\subsection{HI data}
\label{subsec:HI_data}

In \citet{wolfinger2013} the first blind HI survey of the Ursa Major region (covering 480 deg$^2$ and a heliocentric velocity range from 300 to 1900~km~s$^{-1}$) is presented as part of the HI Jodrell All Sky Survey (HIJASS; \citealt{lang2003}). The measured HI parameters are in good agreement with previous values in the literature (where available) for HIJASS detections with \textit{one, clear} associate counterpart. For confused HIJASS detections, i.e. numerous optical counterparts within the beam which may all contribute a significant amount of flux to the HI detection, high-resolution interferometric datasets by \citet{kovac2009} and Verheijen \& Sancisi (2001; both using the Westerbork Synthesis Radio Telescope) are used preferentially over the Jodrell single-dish observations in the following analysis. Furthermore, HI data presented in \citet{haynes2011} as part of the ALFALFA survey using the Arecibo telescope (covering a strip in the southern part of the region: $+24^{\circ} \leq \delta \leq +28^{\circ}$) as well as a compilation of HI properties of a homogenous sample of more than 9000 galaxies from single-dish observations presented in \citet{springob2005} are utilised.  

The HI mass distribution for 294 galaxies is shown in Figure~\ref{fig:hist_mass} (\textit{bottom}) wherein bulge dominated galaxies are shown in red and disk dominated galaxies are shown in blue. There are very few gas-rich early-type galaxies (bulge dominated) in the Ursa Major region (with $\mathrm{M_{HI}} \geq 10^{9.5}~\mathrm{M}_{\odot}$). On average disk dominated galaxies tend to have larger HI masses than bulge dominated galaxies (blue dashed line at $\mathrm{M_{HI}} = 10^9~\mathrm{M}_{\odot}$ as opposed to the red dashed line at $\mathrm{M_{HI}} = 10^{8.8}~\mathrm{M}_{\odot}$).

\begin{figure}
\centering
\includegraphics[trim=0cm 0cm 1cm 0cm, clip=true, width=7cm]{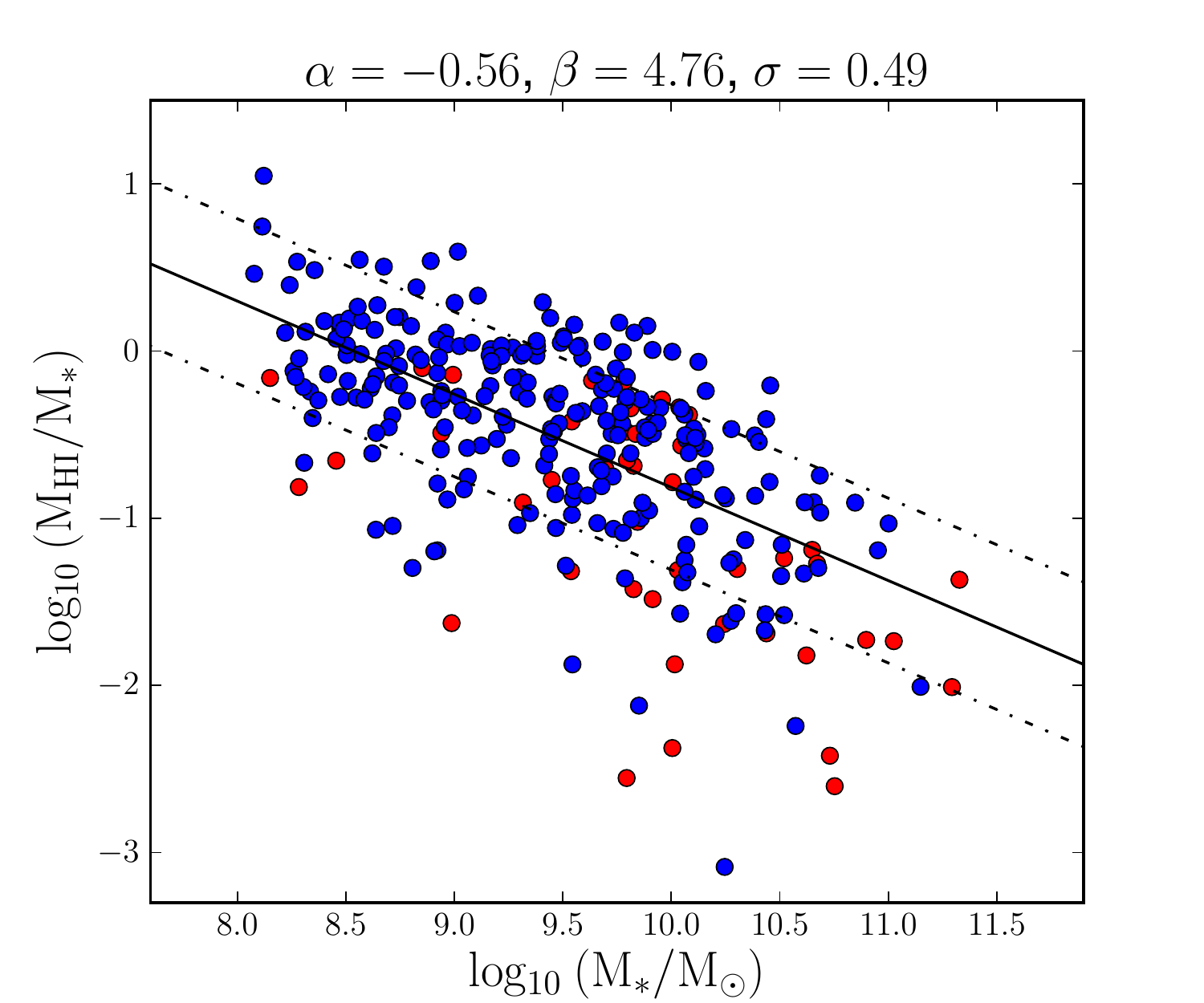}
\includegraphics[trim=0cm 0cm 1cm 0cm, clip=true, width=7cm]{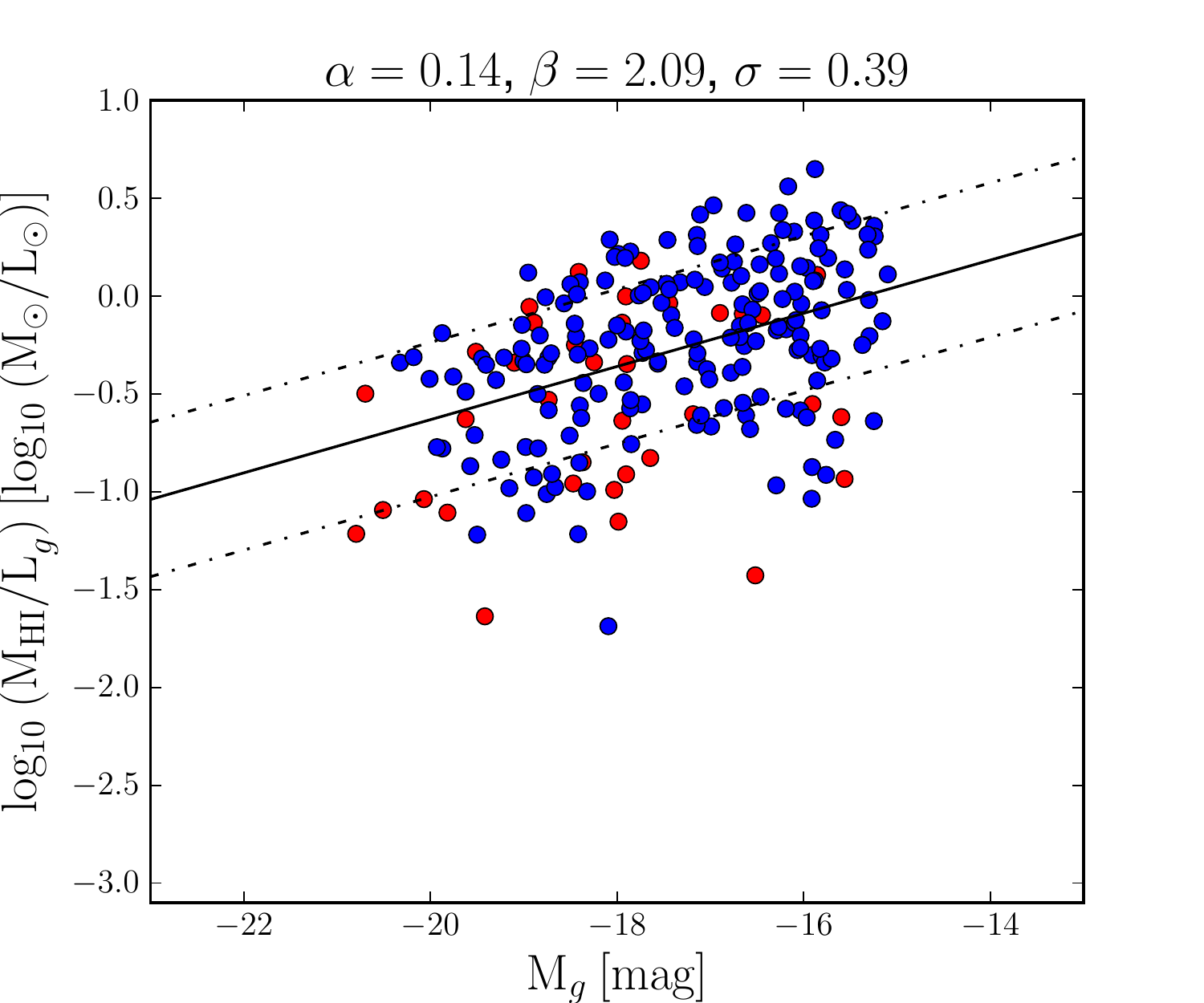}
\includegraphics[trim=0cm 0cm 1cm 0cm, clip=true, width=7cm]{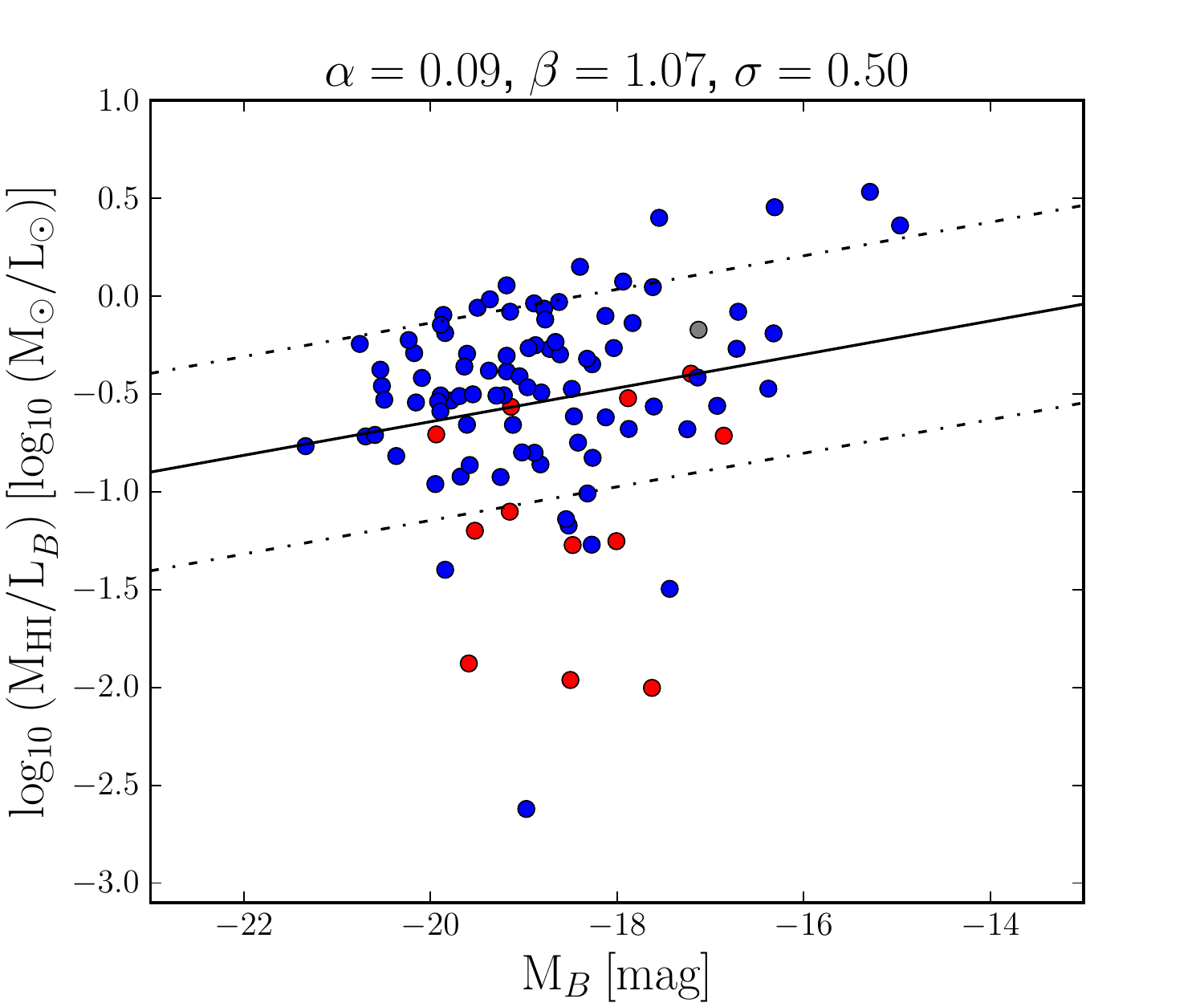}
\caption[HI-to-stellar mass fraction as a function of stellar mass and HI mass-to-light ratio as a function of luminosity.] {(\textit{top:}) HI-to-stellar mass fraction as a function of stellar mass. Bulge dominated galaxies are shown in red and disk dominated galaxies are shown in blue. Galaxies with large stellar masses ($\mathrm{M_{\ast}} \geq 10^{9.5}~\mathrm{M}_{\odot}$) and small HI-to-stellar mass fractions tend to be bulge dominated galaxies, whereas disk dominated galaxies are predominantly gas-rich objects. (\textit{middle} and \textit{bottom:}) HI mass-to-light ratio as a function of absolute magnitude for SDSS ($g-$band \textit{middle}) and RC3 ($B-$band \textit{bottom}) data. The HI mass-to-light ratios are higher for faint galaxies (towards $-14$ in absolute magnitude, which are mostly disk dominated galaxies) than they are for bright galaxies (towards $-22$ in absolute magnitude). Outliers with $\mathrm{log_{10} (M_{HI} / M_{\ast})} \leq 1.8$ reside in galaxy groups as determined in Section~\ref{sec:structures}.}
\label{fig:StellarFracmass}
\end{figure}

The HI-to-stellar mass fraction as a function of stellar mass is shown in Figure~\ref{fig:StellarFracmass} (\textit{top}). Bulge dominated and disk dominated galaxies are colour coded in red and blue, respectively. The HI-to-stellar mass fraction is anti-correlated with stellar mass. Galaxies with large stellar masses ($\mathrm{M_{\ast}} \geq 10^{9.5}~\mathrm{M}_{\odot}$) and small HI-to-stellar mass fractions tend to be bulge dominated galaxies, whereas disk dominated galaxies are predominantly gas-rich objects with higher HI-to-stellar mass fraction and smaller stellar masses ($\mathrm{M_{\ast}} < 10^{9.5}~\mathrm{M}_{\odot}$). The linear regression fit to the data is of the form
\begin{equation} \label{eq:linreg}
\mathrm{log_{10} (\mathrm{M}_{HI} / \mathrm{M}_{\ast})} = \alpha \: \mathrm{log_{10} (M_{\ast})} + \beta 
\end{equation}
with $\alpha$ and $\beta$ as stated above the figure as well as the standard deviation $\sigma$.

The HI mass-to-light ratio as a function of absolute magnitude for SDSS ($g-$band) and RC3 ($B-$band) data are also shown in Figure~\ref{fig:StellarFracmass} (\textit{middle} and \textit{bottom}, similar to the \textit{top}). The HI mass-to-light ratios are higher for faint galaxies (towards $-14$ in absolute magnitude, which are mostly disk dominated galaxies) than the HI mass-to-light ratios for bright galaxies (towards $-22$ in absolute magnitude). The figures show a larger spread in the HI mass-to-light ratio for bright galaxies (outliers towards negative HI mass-to light ratios, which are mostly bulge dominated galaxies) suggesting that bright galaxies lack quantities of HI in comparison to their stellar content, which may have been consumed in star formation or being removed from the galaxies. Note that the outliers with $\mathrm{log_{10} (\mathrm{M}_{HI} / \mathrm{M}_{\ast})} \leq 1.8$ reside in galaxy groups as determined in Section~\ref{sec:groups}.

\section{Defining structures}
\label{sec:structures}

In this Section, we describe the friends-of-friends (FoF) algorithm that has been developed in order to identify bound structures in the Ursa Major region. A similar approach as in \citet{huchra1982} is followed to implement the FoF algorithm. In Section~\ref{sec:lit_comp}, the identified FoF groups are compared to a recently presented all-sky catalogue of 395 nearby galaxy groups with $v_{\mathrm{LG}} \leq 3500$~km~s$^{-1}$ \citep{makarov2011} as well as the original cluster definition of \citet{tully1996}.

\subsection{Friends-of-friends algorithm}
\label{subsec:fof}

To identify galaxy groups or clusters whose members are likely to be physically and dynamically associated, we use a standard FoF algorithm \citep{huchra1982}. Galaxies are associated as `friends' within certain projected and line-of-sight linking lengths. The projected separation (in Mpc) between two galaxies is given by 
\begin{equation} \label{eq:D_12}
D_{12} = \mathrm{sin}(\theta/2) \: \frac{v_1+v_2}{H_0}
\end{equation}
where $\theta$ is the angular separation of the two galaxies and $v_1$ and $v_2$ refer to their velocities \citep{huchra1982}. The radial velocity difference is given by 
\begin{equation} \label{eq:V_12}
V_{12} = |v_1-v_2|
\end{equation}
\citep{huchra1982}.

Two galaxies are associated as `friends' if their projected separation and radial difference are smaller than certain linking lengths:
\begin{equation} \label{eq:D_0}
 D_{12} \leq D_0
\end{equation}
and
\begin{equation} \label{eq:V_0}
 V_{12} \leq V_0
\end{equation}
The galaxy groups consist of `friends' and their `friends'. Hence galaxies that are not directly associated with one another can still be linked together by a chain of common `friends' (see Fig.1 in \citealt{robotham2011} for a schematic and Fig.1 in \citealt{huchra1982} for a flow chart of the grouping algorithm). Only groups with N$_{\mathrm{m}} \geq 4$ are considered as smaller ones are likely to be false positives (e.g. \citealt{ramella1995}).

Note that \citet{robotham2011} introduce further parameters to allow the linking lengths to scale as a function of the observed density contrast and to allow for larger linking lengths for bright galaxies. These parameters are shown to cause only minor perturbations to the grouping and may even be completely negligible. Therefore we only consider the dominant parameters $D_0$ and $V_0$ in our grouping algorithm, which are determined in the following section using mock catalogues.

\subsection{Optimizing linking lengths using mock catalogues}
\label{subsec:mill}

To determine the optimal linking lengths for the FoF algorithm ($D_0$ and $V_0$) we analyze mock galaxy catalogues generated with semi-analytic models \citep{croton2006}, that are based on N-body simulations. The Croton model uses the Millennium Simulation \citep{springel2005}, which is a large $N$-body simulation of dark matter structure in a cosmological volume -- tracing 10 billion dark matter particles in a cubic box of 500$h^{-1}$~Mpc on a side through cosmic time with a halo mass resolution of $\sim 5 \times 10^{10} h^{-1}$~M$_{\odot}$. The cosmological model in Millennium simulation is $ \Omega_M=0.25$, $ \Omega_{\lambda} = 0.75$ and $H_0 = h \: 100$~km~s$^{-1}$~Mpc$^{-1}$ with $h=0.73$. Semi-analytic prescriptions are coupled with the output of the dark matter simulation to describe the baryonic physics, i.e. the formation and evolution of galaxies.

The mock catalogues are obtained from the Theoretical Astrophysical Observatory (TAO; \citealt{bernyk2014}). TAO enables the conversion of the simulation output into mock observables, i.e. lightcones. It also allows us to compute the magnitudes of each galaxy in various filters, including those we need in this paper. The SDSS magnitudes are estimated based on the galaxy model by \citet{croton2006} and stellar population synthesis models by \citet{conroy2009}. We analyze fifty mock lightcones, which do not spatially overlap. To test the group finder on realistic mock galaxy catalogues, the same limitations as in the real data are applied, i.e. the same volume, absolute magnitude limit and stellar mass cut, which lies above the mass resolution limit of the Millennium simulation. The number of galaxies in the mock lightcones vary from 70 to 1200 representing a wide range of environments.

We determine the appropriate linking lengths $D_0$ and $V_0$ from the two dimensional parameter space, which is mapped out for groups with N$_{\mathrm{m}} \geq 10$. Smaller groups show systematic deviations in abundance, projected sizes and velocity dispersions \citep{berlind2006}. The aim is to maximize the correspondence between real space groups and redshift space groups, i.e. maximize the group detection rate with adequate group properties (velocity dispersion, maximal radial extent and number of group members) whilst keeping the fraction of spurious groups in redshift-space low.

First, the FoF algorithm with a single linking length is applied to the real-space mock catalogues (starting at 0.10~Mpc up to 0.50~Mpc in steps of 0.01), i.e. the linking volume is a sphere. Secondly the FoF algorithm with two linking lengths (projected separation between 0.13 to 0.50~Mpc and radial velocity differences between 40 to 400~km~s$^{-1}$) is applied to the redshift-space mock catalogues, which searches for `friends' within a cylinder. 

\begin{figure*}
\centering
\includegraphics[trim=0cm 0cm 0cm 0cm, clip=true, width=18cm]{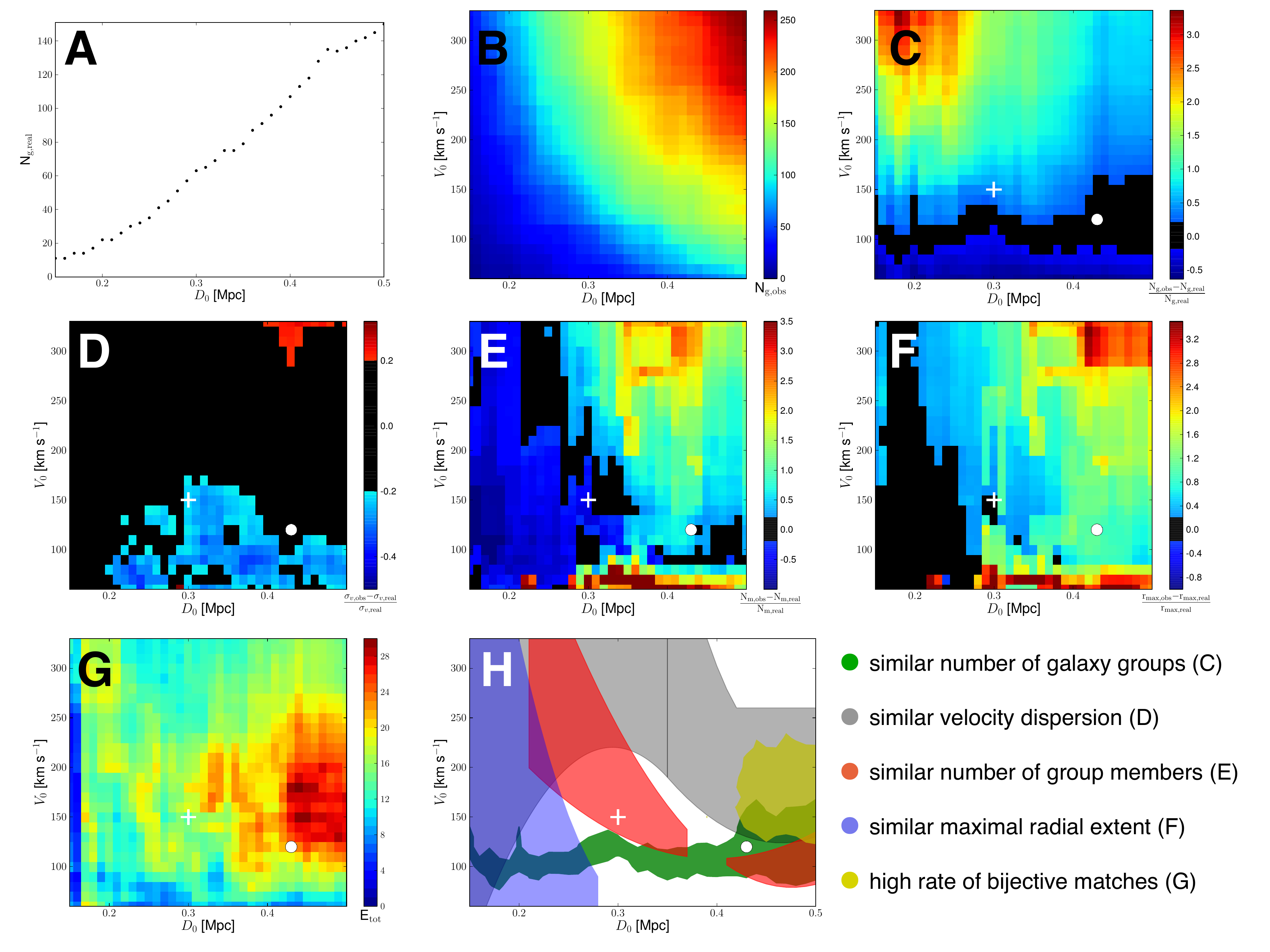}
\caption{Evaluation of the FoF algorithm using mock galaxy lightcones. The number of galaxy groups recovered in real-space (\textit{A}), redshift-space (\textit{B}) and their relative difference (\textit{C}). The region where the FoF algorithm recovers a similar number of galaxy groups to that expected from the mock catalogues is highlighted in black. The next three panels show a comparison of the group properties in real-space to the properties of the bijective or `best match' in redshift-space: the relative differences in velocity dispersion (\textit{D}), the number of group members (\textit{E}) and the maximal radial extent (\textit{F}). The regions where the FoF algorithm recovers groups with similar properties are highlighted in black. (\textit{G}) The percentage of galaxy groups in real- and redshift-space that are bijective matches, i.e. groups that contain more than 50~per~cent of their respective group members. (\textit{H}) Schematic of regions with a high rate of bijective matches (in yellow) and a similar number of galaxy groups in real- and redshift-space (in green). Regions where the FoF algorithm recovers groups with similar properties to their respective groups in real-space: the relative differences in (\textit{i}) velocity dispersion in grey, (\textit{ii}) the maximal radial extent in blue and (\textit{iii}) the number of group members in red. As the shaded regions do not overlap, the choice of linking parameters is not obvious. The linking parameters adopted in this work are indicated by the white cross. For comparison, the large-scale structure of the Ursa Major region found using `looser' linking lengths are indicated by the white circle (which are also discussed in some of the following sections).}
\label{fig:mill}
\end{figure*}

In Figure~\ref{fig:mill} the total number of galaxy groups recovered in real-space ($\mathrm{N_{g,real}}$; \textit{A}) and redshift-space ($\mathrm{N_{g,obs}}$; \textit{B}) are shown. The number of galaxy groups increases with linking lengths. However, it is not a monotonic increase as galaxy groups can merge to from larger structures. The relative difference between the number of galaxy groups is given by 
\begin{equation} \label{eq:delta_Ng}
\Delta \mathrm{N_g} = \frac {\mathrm{N_{g,obs} - N_{g,real}}} {\mathrm{N_{g,real}}}
\end{equation}
and is shown in panel \textit{C}. The region where the FoF algorithm recovers a similar number of galaxy groups ($|\Delta \mathrm{N_g}| \leq 0.2$) is highlighted in black. The radial velocity difference $V_0$ seems to be most adequate in a narrow range between $90-150$~km~s$^{-1}$, which is of the order of the velocity dispersions of groups. The parameter space $D_0 \lesssim 0.3$~Mpc and $V_0 \gtrsim 250$~km~s$^{-1}$ is not appropriate as up to 4 times as many galaxy groups as expected are found, i.e. the number of spurious groups is high.

To compare the properties of the `real-space groups' to the `redshift-space groups' the groups must be matched in a one-to-one way. Here, the galaxy groups are matched when they contain more than 50~per~cent of their respective group members (bijective match; \citealt{robotham2011}). If no bijective match is found, the `best match' is identified, which is the group with the largest product for the relative membership fractions as described in \citet{robotham2011}. Other matching scenarios are the group that contains the central galaxy, or the group whose centroid is closest to the real centre. 

The average relative difference of a group property, e.g. velocity dispersion, is calculated using
\begin{equation} \label{eq:delta_prop}
\Delta \sigma_{\mathrm{v}} = \frac {\sum_{N_{match}} \frac {\sigma_{\mathrm{v,obs}} - \sigma_{\mathrm{v,real}}} {\sigma_{\mathrm{_{v,real}}}}} {N_{match}}
\end{equation}
Similarly, the average relative difference in the number of group members and radial extent are calculated. The regions where the FoF algorithm recovers groups with similar properties are highlighted in black ($|\Delta \mathrm{prop}| \leq 0.2$) in Figure~\ref{fig:mill}. 

The group velocity dispersions in redshift-space are in good agreement with the group velocity dispersions in real-space for most of the parameter space (\textit{D}). Where $D_0 \gtrsim 0.4$~Mpc and $V_0 \gtrsim 260$~km~s$^{-1}$ the group velocity dispersions in redshift-space are overestimated and for $D_0 \gtrsim 0.25$~Mpc and $V_0 \lesssim 120$~km~s$^{-1}$ the group velocity dispersions in redshift-space are underestimated by more than 20~per~cent.

The black region in panel \textit{E} in Figure~\ref{fig:mill} shows linking lengths that pass the number of groups members test. Below $D_0 \lesssim 0.2$~Mpc not enough group members are found. The parameter space $D_0 \gtrsim 0.35$~Mpc and $V_0 \gtrsim 200$~km~s$^{-1}$ is not appropriate as up to 4 times as many group members as expected are found, i.e. the number of interlopers is high. A similar result can be seen in panel \textit{F} where the relative difference in radial extent is shown. The average radial extent becomes up to 4 times as large as expected for $D_0 \gtrsim 0.35$~Mpc and $V_0 \gtrsim 200$~km~s$^{-1}$.

In panel \textit{G}, the percentages of bijective matches (E$_{\mathrm{tot}}$) are shown as defined in \citet{robotham2011}. The percentages range from 0 (no bijective matches) up to 30~per~cent. As one might expect, the bijective matches increase with linking length as groups merge and become less ambiguous. However, where linking lengths $D_0 \gtrsim 0.4$~Mpc, the percentage of bijective matches decreases for larger radial velocity differences ($V_0 \gtrsim 250$~km~s$^{-1}$) as the fraction of interlopers increases.

Figure~\ref{fig:mill} also shows a schematic of the regions that pass the various tests (\textit{H}). The regions with highest rates of bijective matches are shown in yellow, the regions with a similar number of galaxy groups in real- and redshift-space are shown in green. Regions where the FoF algorithm recovers groups in velocity space with similar properties to their respective groups in real-space are shown in grey (velocity dispersion), blue (maximal radial extent) and red (number of group members). As the shaded regions do not overlap, there is no obvious choice of linking parameters. However, some parameter spaces can be ruled out: $V_0 \gtrsim 160$~km~s$^{-1}$ is not appropriate as the number of spurious groups increases, whereas $V_0 \lesssim 100$~km~s$^{-1}$ underestimates the velocity dispersions of the recovered groups. $D_0 \lesssim 0.2$~Mpc does not recover enough group members, whereas $D_0 \gtrsim 0.33$~Mpc overestimates the radial extent of the groups. Therefore, the linking lengths adopted in this work are $D_0=0.3$~Mpc and $V_0=150$~km~s$^{-1}$ (marked with a white cross in Figure~\ref{fig:mill}). The choice of $D_0$ is constraint by the relative differences in the number of group members (E) and the maximal radial extent (F), whereas $V_0$ is constraint by the relative differences in the number of galaxy groups recovered (C) and the velocity dispersion (D). For comparison we sometimes discuss `looser' linking lengths $D_0=0.43$~Mpc and $V_0=120$~km~s$^{-1}$ in the following analysis of the galaxy groups, which recover a structure close to the `Ursa Major cluster' as previously defined in the literature (marked with a white circle in Figure~\ref{fig:mill}). The radial extent of these `looser' groups will be overestimated, but the number of bijective matches is high as galaxy groups merge and become less ambiguous. Note that in comparison to previous work, e.g. \citet{brough2006} our adopted line-of-sight linking length used here is rather tight. \citet{brough2006} use a FoF algorithm to examine substructures in the Eridanus region (velocity range 500-2500~km~s$^{-1}$) with linking parameters $D_0=0.29$~Mpc and $V_0=347$~km~s$^{-1}$. If we used this `loose' radial linking length, there would be a high number of interlopers in the groups.

To test for the fidelity of the FoFs group catalogue with respect to the halo distribution, we compare their halo multiplicity functions. The halo multiplicity function describes the distribution of mass as a function of halo mass (e.g. \citealt{springel2005}). If the multiplicity functions are sufficiently similar the FoFs groups will faithfully reflect the correct dark matter halo distribution. To justify that our choice of linking lengths produces FoFs groups that reflect the underlying dark matter distribution we turn to the tests presented in \citet{berlind2006}. These authors show that the halo multiplicity function of FoFs groups is within 10 per cent of the dark matter distribution for range of linking lengths around $b_{\perp}=0.14$, $b_{\parallel}=0.75$, and is particularly good for groups with $N>10$. In these units our linking lengths are withing 10 per cent, $b_{\perp}=0.12$, $b_{\parallel}=0.83$. Berlind et al. find that halos with $N>5$ have a completeness of more than 95 per cent, and the fraction of spurious groups with $N>5$ is less than 5 per cent.

\onecolumn
\begin{landscape}
\begin{scriptsize}
\begin{longtable}{lcrrrrrrrrrrrr}
\caption[The galaxy groups and their properties, which have been found in the Ursa Major region.]{The galaxy groups and their properties, which have been found in the Ursa Major region.}
\label{tab:group_prop} \\
  
  \hline\hline\\[-2ex]
  & 
  & 
  \multicolumn{1}{c}{RA} &  
  \multicolumn{1}{c}{Dec.} & 
  \multicolumn{1}{c}{$v_{c}$} & 
  \multicolumn{1}{c}{$\sigma_v$} & 
  \multicolumn{1}{c}{$r_{\mathrm{500}}$} & 
  \multicolumn{1}{c}{$\mathrm{M}_v$} & 
  \multicolumn{1}{c}{$t_c$} & 
  \multicolumn{1}{c}{$\mathrm{L}_t$} & 
  \multicolumn{1}{c}{$\mathrm{M}_v/\mathrm{L}_t$} &
  \multicolumn{1}{c}{$\mathrm{log \: (M_{HI,min})}$} & 
  & 
  \multicolumn{1}{c}{$r_{max}$}  \\

ID & 
  \multicolumn{1}{c}{N$_m$ (N$_c$)} & 
  \multicolumn{1}{c}{(J2000.0)} &  
  \multicolumn{1}{c}{(J2000.0)} & 
  \multicolumn{1}{c}{(km~s$^{-1}$)} & 
  \multicolumn{1}{c}{(km~s$^{-1}$)} & 
  \multicolumn{1}{c}{(Mpc)} & 
  \multicolumn{1}{c}{(10$^{13}$ M$_{\odot}$)} & 
  \multicolumn{1}{c}{($H_0^{-1}$)} & 
  \multicolumn{1}{c}{(10$^{11}$ L$_{\odot}$)} & 
  \multicolumn{1}{c}{(M$_{\odot}$/L$_{\odot}$)} &
  \multicolumn{1}{c}{(M$_{\odot}$)} &    
  \multicolumn{1}{c}{$f_{early}$} & 
  \multicolumn{1}{c}{(Mpc)} \\

(A1) & 
  \multicolumn{1}{c}{(A2)} & 
  \multicolumn{1}{c}{(A3)} &  
  \multicolumn{1}{c}{(A4)} & 
  \multicolumn{1}{c}{(A5)} & 
  \multicolumn{1}{c}{(A6)} & 
  \multicolumn{1}{c}{(A7)} & 
  \multicolumn{1}{c}{(A8)} & 
  \multicolumn{1}{c}{(A9)} & 
  \multicolumn{1}{c}{(A10)} & 
  \multicolumn{1}{c}{(A11)} & 
  \multicolumn{1}{c}{(A12)} & 
  \multicolumn{1}{c}{(A13)} & 
  \multicolumn{1}{c}{(A14)} \\[0.5ex] \hline
  \\[-1.8ex]
  \endfirsthead
  
  \multicolumn{13}{c}{{\tablename} \thetable{} -- Continued} \\[0.5ex]
  \hline\hline\\[-2ex]
  & 
  & 
  \multicolumn{1}{c}{RA} &  
  \multicolumn{1}{c}{Dec.} & 
  \multicolumn{1}{c}{$v_{c}$} & 
  \multicolumn{1}{c}{$\sigma_v$} & 
  \multicolumn{1}{c}{$r_{\mathrm{500}}$} & 
  \multicolumn{1}{c}{$\mathrm{M}_v$} & 
  \multicolumn{1}{c}{$t_c$} & 
  \multicolumn{1}{c}{$\mathrm{L}_t$} & 
  \multicolumn{1}{c}{$\mathrm{M}_v/\mathrm{L}_t$} & 
  \multicolumn{1}{c}{$\mathrm{log \: (M_{HI,min})}$} & 
  & 
  \multicolumn{1}{c}{$r_{max}$}  \\

ID & 
  \multicolumn{1}{c}{N$_m$ (N$_c$)} & 
  \multicolumn{1}{c}{(J2000.0)} &  
  \multicolumn{1}{c}{(J2000.0)} & 
  \multicolumn{1}{c}{(km~s$^{-1}$)} & 
  \multicolumn{1}{c}{(km~s$^{-1}$)} & 
  \multicolumn{1}{c}{(Mpc)} & 
  \multicolumn{1}{c}{(10$^{13}$ M$_{\odot}$)} & 
  \multicolumn{1}{c}{($H_0^{-1}$)} & 
  \multicolumn{1}{c}{(10$^{11}$ L$_{\odot}$)} & 
  \multicolumn{1}{c}{(M$_{\odot}$/L$_{\odot}$)} &
  \multicolumn{1}{c}{((M$_{\odot}$)} & 
  \multicolumn{1}{c}{$f_{early}$} & 
  \multicolumn{1}{c}{(Mpc)} \\

(F1) &  
  \multicolumn{1}{c}{(F2)} &  
  \multicolumn{1}{c}{(F3)} &  
  \multicolumn{1}{c}{(F4)} & 
  \multicolumn{1}{c}{(F5)} & 
  \multicolumn{1}{c}{(F6)} & 
  \multicolumn{1}{c}{(F7)} & 
  \multicolumn{1}{c}{(F8)} & 
  \multicolumn{1}{c}{(F9)} & 
  \multicolumn{1}{c}{(F10)} & 
  \multicolumn{1}{c}{(F11)} & 
  \multicolumn{1}{c}{(F12)} & 
  \multicolumn{1}{c}{(F13)} & 
  \multicolumn{1}{c}{(F14)} \\[0.5ex] \hline
  \\[-1.8ex]
   \endhead

MESSIER106 & 6 (4) & 12:17:03.99 & 47:01:17.2 & 454$\pm$91 & 56$\pm$23 & 0.07$\pm$0.03 & 0.21$\pm$0.17 & 0.11$\pm$0.057 & 0.16$\pm$0.006 & 130$\pm$107 & 9.88 & 0.0 & 0.21 \\ 
NGC4449 & 15 (6) & 12:29:36.80 & 42:57:21.0 & 489$\pm$40 & 81$\pm$19 & 0.11$\pm$0.02 & 0.46$\pm$0.17 & 0.07$\pm$0.04 & 0.14$\pm$0.007 & 319$\pm$118 & 9.83 & 0.0 & 0.38 \\ 
NGC4278 & 10 (5) & 12:22:01.90 & 29:09:21.2 & 498$\pm$27 & 47$\pm$8 & 0.06$\pm$0.01 & 0.11$\pm$0.14 & 0.11$\pm$0.042 & 0.05$\pm$0.01 & 235$\pm$288 & 8.47 & 0.4 & 0.33 \\ 
NGC3972 & 4 (4) & 11:55:01.04 & 55:15:55.3 & 738$\pm$32 & 23$\pm$25 & 0.03$\pm$0.03 & 0.08$\pm$0.2 & 0.21$\pm$0.249 & 0.03$\pm$0.007 & 270$\pm$675 & 8.75 & 0.25 & 0.11 \\ 
NGC4026 & 30 (12) & 11:54:47.33 & 49:43:22.9 & 820$\pm$36 & 111$\pm$12 & 0.15$\pm$0.02 & 0.76$\pm$0.19 & 0.09$\pm$0.012 & 0.19$\pm$0.012 & 397$\pm$97 & 10.16 & 0.5 & 0.5 \\ 
NGC3938 & 29 (16) & 12:01:34.69 & 43:40:49.1 & 891$\pm$80 & 152$\pm$25 & 0.2$\pm$0.03 & 1.16$\pm$0.41 & 0.05$\pm$0.012 & 0.36$\pm$0.022 & 327$\pm$115 & 10.07 & 0.31 & 0.45 \\ 
NGC4274 & 13 (7) & 12:20:41.42 & 29:39:31.4 & 907$\pm$73 & 121$\pm$22 & 0.16$\pm$0.03 & 0.6$\pm$0.22 & 0.06$\pm$0.013 & 0.19$\pm$0.013 & 313$\pm$115 & 7.5 & 0.29 & 0.27 \\ 
NGC5033 & 18 (7) & 13:12:19.02 & 36:42:33.0 & 923$\pm$54 & 155$\pm$24 & 0.2$\pm$0.03 & 0.95$\pm$0.4 & 0.04$\pm$0.011 & 0.28$\pm$0.037 & 341$\pm$142 & 10.06 & 0.0 & 0.57 \\ 
NGC4725 & 8 (6) & 12:50:17.48 & 25:35:09.4 & 1044$\pm$61 & 148$\pm$52 & 0.19$\pm$0.07 & 1.34$\pm$0.96 & 0.07$\pm$0.031 & 0.33$\pm$0.008 & 400$\pm$288 & 9.92 & 0.33 & 0.48 \\ 
NGC3998 & 29 (14) & 11:55:07.36 & 55:35:32.3 & 1065$\pm$202 & 188$\pm$15 & 0.25$\pm$0.02 & 2.45$\pm$0.66 & 0.05$\pm$0.009 & 0.22$\pm$0.009 & 1116$\pm$299 & 8.75 & 0.21 & 0.47 \\ 
NGC3301 & 5 (5) & 10:36:55.72 & 21:45:37.8 & 1093$\pm$47 & 61$\pm$36 & 0.08$\pm$0.05 & 0.25$\pm$0.15 & 0.16$\pm$0.111 & 0.08$\pm$0.003 & 335$\pm$195 & 8.73 & 0.4 & 0.47 \\ 
NGC3079 & 5 (5) & 10:01:53.03 & 55:40:16.0 & 1118$\pm$181 & 121$\pm$54 & 0.16$\pm$0.07 & 0.36$\pm$0.7 & 0.02$\pm$0.015 & 0.1$\pm$0.006 & 354$\pm$683 & 9.94 & 0.2 & 0.14 \\ 
NGC3992 & 5 (4) & 11:57:38.07 & 53:22:30.8 & 1267$\pm$54 & 43$\pm$17 & 0.06$\pm$0.02 & 0.04$\pm$0.02 & 0.08$\pm$0.034 & 0.28$\pm$0.003 & 13$\pm$9 & 9.82 & 0.25 & 0.11 \\ 
NGC3248 & 8 (6) & 10:28:06.76 & 22:58:55.8 & 1270$\pm$33 & 67$\pm$22 & 0.09$\pm$0.03 & 0.29$\pm$0.17 & 0.18$\pm$0.066 & 0.06$\pm$0.001 & 443$\pm$257 & 8.4 & 0.5 & 0.39 \\ 
NGC3245 & 9 (6) & 10:29:26.43 & 28:49:34.0 & 1328$\pm$34 & 101$\pm$36 & 0.13$\pm$0.05 & 0.36$\pm$0.22 & 0.07$\pm$0.037 & 0.25$\pm$0.009 & 144$\pm$87 & 9.85 & 0.5 & 0.38 \\ 
NGC3003 & 6 (6) & 09:50:16.61 & 33:10:26.1 & 1346$\pm$27 & 62$\pm$33 & 0.08$\pm$0.04 & 0.27$\pm$0.17 & 0.22$\pm$0.124 & 0.11$\pm$0.003 & 238$\pm$147 & 0.0 & 0.5 & 0.51 \\ 
NGC3430 & 6 (5) & 10:51:00.87 & 32:55:58.2 & 1373$\pm$22 & 13$\pm$6 & 0.02$\pm$0.01 & 0.01$\pm$0.04 & 0.18$\pm$0.287 & 0.16$\pm$0.013 & 5$\pm$25 & 10.11 & 0.4 & 0.14 \\ 
NGC3193 & 12 (8) & 10:18:02.96 & 21:47:34.2 & 1383$\pm$53 & 84$\pm$8 & 0.11$\pm$0.01 & 0.37$\pm$0.19 & 0.09$\pm$0.018 & 0.27$\pm$0.006 & 140$\pm$70 & 8.8 & 0.38 & 0.34 \\ 
NGC5582 & 4 (4) & 14:21:11.60 & 39:41:27.4 & 1385$\pm$40 & 17$\pm$11 & 0.02$\pm$0.01 & 0.11$\pm$0.14 & 0.46$\pm$0.459 & 0.12$\pm$0.004 & 94$\pm$116 & 0.0 & 0.5 & 0.27 \\ 
NGC3414 & 6 (5) & 10:51:17.49 & 27:59:33.2 & 1458$\pm$169 & 177$\pm$71 & 0.23$\pm$0.09 & 0.47$\pm$0.27 & 0.01$\pm$0.006 & 0.14$\pm$0.006 & 334$\pm$192 & 8.57 & 0.2 & 0.06 \\ 
NGC3683 & 22 (14) & 11:36:00.83 & 57:56:30.0 & 1478$\pm$266 & 87$\pm$20 & 0.11$\pm$0.03 & 0.69$\pm$0.2 & 0.13$\pm$0.037 & 0.22$\pm$0.003 & 317$\pm$93 & 9.78 & 0.21 & 0.6 \\ 
NGC5473 & 20 (15) & 14:04:55.15 & 55:13:26.7 & 1697$\pm$88 & 123$\pm$15 & 0.16$\pm$0.02 & 0.85$\pm$0.3 & 0.07$\pm$0.02 & 0.45$\pm$0.006 & 189$\pm$67 & 0.0 & 0.33 & 0.39 \\ 
NGC5631 & 6 (5) & 14:26:21.60 & 56:33:51.1 & 1743$\pm$51 & 66$\pm$30 & 0.09$\pm$0.04 & 0.12$\pm$0.11 & 0.08$\pm$0.068 & 0.11$\pm$0.003 & 111$\pm$95 & 0.0 & 0.2 & 0.26 \\ 
UGC06570 & 4 (4) & 11:35:14.96 & 35:19:06.1 & 1892$\pm$21 & 36$\pm$25 & 0.05$\pm$0.03 & 0.13$\pm$0.13 & 0.38$\pm$0.282 & 0.12$\pm$0.001 & 109$\pm$109 & 9.14 & 0.75 & 0.4 \\ 
NGC3613 & 21 (16) & 11:20:00.84 & 58:29:11.5 & 1893$\pm$54 & 129$\pm$18 & 0.17$\pm$0.02 & 1.45$\pm$0.46 & 0.08$\pm$0.02 & 0.75$\pm$0.009 & 194$\pm$62 & 10.19 & 0.19 & 0.63 \\ 
NGC5322 & 17 (16) & 13:49:54.56 & 60:17:02.3 & 1923$\pm$286 & 93$\pm$22 & 0.12$\pm$0.03 & 1.32$\pm$0.38 & 0.11$\pm$0.032 & 0.66$\pm$0.01 & 200$\pm$57 & 0.0 & 0.31 & 0.54 \\ 
NGC3665 & 20 (15) & 11:24:13.44 & 38:31:25.7 & 1944$\pm$166 & 197$\pm$17 & 0.26$\pm$0.02 & 2.13$\pm$0.69 & 0.05$\pm$0.01 & 0.38$\pm$0.011 & 559$\pm$181 & 9.74 & 0.27 & 0.43 \\ 
UGC09112 & 5 (4) & 14:12:46.90 & 47:44:48.4 & 2136$\pm$68 & 117$\pm$63 & 0.15$\pm$0.08 & 1.14$\pm$0.59 & 0.11$\pm$0.066 & 0.02$\pm$0.001 & 4619$\pm$2372 & 0.0 & 0.0 & 0.39 \\ 
UGC09071 & 6 (5) & 14:10:37.19 & 54:05:00.7 & 2160$\pm$58 & 71$\pm$35 & 0.09$\pm$0.05 & 0.73$\pm$0.47 & 0.18$\pm$0.091 & 0.05$\pm$0.001 & 1435$\pm$917 & 0.0 & 0.0 & 0.37 \\ 
NGC3182 & 6 (6) & 10:20:35.57 & 58:11:49.6 & 2323$\pm$101 & 87$\pm$44 & 0.11$\pm$0.06 & 0.74$\pm$0.72 & 0.14$\pm$0.074 & 0.22$\pm$0.001 & 340$\pm$328 & 9.68 & 0.33 & 0.37 \\ 
SBS1144+591 & 4 (4) & 11:47:05.81 & 58:53:51.1 & 2349$\pm$23 & 47$\pm$30 & 0.06$\pm$0.04 & 0.03$\pm$0.01 & 0.06$\pm$0.067 & 0.01$\pm$0.001 & 298$\pm$128 & 0.0 & 0.25 & 0.1 \\ 
NGC5371 & 39 (31) & 13:53:01.51 & 40:12:35.3 & 2350$\pm$94 & 186$\pm$20 & 0.24$\pm$0.03 & 2.67$\pm$0.67 & 0.07$\pm$0.013 & 1.94$\pm$0.022 & 138$\pm$35 & 10.28 & 0.48 & 0.81 \\ 
NGC5289 & 9 (8) & 13:45:11.39 & 41:35:42.4 & 2382$\pm$136 & 84$\pm$27 & 0.11$\pm$0.04 & 0.41$\pm$0.4 & 0.08$\pm$0.039 & 0.06$\pm$0.014 & 677$\pm$656 & 9.67 & 0.0 & 0.55 \\ 
UGC08603 & 6 (5) & 13:36:31.37 & 44:39:43.4 & 2480$\pm$93 & 102$\pm$51 & 0.13$\pm$0.07 & 0.36$\pm$0.26 & 0.05$\pm$0.032 & 0.1$\pm$0.001 & 367$\pm$264 & 0.0 & 0.2 & 0.14 \\ 
NGC5198 & 4 (4) & 13:29:42.84 & 46:37:56.0 & 2481$\pm$50 & 48$\pm$31 & 0.06$\pm$0.04 & 0.45$\pm$0.58 & 0.2$\pm$0.143 & 0.22$\pm$0.004 & 205$\pm$265 & 9.25 & 0.25 & 0.24 \\ 
CGCG293-042 & 5 (4) & 12:54:45.96 & 58:51:31.5 & 2490$\pm$45 & 70$\pm$34 & 0.09$\pm$0.04 & 0.07$\pm$0.18 & 0.05$\pm$0.029 & 0.07$\pm$0.001 & 100$\pm$260 & 9.53 & 0.5 & 0.28 \\ 
NGC5336 & 5 (5) & 13:51:17.43 & 43:21:57.2 & 2513$\pm$61 & 56$\pm$22 & 0.07$\pm$0.03 & 0.79$\pm$0.63 & 0.18$\pm$0.115 & 0.11$\pm$0.001 & 741$\pm$595 & 9.9 & 0.0 & 0.32 \\ 
MCG+11-16-005 & 4 (4) & 12:44:08.26 & 62:15:32.0 & 2676$\pm$86 & 50$\pm$53 & 0.07$\pm$0.07 & 0.5$\pm$0.56 & 0.07$\pm$0.08 & 0.02$\pm$0.001 & 2501$\pm$2817 & 0.0 & 0.0 & 0.13 \\ 
NGC5320 & 4 (4) & 13:50:16.81 & 41:24:33.1 & 2762$\pm$57 & 19$\pm$16 & 0.02$\pm$0.02 & 0.51$\pm$0.53 & 0.74$\pm$0.675 & 0.16$\pm$0.001 & 321$\pm$337 & 9.94 & 0.25 & 0.43 \\ 
NGC5346 & 6 (6) & 13:53:12.04 & 39:40:30.8 & 2773$\pm$67 & 72$\pm$33 & 0.09$\pm$0.04 & 0.25$\pm$0.17 & 0.11$\pm$0.064 & 0.1$\pm$0.001 & 243$\pm$165 & 9.22 & 0.0 & 0.34 \\ 
UGC07017 & 5 (4) & 12:02:31.92 & 29:50:05.5 & 2925$\pm$61 & 83$\pm$46 & 0.11$\pm$0.06 & 0.42$\pm$0.7 & 0.08$\pm$0.071 & 0.27$\pm$0.001 & 159$\pm$264 & 9.94 & 0.5 & 0.35 \\ 
\hline\\[-2ex]
\hline
\end{longtable}
\end{scriptsize}
\end{landscape}
\twocolumn

\section{Results}
\label{sec:results}

\subsection{Group properties}
\label{sec:est_prop}

The FoF algorithm with linking parameters $D_0=0.3$~Mpc and $V_0=150$~km~s$^{-1}$ was implemented on the complete sample of 1209 galaxies in the Ursa Major region (see Section~\ref{sec:sample}). We find 41 galaxy groups in the Ursa Major region. After defining the luminosity-weighted centroid (RA, Dec. and recession velocity), the maximal radial extent and the velocity dispersion, faint galaxies within each cylindrical volume (that are not part of the complete sample) were added as group members and the group properties were re-calculated. The groups and their properties are listed in Table~\ref{tab:group_prop} sorted with increasing central recession velocity. 

The columns are as follows: column~(A1) contains the group ID used in this work, which is the name of the brightest group galaxy; column~(A2) gives the number of group members in the full and complete sample (in brackets, with $\mathrm{M}_r \leq -15.3$~mag and $\mathrm{M_{\ast}} \geq 10^8~\mathrm{M}_{\odot}$); columns~(A3) and (A4) list the group luminosity-weighted centroid [RA (J2000), Dec. (J2000)]; column~(A5) contains the luminosity-weighted central velocity in the Local Group rest frame (km~s$^{-1}$) and its error; column~(A6) gives the group's velocity dispersion and its error (km~s$^{-1}$); column~(A7) lists the radius $r_{\mathrm{500}}$ with its error (Mpc); column~(A8) contains the virial mass estimate and its error (M$_{\odot}$); column~(A9) gives the crossing time as a function of Hubble time ($H_0^{-1}$) and its error; columns~(A10) and (A11) list the total group luminosity and mass-to-light ratio with their errors; column~(A12) gives a lower limit of the total HI mass in the group (all members with HI masses available are summed; M$_{\odot}$); column~(A13) contains the fraction of bulge dominated galaxies in the group and column~(A14) lists the maximal radial extent of the group (Mpc).

The velocity dispersion of the galaxy groups is estimated as discussed in \citet{beers1990} wherein the gaps between the ordered velocities are calculated $g_i = v_{i+1} - v_i$ for $i = 1,2,...,N-1$. With weights defined as $w_i = i(N-i)$ the velocity dispersion is given by 
\begin{equation} 
\sigma_v  = \frac{\sqrt{\pi}}{N(N-1)} \: \sum\limits_{i=1}^{N-1} w_ig_i.
\end{equation}

This is known as the gapper technique, which gives robust estimates of the velocity dispersions for small groups. The errors are calculated using the jackknife method. The radius $r_{500}$ is calculated as in \citet{osmond2004} 
\begin{equation} 
r_{500} = \frac{0.096 \sigma_v}{H_0} 
\end{equation}
corresponding to an overdensity of 500~times the critical density with errors estimated by standard error propagation. The crossing time as a function of Hubble time ($H_0^{-1}$) is given by
\begin{equation} 
t_c = \frac{3r_H}{5^{3/2}\sigma_v} 
\end{equation}
\citep{huchra1982}, with 
\begin{equation} 
r_H=\pi D \mathrm{sin}\left[\frac{N(N-1)}{4\sum\limits_i\sum\limits_{j>i} \theta_{ij}^{-1}}\right]
\end{equation} 
being the harmonic radius where $D$ is the distance of the group in Mpc, $N$ is the number of members and $\theta_{ij}$ is the angular separation of the group members. The error on $r_H$ is estimated using the jackknife algorithm and the error on the crossing time is calculated with standard error propagation. Note that galaxy groups with $t_c > 0.09 H_0^{-1}$ may not have had time to virialize yet \citep{nolthenius1987}.

For virialized systems, a dynamical mass can be estimated via 
\begin{equation} 
\mathrm{M}_v = \frac{3\pi N}{2G} \frac{\sum\limits_i V_i^2}{\sum\limits_{i<j} 1/r_i}
\end{equation}
where $V_i$ is the observed radial component of the velocity of the galaxy \textit{i} with respect to the mean group velocity and $r_i$ is the projected separation from the group centre \citep{heisler1985}. Again, the error is estimated using the jackknife method.

In general, galaxy groups with numerous group members have a larger maximal radial extent ($r_{max}$), higher velocity dispersion ($\sigma_v$), larger virial mass ($\mathrm{M}_v$) and higher total luminosity ($\mathrm{L}_t$) than galaxy groups with a small number of members. The fraction of early-type galaxies in the groups is on average 27~per~cent, which is low in comparison to other studies -- e.g. \citet{brough2006} find 90 to 100~per cent early-type galaxies in the densest regions of the Eridanus supergroup. The average group properties are $N_{m,F}$=11, $N_{m,C}$=8, $\sigma_v = 90$~km~s$^{-1}$, $r_{max} =360$~kpc, $\mathrm{M}_v = 0.6 \times 10^{13}~\mathrm{M}_{\odot}$, $\mathrm{L}_t = 0.23 \times 10^{11}~\mathrm{L}_{\odot}$, a mass-to-light ratio of $\mathrm{M}_v/\mathrm{L}_t = 480$ and a crossing time of 0.13~$H_0^{-1}$. The group properties, such as (\textit{i}) the small number of group members, (\textit{ii}) the small fraction of early-type galaxies, (\textit{iii}) the low velocity dispersions and (\textit{iv}) the virial masses $< 10^{14}~\mathrm{M}_{\odot}$  emphasize the picture of Ursa Major being composed of galaxy groups as opposed to an `Ursa Major cluster'. 

Only 26~per~cent of galaxies in the complete sample reside in groups (with $N_m \geq 4$), i.e. 74~per~cent of galaxies are left ungrouped. The ungrouped fraction is rather high in comparison to e.g. Giuricin et al. (2000), wherein only 40~per~cent of galaxies in a complete, distance-limited ($cz<=6000$~km~s$^{-1}$) and magnitude-limited ($B \leq$14) sample of $\sim$7000 galaxies are field galaxies. However, Giuricin et al. (2000) also includes galaxy pairs and groups with three members.

\subsection{Structures in the Ursa Major region - An overview}
\label{sec:groups}

\begin{figure*}
\begin{center}
\includegraphics[bb = 20 250 550 580, width=15cm]{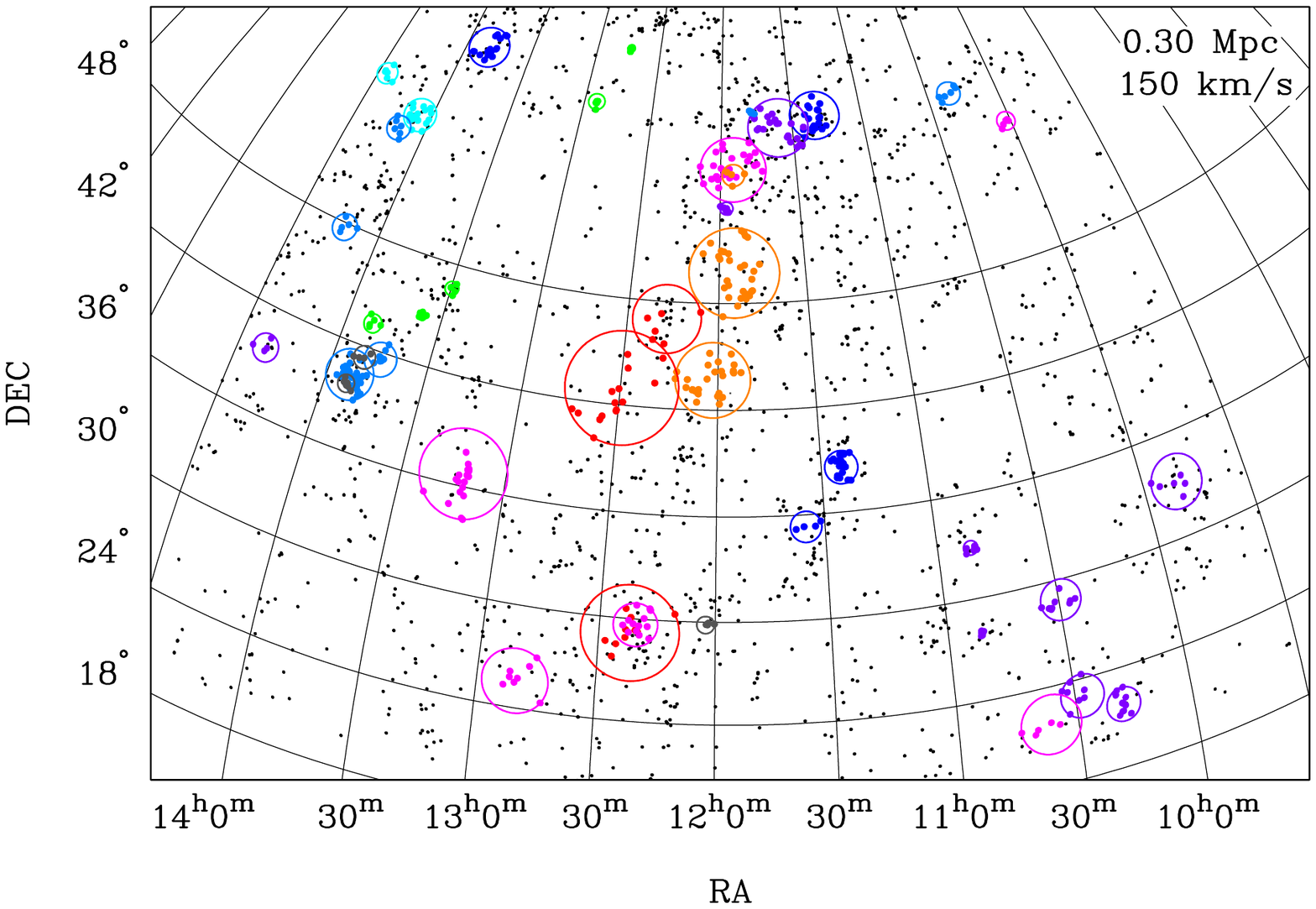}
\includegraphics[bb = 20 200 550 600, width=15cm]{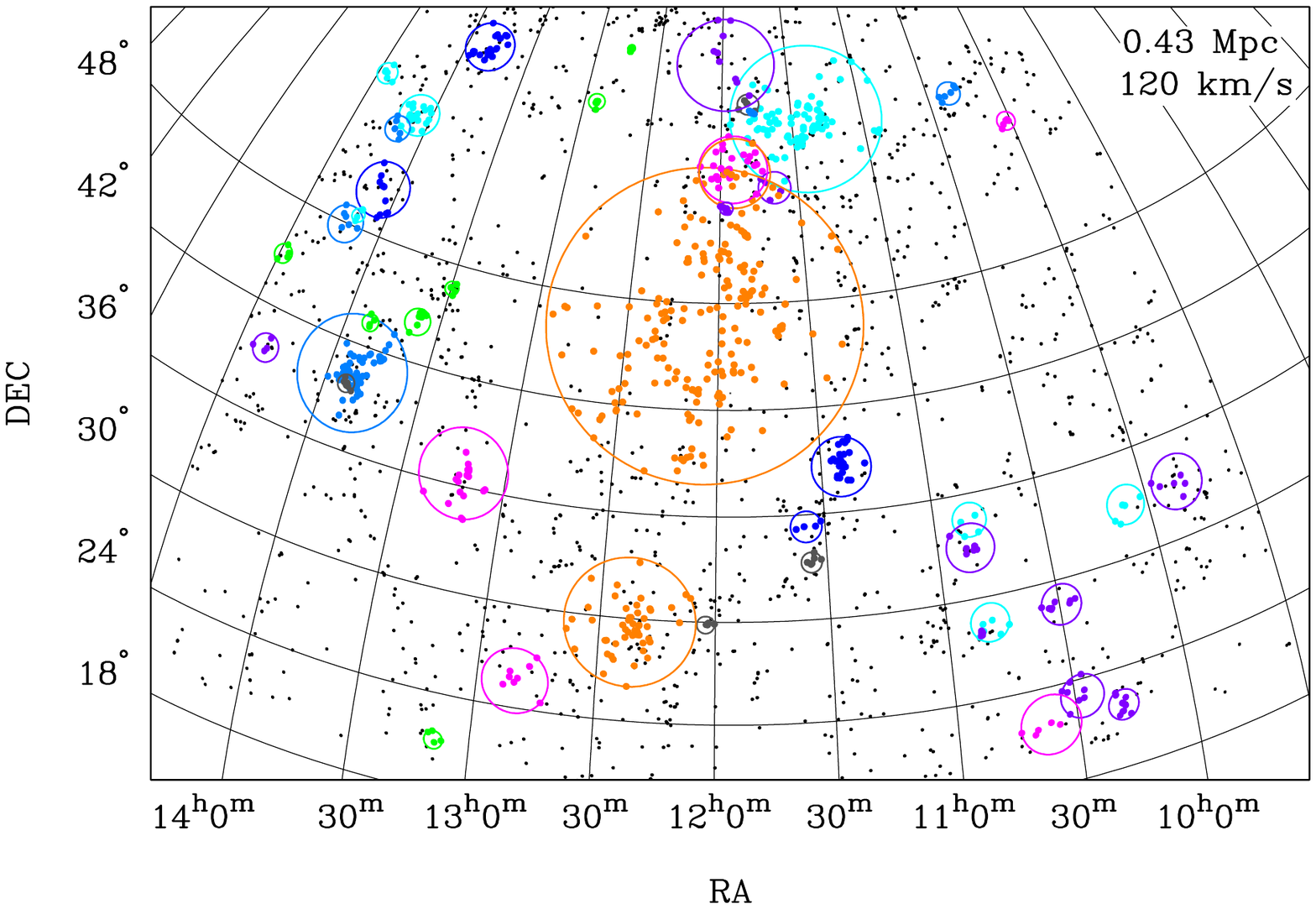} 
\caption{Projected overviews of the groups found in the Ursa Major region using the FoF algorithm. The linking lengths are labelled in the top right corner. The group's maximum radial extent is shown by a circle and the colour indicates the central group velocities -- red ($v_c \leq 600$~km~s$^{-1}$), orange ($600 < v_c \leq 900$~km~s$^{-1}$), magenta ($900 < v_c \leq 1200$~km~s$^{-1}$), violet ($1200 < v_c \leq 1500$~km~s$^{-1}$), cyan ($1500 < v_c \leq 1800$~km~s$^{-1}$), blue ($1800 < v_c \leq 2100$~km~s$^{-1}$), azure ($2100 < v_c \leq 2400$~km~s$^{-1}$), green ($2400 < v_c \leq 2700$~km~s$^{-1}$) and grey ($v_c > 2700$~km~s$^{-1}$). Group members are indicated by the same colour, whereas galaxies that are not residing in any group are shown in black.}
\label{fig:skymap}
\end{center}
\end{figure*}

To illustrate the galaxy groups identified in the Ursa Major region using the FoF algorithm, projected skymaps are shown in Figure~\ref{fig:skymap} (with linking lengths $D_0=0.3$~Mpc and $V_0=150$~km~s$^{-1}$ \textit{top} and $D_0=0.43$~Mpc and $V_0=120$~km~s$^{-1}$ \textit{bottom}). For illustrative purposes, the group radius is defined as the maximal radial extent, i.e. the radius to the galaxy the furthest away from the group centre. The group members are indicated by the same colour as the group radius, i.e. red ($v_c \leq 600$~km~s$^{-1}$), orange ($600 < v_c \leq 900$~km~s$^{-1}$), magenta ($900 < v_c \leq 1200$~km~s$^{-1}$), violet ($1200 < v_c \leq 1500$~km~s$^{-1}$), cyan ($1500 < v_c \leq 1800$~km~s$^{-1}$), blue ($1800 < v_c \leq 2100$~km~s$^{-1}$), azure ($2100 < v_c \leq 2400$~km~s$^{-1}$), green ($2400 < v_c \geq 2700$~km~s$^{-1}$) and grey ($v_c > 2700$~km~s$^{-1}$).

\begin{figure*}
\begin{center}
\includegraphics[trim=0.5cm 1cm 1cm 2cm, clip=true, width=15cm]{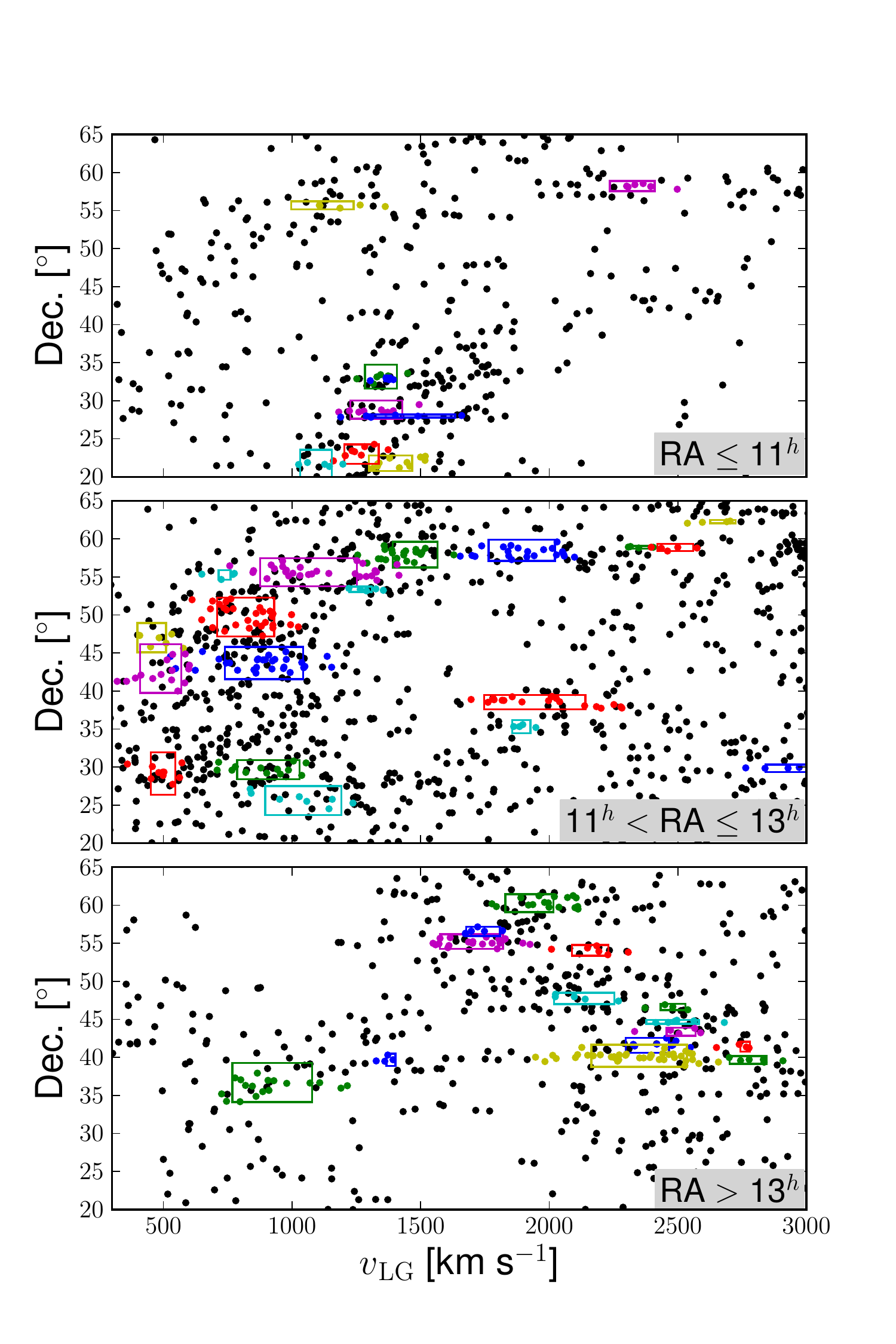}
\caption{Groups identified in the Ursa Major region using the FoF algorithm with linking lengths $D_0 = 0.30$~Mpc and $V_0 = 150$~km~s$^{-1}$ shown in velocity space. The slices in right ascension are stated in the bottom right corner. Galaxy groups and their members are colour-coded. The boxes surrounding the group members are centred on the luminosity-weighted central group velocities with a width of $\pm 2\sigma_v$ (velocity dispersion) and a height of $\pm r_{max}$ (maximal radial extent).}
\label{fig:mapvel30}
\end{center}
\end{figure*}

\begin{figure*}
\begin{center}
\includegraphics[trim=0.5cm 1cm 1cm 2cm, clip=true, width=15cm]{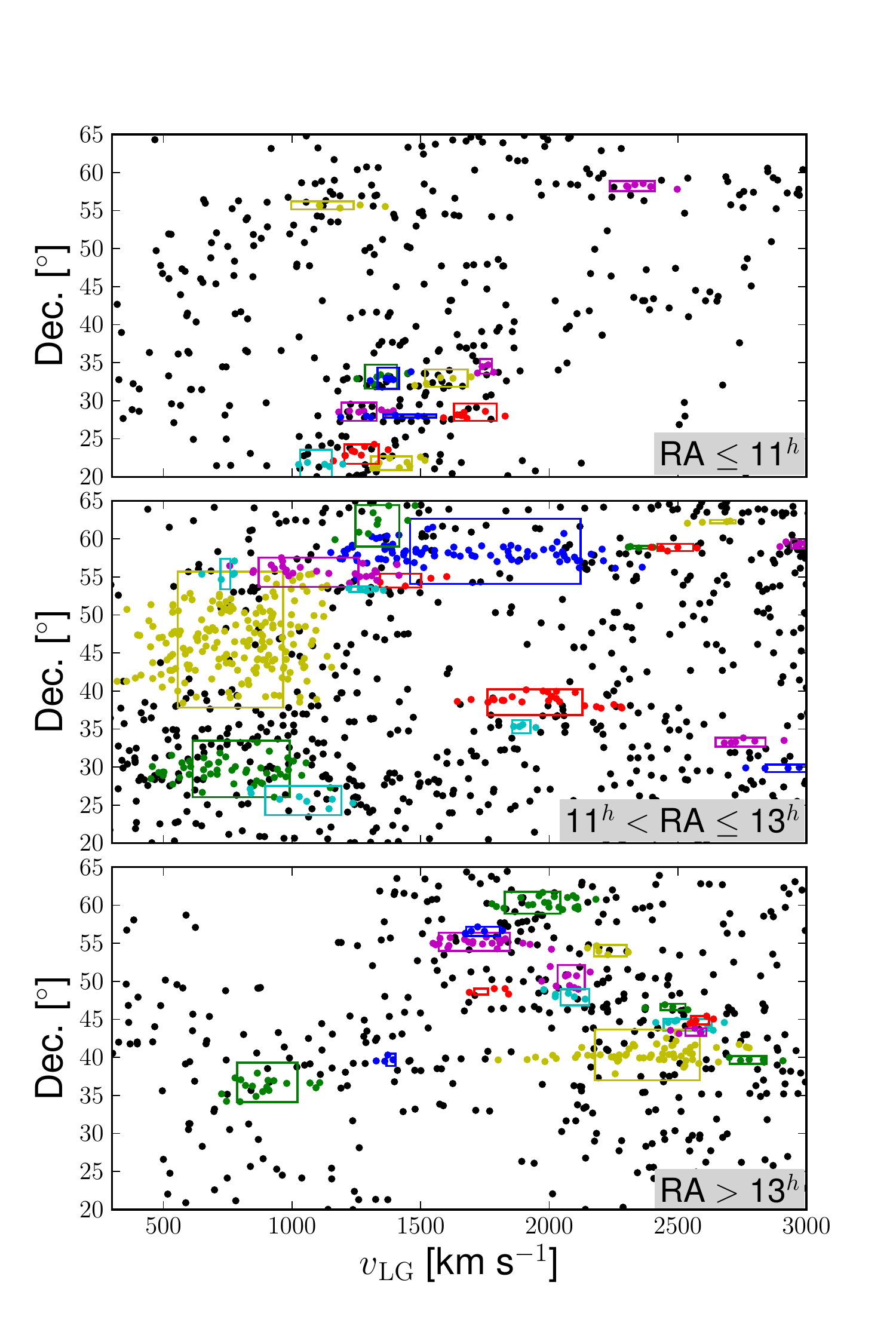}
\caption{Groups identified in the Ursa Major region using the FoF algorithm with linking lengths $D_0 = 0.43$~Mpc and $V_0 = 120$~km~s$^{-1}$ shown in velocity space. Box sizes are detailed in the Figure~\ref{fig:mapvel30}.}
\label{fig:mapvel43}
\end{center}
\end{figure*}

In Figures~\ref{fig:mapvel30} and \ref{fig:mapvel43} the identified galaxy groups are shown in velocity space. Three slices in right ascension are used, which are stated in the bottom right corner. For clarity, the galaxy groups and their members are coloured. The boxes surrounding the group members are centred on the luminosity-weighted central group velocities with a width of $\pm 2\sigma_v$ (velocity dispersion) and a height of $\pm r_{max}$ (maximal radial extent).

The majority of galaxy groups with RA$ \leq 11^h$ are at recession velocities $v_c \approx 1200$~km~s$^{-1}$ (see Figure~\ref{fig:mapvel30} and ~\ref{fig:mapvel43} \textit{top}), with a conglomeration of galaxy groups in the South (see Figure~\ref{fig:skymap} -- bottom right corner).

These galaxy groups are mostly unaffected by the `looser' linking lengths (see Figure~\ref{fig:skymap} \textit{bottom}), i.e. insignificant changes in spatial extent, the group members etc. A few additional galaxy groups appear in cyan (with $1500 < v_c \leq 1800$~km~s$^{-1}$) when using `looser' linking lengths.

In the central region ($11^h < \mathrm{RA} \leq 13^h$), the FoF analysis reveals several groups that constitute to an extended filamentary structure -- the groups are connected in projection from the central (Dec.$\approx 40^{\circ}$) to the northern part (Dec.$\approx 60^{\circ}$ see Figure~\ref{fig:skymap} \textit{top}). Three sizable sub-structures sit further to the South, which contain between 8 to 13 galaxies each and have similar central group velocities to the South end of the main Ursa Major structure. The recession velocity of the filamentary structure is approximately 700~km~s$^{-1}$ in the South and 1800~km~s$^{-1}$ in the North (see Figure~\ref{fig:mapvel30} \textit{middle}). The main galaxy groups in this filamentary structure are discussed in detail in Section~\ref{sec:IndGrps}. The filamentary structure is connected to the Virgo cluster, which is centred at RA$=12^h30^m$, Dec.$=12^{\circ}20'$ and $v_{\mathrm{LG}}=975$~km~s$^{-1}$ (to the South of the displayed region). Note that behind the filamentary structure is a mostly empty region (see Figure~\ref{fig:mapvel30} and \ref{fig:mapvel43} \textit{middle} at $1200 < v_{\mathrm{LG}} < 1800$~km~s$^{-1}$). When using `looser' linking lengths, several groups in the central Ursa Major region merge (see Figure~\ref{fig:skymap} \textit{bottom}), e.g. the two galaxy groups at Dec.$\approx 30^{\circ}$, the four galaxy groups at Dec.$\approx 46^{\circ}$ and the two galaxy groups at Dec.$\approx 60^{\circ}$ (in Figure~\ref{fig:skymap} \textit{top}).

The majority of galaxy groups with RA$> 13^h$ are at high recession velocities ($v_c \geq 1600$~km~s$^{-1}$). The groups are not scattered throughout the field, they appear to form a filamentary structure from the North (with $v_c \approx 1600$~km~s$^{-1}$) to the central region (with $v_c \approx 2700$~km~s$^{-1}$). When using `looser' linking lengths, some groups grow slightly in size and a few additional galaxy groups appear (Figure~\ref{fig:skymap} \textit{bottom}). However, the `looser' linking lengths only affect the central Ursa Major region where groups are getting merged and grow rapidly -- the majority of groups with RA$ \leq 11^h$ and RA$> 13^h$ stay as individual groups.

\subsection{Galaxy groups -- Individual structures}
\label{sec:IndGrps}

\begin{figure}
\begin{center}
\mbox{\includegraphics[trim=1.8cm 0.5cm 2cm 0.5cm, clip=true, height=8cm]{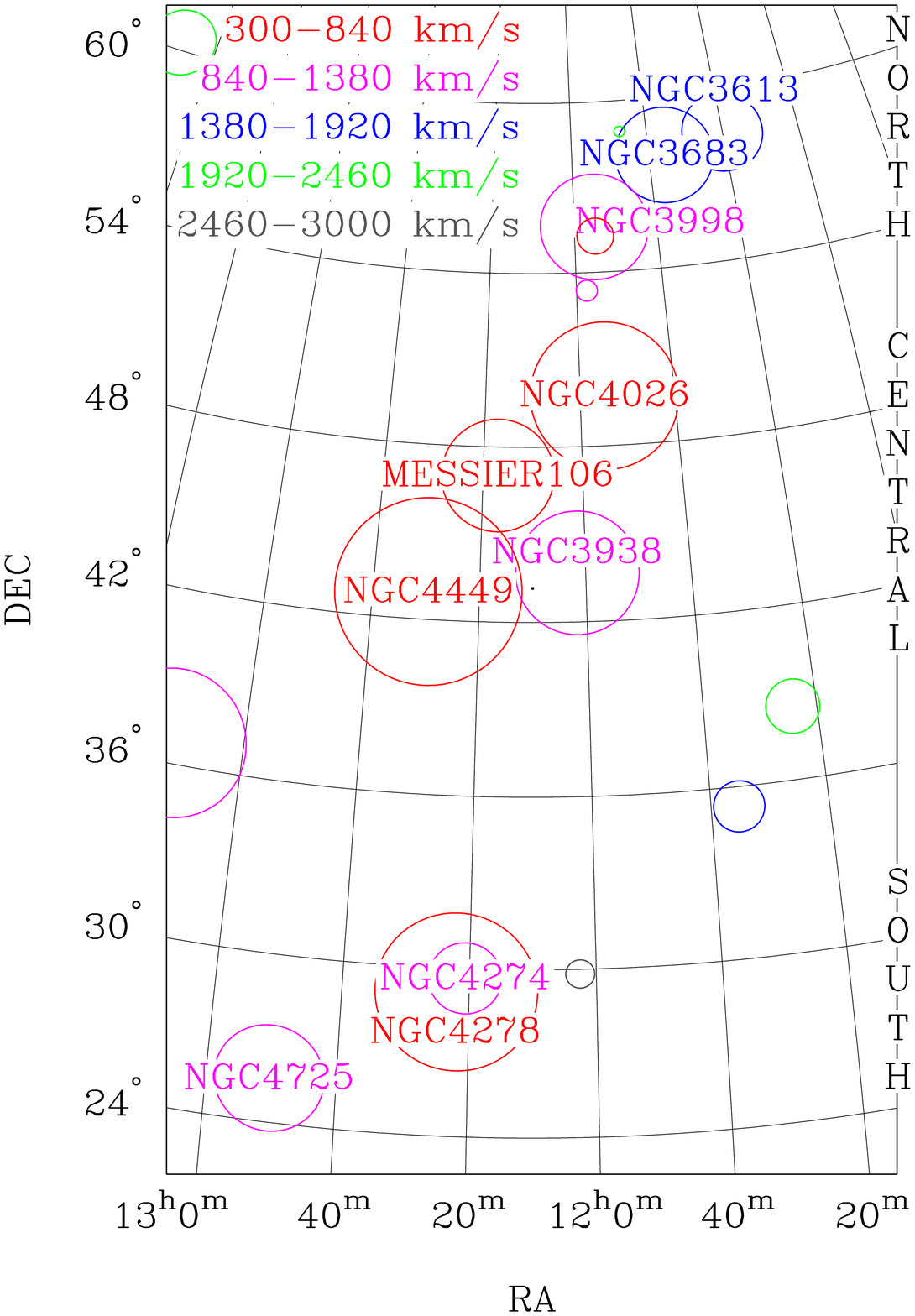}}
\caption{Overview of the main structures identified in the Ursa Major region using linking lengths $D_0=0.30$~Mpc and $V_0=150$~km~s$^{-1}$. The colour key is described in the top left corner. The groups are indicated by circles showing their maximal radial extent and labelled by their brightest group galaxy (BGG). The main galaxy groups are discussed in Sections~\ref{sec:IndGrps} and the Appendix.}
\label{fig:overview}
\end{center}
\end{figure}

The main structures identified in the central Ursa Major region ($11^{h}\leq\alpha\leq+13^{h}$) using linking lengths $D_0=0.3$~Mpc and $V_0=150$~km~s$^{-1}$ are shown in Figure~\ref{fig:overview}. The galaxy groups are labelled by their brightest group galaxy (BGG). 

The dominant structures in the central part of the Ursa Major region ($35^{\circ}\leq\delta\leq+50^{\circ}$) are the NGC4026, NGC3938, NGC4449 and MESSIER106 groups, which are discussed in the following. For further information on the galaxy groups that are located to the South and North (as labelled in Figure~\ref{fig:overview}) see the Appendix.

\begin{figure*}
\begin{center}
\begin{tabular}{p{4.3cm} p{4.7cm} p{5cm}}
\mbox{\includegraphics[trim=2.5cm 2.5cm 3.0cm 0.0cm, clip=true, height=4.4cm]{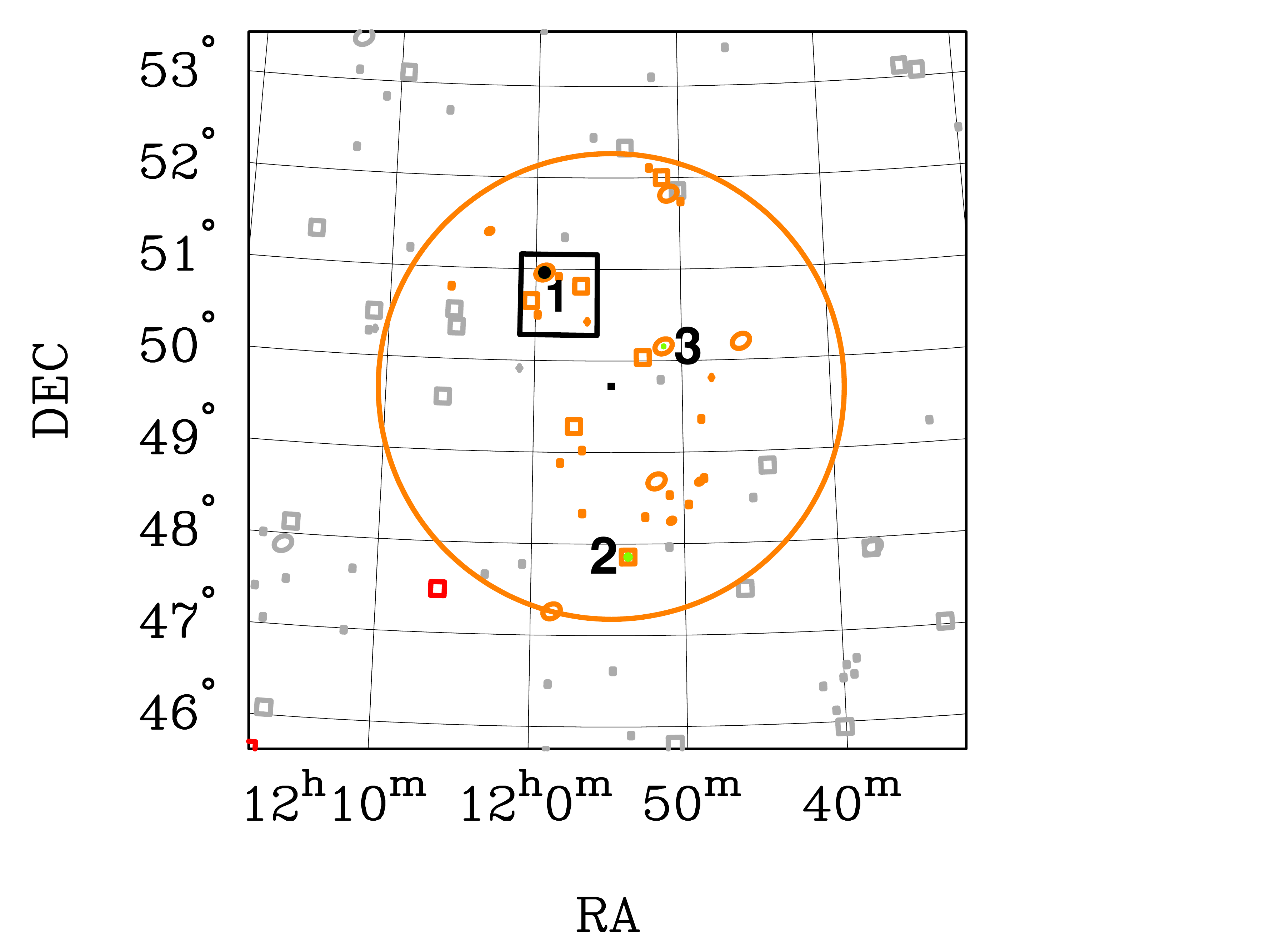}} & 
\mbox{\includegraphics[trim=0cm 0cm 0.5cm 0.5cm, clip=true, height=4.5cm]{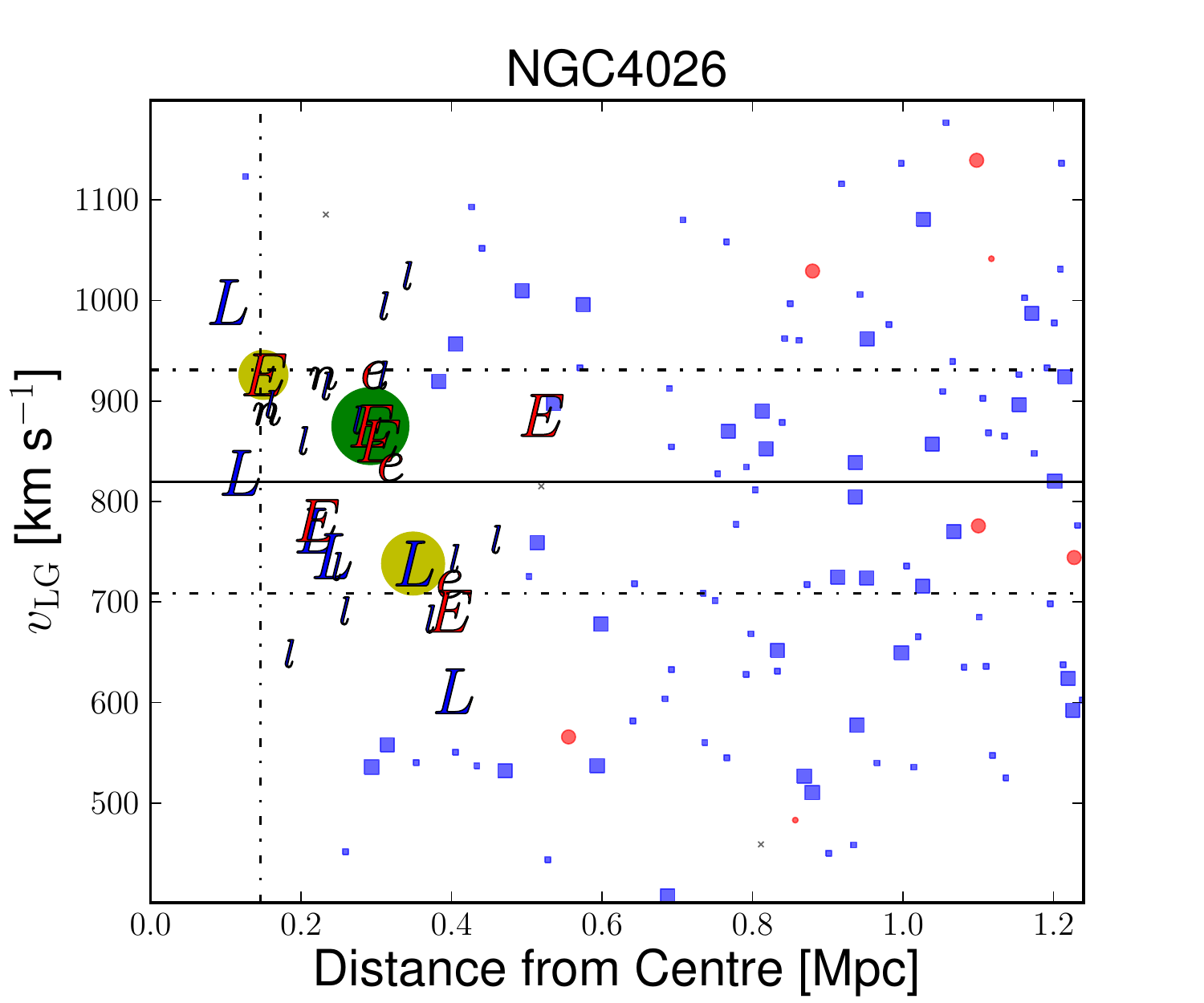}} &
\mbox{\includegraphics[trim=0.5cm 0cm 0.5cm 0.5cm, clip=true, height=4.5cm]{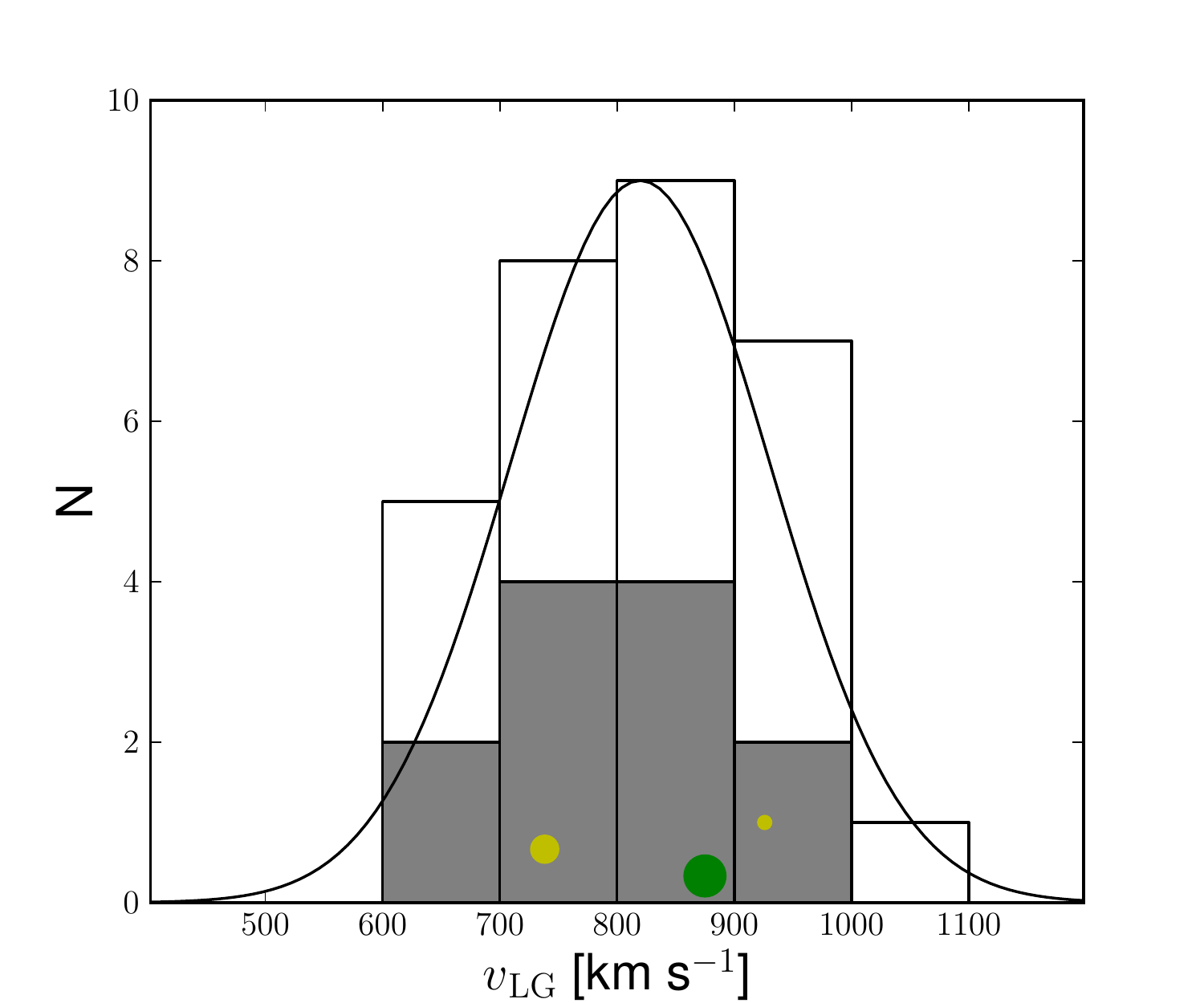}} \\
[3pt]
\mbox{\includegraphics[trim=2.5cm 2.5cm 3.0cm 0.0cm, clip=true, height=4.6cm]{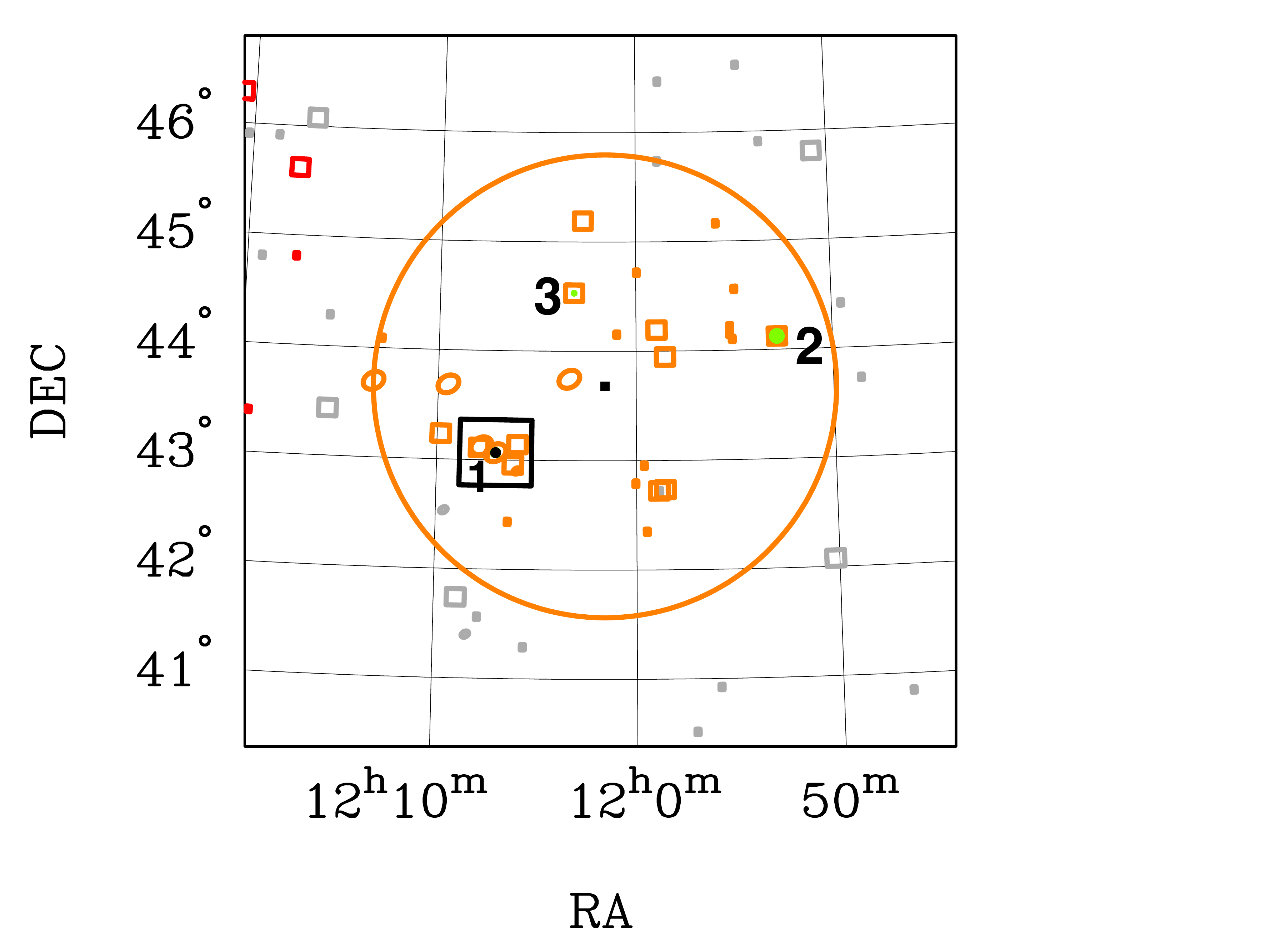}} & 
\mbox{\includegraphics[trim=0cm 0cm 0.5cm 0.5cm, clip=true, height=4.5cm]{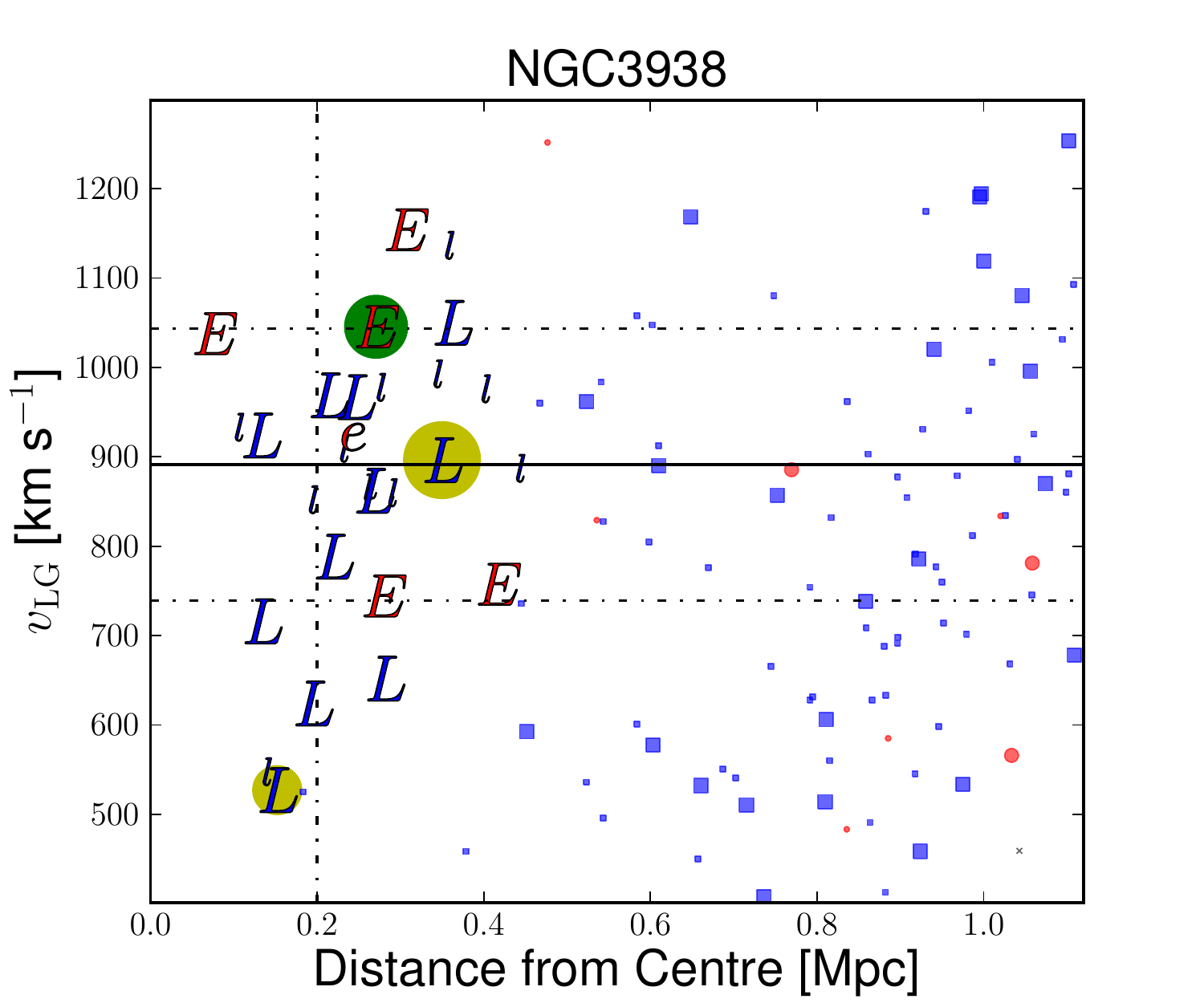}} &
\mbox{\includegraphics[trim=0.5cm 0cm 0.5cm 0.5cm, clip=true, height=4.5cm]{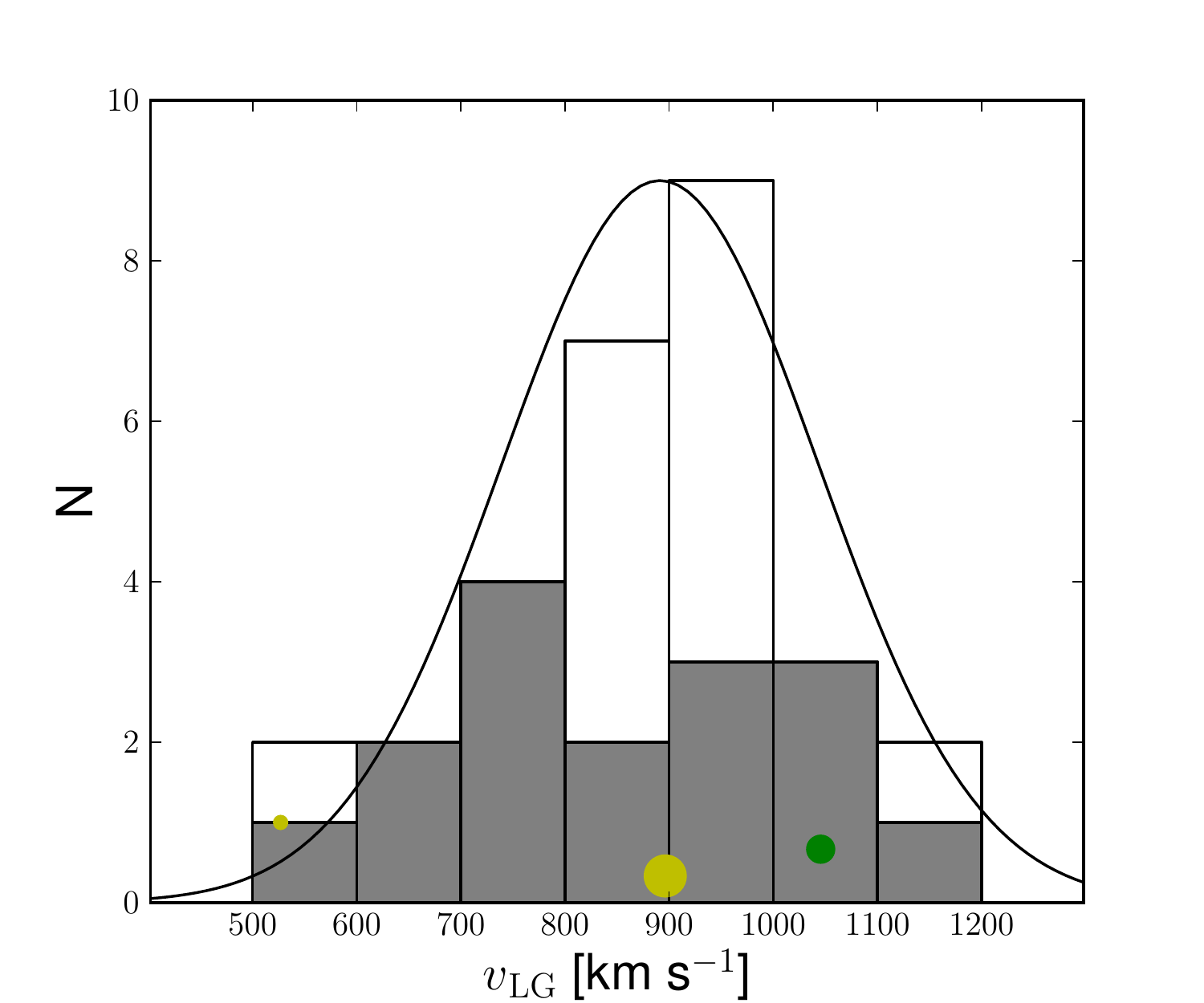}} \\
[3pt]
\mbox{\includegraphics[trim=2.5cm 2.5cm 3.0cm 0.0cm, clip=true, height=4.6cm]{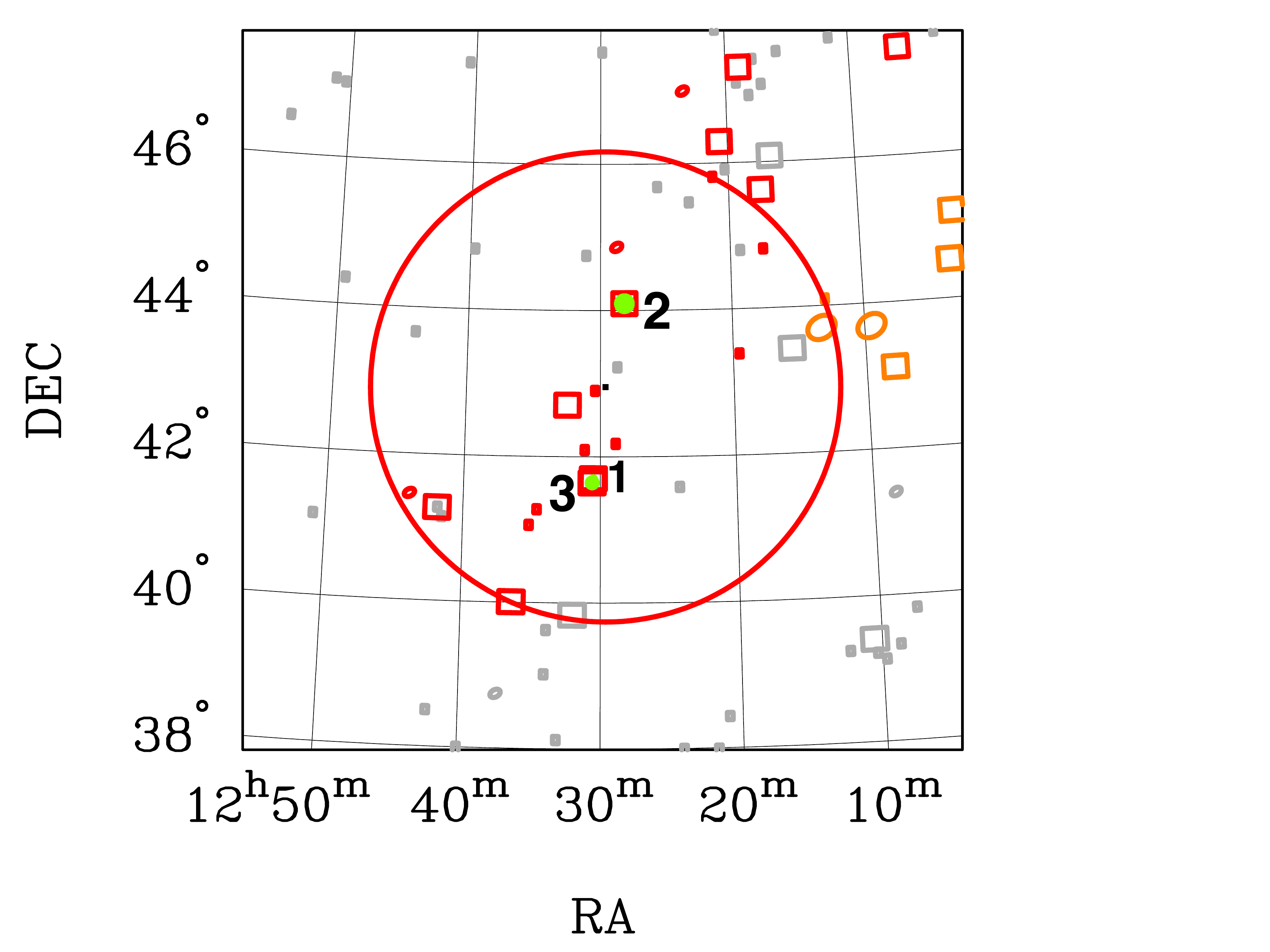}} & 
\mbox{\includegraphics[trim=0cm 0cm 0.5cm 0.5cm, clip=true, height=4.5cm]{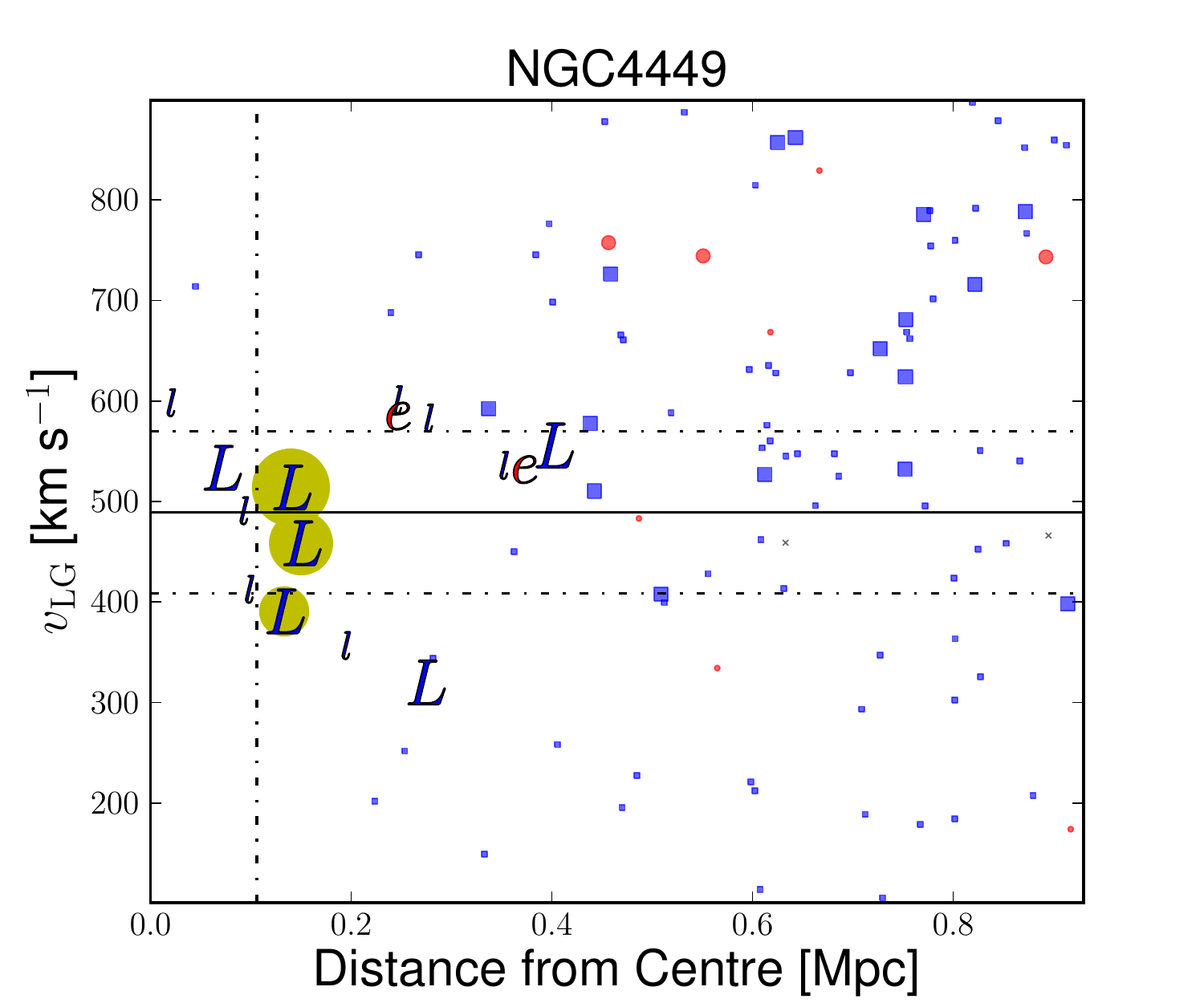}} &
\mbox{\includegraphics[trim=0.5cm 0cm 0.5cm 0.5cm, clip=true, height=4.5cm]{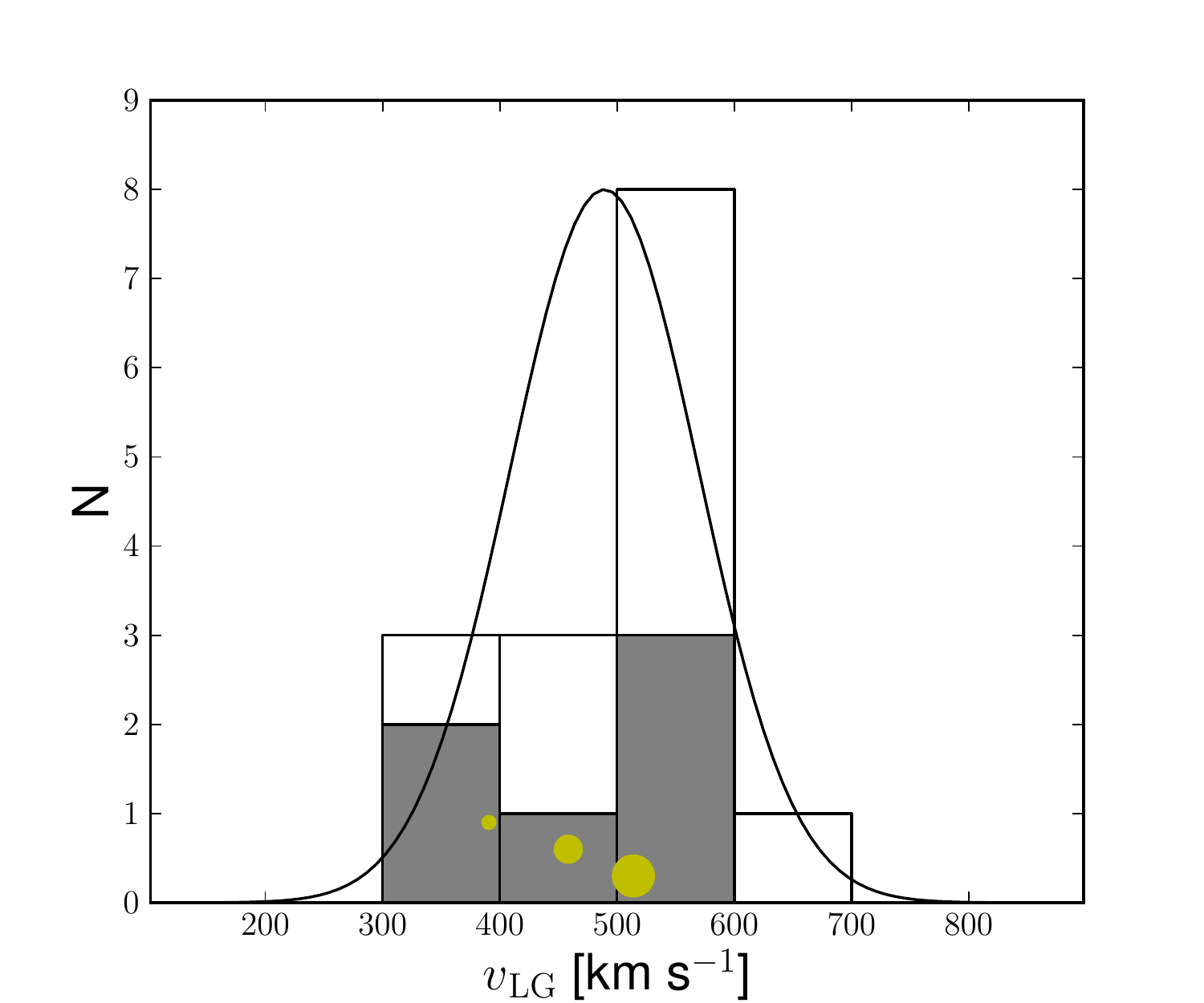}} \\
[3pt]
\mbox{\includegraphics[trim=2.5cm 2.5cm 3.0cm 0.0cm, clip=true, height=4.6cm]{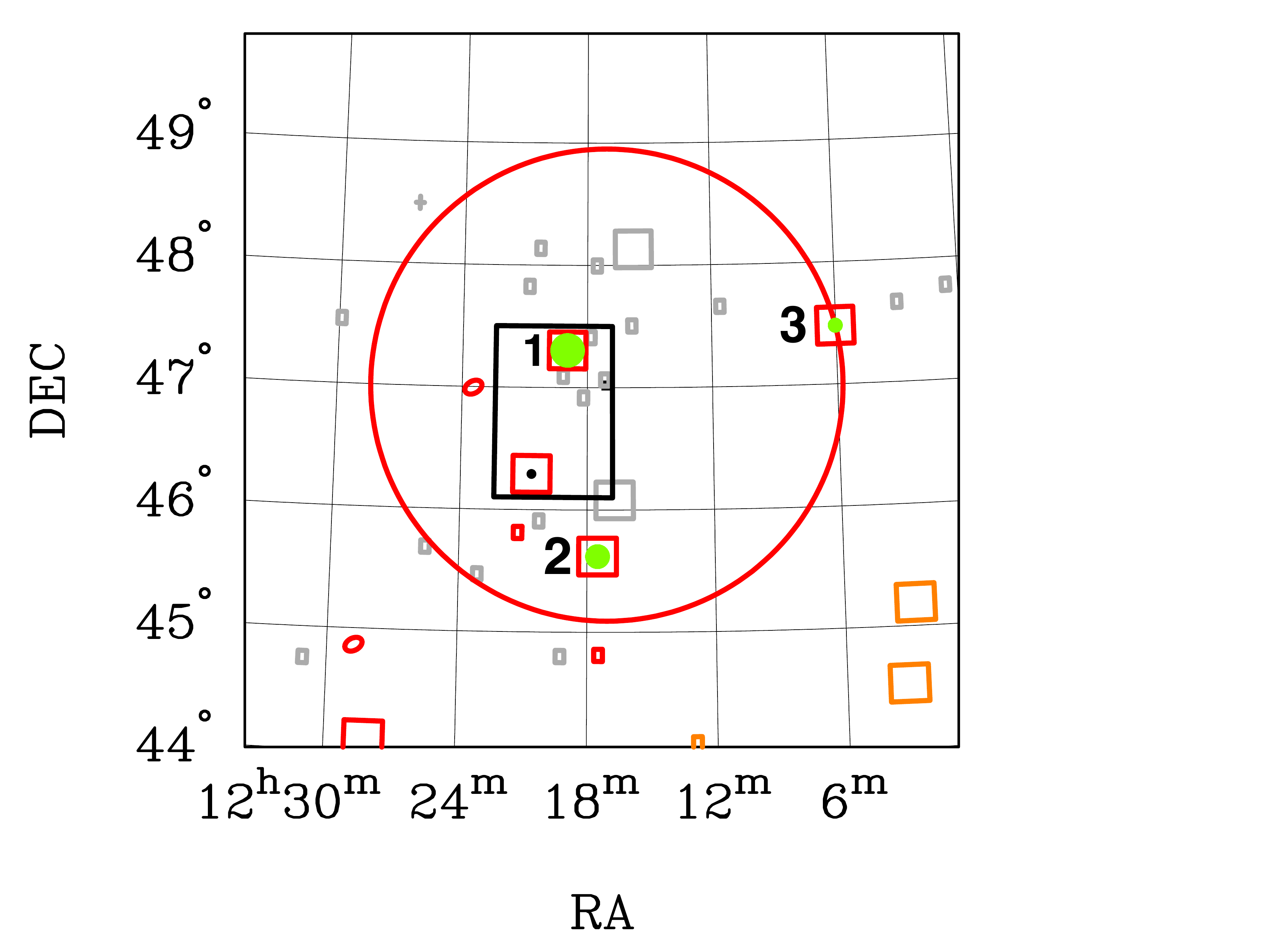}} & 
\mbox{\includegraphics[trim=0cm 0cm 0.5cm 0.5cm, clip=true, height=4.5cm]{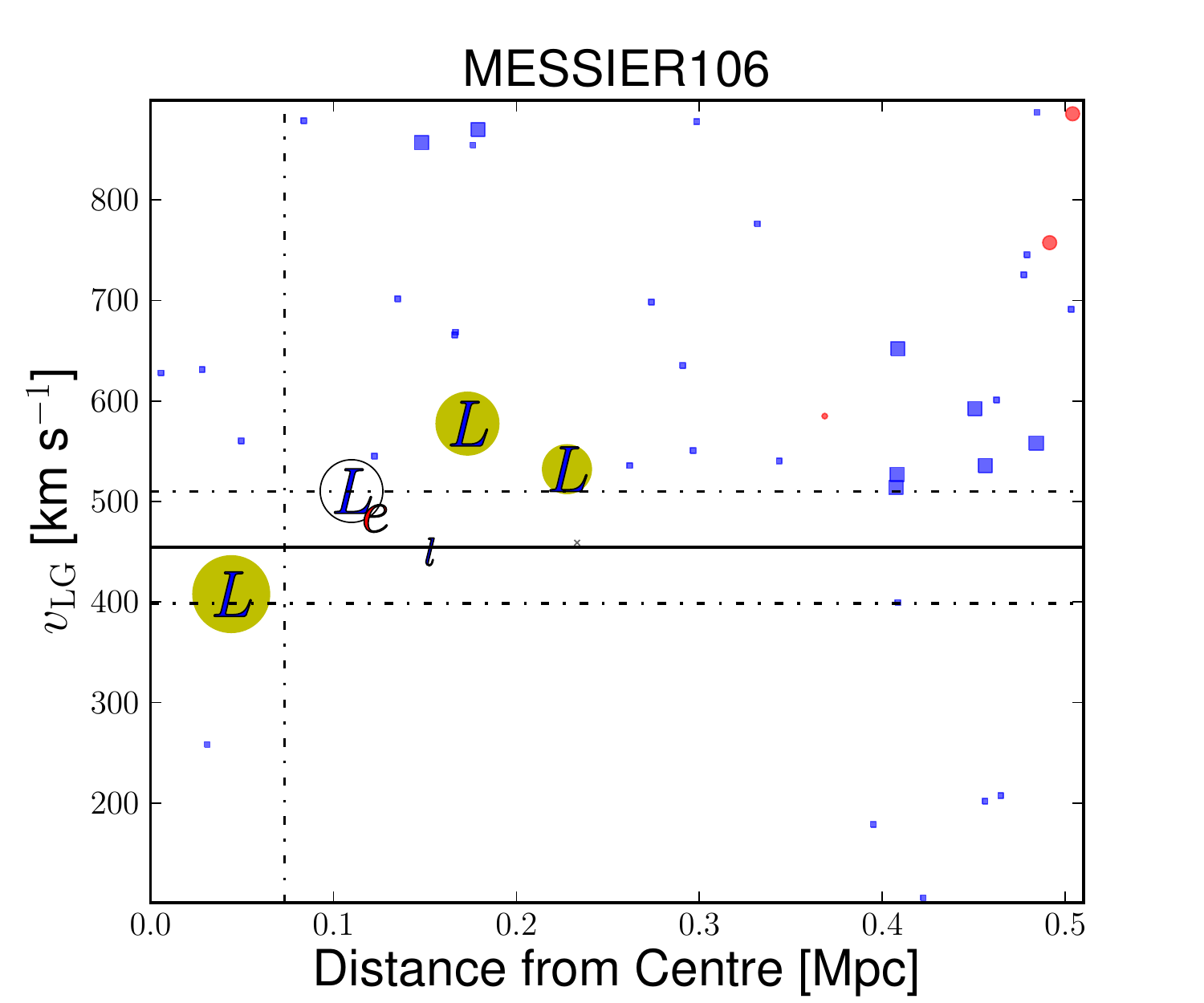}} &
\mbox{\includegraphics[trim=0.5cm 0cm 0.5cm 0.5cm, clip=true, height=4.5cm]{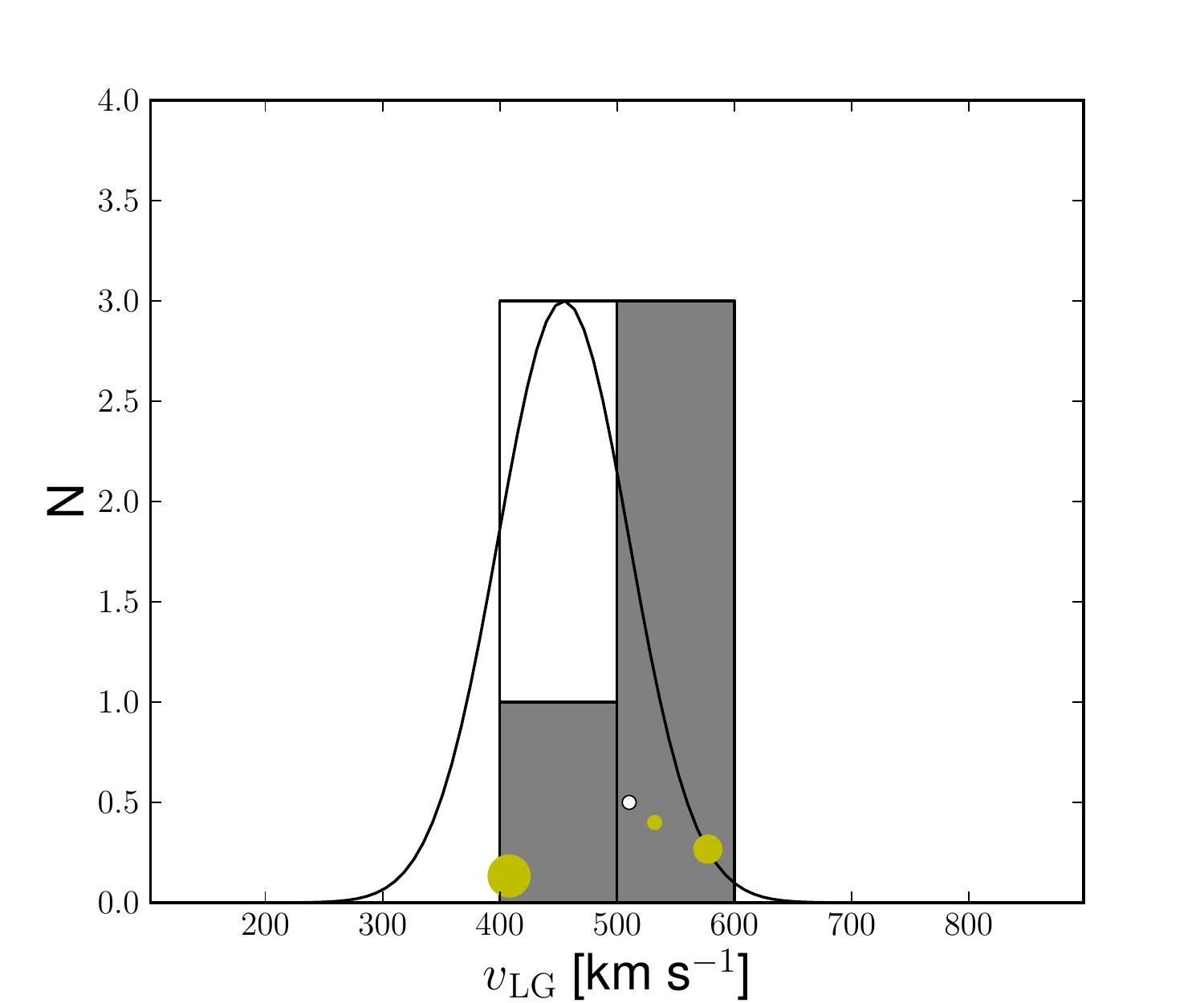}} \\
\end{tabular} 
\caption{The main structures in the central part of the Ursa Major region (found using linking lengths $D_0=0.3$~Mpc and $V_0=150$~km~s$^{-1}$). The group ID (name of the BGG) is given in the middle (\textit{top}). (\textit{left}) Projected map of the group and its members centred on the luminosity-weighted centroid and colour coded as in Figure~\ref{fig:skymap}. The circle indicates the maximal radial extent. Early-type galaxies are marked with ellipses, whereas late-type galaxies are indicated by boxes -- large symbols represent the complete sample and small symbols indicate additional faint galaxies. The BGGs are marked with filled circles, which are decreasing in size. The black squares mark regions, where galaxies are being likely transformed due to galaxy-galaxy interactions -- galaxies in the process of being stripped showing HI tails are marked with a black dot in the skymap and circles in the distance-velocity diagram and velocity histogram (green for BGGs and white for other members). (\textit{middle}) The distance-velocity diagram showing group members marked with letters and nearby galaxies indicated by symbols. Early-type galaxies are shown in red ($E$, $e$ and circles) and late-type galaxies are shown in blue ($L$, $l$ and boxes). Capital letters and large symbols show the complete sample, whereas small letters and symbols indicate additional faint galaxies.The lines mark the luminosity-weighted mean group velocity (solid), the velocity dispersion (dashed horizontal) and $r_{500}$ (dashed vertical). The BGGs are surrounded by yellow circles (sizes as previously discussed). (\textit{right}) Velocity distribution of the group members -- the complete sample is shown in grey and additional faint galaxies are shown in white. The velocities of the BGGs are marked with yellow circles (sizes as previously discussed). Overlaid is a Gaussian with the peak centred at the luminosity-weighted mean group velocity, the height corresponds to the maximum members in a bin and a FWHM of twice the velocity dispersion.}
\label{fig:middle1}
\end{center}
\end{figure*}

To investigate the evolutionary state of a galaxy group as well as its purity (i.e. distinct group members), three diagrams are shown in Figure~\ref{fig:middle1} -- (\textit{i}) a projected skymap, which is centred on the luminosity-weighted centroid of the group, (\textit{ii}) the distance-velocity diagram and (\textit{iii}) the velocity distribution of the group members. The distance-velocity diagram can provide evidence of distinct group membership when other nearby galaxies are separated by a clear gap from the group members. Furthermore dynamically evolved groups are predicted to have the brightest, most massive galaxy residing at the centre of their dark matter halos. Therefore, small offsets of the brightest group galaxy (BGG) from the spatial and kinematic group centres can provide evidence for dynamically evolved groups. Note that the determination of the spatial and kinematic group centres are luminosity-weighted and therefore biased towards the brightest group galaxy. By plotting the velocity histogram of each group we can assess whether the group is unrelaxed or dynamically evolved as dynamically evolved groups tend to show a Gaussian velocity distribution \citep{bird1993}. 

The four galaxy groups presented here are arranged with decreasing number of group members. In the projected skymap of each galaxy group, early-type galaxies are shown with ellipses, whereas late-type galaxies are shown with boxes. Group members are colour-coded and galaxies without group membership are shown in grey. Large symbols indicate the complete galaxy sample, whereas smaller symbols show additional faint galaxies in the region. The three brightest group galaxies are marked with filled circles - the brightest group galaxy (BGG) is indicated by the largest filled circle to the third brightest group galaxy ($3^{rd}$ BGG) marked with the smallest filled circle.

In the distance-velocity diagram, the group members are marked with letters and other nearby galaxies are indicated by symbols. Early-type galaxies are shown in red ($E$, $e$ and circles), whereas late-type galaxies are shown in blue ($L$, $l$ and boxes). Capital letters and large symbols show the complete sample, whereas small letters and symbols indicate additional faint galaxies in the region. The lines mark the luminosity-weighted mean group velocity (solid), the velocity dispersion (dashed horizontal) and $r_{500}$ (dashed vertical). The brightest group galaxies (BGGs) are surrounded by yellow circles (sizes as discussed above). 

In the velocity histogram of the group members, the complete galaxy sample is shown in grey, whereas additional faint galaxies are shown in white. The velocities of the BGGs are marked with yellow circles (sizes as discussed above). Overlaid is a Gaussian with the peak centred at the luminosity-weighted mean group velocity, the height corresponding to the maximum members in a bin and a FWHM of twice the velocity dispersion. 

Galaxies that are known to show HI tails are surrounded by black boxes in the projected skymap and are indicated by a circle in the distance-velocity diagram and the velocity histogram (green for BGGs and white for other group members).

The NGC4026, NGC3938 and NGC5033 groups reside in close vicinity of one another (in projection) together with the smaller MESSIER106 group (6 group members in the full and 4 group members in the complete sample), which are shown in Figure~\ref{fig:middle1}. The luminosity-weighted mean velocities (hereafter referred to as \textit{central} velocity) of this conglomeration range from 454~km~s$^{-1}$ (MESSIER106) to 891~km~s$^{-1}$ (NGC3938).

The NGC4026 group is the richest group in the central part of the Ursa Major region with 30 group members (12 in the complete sample). NGC4026 (BGG) is located 283~kpc and 55~km~s$^{-1}$ from the group centroid -- the projected offset from the centre is well beyond $r_{500}$, but only about half of $r_{max}$. The group has a relatively high early-type fraction of 50~per cent (determined in the full sample) compared to other groups in the Ursa Major region and a velocity dispersion of 111~km~s$^{-1}$. According to quality distance measurements in the Extragalactic Distance Database (hereafter `EDD'; \citealt{tully2009}), the BGG is located at 13.6~Mpc and NGC3928 at 16.9~Mpc, whereas the group distance estimated assuming Hubble flow is only 11.2~Mpc. `Quality distances' are accumulations of multiple sources, e.g. Cepheid period-luminosity relation, the  tip of the red giant branch method and/or surface brightness fluctuations. The BGG is likely being stripped of its HI gas (the galaxy and its surroundings are marked with a black box in Figure~\ref{fig:middle1}; NGC4026 is highlighted with a green circle in the distance-velocity diagram and in the velocity histogram).

The NGC3938 group has a velocity dispersion of 152~km~s$^{-1}$ and an early-type fraction of 31~per cent. NGC3938 (BGG) is at the kinematic centre of the group (the difference is only 5~km~s$^{-1}$). However, the projected spatial offset is 349~kpc, which is beyond $r_{500}$ indicating that the BGG resides in the outskirts of the group. Redshift-independent distances (EDD; \citealt{tully2009}) are available for NGC4111 ($2^{nd}$ BGG) and NGC4138, which are 15.0~Mpc and 13.8~Mpc, respectively. Assuming Hubble flow, the NGC3938 group is at similar distance to the NGC4026 group (recession velocities of 891~km~s$^{-1}$ and 820~km~s$^{-1}$, which correspond to 12.2~Mpc and 11.2~Mpc, respectively). The $2^{nd}$ BGG (NGC4111) shows a large HI tail (the galaxy and its surroundings are marked with a black box in Figure~\ref{fig:middle1}; NGC4111 is highlighted with a green circle in the distance-velocity diagram and in the velocity histogram).

The NGC4449 group resides in the foreground. The central group velocity is 489~km~s$^{-1}$, which corresponds to a distance of 6.7~Mpc (Hubble flow). The quality distance of NGC4460 listed in EDD is 9.6~Mpc \citep{tully2009}. There are no early-type galaxies amongst the 15 group members (6 in the complete sample). The group has a velocity dispersion of 81~km~s$^{-1}$. NGC4449 (BGG) is 136~kpc and 24~km~s$^{-1}$ offset from the group centroid.

The much smaller MESSIER106 group with 6 group members in the full and 4 members in the complete sample resides at a distance of 6.2~Mpc assuming Hubble flow ($v_{\mathrm{LG}} = 454$~km~s$^{-1}$). A quality independent distance measurement of 14.3~Mpc is available for the NGC4346 galaxy placing the group behind the NGC4026, NGC3938 and NGC4449 groups. Large peculiar velocities may explain the inconsistency in quality and redshift distance measurements. The MESSIER106 group may in fact be located at 14.3~Mpc and infalling, which would lead to blue-shifted redshift measurements, i.e. the distances calculated assuming Hubble flow are too small. The MESSIER106 group has a velocity dispersion of 56~km~s$^{-1}$ and the group consists solely of late-type galaxies. The BGG `MESSIER106' is located 47~kpc and 46~km~s$^{-1}$ offset from the group centroid. The NGC4288 galaxy -- located to the South of MESSIER106 -- shows an HI tail (Wolfinger et al., in prep.; the galaxy and its surroundings are marked with a black box in Figure~\ref{fig:middle1}; NGC4288 is highlighted with a white circle in the distance-velocity diagram and in the velocity histogram).

Note that \textit{looser} linking lengths ($D_0=0.43$~Mpc and $V_0=120$~km~s$^{-1}$) result in the NGC4026, NGC3938, NGC4449 and MESSIER106 group being merged to form the `Ursa Major cluster' (as discussed in Section~\ref{sec:lit_comp}) comprising 185 galaxies (52 in the complete sample). This large-scale structure has a velocity dispersion of 205~km~s$^{-1}$ and an early-type fraction of 23~per cent. Assuming a `cluster' distance of 10.4~Mpc (Hubble flow), the Ursa Major `cluster' spans 3.24~Mpc across (projected diameter). The BGG is MESSIER106, which is located 455~kpc and 353~km~s$^{-1}$ from the group centroid indicating an unrelaxed structure. The virial mass of the Ursa Major `cluster' as a whole is $7.96 \times 10^{13}$~M$_{\odot}$, whereas the virial mass estimates of the four substructures sum up to $2.59 \times 10^{13}$~M$_{\odot}$. Previous estimates of the virial mass of Ursa Major are $4.4 \times 10^{13}$~M$_{\odot}$ \citep{tully1987} and $4.3 \times 10^{13}$~M$_{\odot}$ \citep{karachentsev2014}. The latter is the sum of seven groups constituting to the Ursa Major complex. The virial mass estimates are typical for rich groups or poor clusters -- a galaxy cluster usually exceeds a mass of $10^{14}$~M$_{\odot}$.

\subsection{Group dynamics}
\label{sec:dynamics}

\begin{table}
\begin{scriptsize}
\caption[High-probability two-body bound systems ($P_{bound} \geq 0.8$).]{High-probability two-body bound systems ($P_{bound} \geq 0.8$).} 
\label{tab:Pbound}
\begin{tabular}{lcccc}
\hline\hline\\[-2ex] 
& 
\multicolumn{1}{c}{$M_{tot}$} & 
\multicolumn{1}{c}{$V_r$} & 
\multicolumn{1}{c}{$R_p$} & 
\multicolumn{1}{c}{$P_{bound}$} \\ 

\multicolumn{1}{c}{Group pair} & 
\multicolumn{1}{c}{($10^{13} M_{\odot}$)} & 
\multicolumn{1}{c}{(km~s$^{-1}$)} & 
\multicolumn{1}{c}{(Mpc)} & 
\multicolumn{1}{c}{per cent} \\

\multicolumn{1}{c}{(B1)} & 
\multicolumn{1}{c}{(B2)} & 
\multicolumn{1}{c}{(B3)} & 
\multicolumn{1}{c}{(B4)} &
\multicolumn{1}{c}{(B5)} \\[0.5ex] \hline
\\[-1.8ex]

NGC4278, NGC4449 & 0.6 & 9 & 1.7 & 95.4  \\
NGC4274, NGC5033 & 1.6 & 16 & 2.9 & 94.0  \\
NGC3938, NGC4274 & 1.8 & 16 & 3.2 & 93.9  \\
MESSIER106, NGC4449 & 0.7 & 35 & 0.5 & 91.6  \\
NGC3193, NGC3430 & 0.4 & 10 & 4.5 & 90.8  \\
NGC3003, NGC3245 & 0.6 & 18 & 3.0 & 89.2  \\
NGC3938, NGC5033 & 2.1 & 32 & 3.4 & 89.0  \\
NGC5473, NGC5631 & 1.0 & 45 & 1.4 & 85.1  \\
NGC3998, NGC3972 & 2.5 & 327 & 0.1 & 84.8  \\
NGC4026, NGC3938 & 1.9 & 72 & 1.3 & 83.8  \\
NGC3079, NGC3998 & 2.8 & 54 & 4.2 & 81.2  \\
\hline\hline\\[-2ex]
\end{tabular}
\end{scriptsize}
\end{table}

To further investigate the galaxy groups in the Ursa Major region and to determine if the individual structures are bound to one another and will eventually merge to form larger structures, the Newtonian binding criterion is used. \citet{brough2006} carried out a similar analysis for the Eridanus supergroup. A two-body system is bound if the kinetic energy is less or equal to the potential energy of the system, which is expressed in mathematical term as
\begin{equation} 
V_r^2 R_p \leq 2GM_{tot} \: sin^2{\alpha} \: cos{\alpha}
\end{equation}
with $M_{tot}$ being the total mass of the system, $V_r$ the relative velocity difference, $R_p$ the projected distance between the groups and $G$ the gravitational constant (e.g. \citealt{beers1982}; \citealt{cortese2004}; \citealt{brough2006}). $V_r$ and $R_p$ are corrected for projection effects by varying the projection angle $\alpha$ between $0-90^{\circ}$. Given the total mass of the system, the relative velocity, the projected distance and the gravitational constant $G$, the probability for each two-body system to be bound is given by integrating over the solid angle $\alpha$:
\begin{equation} 
P_{bound} = \int_{\alpha_1}^{\: \alpha_2} cos(\alpha) \: \mathrm{d}\alpha
\end{equation}
\citep{girardi2005}. In Table ~\ref{tab:Pbound} the probabilities for all two-body systems with $P_{bound} > 80$~per cent are given with the most likely bound two-body system listed first (for galaxy groups found using the FoF algorithm with \textit{strict} linking lengths $D_0=0.30$~Mpc and $V_0=150$~km~s$^{-1}$). The columns are as follows: column (B1) contains the IDs of the galaxy groups considered in the two-body system, column (B2) gives the total mass of the system, column (B3) lists the relative velocity difference $V_r$, column (B4) contains the projected distance between the groups $R_p$ and column (B5) gives the probability for the two-body systems to be bound.

Note that this analysis depends strongly on the relative velocity difference and mass of the system. For binding the potential energy is required to be smaller or equal to the kinetic energy and the former strongly depends on the relative velocity ($V_r^2$) and to a minor degree on the projected separation ($R_p$). Over large separations (8~Mpc or more) but with small relative velocities (tens of km~s$^{-1}$) galaxy groups can be bound to one another. However, back-of-the-envelope calculations show that the merging times for these systems are too large to be considered a bound system (8~Gyr to more than a Hubble time). Therefore, we do not consider two-body systems with projected distances $R_p \geq 8.8$~Mpc to be bound.

In Figures~\ref{fig:skymap300} to \ref{fig:skymap1000} the high-probability two-body bound systems are indicated by a connecting line (shown in projection). For clarity, the Ursa Major region is divided into velocity slices of 250 to 400~km~s$^{-1}$ depth to avoid likely bound systems being broken up in two figures. Note that group members can be broken up into multiple figures depending on their $v_{LG}$. The main structures in the Ursa Major region and their dynamics are as follows:

\begin{figure}
\includegraphics[trim=1.1cm 7cm 1cm 7.5cm, clip=true, width=8.4cm]{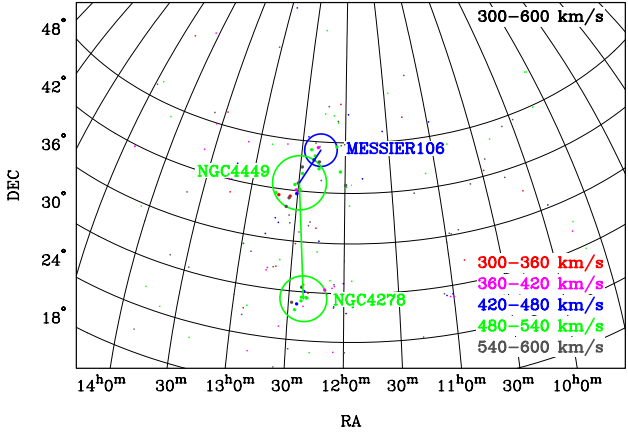}
\caption{High-probability two-body bound systems (with $P_{bound} \geq 80$~per cent) in the velocity range 300 to 600~km~s$^{-1}$, which are indicated by a connecting line. The colour represents the recession velocity as stated in the key. Galaxy groups with central group velocities within the displayed velocity range are marked by large, open circles with the BGG labelled in the figure, whereas the individual galaxies are indicated by filled circles -- large, filled circles for group members and small, filled circles for non-group members.} 
\label{fig:skymap300}
\end{figure}

\begin{figure}
\includegraphics[trim=1.1cm 7cm 1cm 7.5cm, clip=true, width=8.4cm]{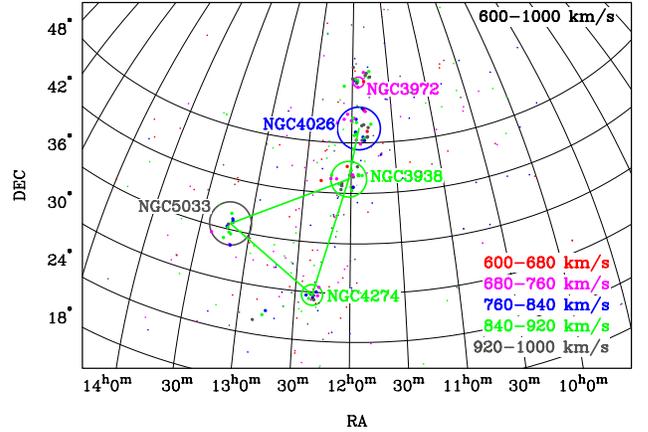} 
\caption{High-probability two-body bound systems in the velocity range 600 to 1000~km~s$^{-1}$. Similar to Figure~\ref{fig:skymap300}. Note that the NGC3998 (shown in Figure~\ref{fig:skymap1000}) and NGC3972 groups are likely to be bound at the 84.8~per~cent level. The connecting line to mark high-probability bound systems is not displayed.} 
\label{fig:skymap600}
\end{figure}

Figure~\ref{fig:skymap300} displays the velocity range from 300 to 600~km~s$^{-1}$. An angular separation of 9.3$^{\circ}$ is approximately a projected physical distance of 1~Mpc (determined at an intermediate velocity of 450~km~s$^{-1}$, assuming Hubble flow). The central galaxy group `NGC4449' has a high-probability to be bound to the NGC4278 group (at the 95.4~per cent level) and the MESSIER106 group (at the 91.6~per cent level). The structures `MESSIER106' and `NGC4278' have a probability of 63.9~per cent to be bound to one another. The probability that all three groups in the velocity range 300-600~km~s$^{-1}$ are bound and will eventually merge can be estimated from the minimum probability of the three pairs, which is 63.9~per cent. The NGC4278 group can be linked to a structure at slightly higher central velocity, the NGC4274 group (shown in Figure~\ref{fig:skymap600}). However, this two-body system has a low probability of being bound (42.4~per cent). 

In the velocity range 600 to 1000~km~s$^{-1}$, an angular separation of 5.2$^{\circ}$ approximates to a physical distance of 1~Mpc, which can be used as a rule of thumb when viewing Figure~\ref{fig:skymap600}. The eye-catching feature is the green triangle, which constitutes of the NGC4274, NGC5033 and NGC3938 groups. Given that lines are drawn between all group pairs, these three pairs are likely to be bound ($P_{bound} \geq 80$~per cent). In fact the minimum probability of the three pairs to be bound is 89~per cent (the `NGC3938' and `NGC5033' system, which are separated by 3.4~Mpc and 32~km~s$^{-1}$) suggesting a high-probability of the groups to be  a `supergroup', i.e. a group of groups that will merge and form a larger structure. Furthermore, the NGC3938 group is likely to be bound to the NGC4026 group (83.8~per cent level), which is a numerous conglomeration of galaxies. At lower probabilities, the NGC4026 group itself can be linked to structures in the North -- the much smaller NGC3972 group (72.5~per cent) and the NGC3998 group (38.0~per cent; shown in Figure~\ref{fig:skymap1000}). 

We note that the NGC4274 group is likely bound to the Virgo cluster (centred at $\alpha=187.70$, $\delta=12.34$ and $v_{\mathrm{LG}}=975$~km~s$^{-1}$ with a mass of $M_T=8.0 \times 10^{14}$~M$_{\odot}$; \citealt{karachentsev2014}) at the 96~per~cent level (also see Section~\ref{subsec:UMa_Virgo}). 

\begin{figure}
\includegraphics[trim=1.1cm 7cm 1cm 7.5cm, clip=true, width=8.4cm]{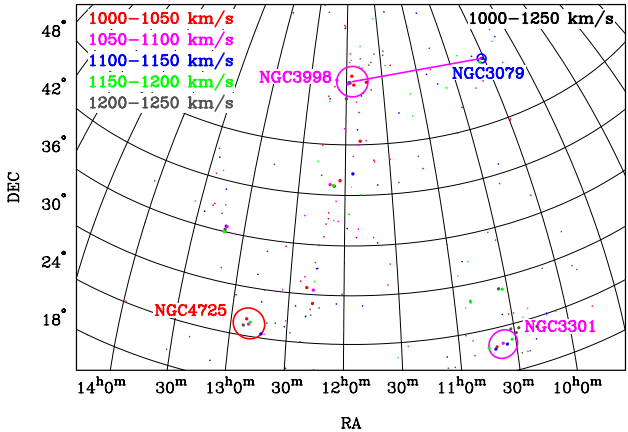} 
\caption{High-probability two-body bound systems in the velocity range 1000 to 1250~km~s$^{-1}$. Similar to Figure~\ref{fig:skymap300}. Note that the NGC3972 (shown in Figure~\ref{fig:skymap600}) and NGC3998 groups are likely to be bound at the 84.8~per~cent level. The connecting line to mark high-probability bound systems is not displayed.} 
\label{fig:skymap1000}
\end{figure}

In the velocity range 1000 to 1250~km~s$^{-1}$ shown in Figure~\ref{fig:skymap1000} (3.7$^{\circ}$ approximates to 1~Mpc) groups seem to be centrally bound to the NGC3998 group (at 1065~km~s$^{-1}$ colour-coded in magenta). This group exceeds the 80~per cent level with the NGC3079 group (81.2~per cent) and the NGC3972 group (which is overlapping and the projected separation is only 0.1~Mpc; at the 84.8~per cent level; see Figure~\ref{fig:skymap600}). The structures to the South, `NGC3938' and `NGC4026' (see Figure~\ref{fig:skymap600}) are unlikely to be bound to the NGC3998 group as the probabilities are only $P_{bound} = 40.8$~per cent and $P_{bound} = 38.0$~per cent, respectively.
 
At higher velocities, the displayed area is sparsely populated with galaxies and groups, in particular in the central region ($11^{h}\leq\alpha\leq+13^{h}$). This can be interpreted as an almost empty region behind the filamentary structure seen in Figures~\ref{fig:skymap300} to \ref{fig:skymap1000} stretching from South to North (also see Figure~\ref{fig:mapvel30}).

The regions beyond 1800~km~s$^{-1}$ are not further discussed in this paper. We conclude that the `Ursa Major cluster' as defined in the literature \citep{tully1996} is a conglomerate of galaxy groups identified using the \textit{strict} linking lengths ($D_0=0.30$~Mpc and $V_0=150$~km~s$^{-1}$). Several of these galaxy groups are likely bound to one another and in the process of merging. Beyond the Ursa Major structures ($v_{LG} \geq$ 1500~km~s$^{-1}$) is a sparsely populated region lacking galaxies and groups defining Ursa Major from the background.

\subsection{Brightest group galaxies (BGG)}
\label{sec:BGG}

Previous studies of galaxy clusters and evolved groups (with extended X-ray emission) agree with the theoretical prediction that the brightest, most massive galaxy resides at the centre of its dark matter halo. 90~per cent of BCGs lie within $0.38 \times r_{200}$ of a cluster's spatial centre \citep{lin2004} and within the mean velocity dispersion of a cluster (\citealt{zabludoff1990}; \citealt{oegerle2001}). \citet{brough2006GEMS} study the less dense group environment finding early-type BGGs within $0.3 \times r_{500}$ and $\pm 0.6 \sigma_v$ of the group centre in dynamically mature groups.

\begin{figure}
\centering
\includegraphics[trim=0.1cm 0.1cm 9cm 0.1cm, clip=true, width=8.0cm]{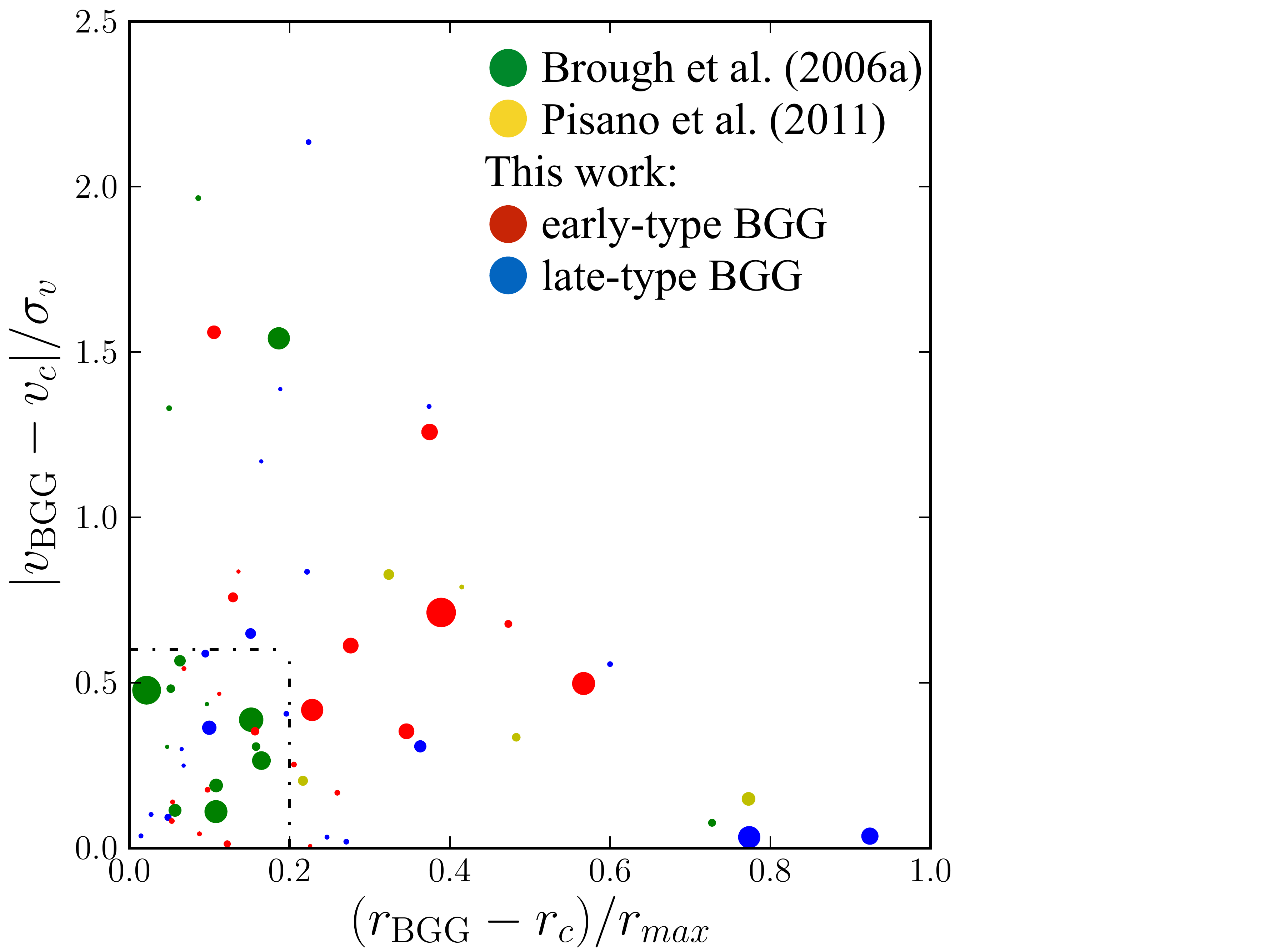}
\caption{Offsets of the BGG from the spatial and kinematic group centres. Early-type BGG are shown in red, whereas late-type BGG are shown in blue. For comparison, the BGG in \citet{brough2006GEMS} and \citet{pisano2011} are shown in green and yellow, respectively. The marker sizes scale with the number of group members. The dashed lines indicate the approximate limitations determined by \citet{brough2006GEMS} to distinguish between dynamically evolved (ranging up to $0.2 \times \frac{r_{\mathrm{BGG}}-r_c}{r_{max}}$ and up to $0.6 \times \frac{|v_{\mathrm{BGG}}-v_c|}{\sigma_{v}}$) and dynamically immature groups.} 
\label{fig:BGG_prop}
\end{figure}

The spatial and kinematic offsets from the group centres of the BGGs in the Ursa Major region (see Figure~\ref{fig:overview}) are shown in Figure~\ref{fig:BGG_prop}. Note that the determination of the group centre is luminosity-weighted and therefore biased towards the brightest group galaxy. Early-type BGGs are shown in red, whereas late-type BGG are shown in blue. For comparison, the offset of BGGs in \citet{brough2006GEMS} and \citet{pisano2011} are shown in green and yellow, respectively.  The marker sizes scale with the number of group members. The dashed lines indicate the approximate limitations determined by \citet{brough2006GEMS} to distinguish between dynamically evolved (ranging up to $0.2 \times \frac{r_{\mathrm{BGG}}-r_c}{r_{max}}$ and up to $0.6 \times \frac{|v_{\mathrm{BGG}}-v_c|}{\sigma_{v}}$) and dynamically immature groups in the process of forming. There are only four dynamically unevolved groups amongst the 16 nearby GEMS groups \citep{brough2006GEMS}, whereas all the BGGs in the \citet{pisano2011} study of loose groups are found beyond the dashed lines.

The majority of structures in the Ursa Major region are dynamically immature groups most similar to Pisano's `loose groups' -- 61 per cent of all groups ($N_m \geq 4$), which increases to 86~per cent for rich groups with $N_m \geq 10$. The two groups with the largest offsets of the BGG from the spatial group centre are the NGC3683 and NGC3938 groups (see Section~\ref{sec:IndGrps}). Both groups have a late-type BGG lying close to the kinematic centre of the group ($|v_{BGG} - v_c| < 0.04 \sigma_v$), but residing in the outskirts of the group ($> 0.7 r_{max}$). 

The galaxy groups with the largest offsets of the BGG from the kinematic group centre are the UGC09071, the NGC5322, the NGC3972, the NGC5336, the NGC3613 and SBS1144+591 groups ($> 1.0 \sigma_v$). Two dynamically immature groups with numerous group members are the NGC5322 (17 members) and NGC3613 (21 members; see Appendix) groups. 

It is remarkable that (\textit{i}) the main structures in the northern part of the Ursa Major region are all dynamically unevolved -- the NGC3613 and NGC3683 groups are discussed in the Appendix as well as the NGC3998 group whose BGG is located just beyond the dashed line at $0.23 r_{max}$ and $0.4 \sigma_v$; (\textit{ii}) the main structures in the central Ursa Major region tend to be dynamically unevolved -- the NGC3938 group as discussed above as well as the NGC4026 (with the BGG at $0.57 r_{max}$ and $0.5 \sigma_v$ ) and the NGC4449 groups (with the BGG at $0.36 r_{max}$ and $0.3 \sigma_v$); there is one exception of a dynamically mature group, which is NGC5033 (with the BGG at $0.10 r_{max}$ and $0.4 \sigma_v$); whereas (\textit{iii}) the main structures in the southern part of the Ursa Major region tend to be dynamically evolved as discussed in the Appendix -- the NGC4278 group (with the BGG at $0.16 r_{max}$ and $0.4 \sigma_v$); the NGC4725 group (with the BGG at $0.05 r_{max}$ and $0.10 \sigma_v$) and the NGC4274 group, whose BGG is just beyond the dashed lines at $0.15 r_{max}$ and $0.6 \sigma_v$. The southern part of the Ursa Major region is near the Virgo cluster. The galaxy groups form a filamentary structure connecting the Virgo cluster and the Ursa Major region, which suggests a change in group properties with distance from the Virgo cluster. Note that the galaxy groups in the southern part of the Ursa Major region have on average less members than the galaxy groups in the central and northern part. The small spatial and kinematic offsets from the group centres of the BGG may result from calculating the luminosity-weighted centroid, which is severely biased towards the BGG for low number statistics.

\section{Galaxy properties}
\label{sec:GalProp}

We now study the properties of galaxies residing in the Ursa Major region as a function of group membership and local projected surface density. Morphologies, colours and luminosities (obtained from SDSS and RC3) combined with neutral hydrogen observations (\citealt{wolfinger2013}; \citealt{kovac2009}; \citealt{verheijen2001}; \citealt{haynes2011} and \citealt{springob2005}) can provide evidence of galaxy transformation occurring in the denser environments when a change in galaxy properties is observed with local projected surface density.

The local projected surface density of each galaxy in the complete sample is calculated using the nearest-neighbour measure, which is based on the principle that galaxies in denser environments have closer neighbours. The local projected surface density to the fifth nearest neighbour within 800~km~s$^{-1}$ is calculated using
\begin{equation} 
 \Sigma_5 = \frac{5}{(4/3)\pi r_n^3}
\end{equation}
The velocity cut of 800~km~s$^{-1}$ minimizes chance alignments, i.e. galaxies may appear in close vicinity (in projection) when they are in fact separated by a large distance in velocity space. The velocity cut is of the order of the velocity dispersion of the Virgo cluster, which is stricter than the cut in \citet{brough2006} allowing up to 1000~km~s$^{-1}$ between the nearest neighbours. To avoid edge effects (edges in the plane of the sky as well as in the velocity range), the analysis is restricted to $10^{h}\leq\alpha\leq+14^{h}$, $22^{\circ}\leq\delta\leq+63^{\circ}$ and $v_{\mathrm{LG}} \leq 2000$~km~s$^{-1}$. The subsample of galaxies with local projected surface density measurements contains 472 galaxies.

\subsection{Morphology}
\label{subsubsec:morph}

\begin{figure}
\mbox{\includegraphics[trim=0.25cm 0cm 2.cm 0cm, clip=true, width=8.2cm]{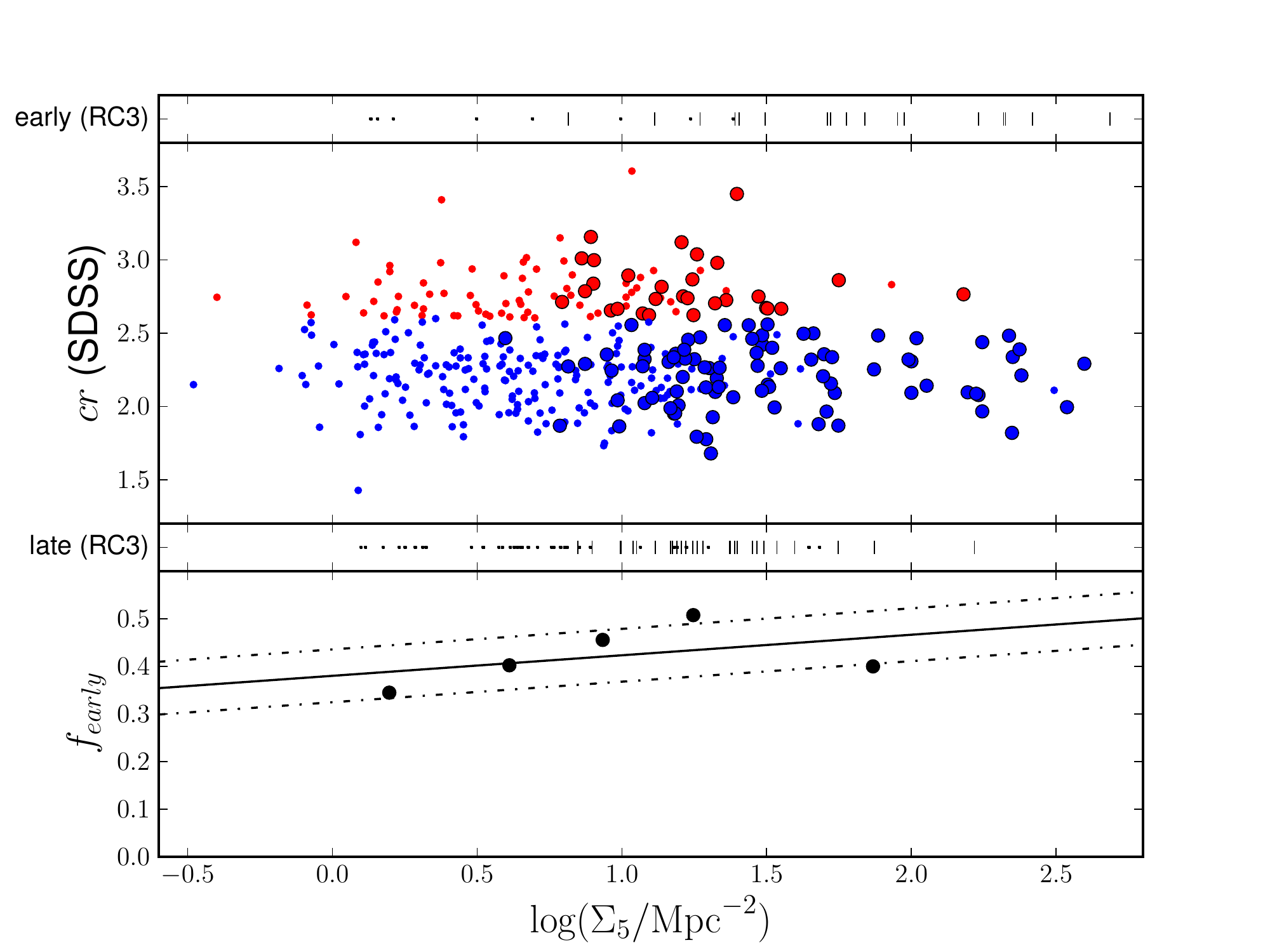}}
\caption[Fraction of early-type galaxies as a function of local projected surface density.]{(\textit{top}) Morphological type as a function of local projected surface density. Morphological types are obtained from SDSS (bulge-dominated galaxies with $cr \geq 2.6$ are shown in red, whereas disk-dominated galaxies are shown in blue) or RC3 (early-type galaxies with $-5 \leq \mathrm{T-type} \leq 0$ are shown in grey above the SDSS data, whereas late-type galaxies with $ 0< \mathrm{T-type} \leq 10$ are shown in grey below the SDSS data). Galaxies residing in groups are shown with large markers, whereas galaxies without group membership are indicated by small symbols. (\textit{bottom}) Fraction of early-type galaxies as a function of local projected surface density. The galaxies are binned per 100 data points.} 
\label{fig:morph_dens}
\end{figure}

The morphology-density relation for the Ursa Major region is shown in Figure~\ref{fig:morph_dens}. The morphology of SDSS galaxies are determined using the concentration index ($cr$). A separator of 2.6 is adopted to define bulge dominated ($cr \geq 2.6$; shown in red) and disk dominated galaxies ($cr < 2.6$; shown in blue) as previously used within the SDSS collaboration (\citealt{strateva2001}; \citealt{shimasaku2001}). For galaxies with morphological T-types from RC3, $-5 \leq \mathrm{T-type} \leq 0$ represent early-type galaxies (shown in grey above the SDSS data), whereas $ 0< \mathrm{T-type} \leq 10$ represents late-type galaxies (shown in grey below the SDSS data). Galaxies residing in groups are indicated by larger symbols than non-group galaxies allowing a consistency check between the local projected surface density and group membership (as determined in Section~\ref{sec:est_prop}). As one expects, group members have higher $\mathrm{log (} \Sigma_5 \mathrm{)}$ than non-group galaxies. There are a few non-group interlopers at high $\mathrm{log (} \Sigma_5 \mathrm{)}$, which may be chance alignments with galaxy groups within 800~km~s$^{-1}$.

The fraction of early-type galaxies (see \textit{bottom} panel in Figure~\ref{fig:morph_dens}) is determined per 100 galaxies (binned data). A linear regression fit to the binned data shows an increase in the fraction of early-type galaxies with local projected surface density. The $1~\sigma$ standard deviation from the fit is indicated by dashed lines ($\pm 6$~per cent). 

Compared to previous studies, e.g. galaxy groups in the Eridanus region as discussed in \citet{brough2006}, the increase in the fraction of early-type galaxies with local density is rather shallow. Brough et al. (2006) measure a similar fraction of early-type galaxies in the less dense environment ($ f_{early} \sim 35$~per cent at $\mathrm{log (} \Sigma_5 \mathrm{)} = 0$), but a significantly higher fraction in the denser regions ($f_{early} > 90$~per cent at $\mathrm{log (} \Sigma_5 \mathrm{)} = 2$). \citet{dressler1980} also shows that the fraction of ellipticals and S0 galaxies are $\sim 20$~per cent in the field and increase to $\sim 90$~per cent in the dense core of rich clusters. Therefore, the galaxy groups in the Ursa Major region are comparatively unevolved by this measure, consisting of a high fraction of late-type galaxies.

\subsection{Colour}
\label{subsubsec:col}

\begin{figure*}
\centering
\begin{tabular}{p{7cm} p{0.3cm} p{7cm}}
\mbox{\includegraphics[trim=1.1cm 0cm 2cm 0cm, clip=true, width=7cm]{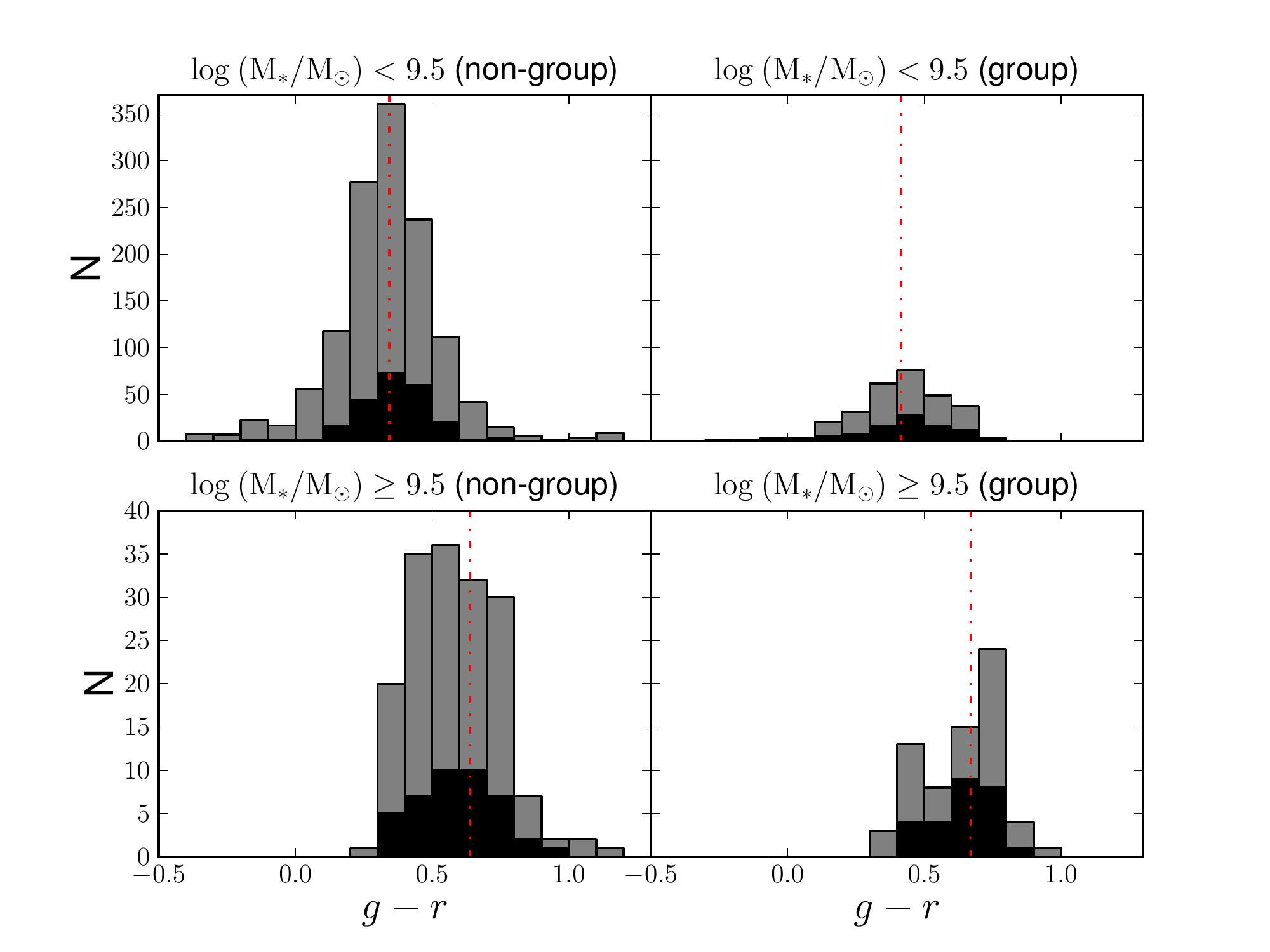}} & & 
\mbox{\includegraphics[trim=1cm 0cm 1cm 0cm, clip=true, width=7cm]{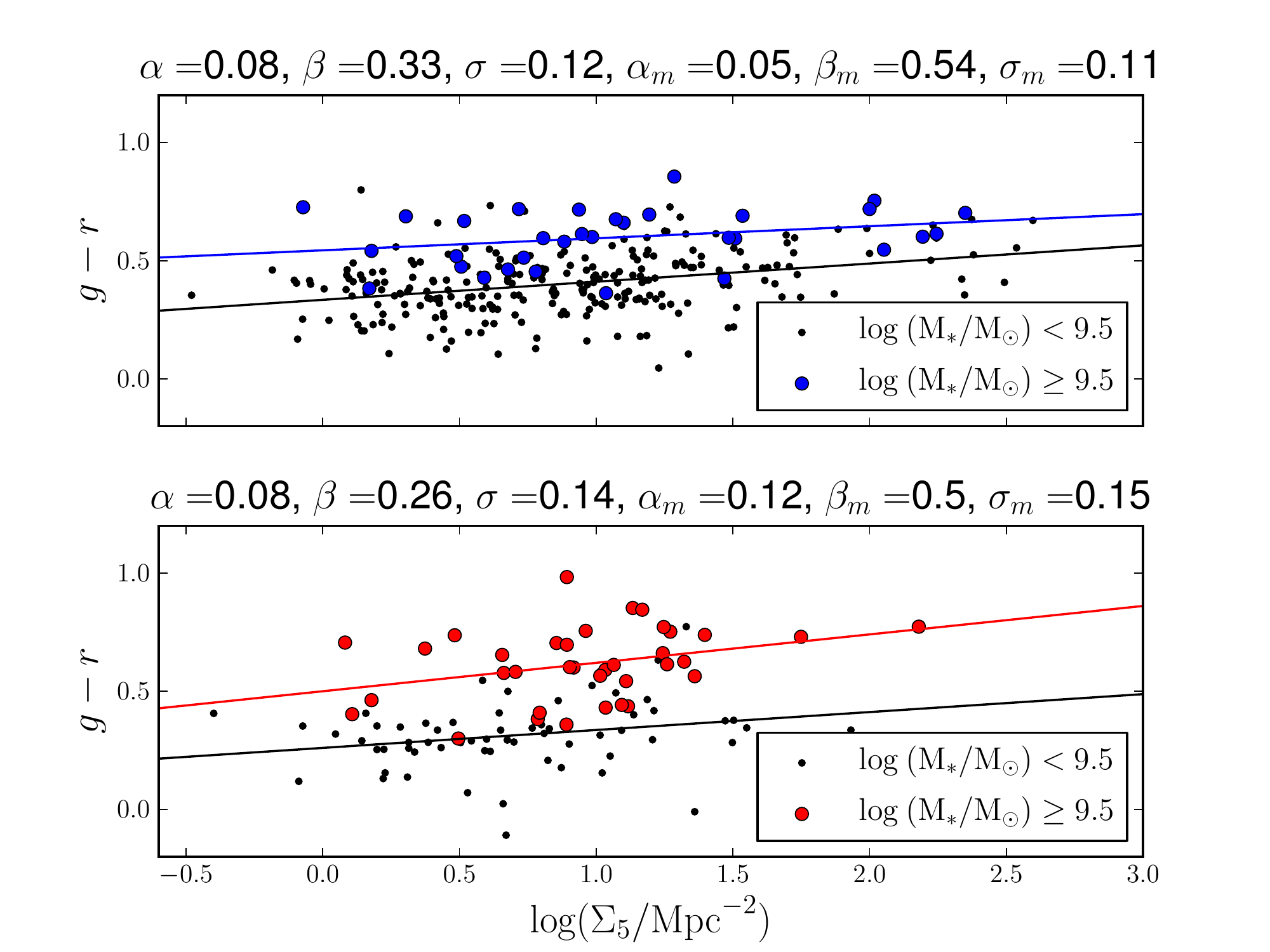}} \\ 
[12pt]
\mbox{\includegraphics[trim=1.1cm 0cm 2cm 0cm, clip=true, width=7cm]{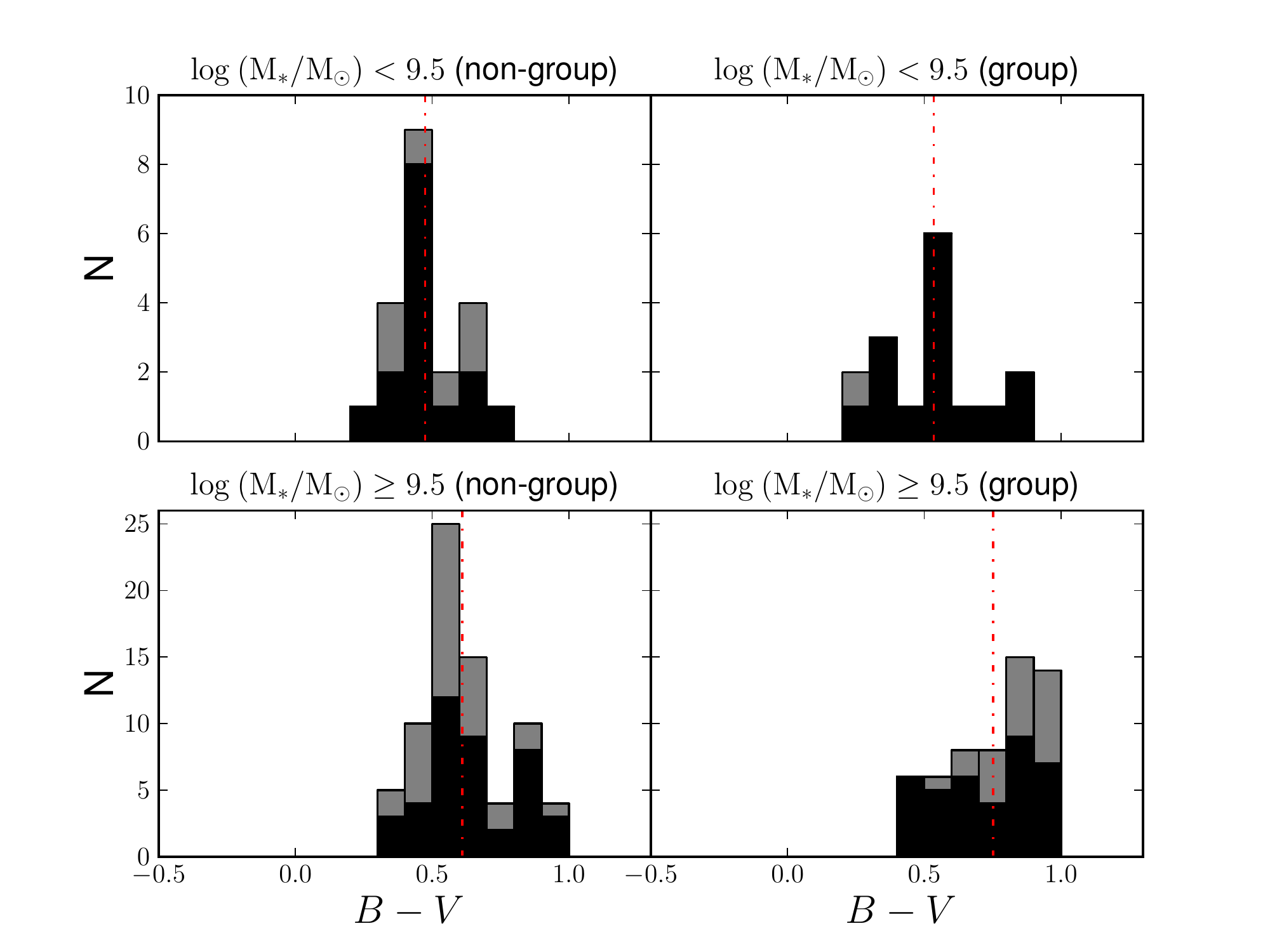}} & &
\mbox{\includegraphics[trim=1cm 0cm 1cm 0cm, clip=true, width=7cm]{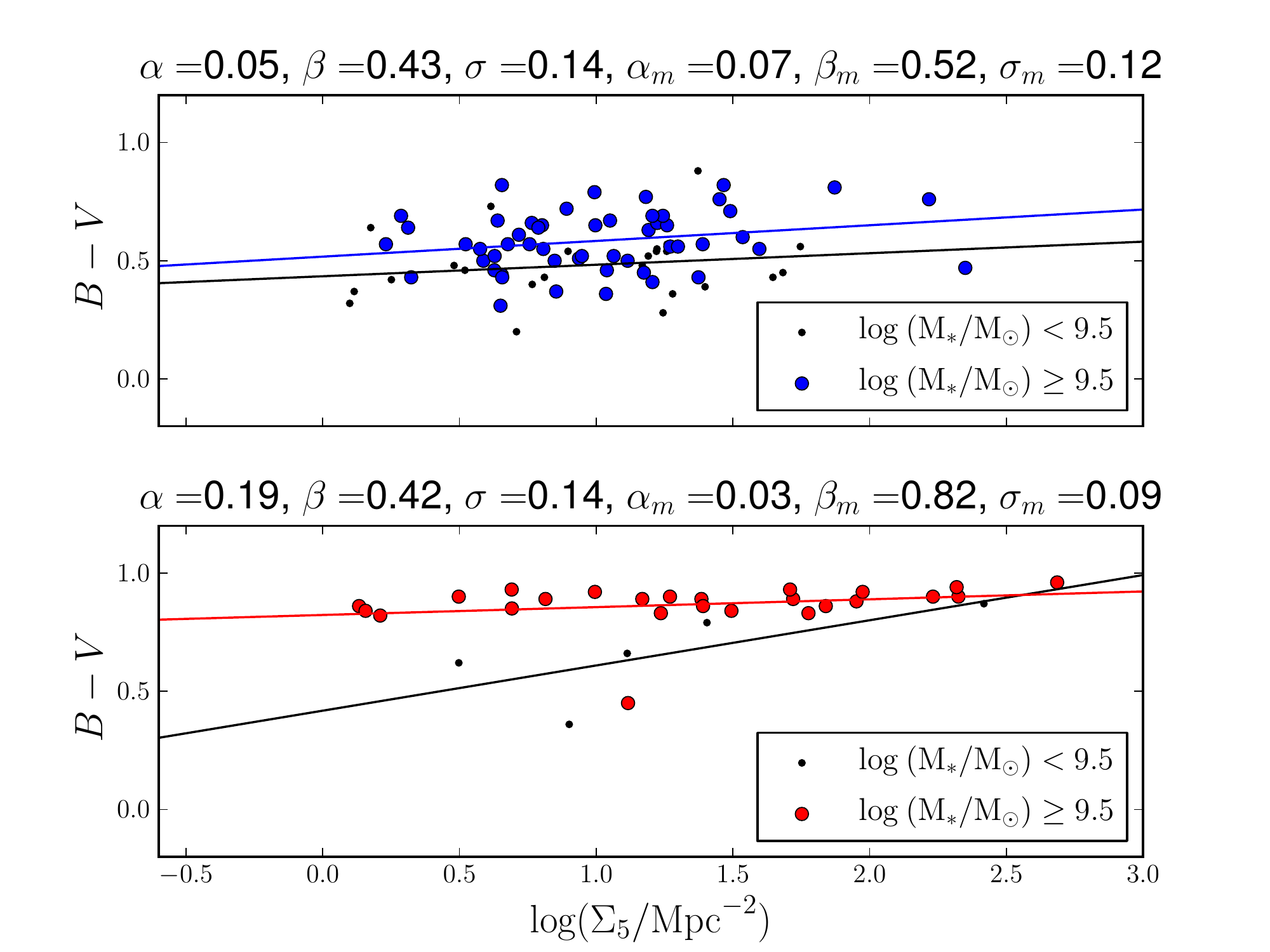}} \\ 
\end{tabular}
\caption{Colour distribution for group/non-group galaxies and galaxy colour as a function of local projected surface density. The $g-r$ colour is shown in the \textit{top} panel (as obtained from SDSS) and the $B-V$ colour in the \textit{bottom} panel (RC3 data). (\textit{left}) Colour distribution for group/non-group galaxies divided by their stellar mass. Galaxies with local projected surface density available are shown in black, whereas faint galaxies and galaxies residing near the edge of the studied area are shown in grey. The average colour is marked with a dashed line (red). (\textit{right}) Galaxy colour as a function of local projected surface density divided into late-type (\textit{top}, blue) and early-type (\textit{bottom}, red) galaxies. The small (black) circles mark galaxies with  $\mathrm{M_{\ast}} < 10^{9.5}~\mathrm{M}_{\odot}$ (linear regression fit with $\alpha$, $\beta$ and $\sigma$), whereas the larger circles represent more massive galaxies with $\mathrm{M_{\ast}} \geq 10^{9.5}~\mathrm{M}_{\odot}$ (linear regression fit with $\alpha_m$, $\beta_m$ and $\sigma_m$).} 
\label{fig:col_dens}
\end{figure*}

To investigate the relationship between galaxy colour and environment, the galaxies are divided by their stellar mass (at a mass of $\mathrm{M_{\ast}} = 10^{9.5}~\mathrm{M}_{\odot}$) and by their morphological type (into late-type and early-type galaxies). $B-V$ (RC3) and $g-r$ (SDSS) colours are discussed separately in the following analysis. Note that the SDSS fragments bright galaxies of large angular extent on the sky. Therefore, RC3 data is utilized for those affected galaxies. In Figure~\ref{fig:col_dens} (\textit{left}), the colour distribution is shown separated by stellar mass and group membership. One can see that massive galaxies (in the galaxy groups as well as the less dense environments) are on average redder than less massive ones. The average (dashed line) colour is redder for group members than non-group galaxies at the same stellar mass. When plotting the galaxy colour versus the local projected surface density (Figure~\ref{fig:col_dens} \textit{right}), the galaxies are redder with increasing density for both morphological types (\textit{top} late-type galaxies in blue for massive and black for less massive ones; \textit{bottom} early-type galaxies in red for massive and black for less massive ones). The linear regression fit to the data is of the form
\begin{equation} \label{eq:linreg}
\mathrm{log_{10} (\mathrm{M}_{HI} / \mathrm{M}_{\ast})} = \alpha \: \mathrm{log_{10} (M_{\ast})} + \beta 
\end{equation}
with the values of $\alpha$ and $\beta$ stated above each figure as well as the standard deviation $\sigma$ (index \textit{m} for the more massive galaxies). Note that RC3 colours tend to be redder than SDSS colours due to the selection bias of RC3 towards brighter galaxies, which are most likely massive and red galaxies. 

Previous studies have also found that the colours of galaxies belonging to the most massive galaxy groups are redder than those in less massive groups and lower density environments (e.g. \citealt{alpaslan2015}). The colour distribution as a function of environmental density has been shown to be bimodal, wherein the observed colour distribution is commonly referred to as the `blue cloud' and `red sequence' (e.g. \citealt{baldry2004}; \citealt{balogh2004}; \citealt{blanton2005}).

\subsection{Luminosity}
\label{subsubsec:lum}

\begin{figure*}
\centering
\begin{tabular}{p{7cm} p{0.3cm} p{7cm}}
\mbox{\includegraphics[trim=1.1cm 0cm 2cm 0cm, clip=true, width=7cm]{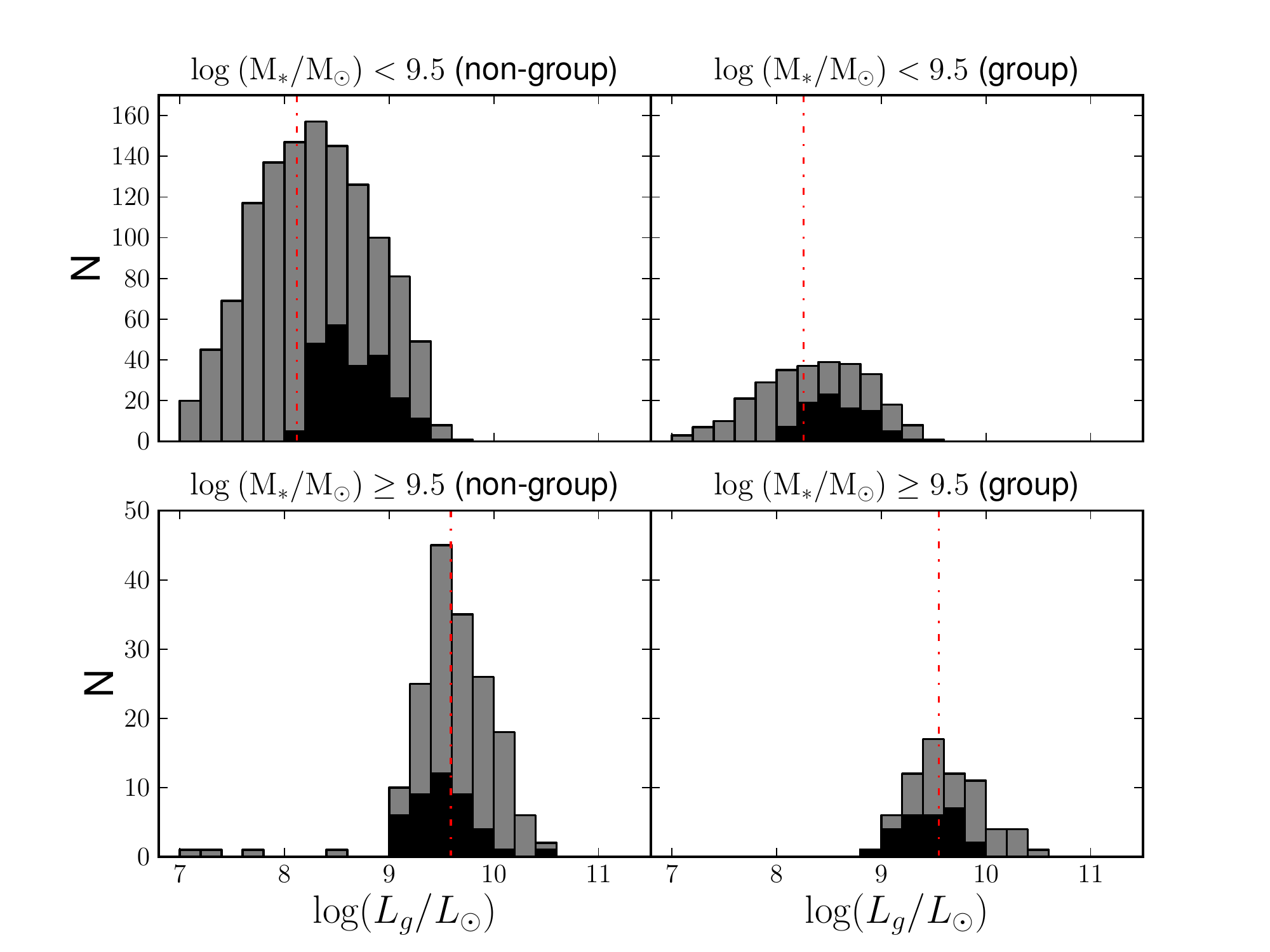}} & & 
\mbox{\includegraphics[trim=1cm 0cm 1cm 0cm, clip=true, width=7cm]{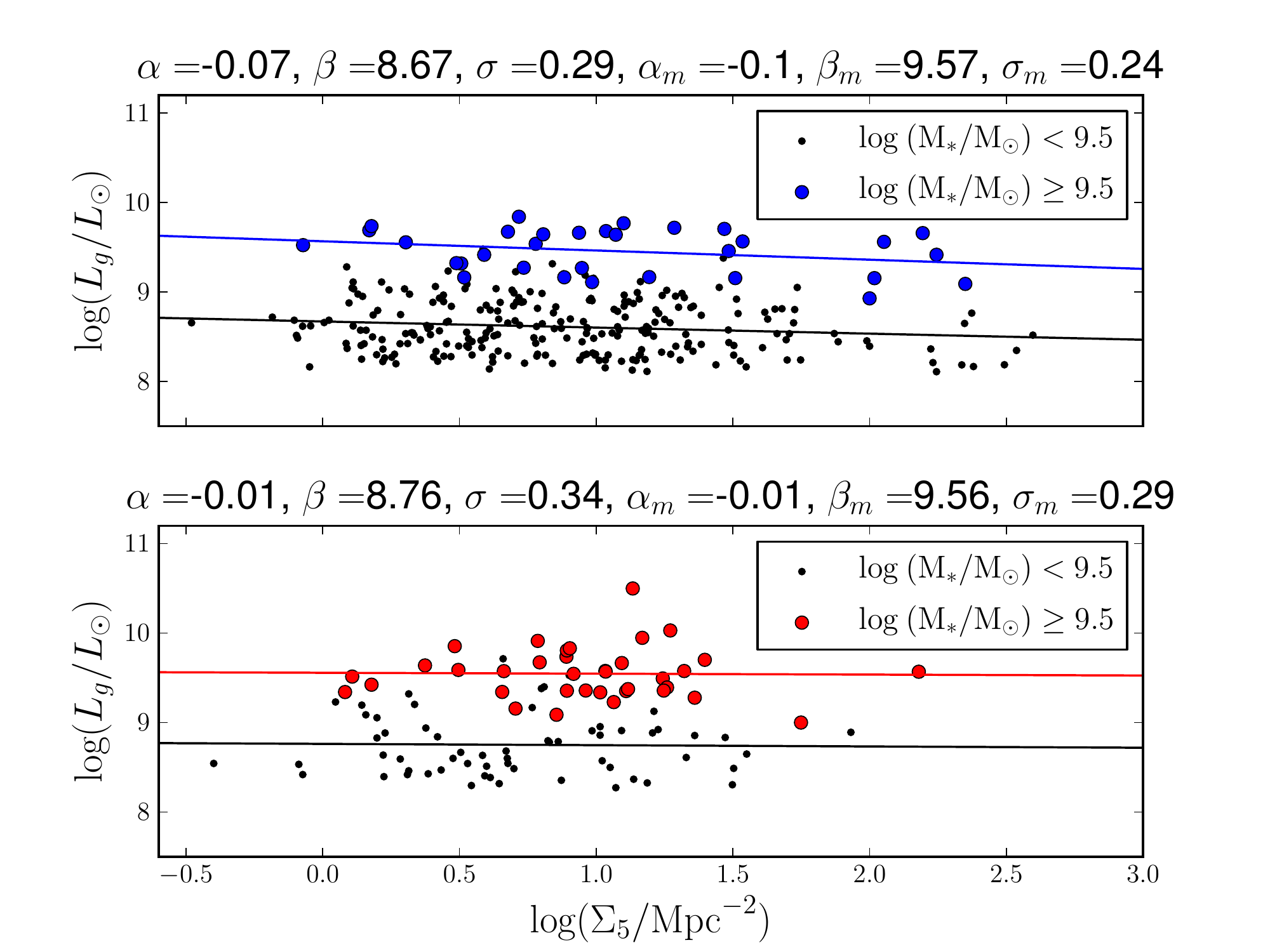}} \\ 
[12pt]
\mbox{\includegraphics[trim=1.1cm 0cm 2cm 0cm, clip=true, width=7cm]{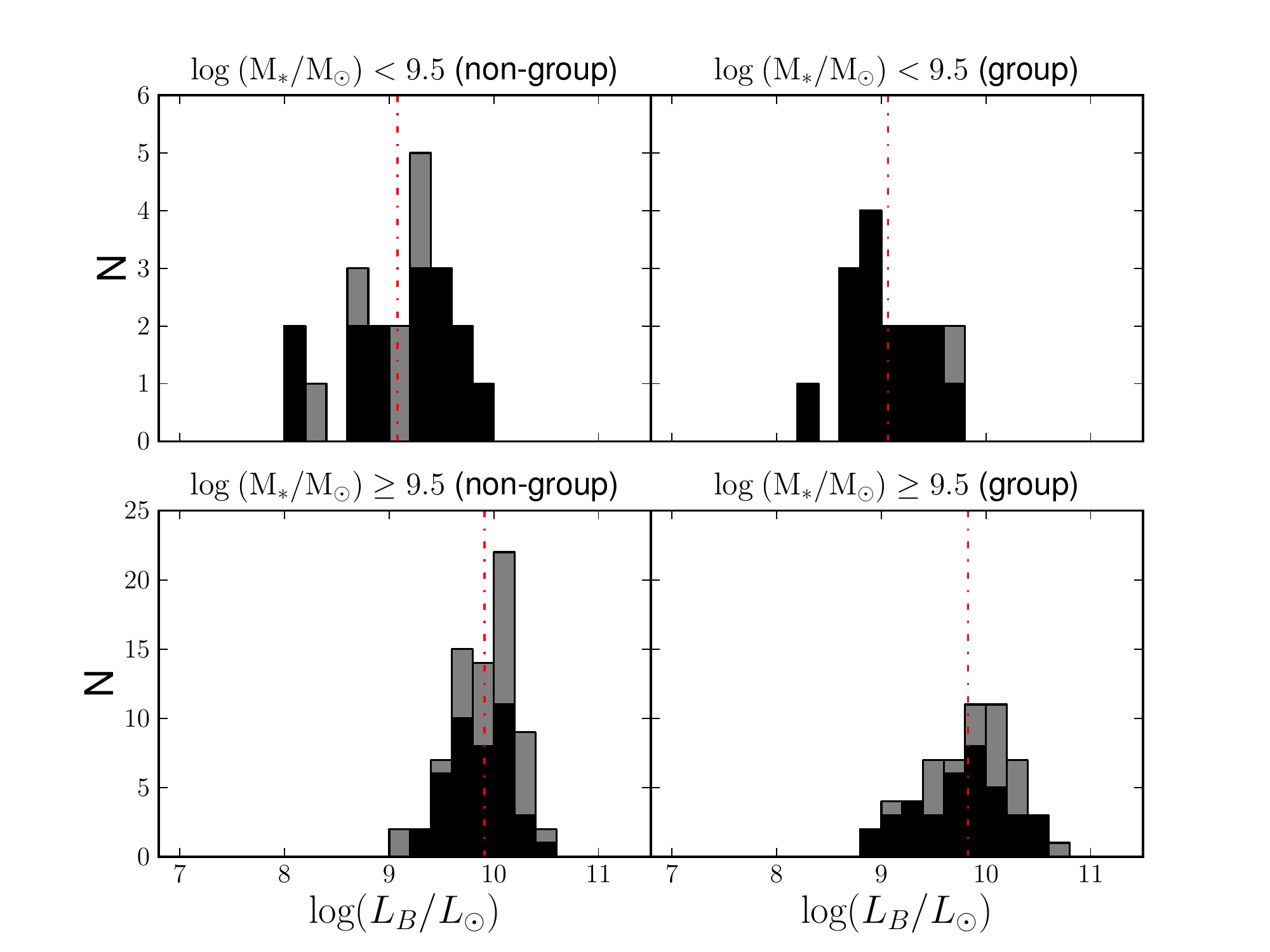}} & &
\mbox{\includegraphics[trim=1cm 0cm 1cm 0cm, clip=true, width=7cm]{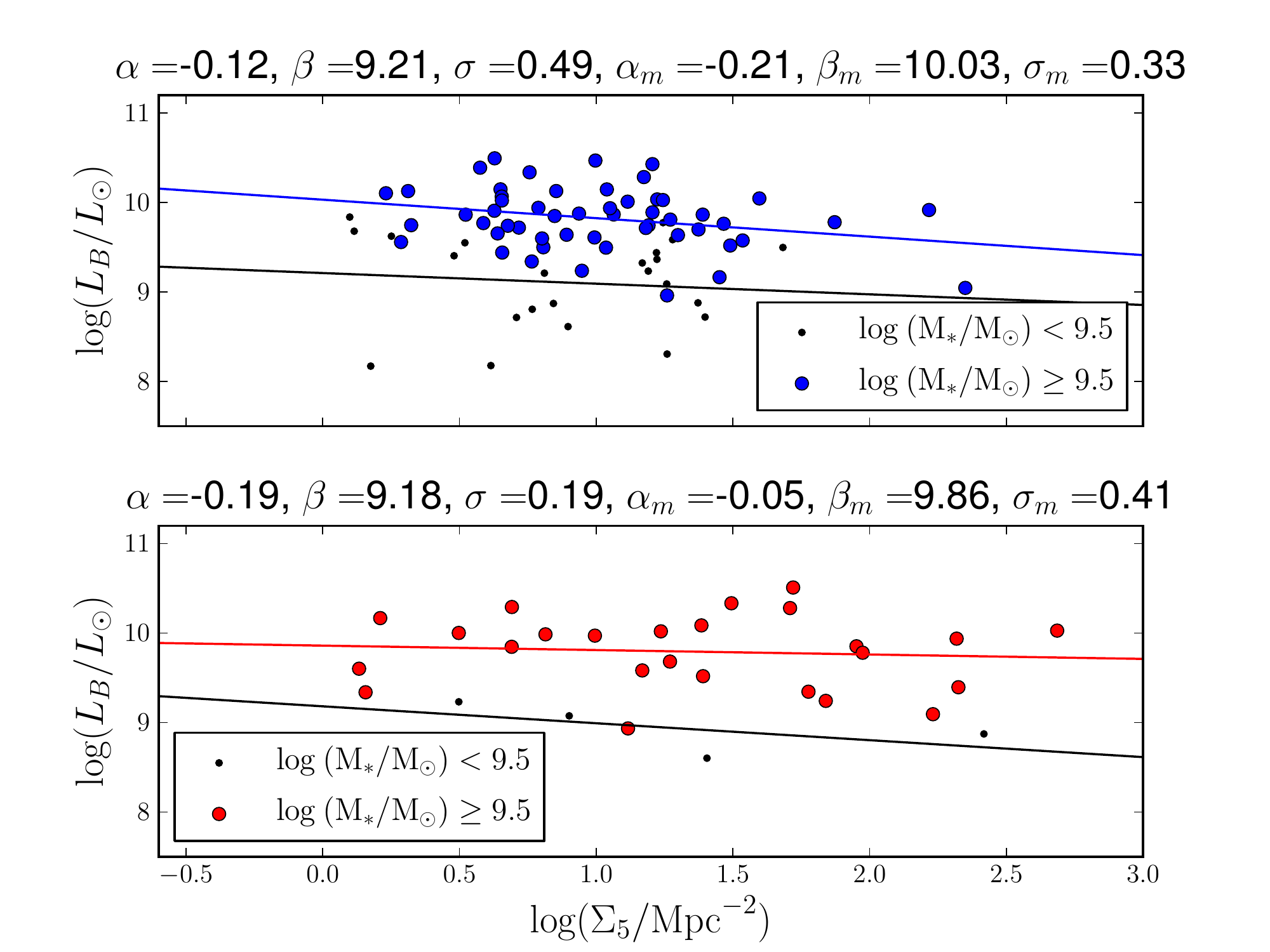}} \\ 
\end{tabular}
\caption{Luminosity distribution for group/non-group galaxies and galaxy luminosity as a function of local projected surface density. Similar to Figure~\ref{fig:col_dens}. The $L_g$ luminosity is shown in the \textit{top} panel (as obtained from SDSS) and the $L_B$ luminosity in the \textit{bottom} panel (RC3 data).} 
\label{fig:lum_dens}
\end{figure*}

The relationship between luminosity and environment is examined in a similar manner to the galaxy colour analysis discussed above. The luminosity distribution for group/non-group galaxies, and the galaxy luminosity as a function of local projected surface density are shown in Figure~\ref{fig:lum_dens} (similar to Figure~\ref{fig:col_dens}). Note that the RC3 catalogue is biased towards bright galaxies as the selection criteria is galaxies with apparent diameters larger than 1~arcmin at the D25 isophotal level and a total $B$-band magnitude limit of 15.5. As a consequence there is a natural offset between the luminosities obtained from SDSS and RC3, which is apparent in Figure~\ref{fig:lum_dens}. SDSS luminosities are an order of magnitude smaller than the luminosities obtained from RC3.

In Figure~\ref{fig:lum_dens} (\textit{left}), one can clearly see that the massive galaxies (in the galaxy groups as well as the less dense environments) are on average brighter than the less massive ones. We find no obvious tendency of the brighter galaxies to reside in galaxy groups. \citet{alpaslan2015} find a weak tendency for brighter galaxies ($\mathrm{log_{10} (\mathrm{M}_{\ast} / \mathrm{M}_{\odot})} > 9.5$) to be associated with the highest mass galaxy groups, but, in agreement with our findings, no trend with the group environment for fainter galaxies ($\mathrm{log_{10} (\mathrm{M}_{\ast} / \mathrm{M}_{\odot})} < 9.5$). The average (dashed line) stays almost unchanged for non-group and group members at a certain stellar mass for SDSS and RC3 luminosities. 

When plotting the galaxy luminosity against the local projected surface density (Figure~\ref{fig:lum_dens} \textit{right}), the galaxy luminosity appears to be unchanged for both morphological types with increasing projected surface density suggesting that the galaxy luminosity does not change with environment. For galaxies where RC3 data is utilized (\textit{bottom} panels, Figure~\ref{fig:lum_dens}), the luminosity-density fits are severely biased towards the bright galaxies due to the selection bias. Therefore, one can not see a trend in galaxy luminosity with density as faint galaxies, which are predominantly less massive galaxies, are missing from the RC3 sample (e.g. only a handful data points for less massive early-type galaxies in the \textit{bottom} panel).

We conclude, that there is no obvious transformation in galaxy luminosities for both morphological types with environment.

\subsection{HI deficiency and excess}
\label{subsec:HI}

\begin{figure*}
\begin{center}
\begin{tabular}{lr}
\mbox{\includegraphics[trim=0cm 0cm 1cm 1cm, clip=true, width=0.47\textwidth]{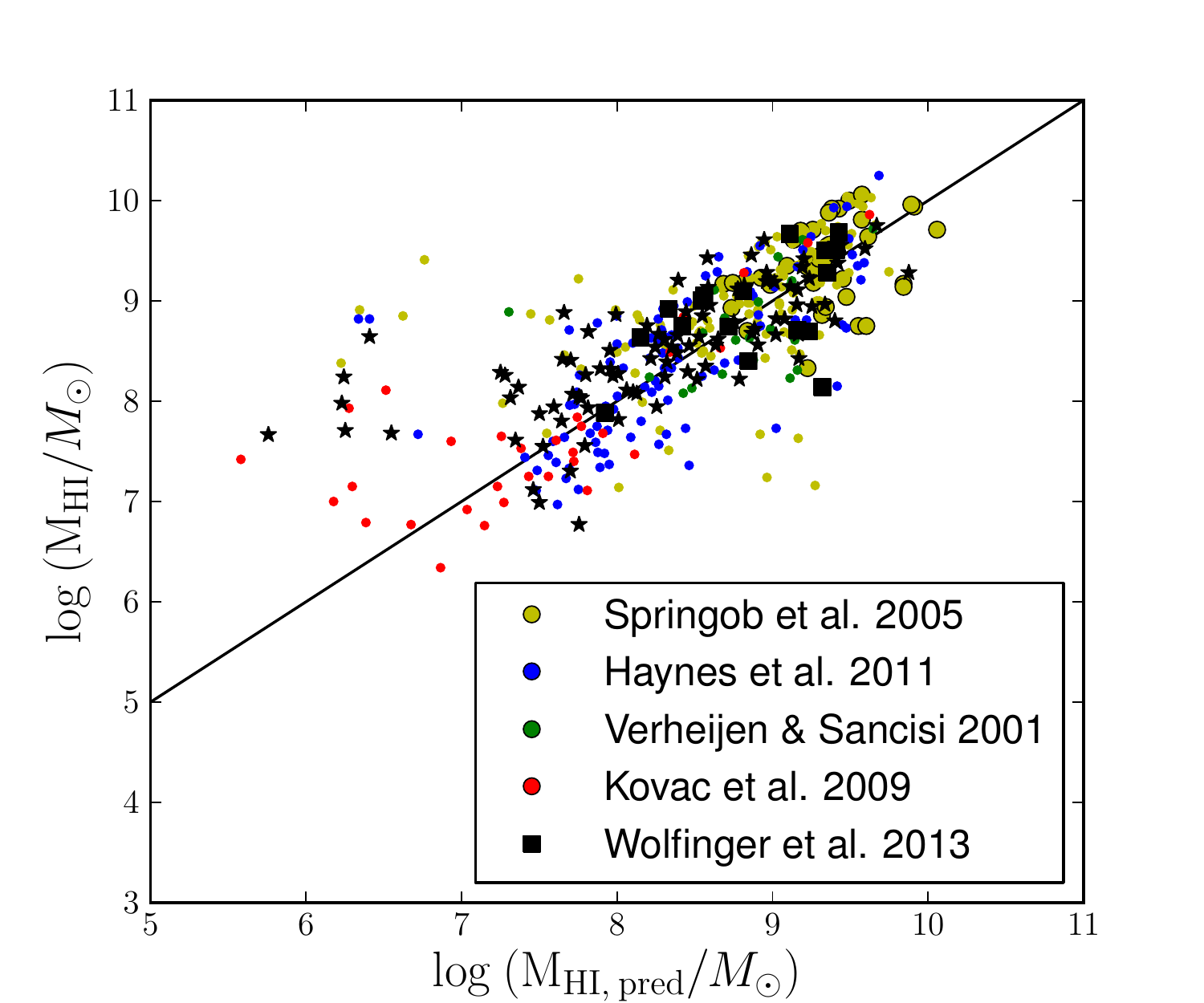}} & 
\mbox{\includegraphics[trim=0cm 0cm 1cm 1cm, clip=true, width=0.47\textwidth]{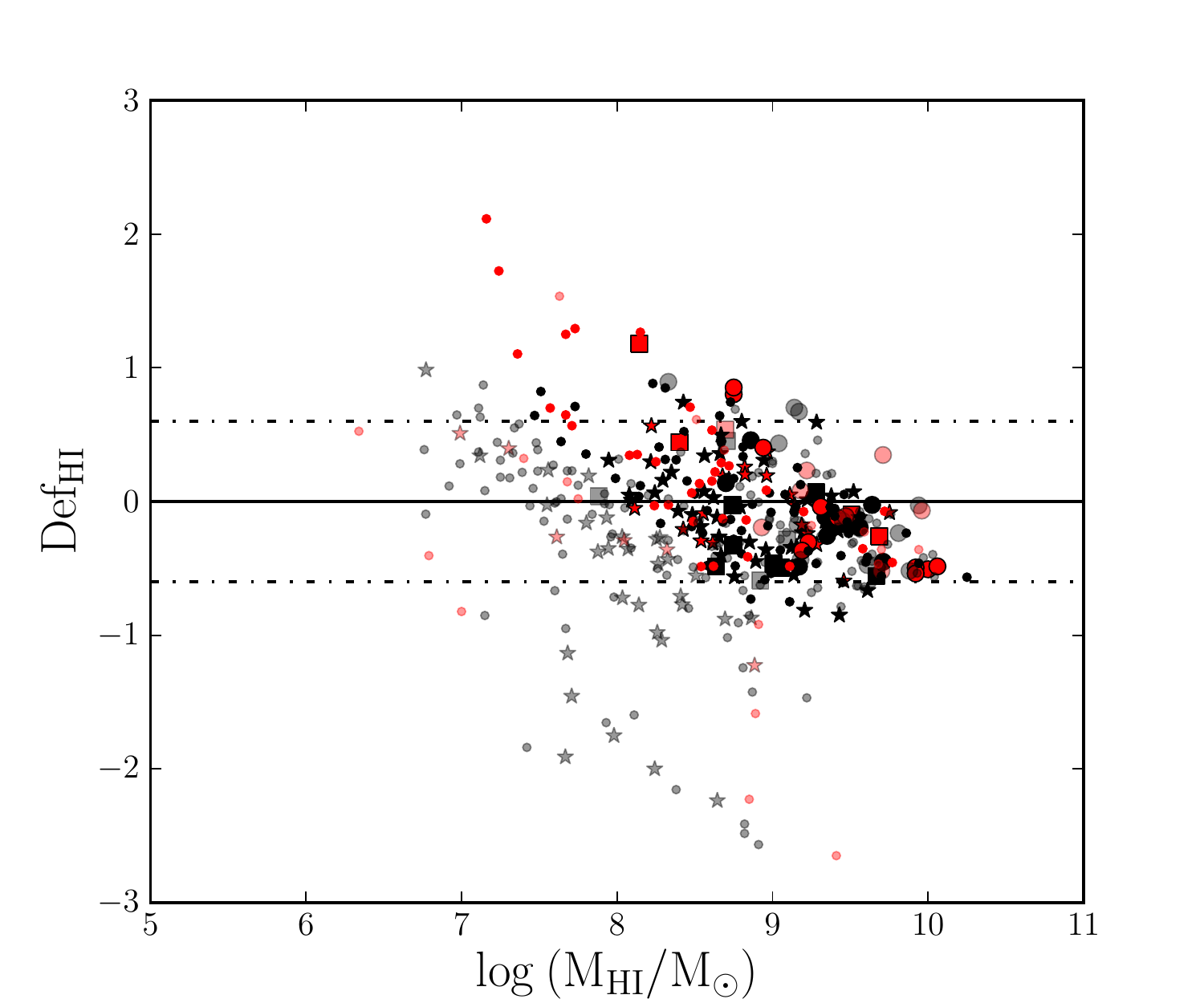}}\\
\end{tabular}
\end{center}
\caption{The measured HI mass plotted against the predicted HI mass and the HI deficiency versus measured HI mass. (\textit{left}) Predicted HI content compared to the measured HI content (the HI observations are referenced in the key). Small circles and stars mark galaxies with \textit{one} optical counterpart within the beam, whereas large circles and square indicate confused HI sources with multiple optical counterparts within the beam. The solid line marks the unity relationship. (\textit{right}) HI deficiency plotted against the measured HI mass. The dashed lines mark the $\mathrm{Def_{HI}} = \pm 0.6$ beyond which we consider galaxies to be HI deficient or with HI excess, respectively. The red markers indicate galaxies, which reside in groups, whereas the black markers show non-group galaxies. Galaxies in the complete sample for which local projected surface densities are available are shown in bright colours, whereas faint galaxies or galaxies towards the edges of the studied region are shown with faded colours.} 
\label{fig:pred_meas}
\end{figure*}

\begin{figure*}
\begin{center}
\begin{tabular}{lr}
\mbox{\includegraphics[trim=0cm 0cm 1cm 0.5cm, clip=true, width=0.47\textwidth]{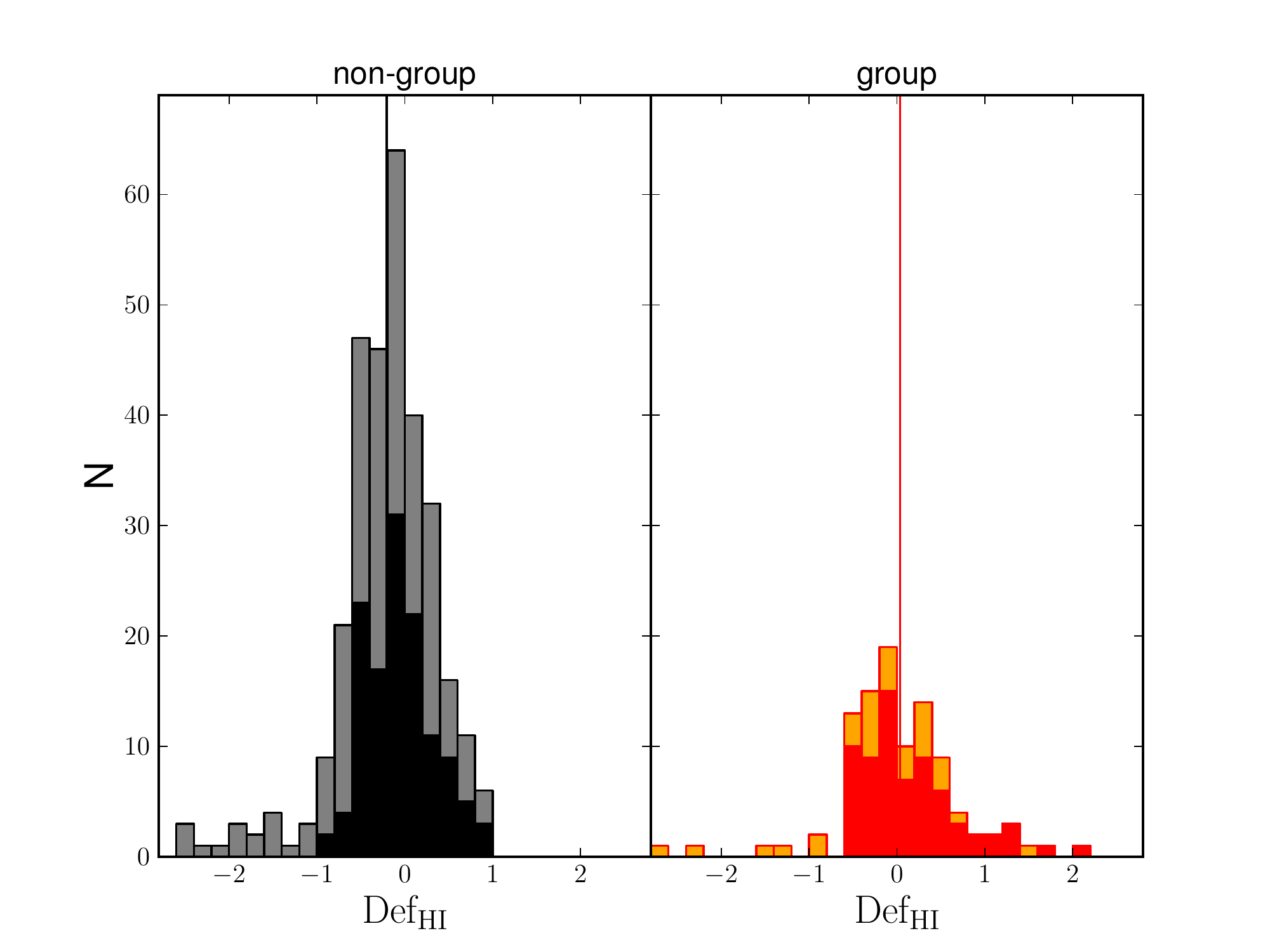}} & 
\mbox{\includegraphics[trim=0cm 0cm 1cm 1cm, clip=true, width=0.47\textwidth]{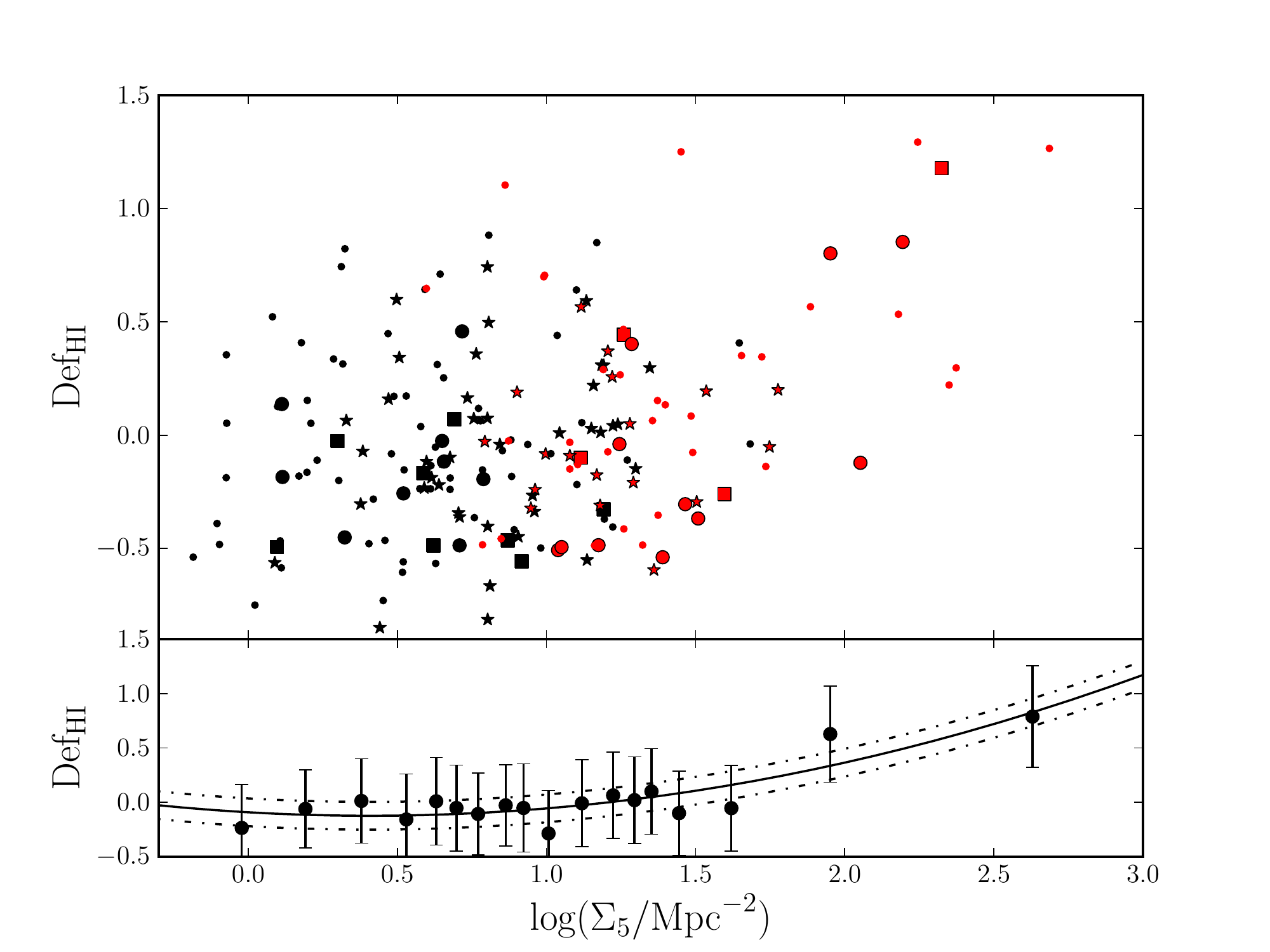}}\\
\end{tabular}
\end{center}
\caption{HI deficiency distributions for group/non-group galaxies and the HI deficiency as a function of local projected surface density. (\textit{left}) HI deficiency distribution for galaxies residing in groups (red and orange) and non-group galaxies (black and grey). The average HI deficiency is indicated by the line. Galaxies in the complete sample for which local projected surface densities are available are shown in red and black, whereas faint galaxies or galaxies towards the edges of the studied region are shown in orange and grey. (\textit{right}) The HI deficiency as a function of local projected surface density. The \textit{bottom panel} shows the binned data (average per 10 data points). Markers and colours are similar to Figure~\ref{fig:pred_meas}.} 
\label{fig:hist_HIdef}
\end{figure*}

\begin{figure*}
\begin{tabular}{p{2.7cm} p{2.5cm} p{2cm} p{2.7cm} p{2.5cm}}
\multicolumn{2}{c}{UGC07698 (J1232+31) -- $\mathrm{Def_{HI}}=-2.34$}
 & &
\multicolumn{2}{c}{UGCA259 (J1158+45) -- $\mathrm{Def_{HI}}=-2.24$}\\
\includegraphics[trim=7.7cm 3.7cm 0cm 0cm, clip=true, height=3cm]{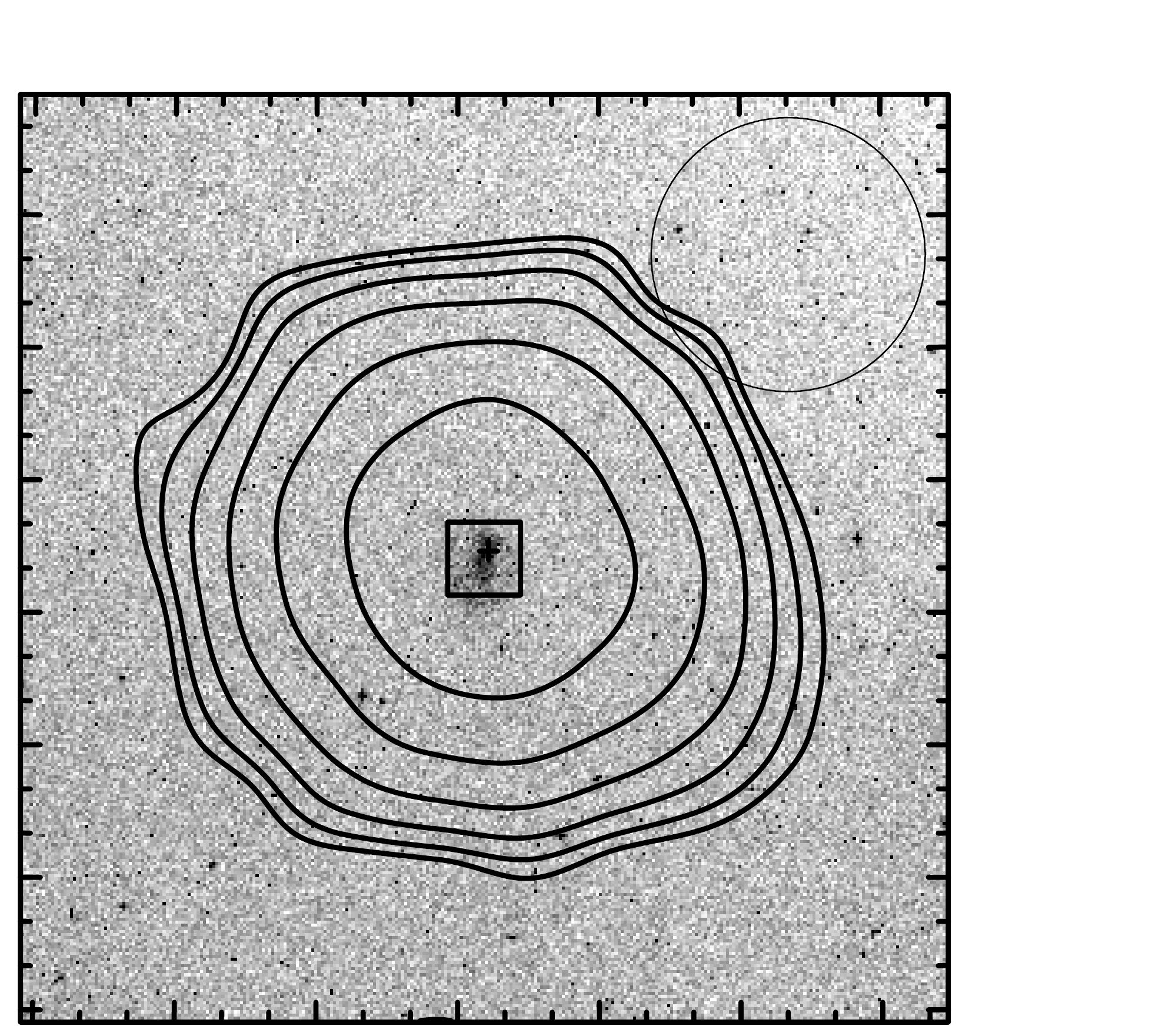} & 
\includegraphics[trim=1cm 1cm 0cm 0cm, clip=true, height=3cm]{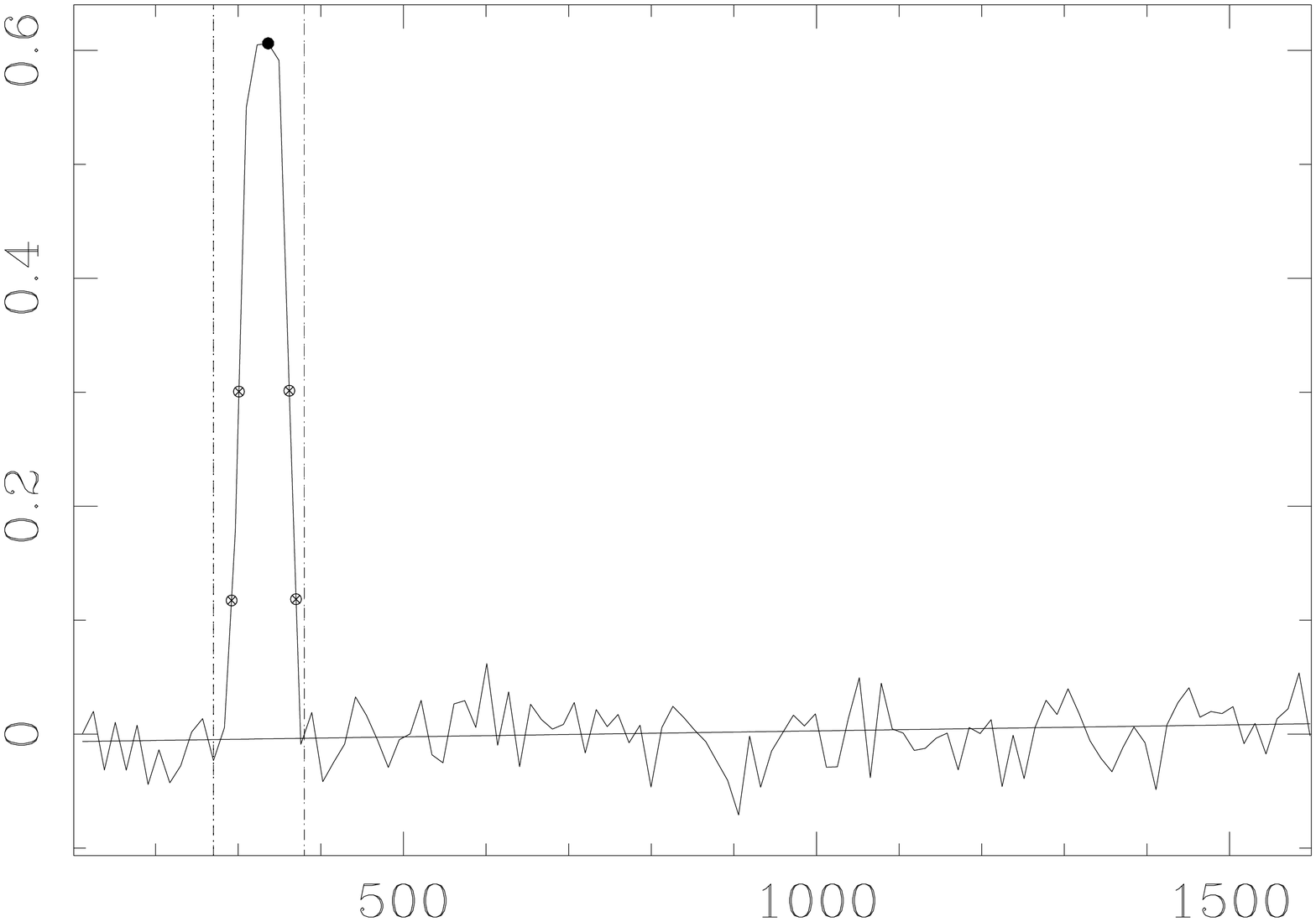} &
&
\includegraphics[trim=7.7cm 3.7cm 0cm 0cm, clip=true, height=3cm]{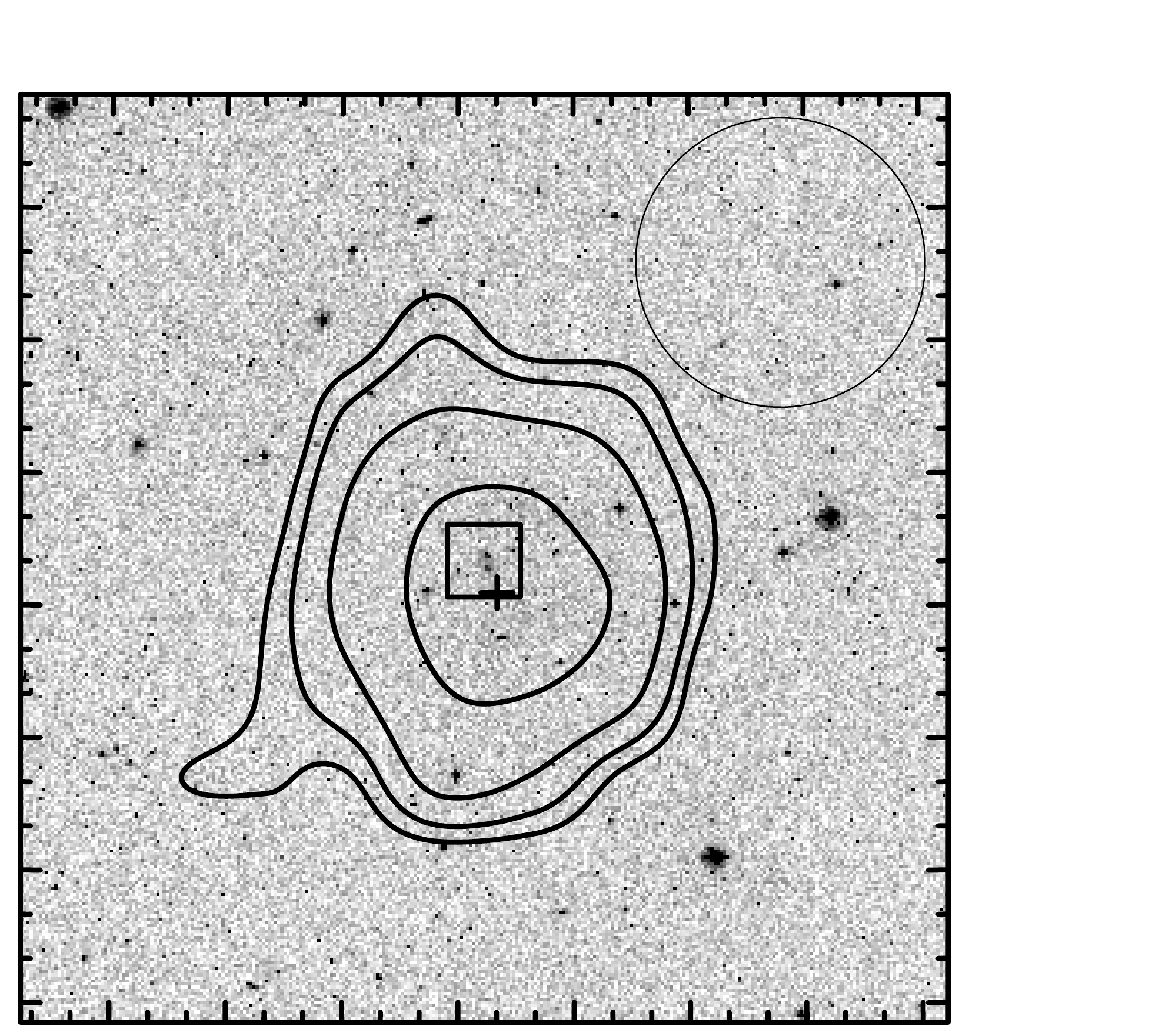} & 
\includegraphics[trim=1cm 1cm 0cm 0cm, clip=true, height=3cm]{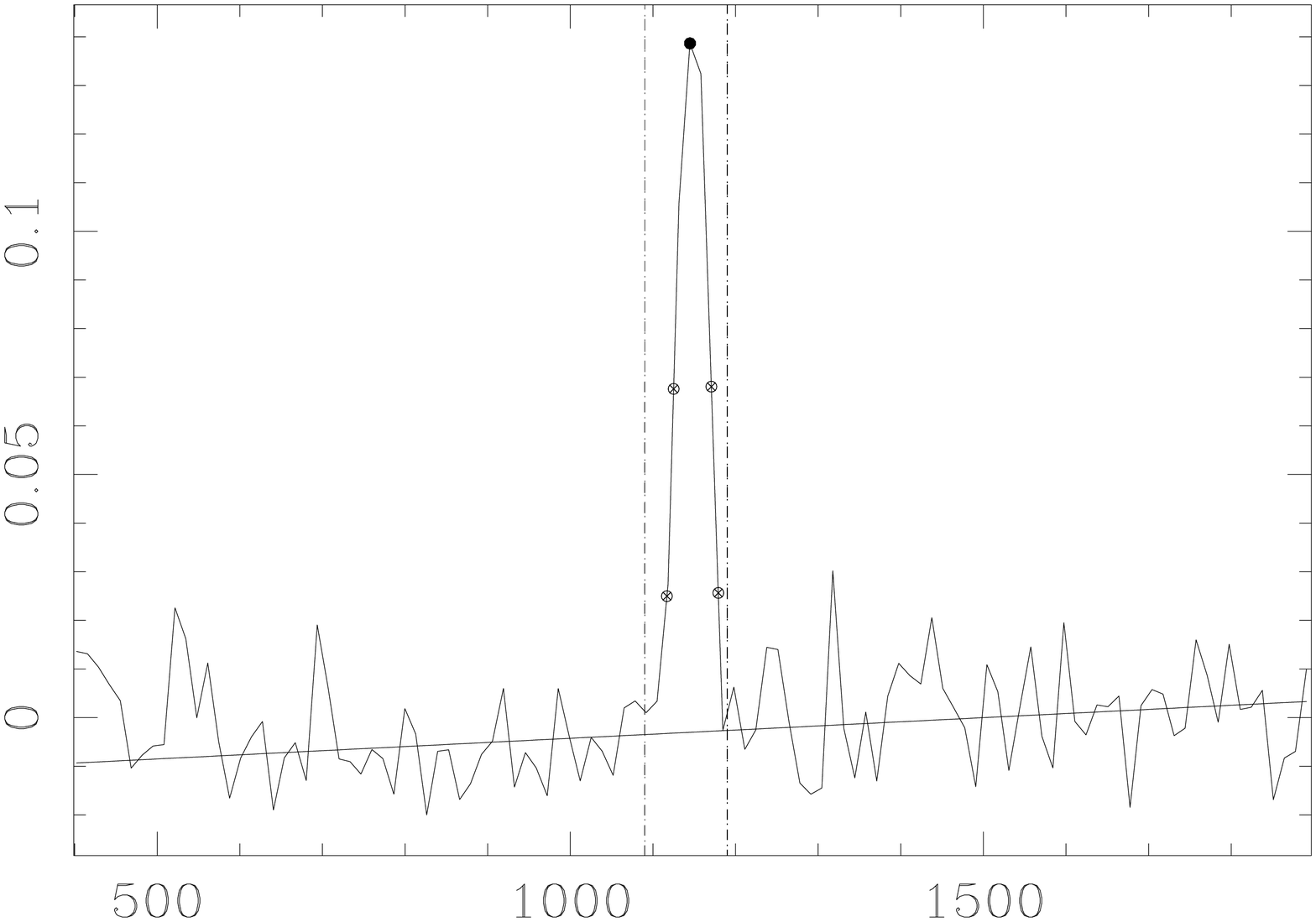} \\
[12pt]
\multicolumn{2}{c}{UGC07236 (J1213+24) -- $\mathrm{Def_{HI}}=-2.00$}
& &
\multicolumn{2}{c}{UGC07559 (J1227+37) -- $\mathrm{Def_{HI}}=-1.91$}\\
\mbox{\includegraphics[trim=7.7cm 3.7cm 0cm 0cm, clip=true,height=3cm]{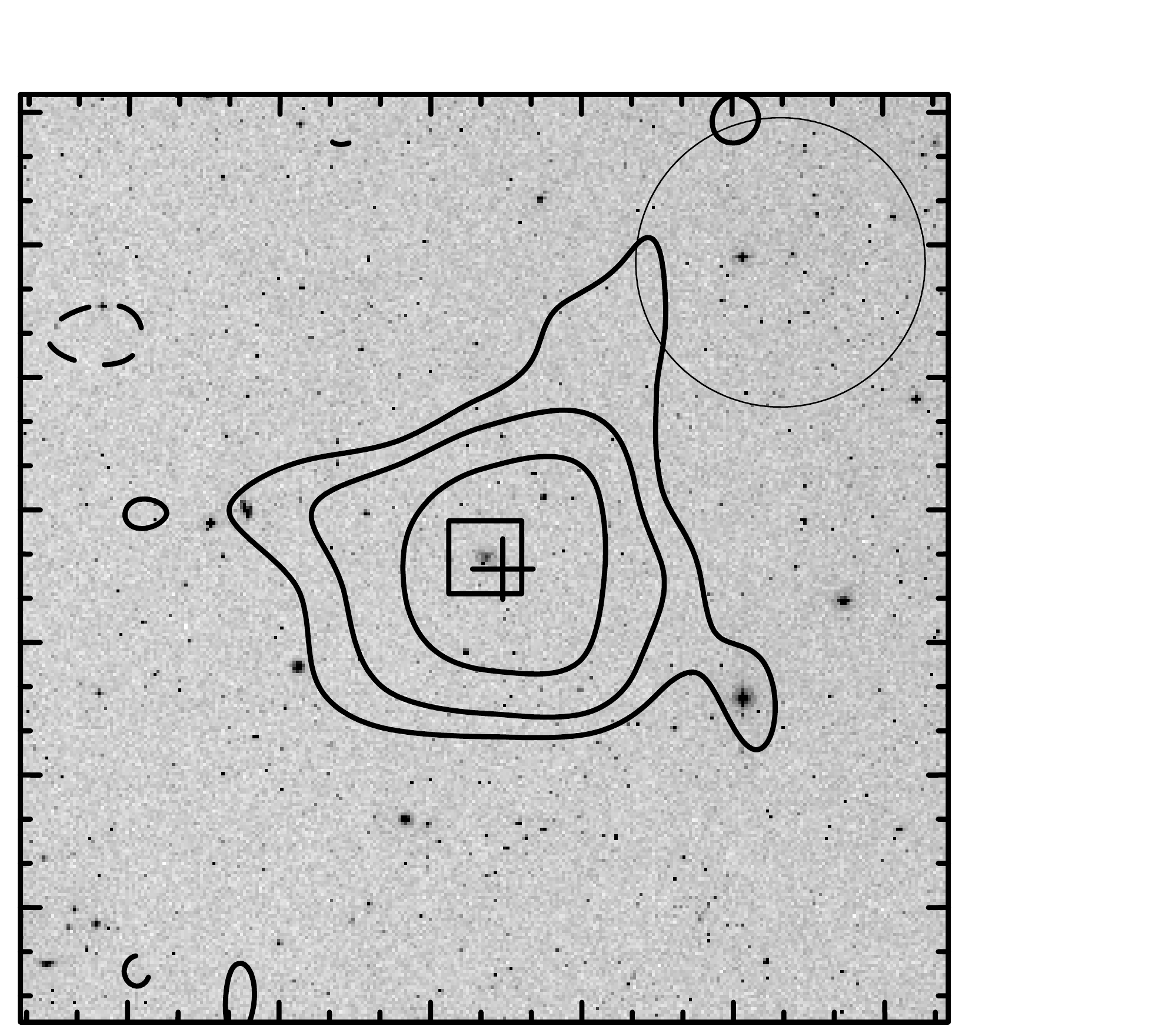}} & 
\mbox{\includegraphics[trim=1cm 1cm 0cm 0cm, clip=true, height=3cm]{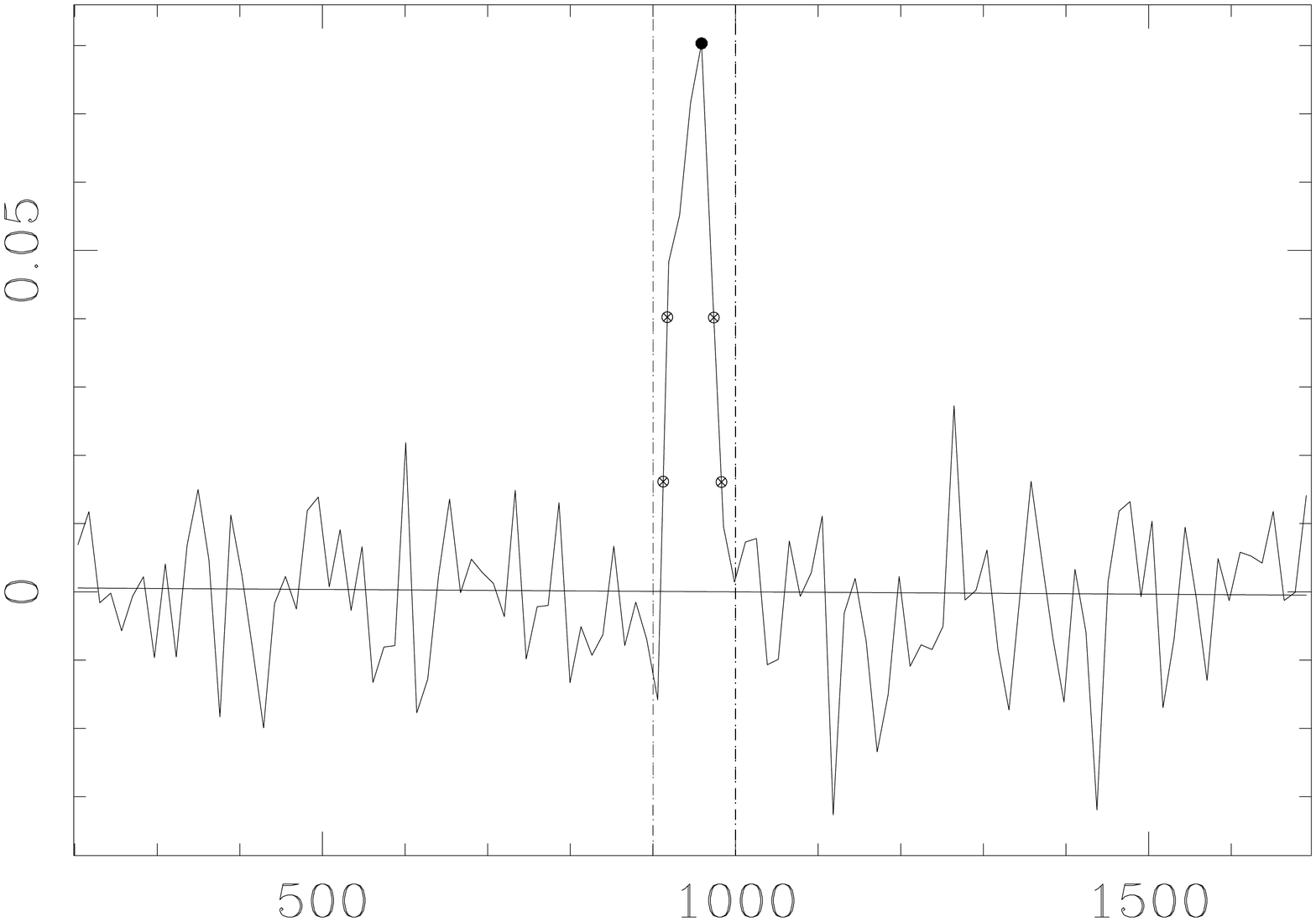}} &
&
\mbox{\includegraphics[trim=7.7cm 3.7cm 0cm 0cm, clip=true,height=2.8cm]{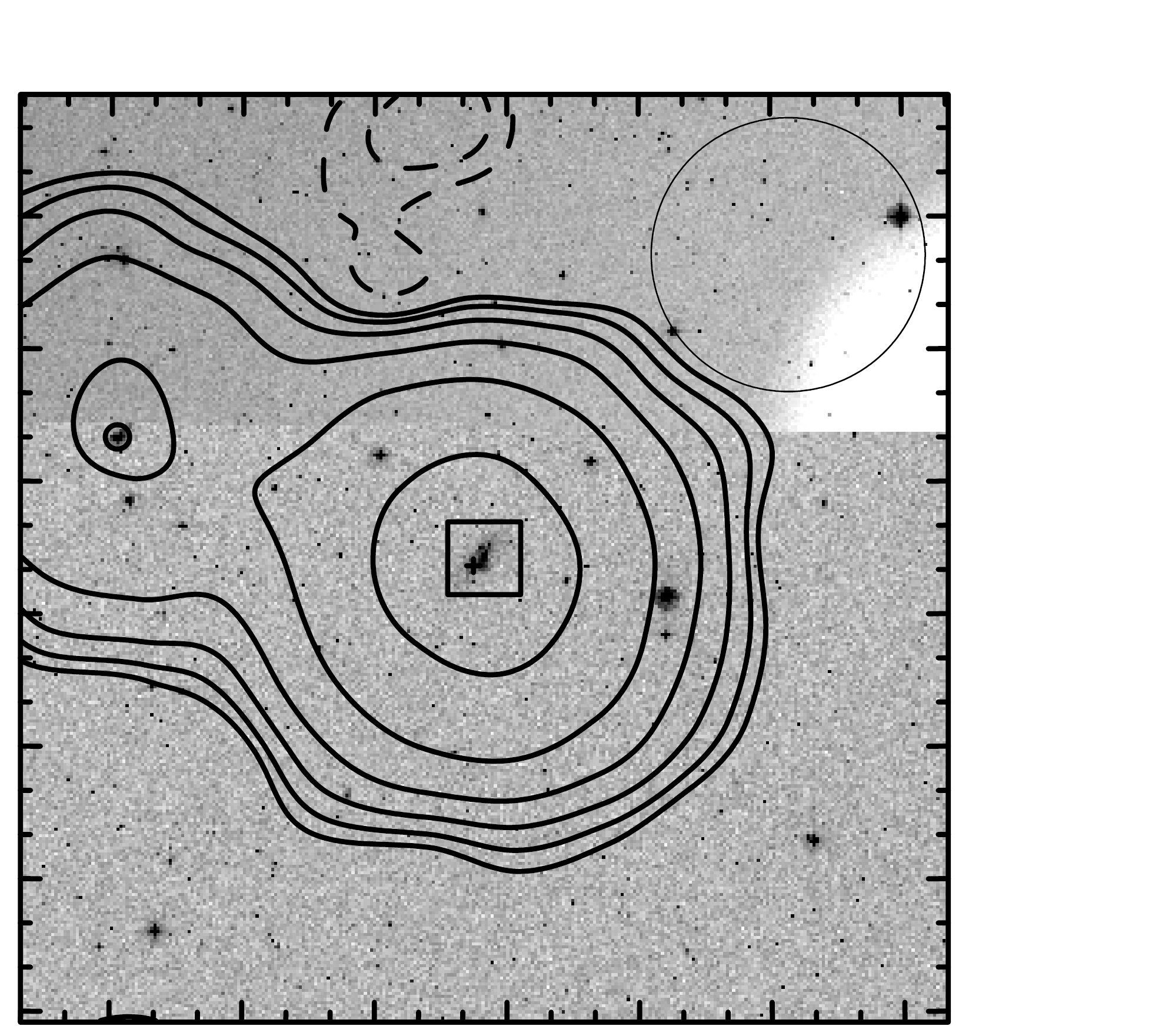}} & 
\mbox{\includegraphics[trim=1cm 1cm 0cm 0cm, clip=true, height=3cm]{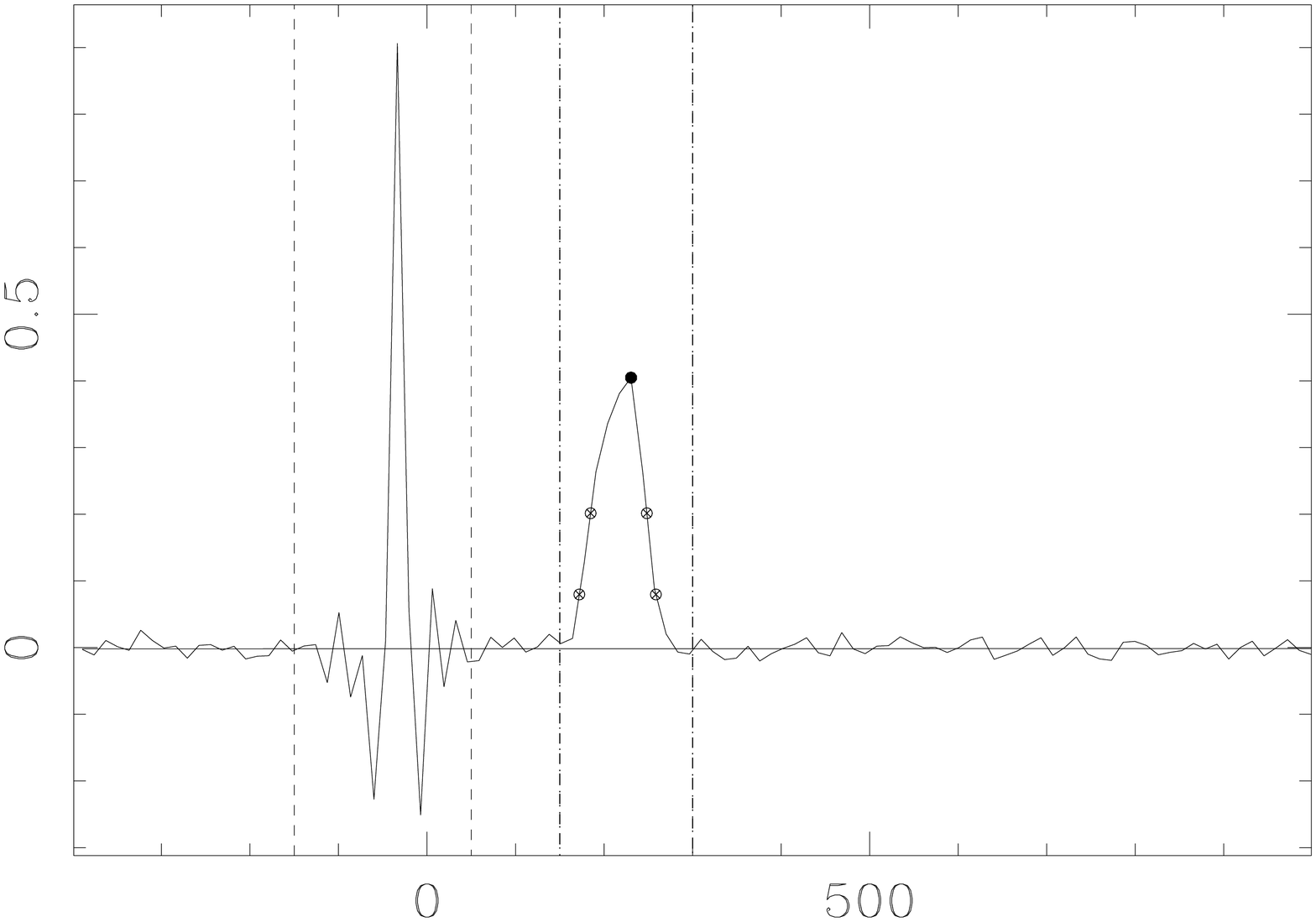}} \\
[12pt]
\multicolumn{2}{c}{UGC06817 (J1150+38) -- $\mathrm{Def_{HI}}=-1.87$}
& &
\multicolumn{2}{c}{UGC05917 (J1048+46) -- $\mathrm{Def_{HI}}=-1.75$}\\
\mbox{\includegraphics[trim=7.7cm 3.7cm 0cm 0cm, clip=true,height=2.8cm]{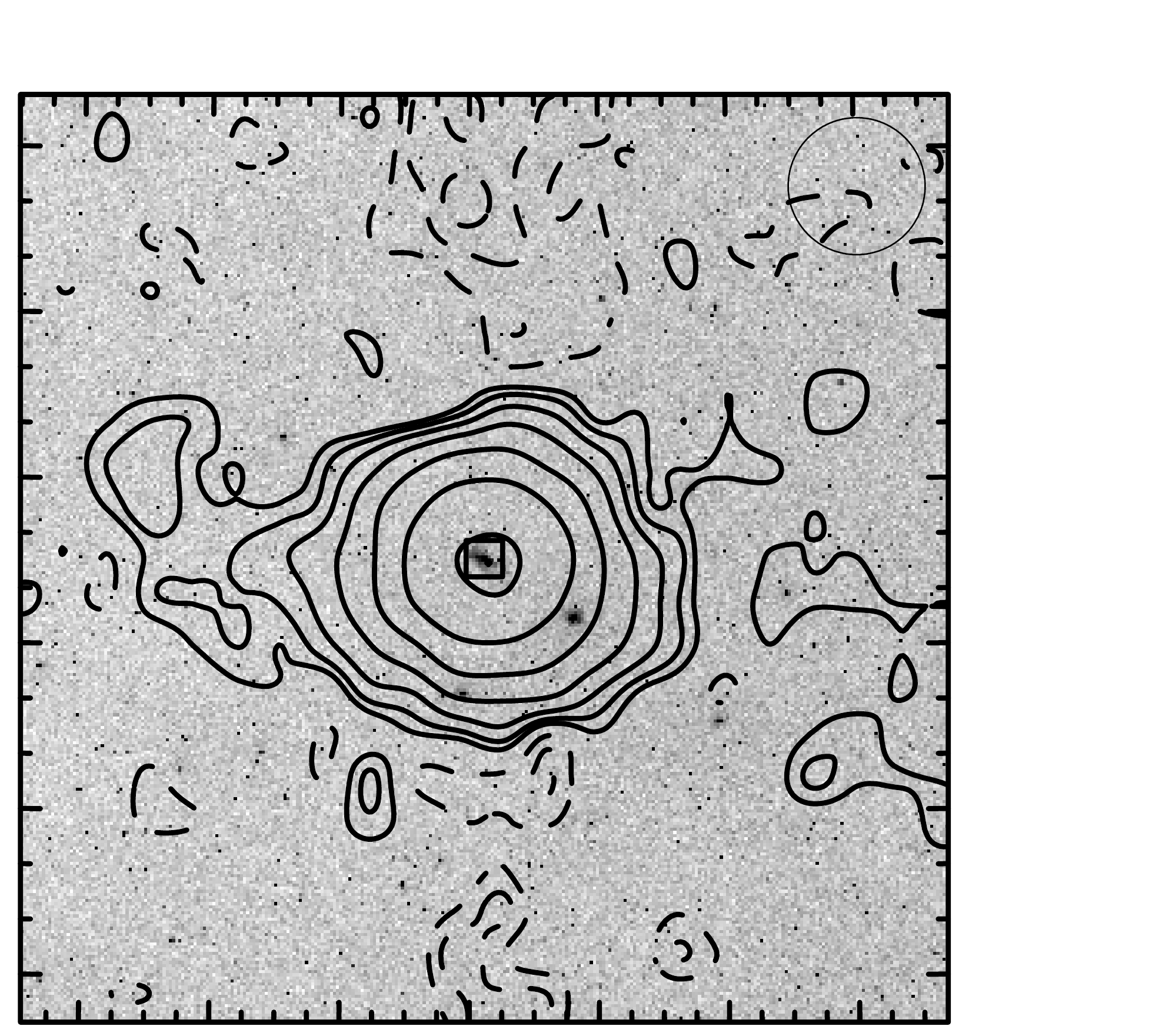}} & 
\mbox{\includegraphics[trim=1cm 1cm 0cm 0cm, clip=true, height=3cm]{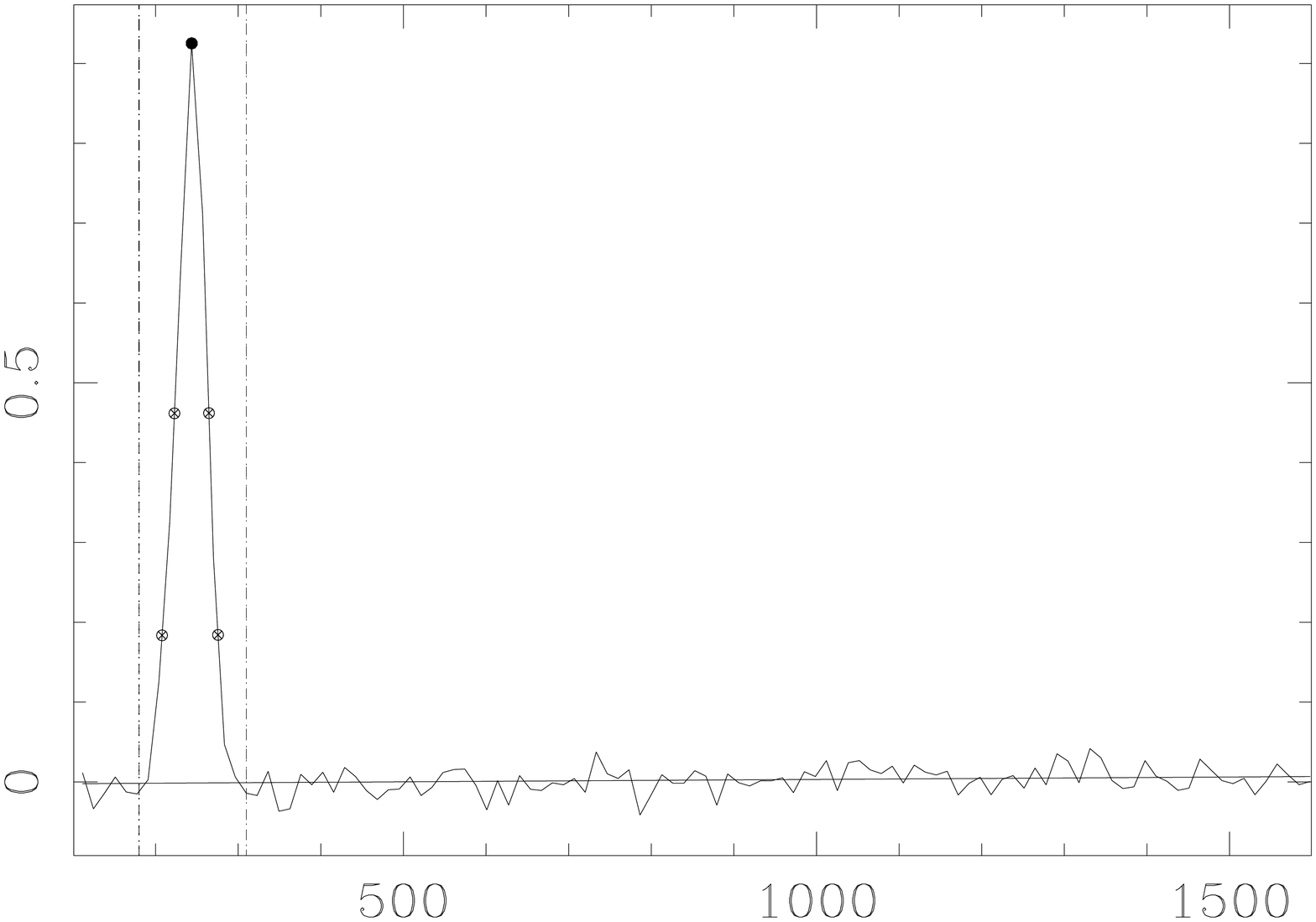}} &
&
\mbox{\includegraphics[trim=7.7cm 3.7cm 0cm 0cm, clip=true,height=3cm]{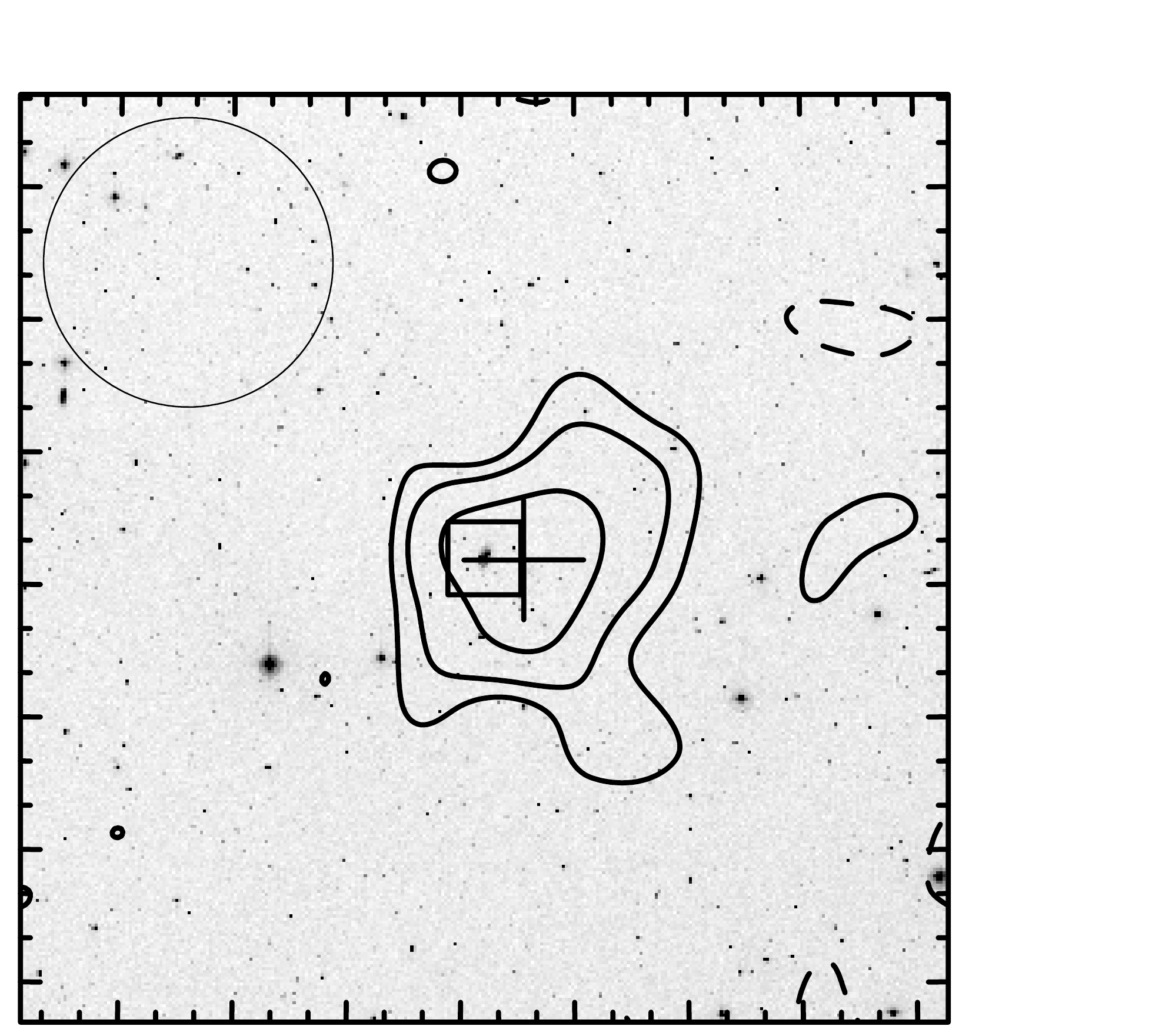}} & 
\mbox{\includegraphics[trim=1cm 1cm 0cm 0cm, clip=true, height=3cm]{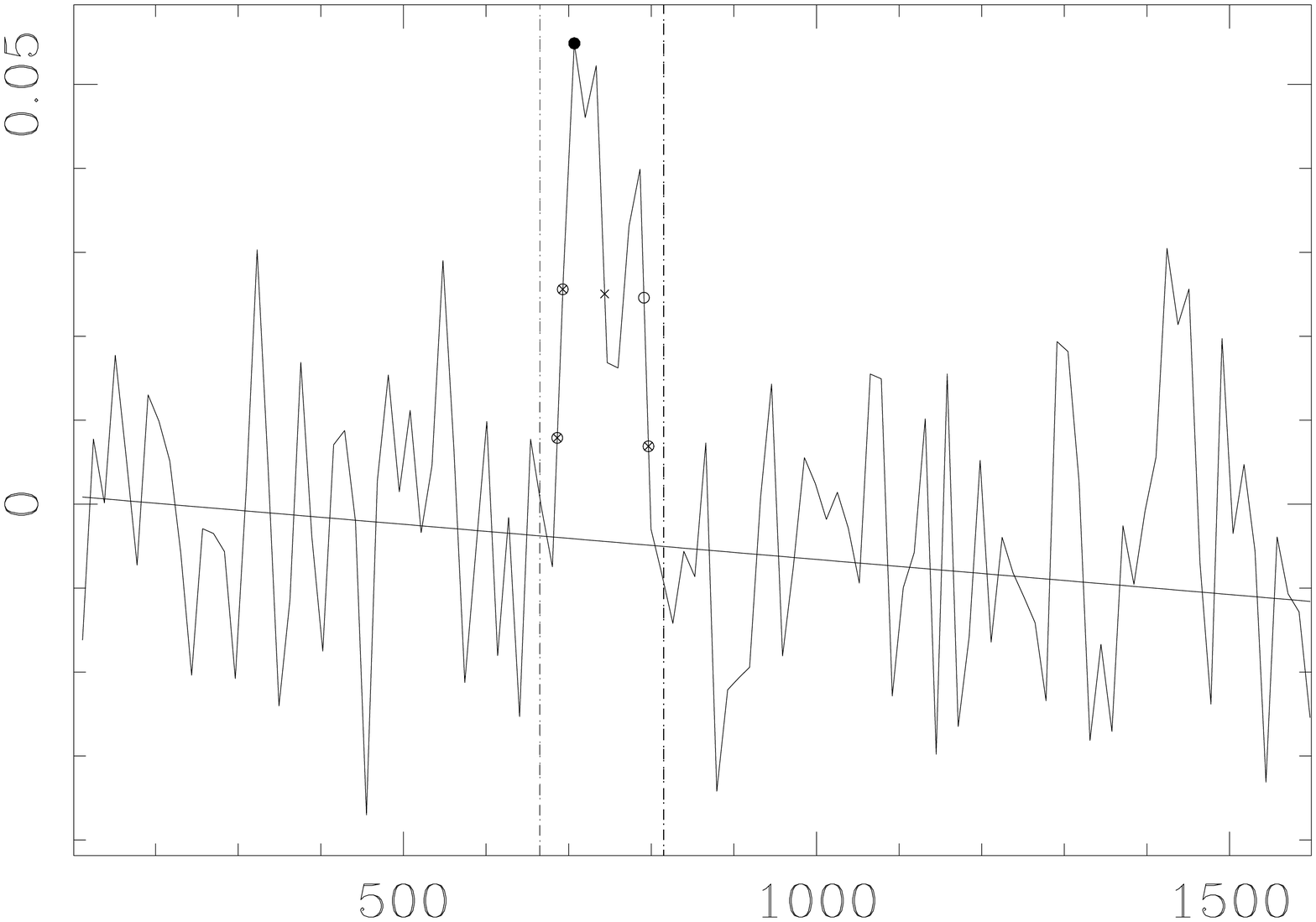}} \\
[12pt]
\multicolumn{2}{c}{NGC4395 (J1225+33) -- $\mathrm{Def_{HI}}=-1.57$}
& & 
\multicolumn{2}{c}{NGC4144 (J1210+46) -- $\mathrm{Def_{HI}}=-1.51$} \\
\mbox{\includegraphics[trim=7.7cm 3.7cm 0cm 0cm, clip=true,height=3cm]{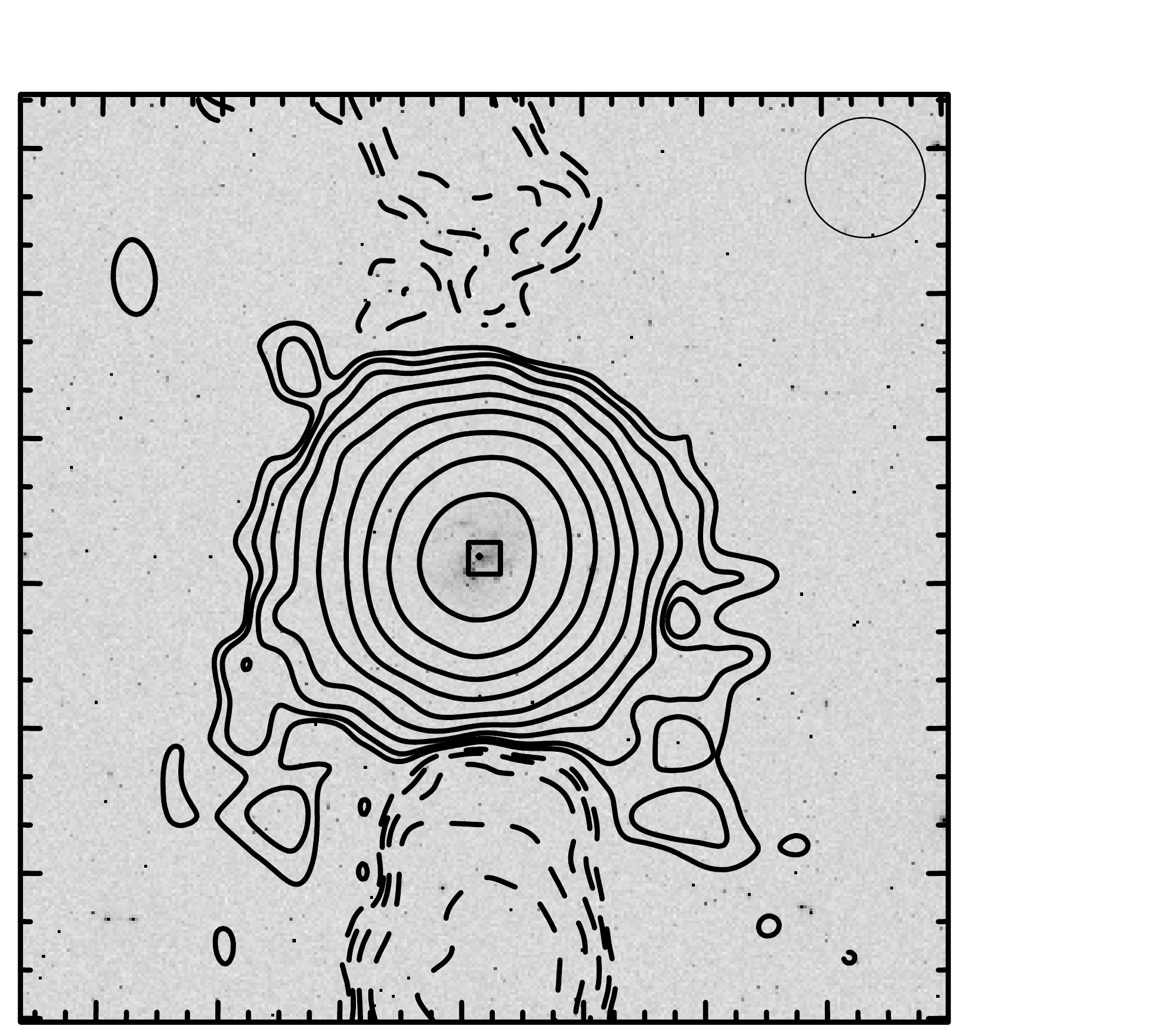}} & 
\mbox{\includegraphics[trim=1cm 1cm 0cm 0cm, clip=true, height=3cm]{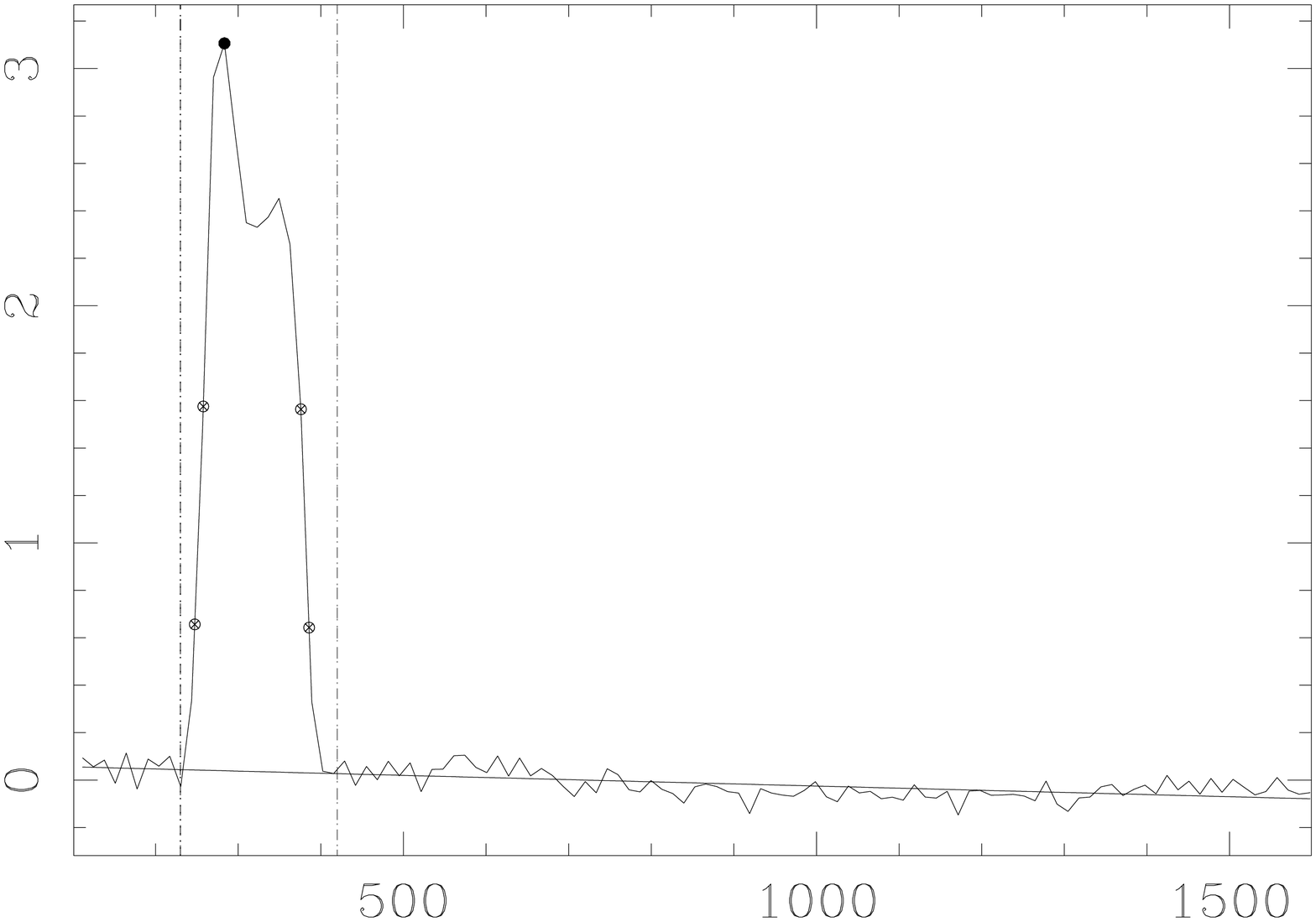}} &
&
\mbox{\includegraphics[trim=7.7cm 3.7cm 0cm 0cm, clip=true,height=3cm]{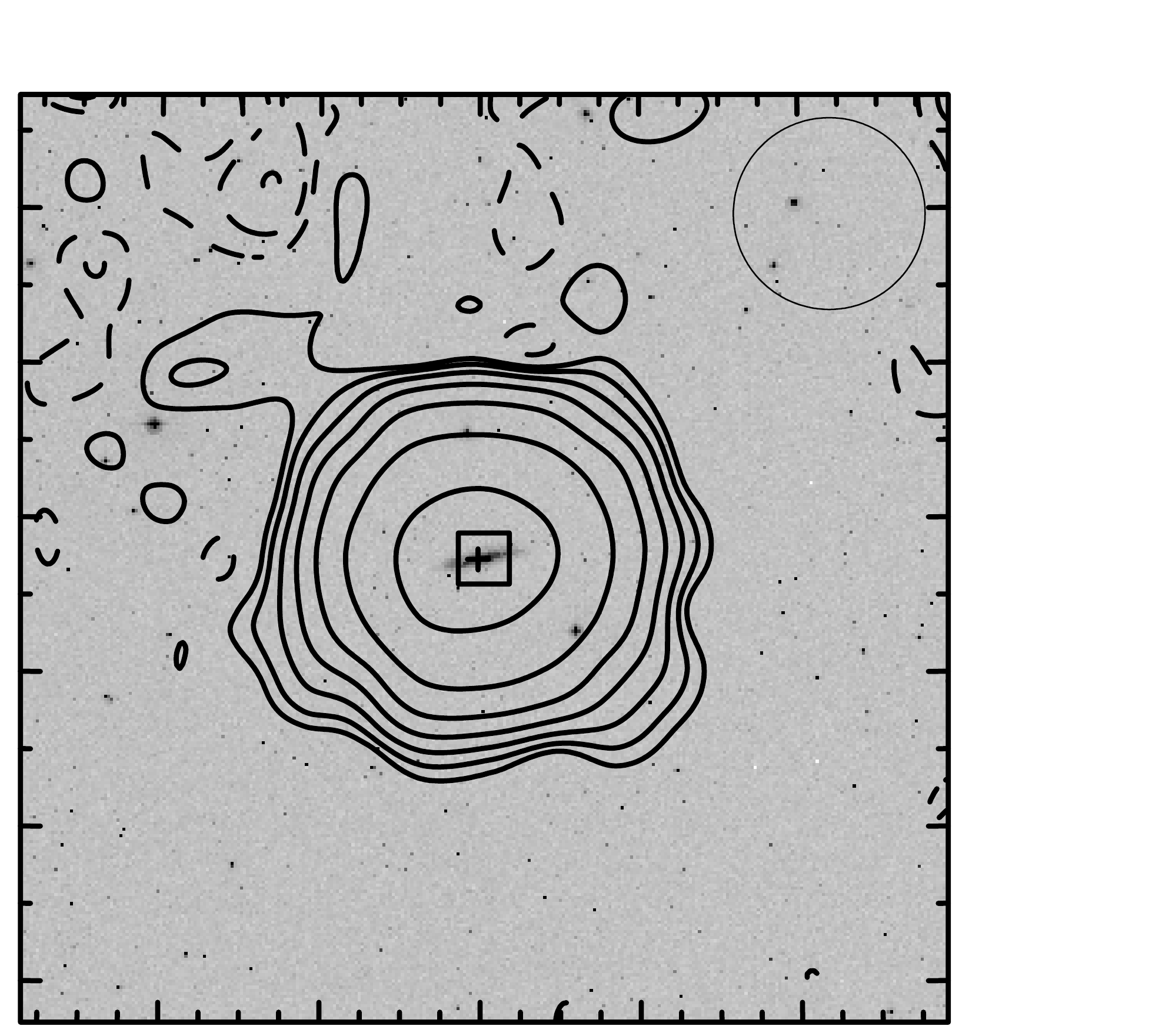}} & 
\mbox{\includegraphics[trim=1cm 1cm 0cm 0cm, clip=true, height=3cm]{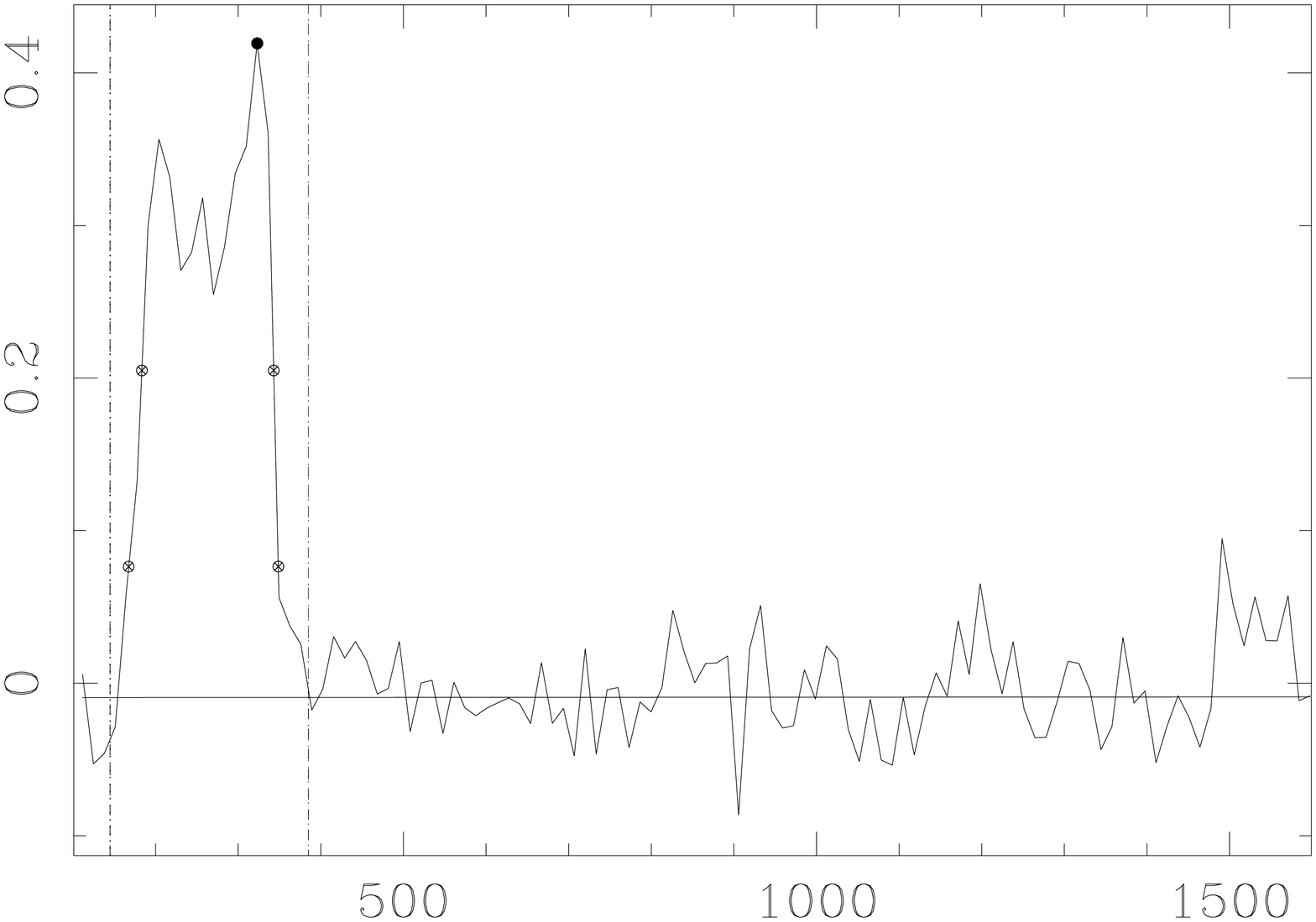}} \\
[12pt]
\multicolumn{2}{c}{KUG1218+387 (J1220+38) -- $\mathrm{Def_{HI}}=-1.46$}
& &
\multicolumn{2}{c}{UGC06840 (J1152+52) -- $\mathrm{Def_{HI}}=-1.23$} \\
\mbox{\includegraphics[trim=7.7cm 3.7cm 0cm 0cm, clip=true,height=3cm]{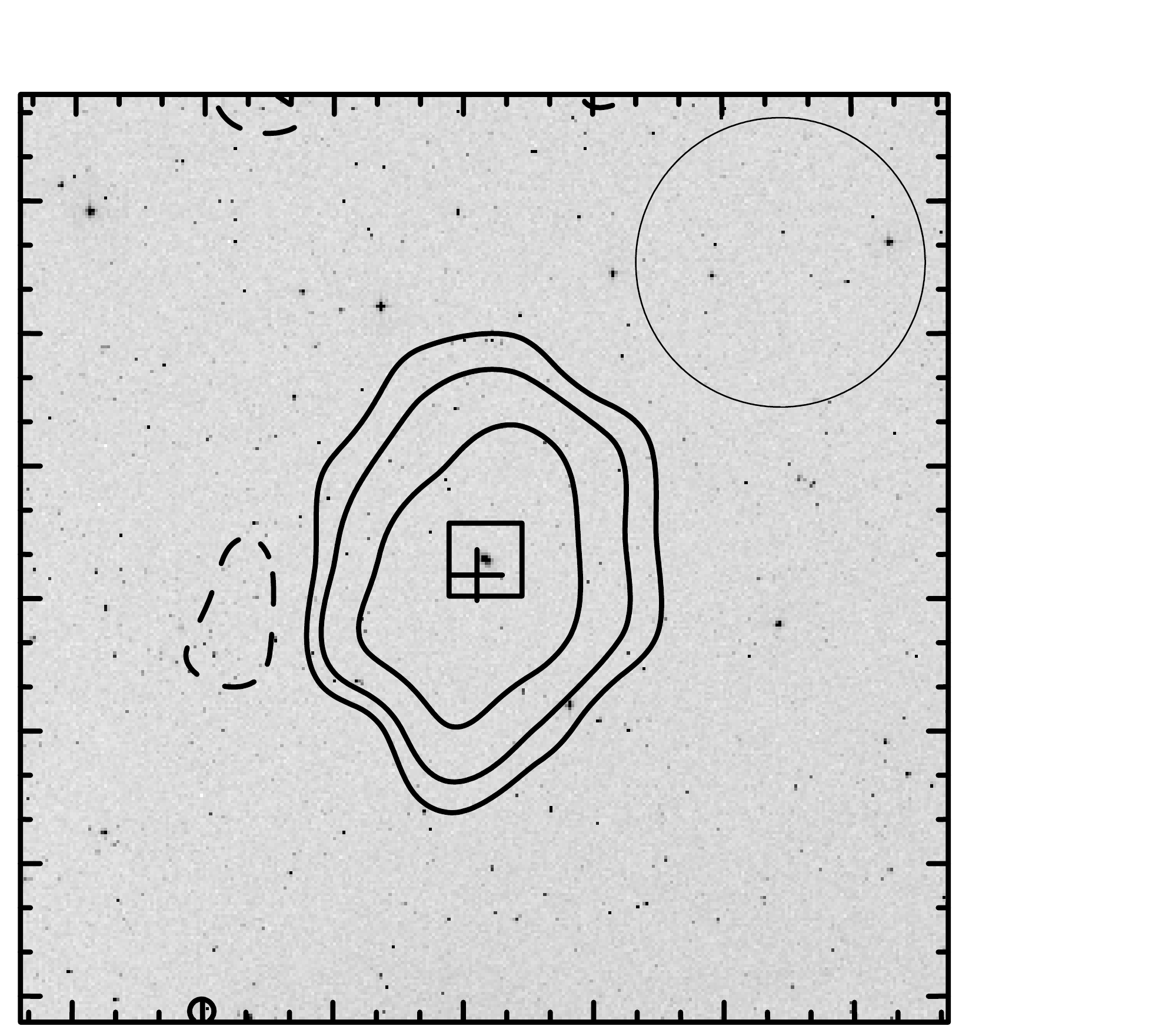}} & 
\mbox{\includegraphics[trim=1cm 1cm 0cm 0cm, clip=true, height=3cm]{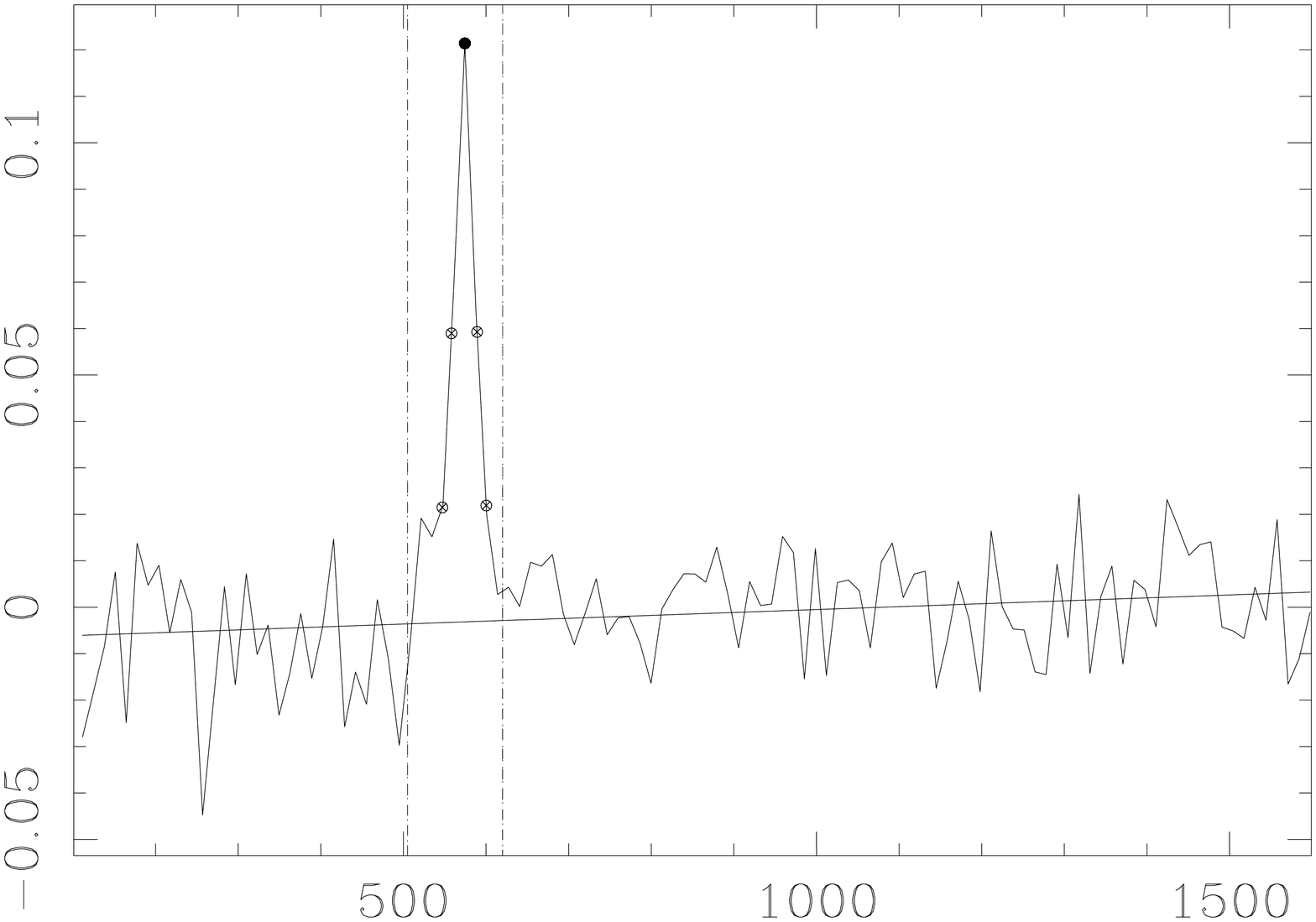}} &
&
\mbox{\includegraphics[trim=7.7cm 3.7cm 0cm 0cm, clip=true,height=3cm]{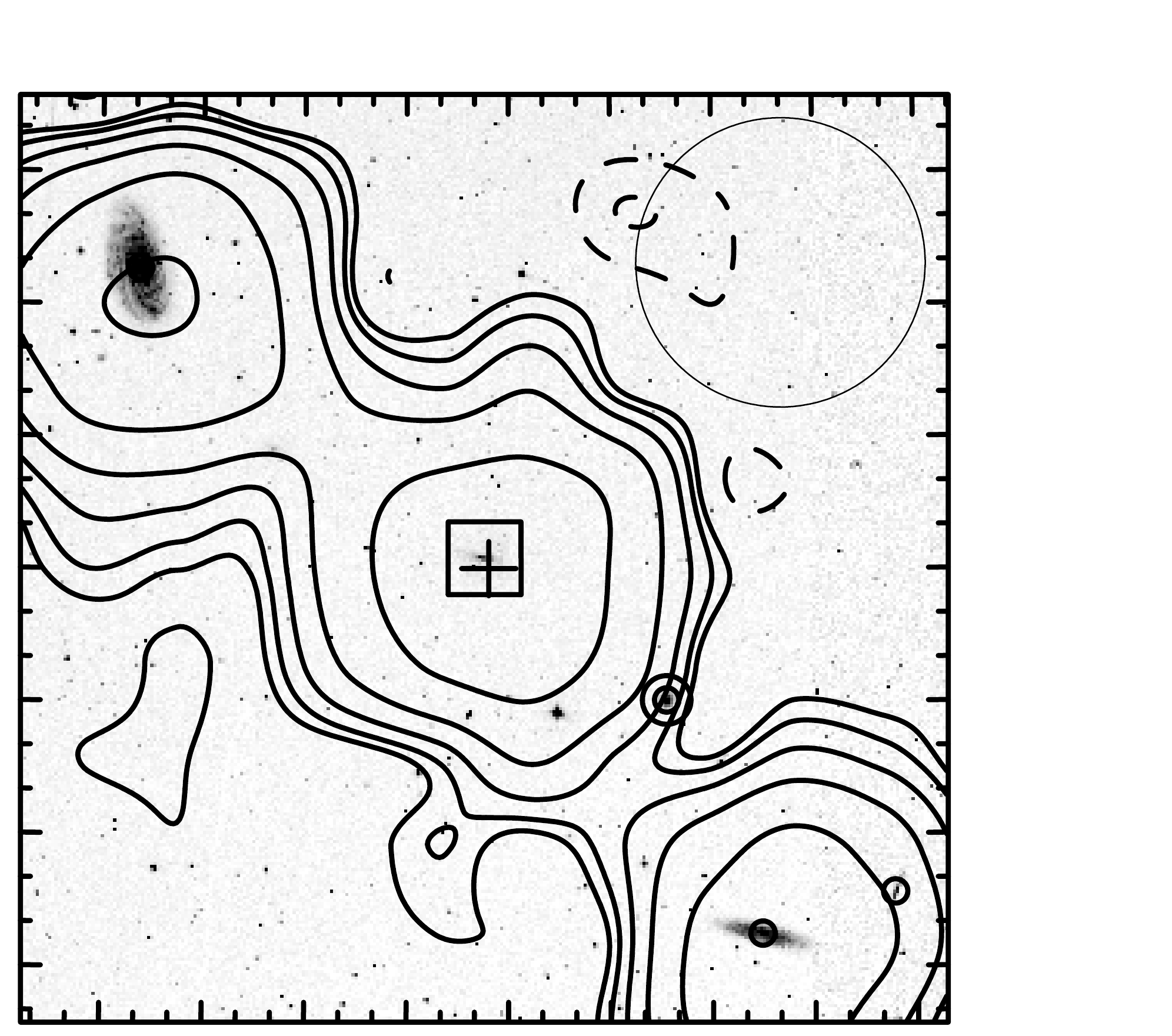}} & 
\mbox{\includegraphics[trim=1cm 1cm 0cm 0cm, clip=true, height=3cm]{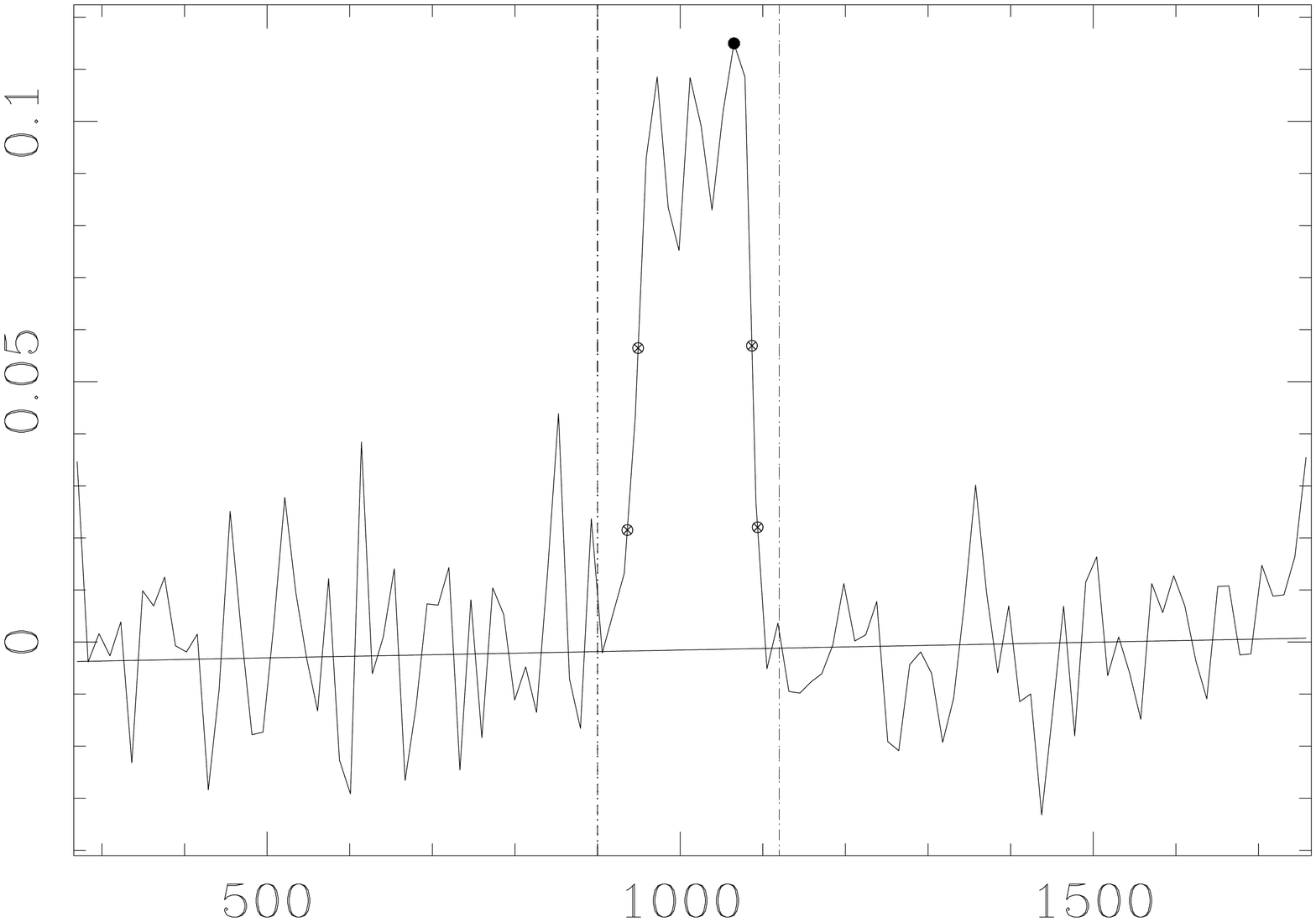}} \\
[12pt]
\end{tabular} 
\caption[HIJASS detections with HI excess.]{HIJASS detections with HI excess, i.e. with HI deficiencies $\leq -1$. (\textit{left}) We show DSS2 B-band images and overlaid integrated flux levels at $\pm0.5 \times 2^n$~Jy~beam$\mathrm{^{-1}}$~km~s$^{-1}$ (contours). The gridded beam is displayed in a corner (13.1~arcmin) and negative contours are shown with dashed lines (e.g. negative bandpass sidelobes). The main optical counterpart within the HIJASS beam is indicated by a box and labelled above the figures along with the HIJASS name (in brackets of the form HIJASS~x, where x is the stated ID) and the $\mathrm{Def_{HI}}$ value. Smaller galaxies which might also contribute to the measured HI flux are indicated by circles. The DSS2 images are centred on the HI position (position uncertainty is indicated by the size of the cross). (\textit{right}) The HI spectra have the optical velocities ($cz$ in km~s$^{-1}$) along the abscissas and the flux densities (in Jy) along the ordinates. The HI line emission analysis is conducted within the velocity range as indicated by the dotted vertical lines. The baseline fits to the spectra are conducted within the displayed velocity range excluding the velocity ranges of HI line emission analysis and Galactic emission (see UGC07559). Note that we show the HI spectra prior to baseline subtraction. The HI excess decreases from the left to the right and from the top to the bottom.}
\label{fig:HIexcI}
\end{figure*}

\begin{figure*}
\begin{tabular}{p{2.7cm} p{2.5cm} p{2cm} p{2.7cm} p{2.5cm}}
\multicolumn{2}{c}{UGCA292 (J1238+32) -- $\mathrm{Def_{HI}}=-1.13$}
& &
\multicolumn{2}{c}{UGC06113 (J1102+52) -- $\mathrm{Def_{HI}}=-1.04$} \\
\mbox{\includegraphics[trim=7.7cm 3.7cm 0cm 0cm, clip=true,height=3cm]{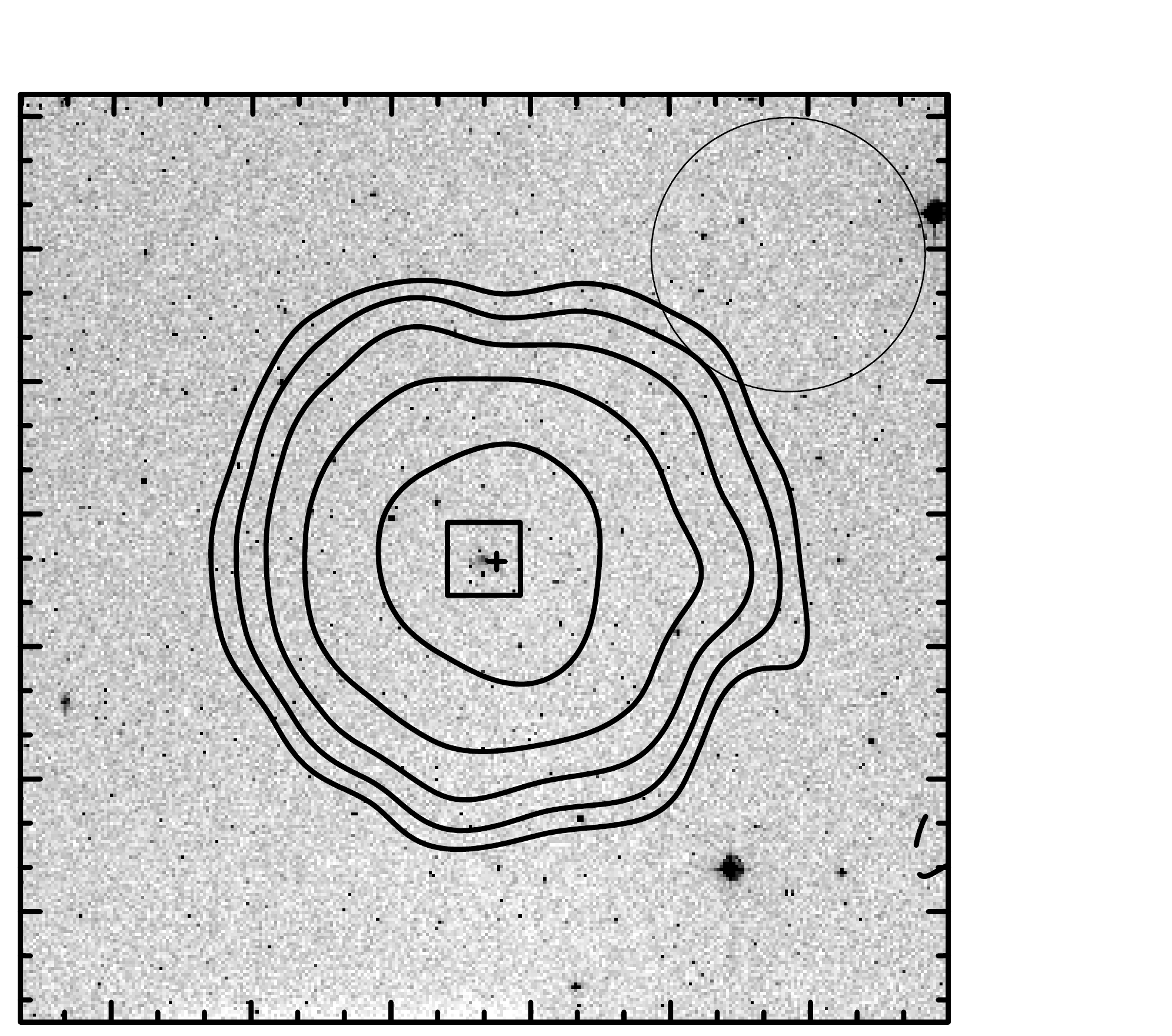}} & 
\mbox{\includegraphics[trim=1cm 1cm 0cm 0cm, clip=true, height=3cm]{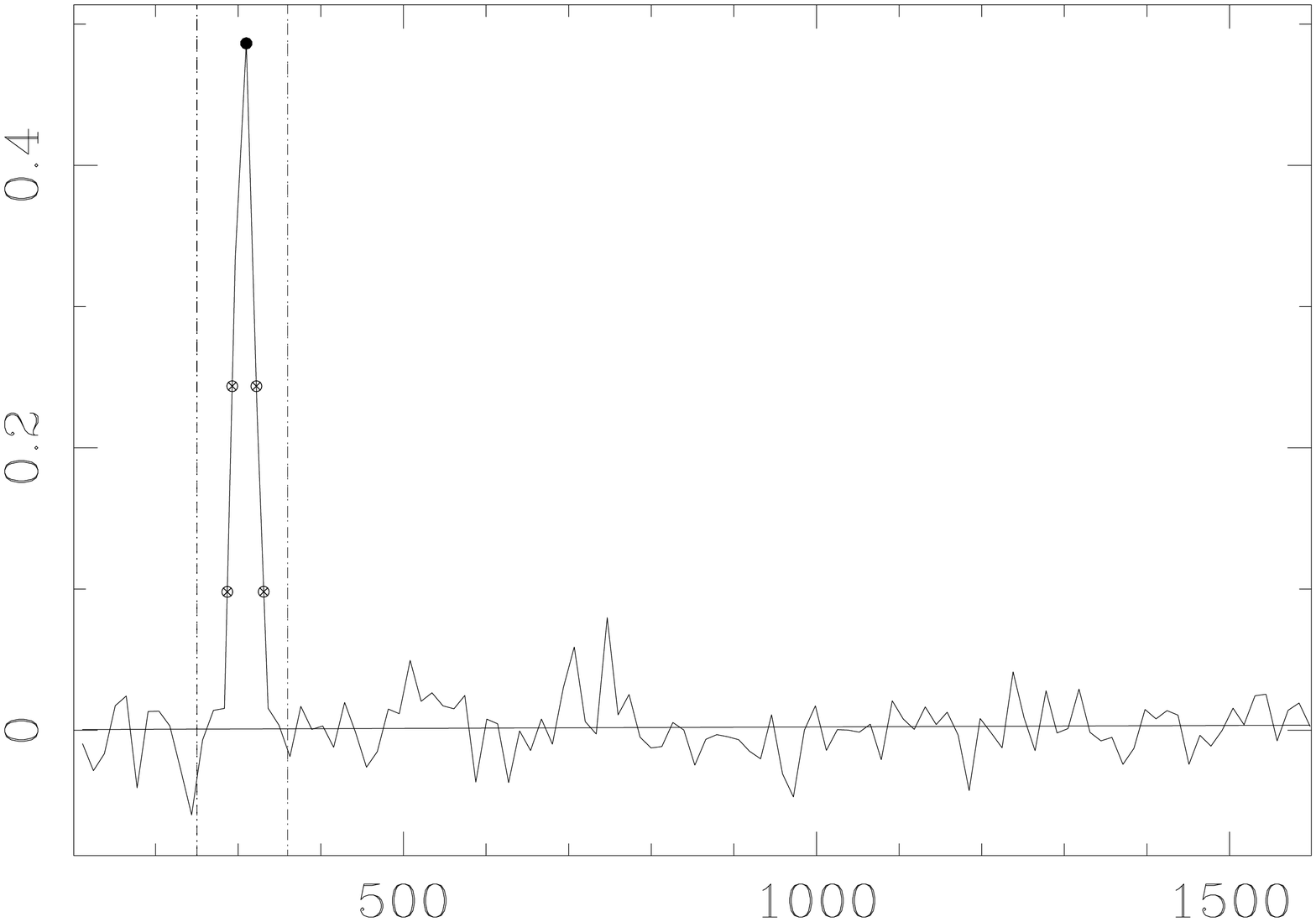}} &
&
\mbox{\includegraphics[trim=7.7cm 3.7cm 0cm 0cm, clip=true,height=3cm]{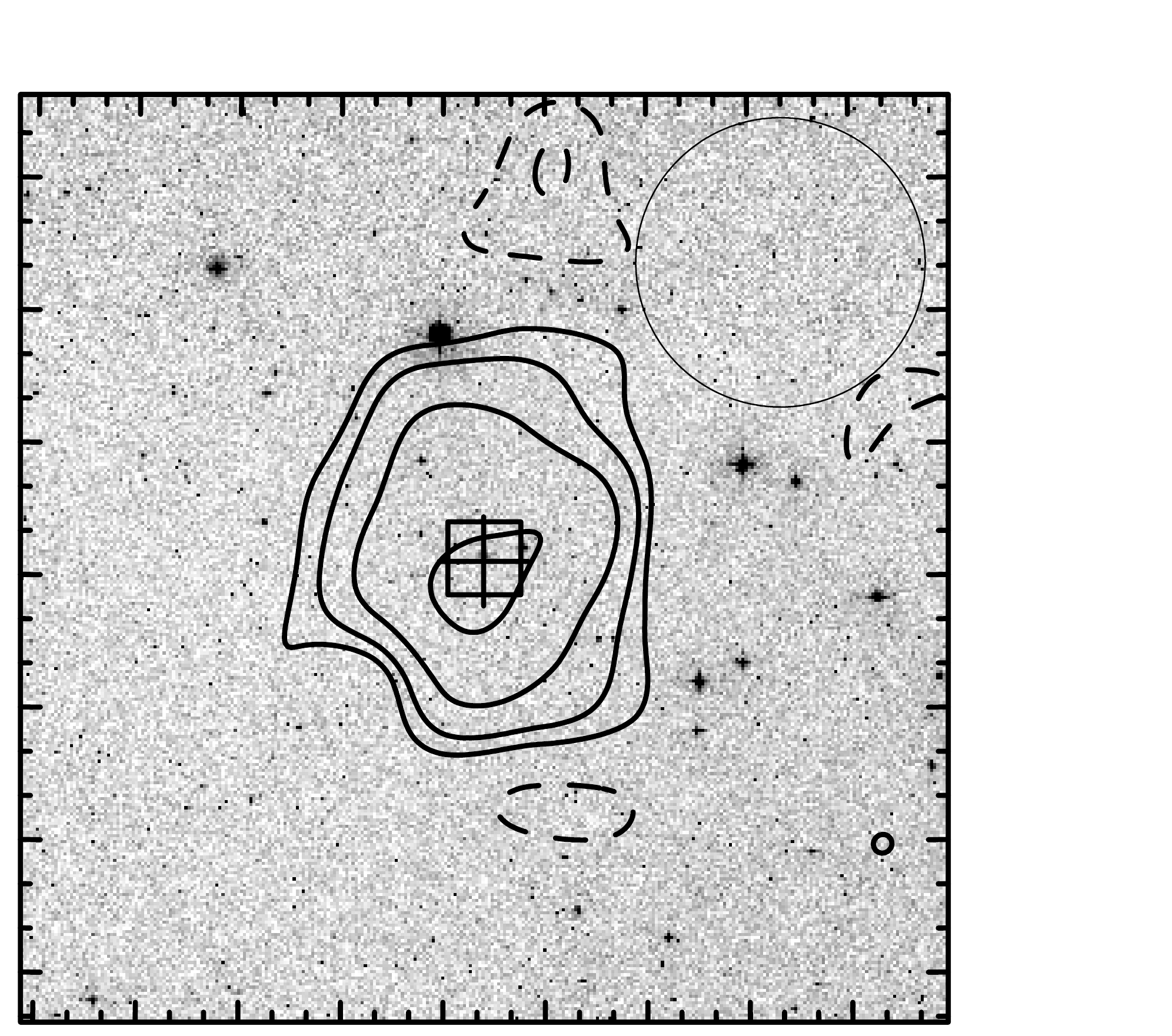}} & 
\mbox{\includegraphics[trim=1cm 1cm 0cm 0cm, clip=true, height=3cm]{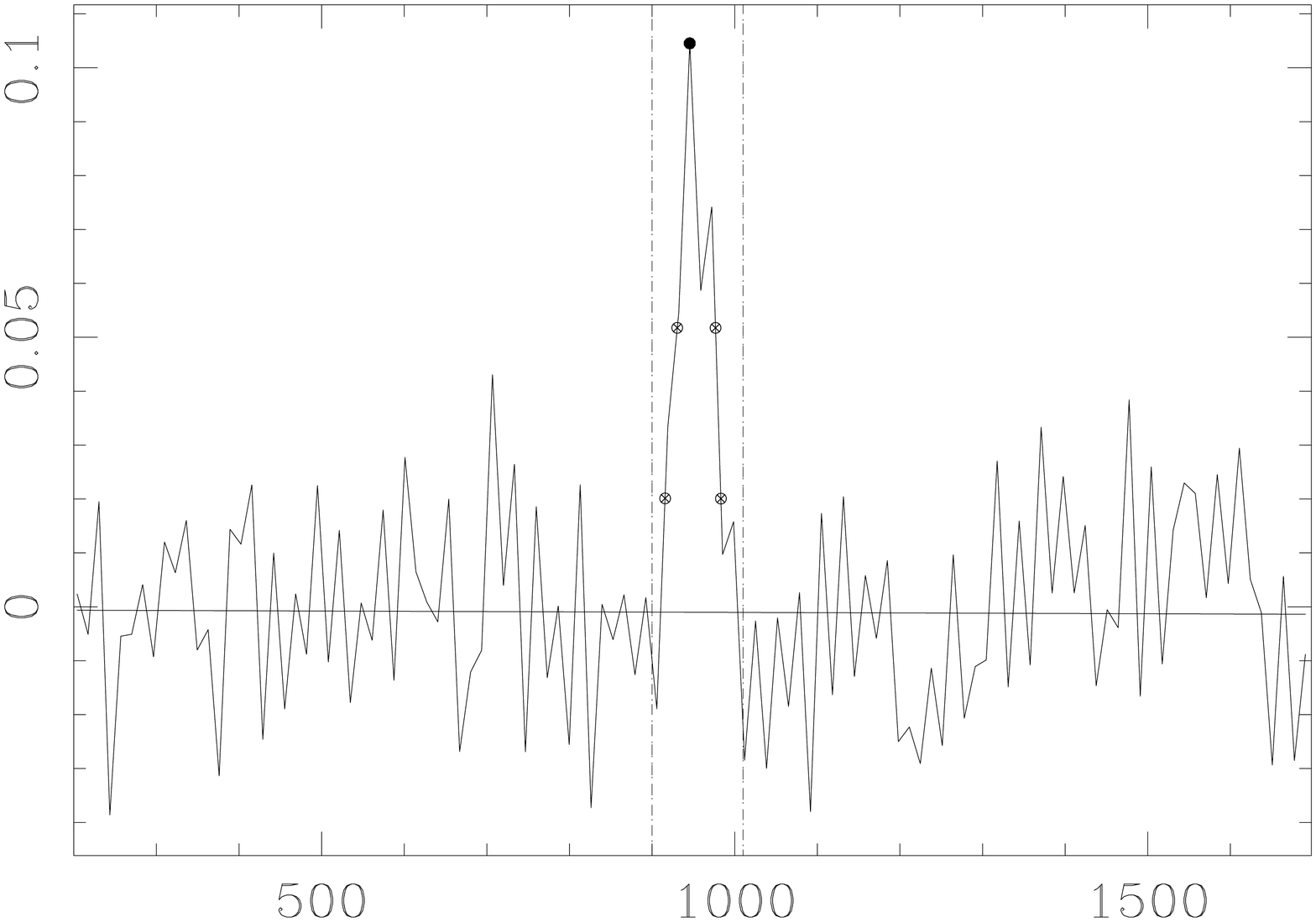}} \\
[12pt]
\end{tabular} 
\caption[HIJASS detections with HI excess --\textit{continued}.]{HIJASS detections with HI excess, i.e. with HI deficiencies $\mathrm{Def_{HI}} \leq -1$ --\textit{continued} (similar to Figure~\ref{fig:HIexcI}).} 
\label{fig:HIexcII}
\end{figure*}

Galaxies with highly disturbed HI content have been found near the cluster cores of e.g. Virgo and Coma. Recent observations have found spiral galaxies in the less dense group environments, which are loosing their HI reservoir. \citet{denes2014} determined scaling relations based on optical and infrared magnitudes and diameters, which can be used to predict the HI content of galaxies. Using the scaling relations in the $B$- and $r$-band, the predicted HI content of Ursa Major galaxies can be compared to the measured HI content (where available), which is shown in Figure~\ref{fig:pred_meas} (\textit{left} the HI observations are referenced in the key). HI measurements presented in \citet{wolfinger2013} are highlighted with stars and squares. Small circles and stars mark galaxies with \textit{one} optical counterpart within the beam, whereas large circles and square indicate confused HI sources with multiple optical counterparts within the beam. When dealing with a confused source, the predicted HI content of all galaxies, which may contribute to the measured HI flux, are summed. The solid line marks the unity relationship -- a number of galaxies with measured HI masses larger than the predicted ones are apparent to the left of the unity relationship. 

The HI deficiency is defined as $$\mathrm{Def_{HI} = log [M_{HI, \: pred}]- log [M_{HI}]}$$ with $\mathrm{M_{HI, \: pred}}$ being the predicted HI mass from scaling relations of isolated galaxies and $\mathrm{M_{HI}}$ the calculated HI mass from the radio observations \citep{haynes1983}. The HI deficiency is plotted against the measured HI mass in Figure~\ref{fig:pred_meas} (\textit{right}). Positive HI deficiency  values indicate galaxies with less HI than predicted (deficient), whereas negative values show galaxies with HI excess relative to their counterparts in the field. We consider galaxies beyond the dashed lines ($\mathrm{Def_{HI}} = \pm 0.6$) to be HI deficient and with HI excess, respectively. Given that HIJASS is a shallow HI survey (HI masses in the range of 10$^{7}$--10$^{10.5}$~M$_{\odot}$) we find more HI excess galaxies than galaxies with an HI deficiency as HI deficient galaxies are more difficult to detect (their HI masses still have to be larger than the detection limit). Red markers indicate galaxies residing in groups, whereas black markers show non-group galaxies. As one might expect, the majority of HI excess galaxies are non-group galaxies and the HI deficient galaxies tend to reside in galaxy groups (some of the individual HIJASS detections are discussed further below). Note that local projected surface densities are only available for galaxies in the complete sample, which are shown in bright colours, whereas faint galaxies or galaxies towards the edges of the studied region are shown with faded colours.

To investigate the dependance of HI deficiency on galaxy environment, the group membership and the local projected surface density are considered. Figure~\ref{fig:hist_HIdef} (\textit{left}) shows the HI deficiency distribution of galaxies residing in groups (red) and galaxies without group membership (black) (galaxies without local projected surface densities available are shown in orange and grey, respectively). The HI deficiency values of non-group galaxies tend to be between -1 and 1, with a few HI excess galaxies lying at negative HI deficiency values. The HI deficiency distribution of galaxies residing in groups shows numerous HI deficient galaxies (positive HI deficiency values). All Ursa Major galaxies with HI measurements available and $\mathrm{Def_{HI}} \geq 1$ reside in galaxy groups. The average HI deficiency of group members is higher than the average HI deficiency of non-group members (red and black line, respectively). Even though these averages reside in adjacent bins, significant different distributions are indicated by the tails.

The HI deficiency as a function of local projected surface density is shown in Figure~\ref{fig:hist_HIdef} (\textit{right}; with markers and colours similar to Figure~\ref{fig:pred_meas}). There is a trend of increasing HI deficiency with increasing local projected surface density, which can be seen in the binned data averaged per 10 data points (\textit{bottom} panel). Up to a local projected surface density of $\mathrm{log (} \Sigma_5 \mathrm{)} = 1$ the HI deficiency values appear to be evenly distributed around 0 followed by a steep increase with higher local projected surface density values suggesting that the HI properties of galaxies change with environment.

\subsubsection{HIJASS detections with HI excess}

The HIJASS detections with the largest HI excesses ($\mathrm{Def_{HI}} < -1$) are shown in Figures~\ref{fig:HIexcI} and \ref{fig:HIexcII}. A DSS2 B-band image with overlaid integrated HI flux contours is shown as well as their HIJASS spectra. The spectra show high signal-to-noise detections making HI excess due to measurement errors unlikely. Some galaxies, such as UGC06817, NGC4395, NGC4144 show signs of HI extension, which can not be parameterized separately. Note that a point source is assumed for UGC07559 and UGC06840, given that the galaxies are part of double/triple detections.

High resolution follow-up observations are required to determine the nature of these HI excess galaxies.

\subsubsection{HI deficient galaxies in HIJASS}

\begin{figure*}
\begin{center}
\begin{tabular}{p{3cm} p{2.5cm}}
\multicolumn{2}{c}{NGC4278 (HIJASS J1220+29) -- $\mathrm{Def_{HI}}=1.18$} \\
\mbox{\includegraphics[trim=7.7cm 3.7cm 0cm 0cm, clip=true, height=3cm]{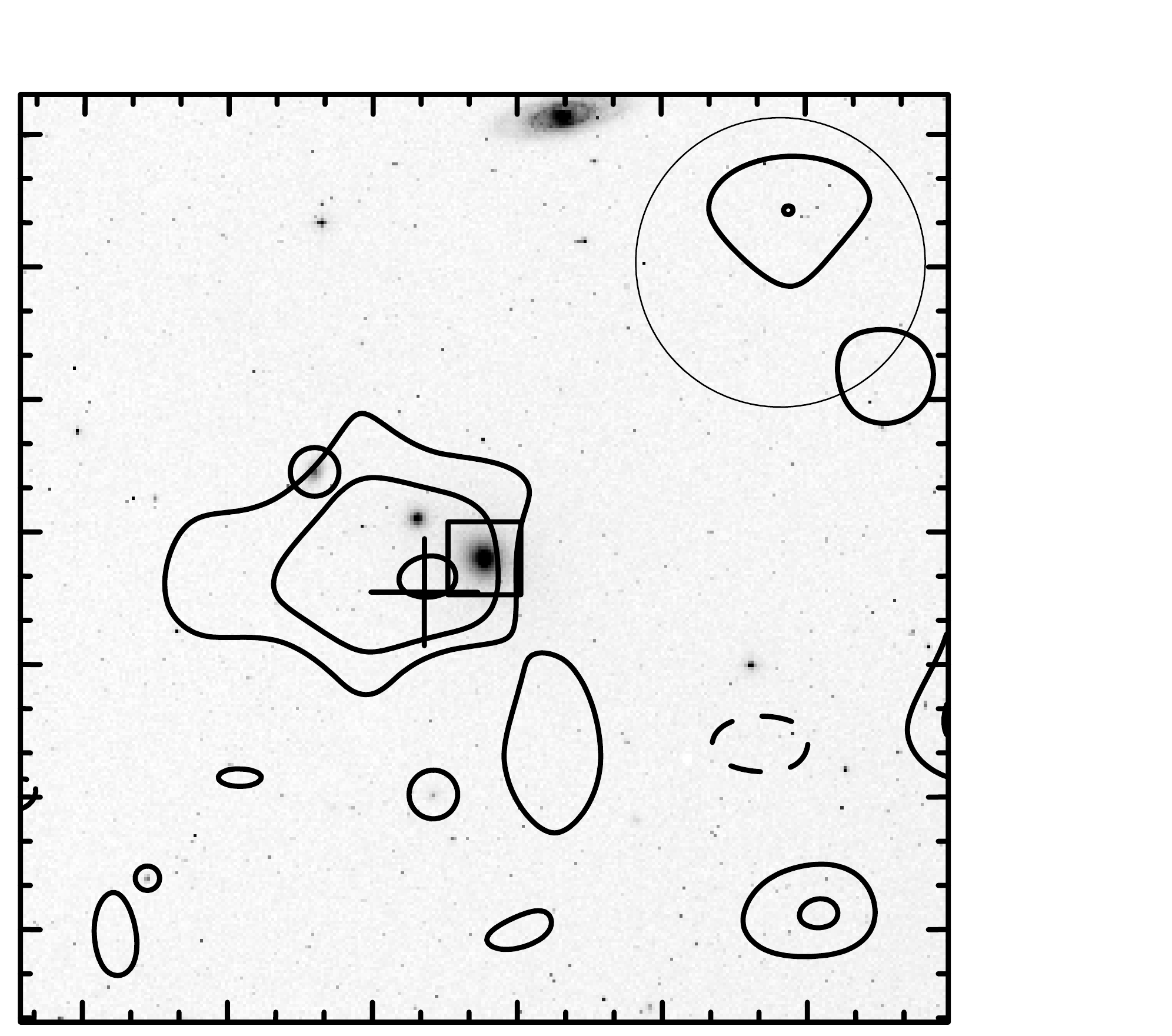}} & 
\mbox{\includegraphics[trim=1cm 1cm 0cm 0cm, clip=true, height=3cm]{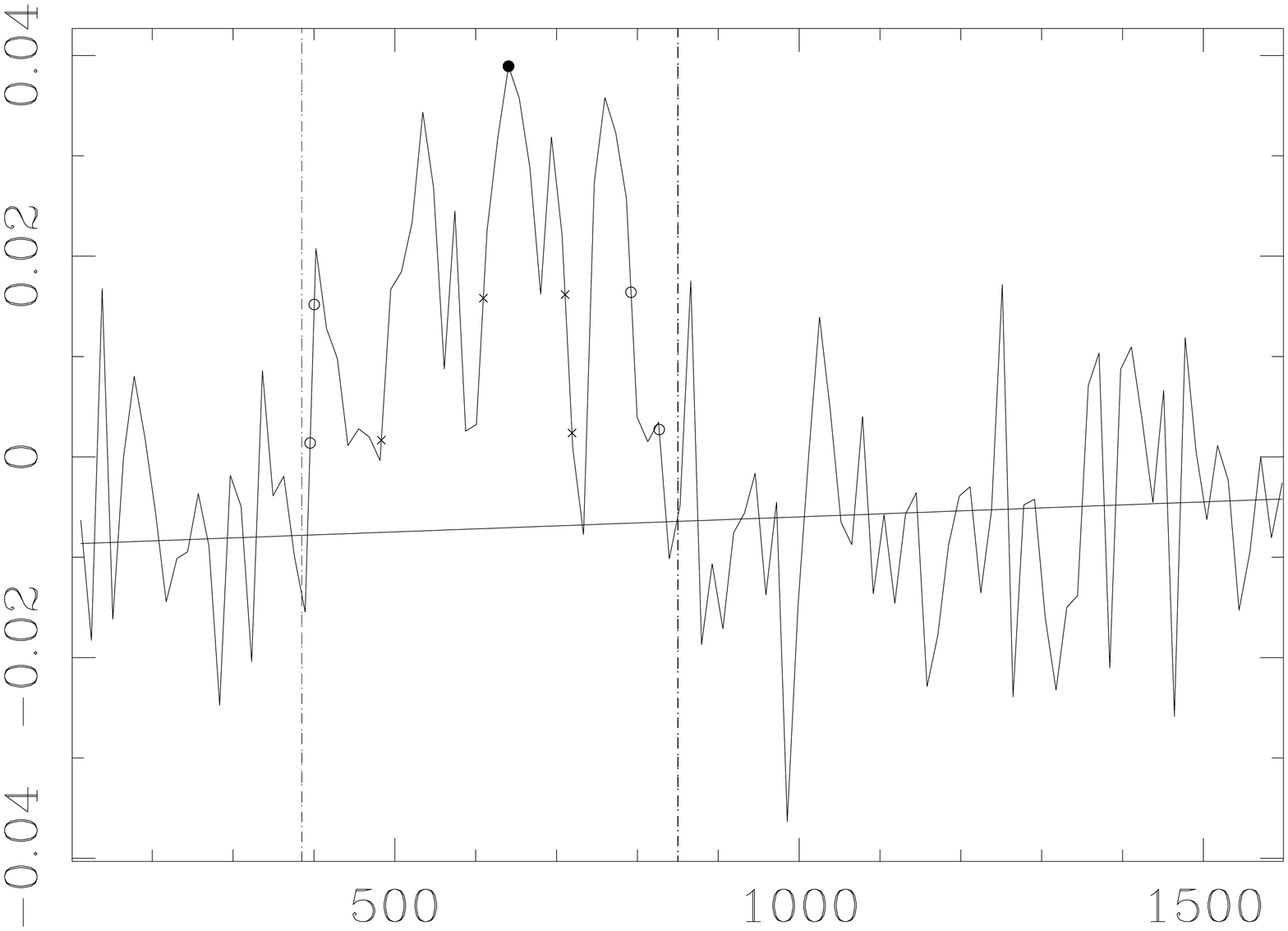}} \\
[12pt]
\end{tabular} 
\caption[HI deficient galaxy detected in HIJASS.]{HI deficient galaxy detected in HIJASS with HI deficiency $\mathrm{Def_{HI}} \geq 1$ (similar to Figure~\ref{fig:HIexcI}). NGC4278 is the most likely optical counterpart to the HIJASS detection (marked by a box). However, HIJASS J1220+29 is a confused HI source as NGC4286 is at matching recession velocity and might also contribute to the measured HI flux (surrounded by a circle). The measured HI flux appears to be between the two galaxies and not evenly distributed around NGC4278.}
\label{fig:HIdefI}
\end{center}
\end{figure*}

\begin{figure}
\includegraphics[trim=0cm 0cm 0cm 0cm, clip=true, width=8cm]{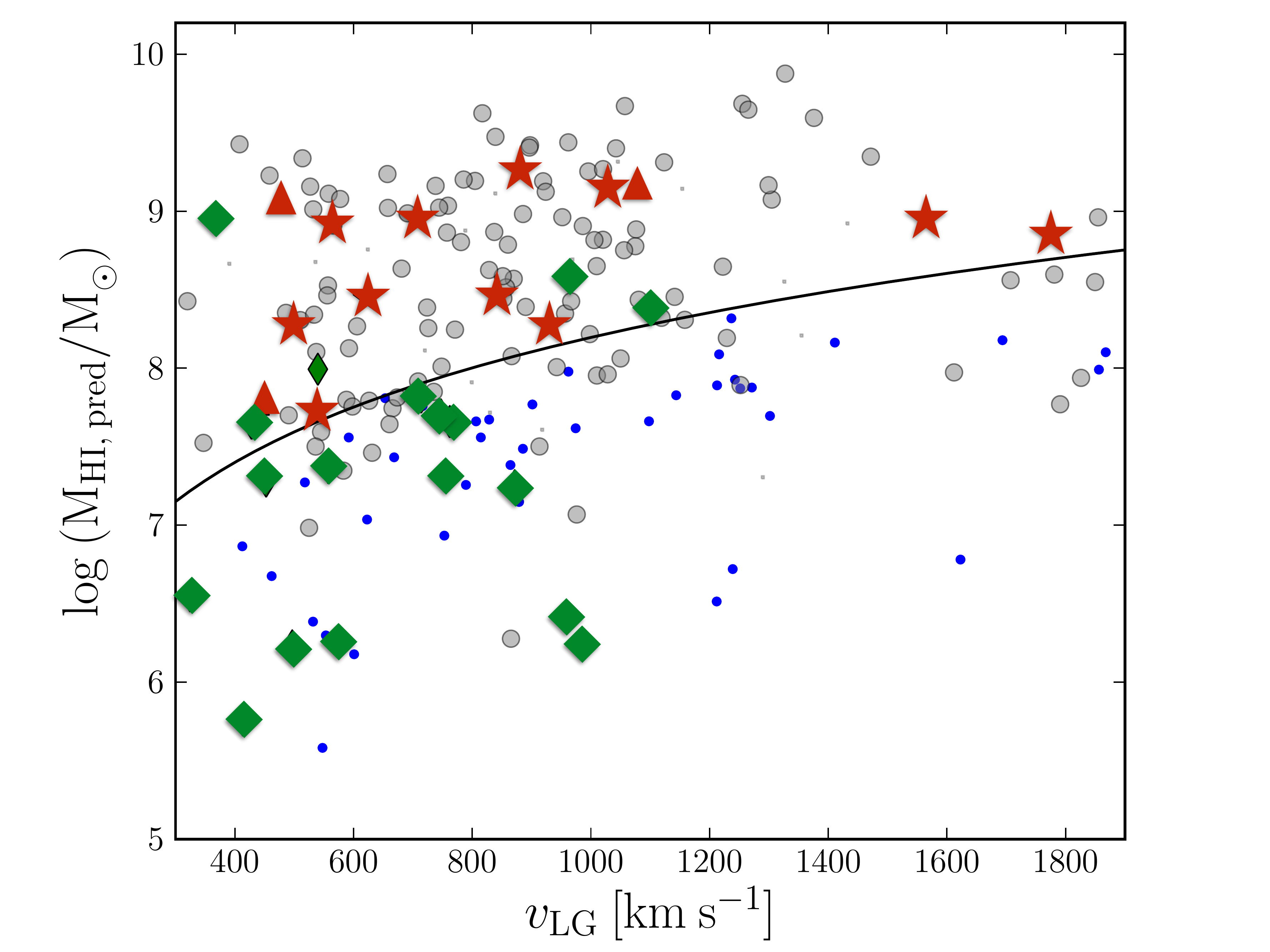}
\caption[Predicted HI masses of galaxies within the HIJASS boundaries plotted against their LG velocities.]{Predicted HI masses of galaxies within the HIJASS boundaries plotted against their LG velocities. Grey circles mark galaxies that are listed in the 5$\sigma$ peak flux HI catalogue -- large circle mark the main detections, whereas small circles indicate galaxies that are part of confused HI detections. The black line indicates the completeness level of the 5$\sigma$ peak flux HI catalogue. Blue points represent galaxies that are not listed in the HIJASS catalogue, which have predicted HI masses below the completeness limit, whereas red stars mark HIJASS non-detections, with predicted HI masses larger than the completeness limit. These galaxies are likely to be HI deficient. The red triangle marks NGC4278 (and NGC4286), which is the only HI deficient galaxy found in the 5$\sigma$ peak flux HIJASS catalogue discussed above. Galaxies with HI excess are marked with green diamonds, which are shown in Figures~\ref{fig:HIexcI} and \ref{fig:HIexcII}.}
\label{fig:HIdefII}
\end{figure}

\begin{figure*}
\begin{tabular}{p{2.5cm} p{3cm} p{1cm} p{2.5cm} p{3cm}}
\multicolumn{2}{c}{NGC4448 ($6.5' \times 6.5'$) -- $\mathrm{Def_{HI}}=1.22$}
 & &
\multicolumn{2}{c}{NGC4314 ($6.5' \times 6.5'$) -- $\mathrm{Def_{HI}}=1.19$}\\
\includegraphics[height=2.5cm]{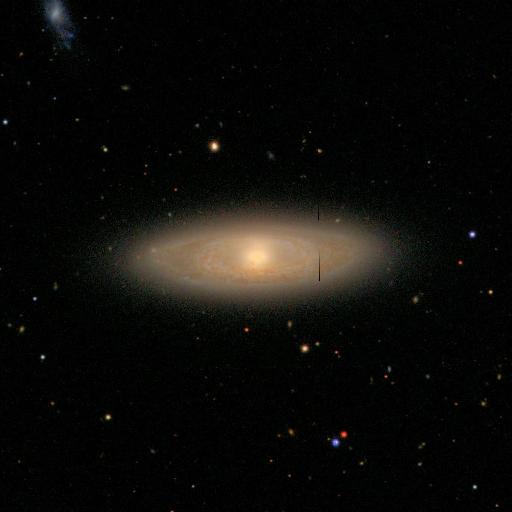} & 
\includegraphics[trim=1cm 1cm 0.5cm 1.5cm, clip=true, height=2.5cm]{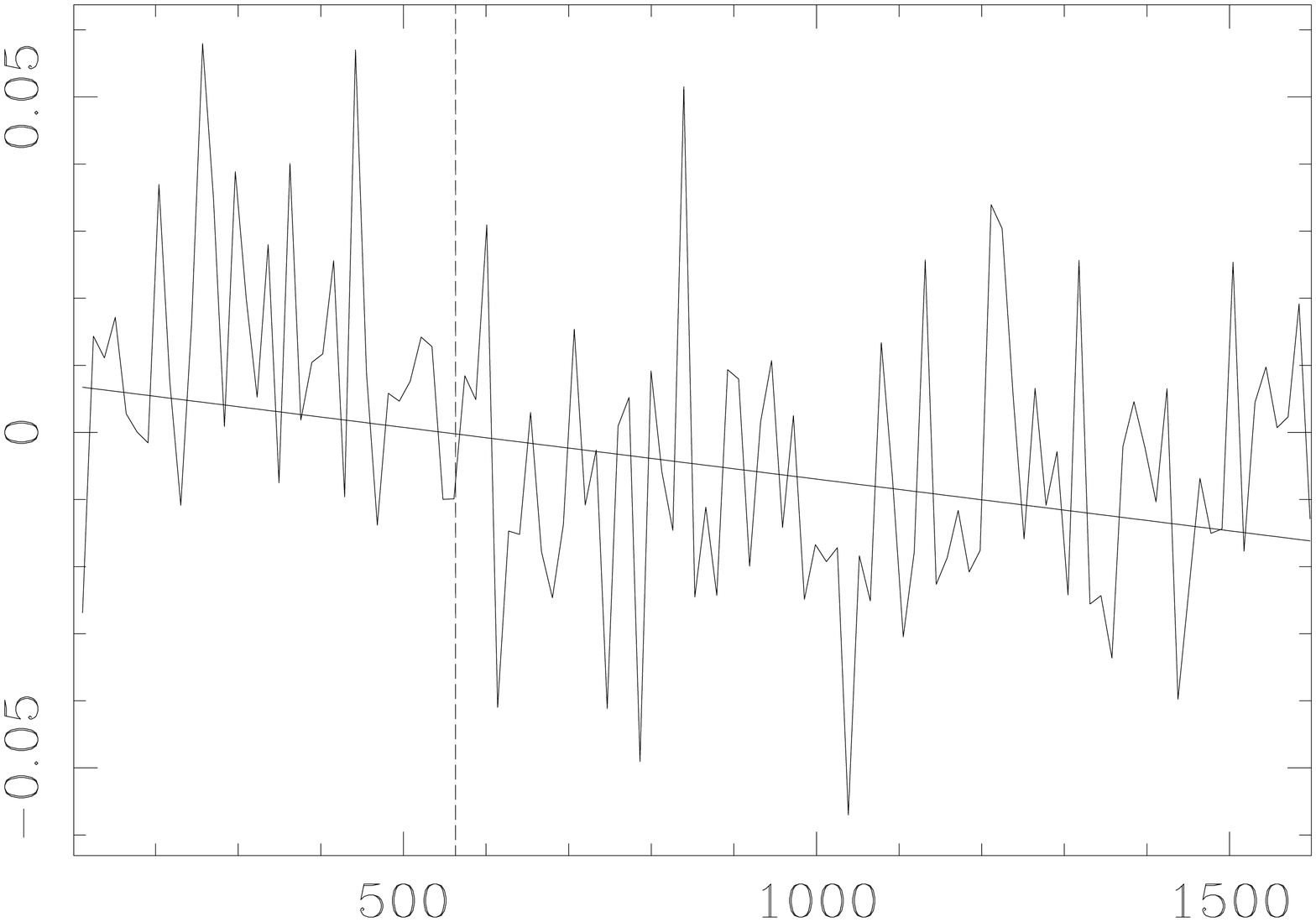} & 
 & 
\includegraphics[height=2.5cm]{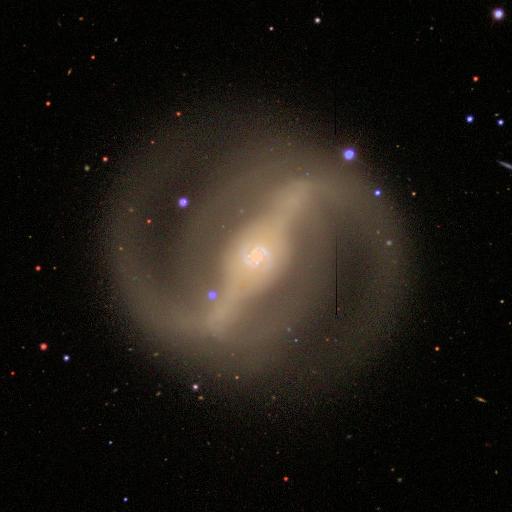} & 
\includegraphics[trim=1cm 1cm 0.5cm 1.5cm, clip=true, height=2.5cm]{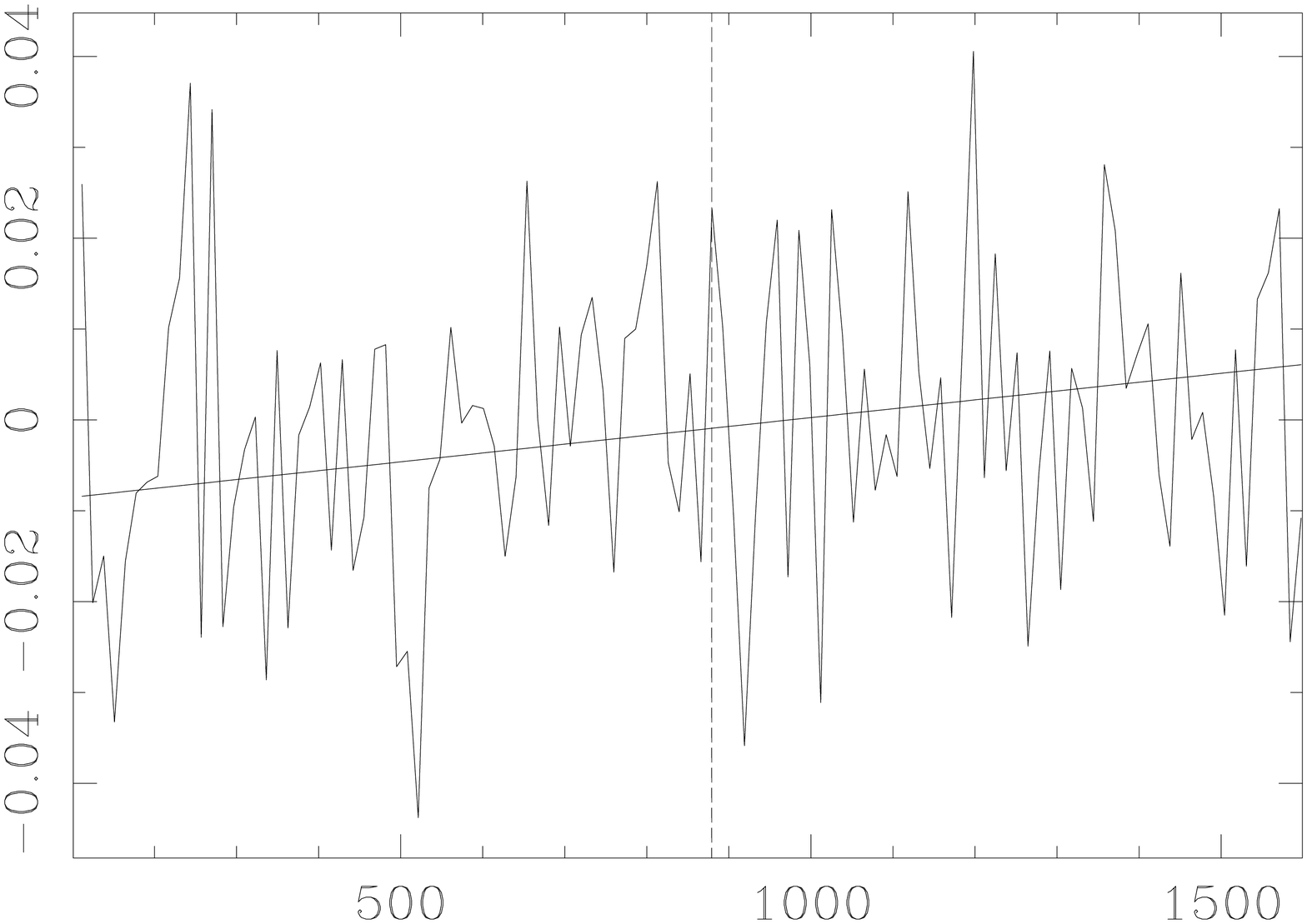}\\
[5pt]
\multicolumn{2}{c}{UGCA298 ($1.6' \times 1.6'$) -- $\mathrm{Def_{HI}}=1.16$}
 & &
 \multicolumn{2}{c}{NGC4080 ($1.6' \times 1.6'$) -- $\mathrm{Def_{HI}}=1.10$} \\
\includegraphics[height=2.5cm]{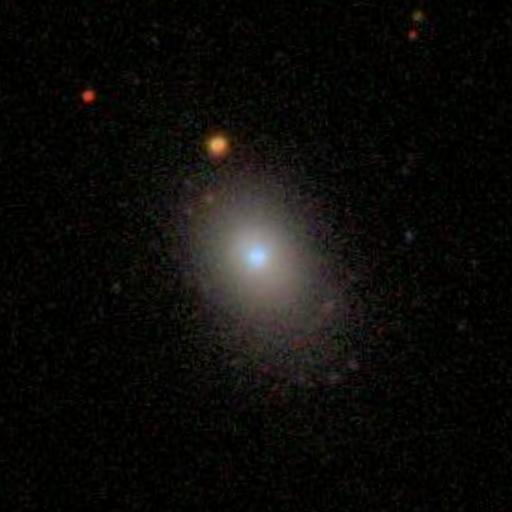} & 
\includegraphics[trim=1cm 1cm 0.5cm 1.5cm, clip=true, height=2.5cm]{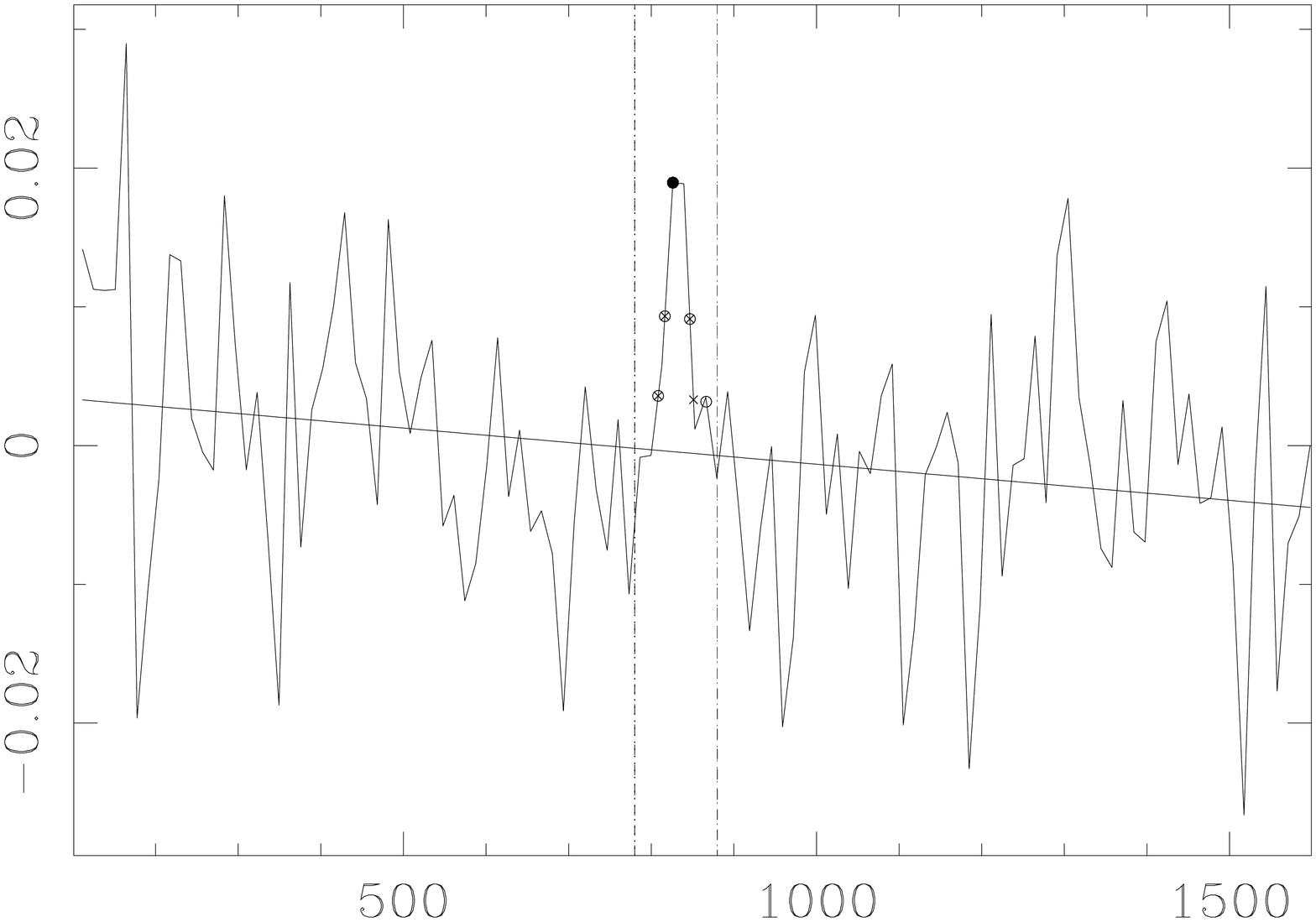} & 
 & 
\includegraphics[height=2.5cm]{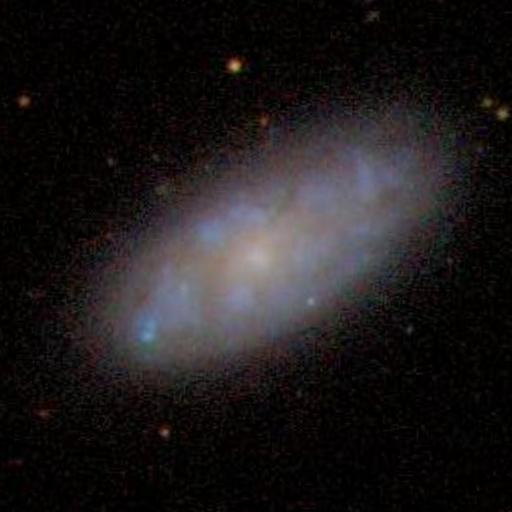} & 
\includegraphics[trim=1cm 1cm 0.5cm 1.5cm, clip=true, height=2.5cm]{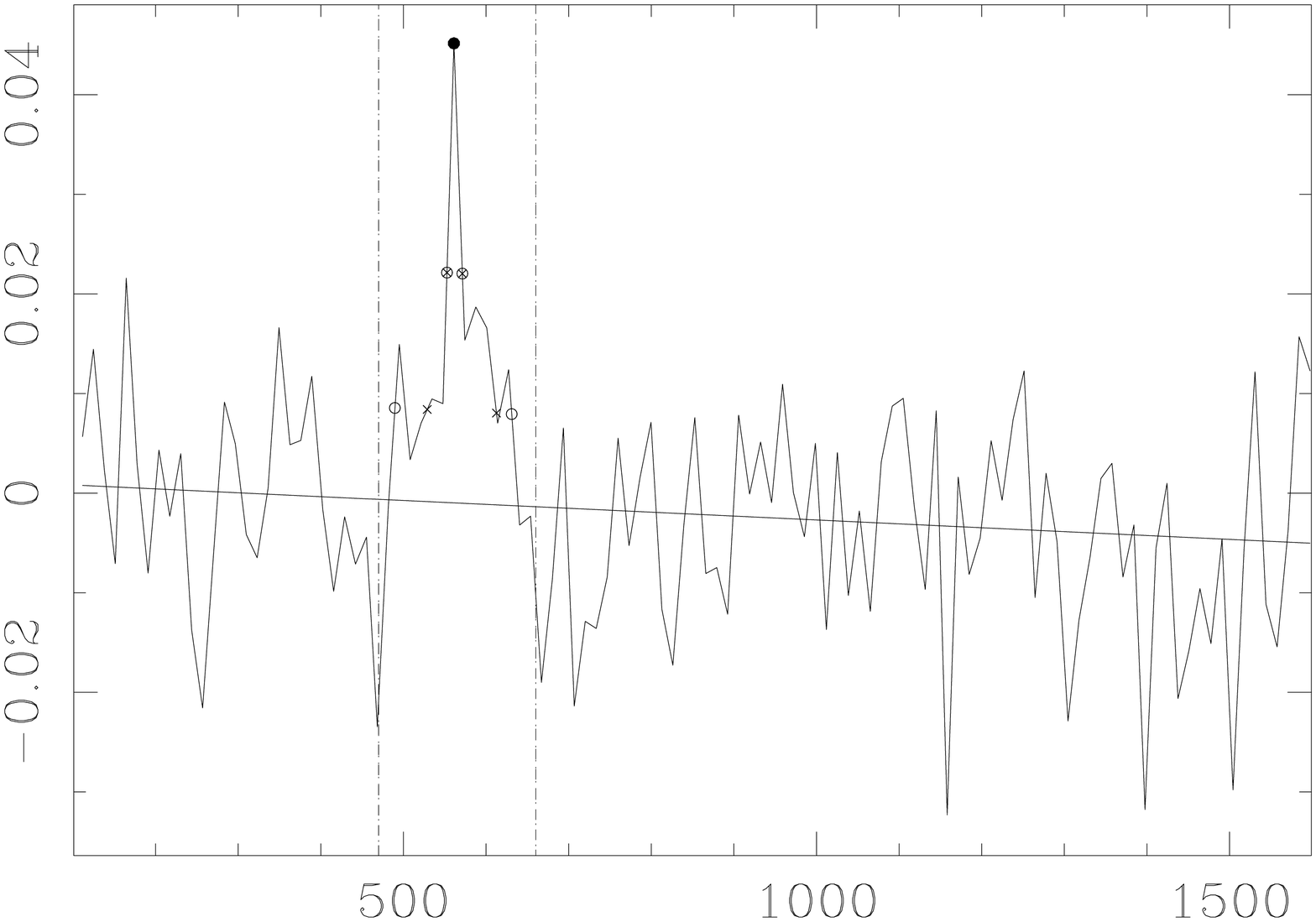}\\
[5pt]
\multicolumn{2}{c}{NGC4245 ($6.5' \times 6.5'$) -- $\mathrm{Def_{HI}}=1.06$}
 & &
 \multicolumn{2}{c}{NGC4220 ($3.3' \times 3.3'$) -- $\mathrm{Def_{HI}}=0.94$} \\
\includegraphics[height=2.5cm]{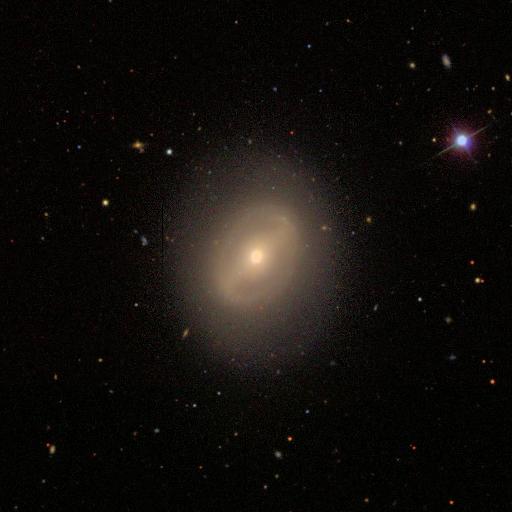} & 
\includegraphics[trim=1cm 1cm 0.5cm 1.5cm, clip=true, height=2.5cm]{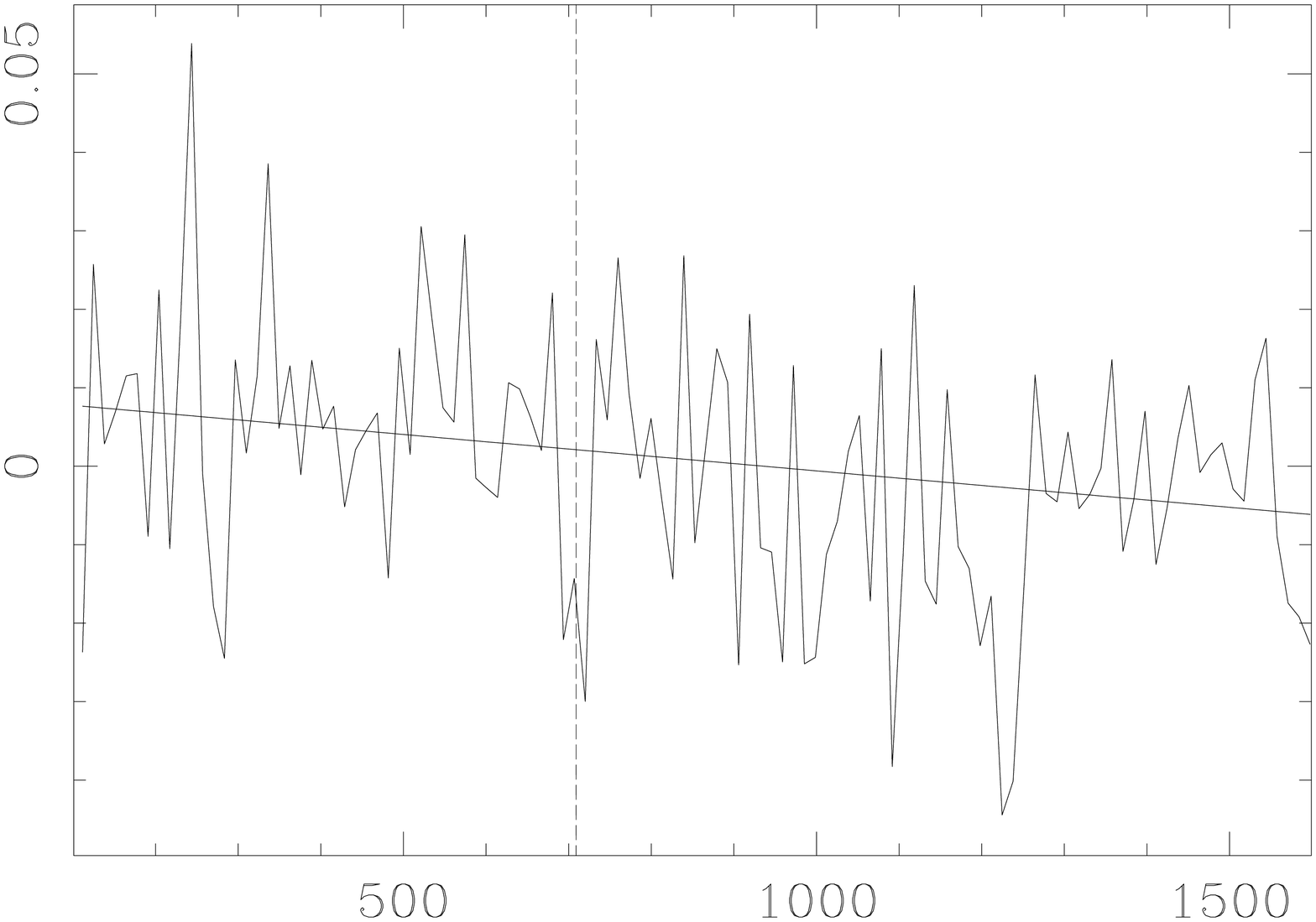} & 
 & 
\includegraphics[height=2.5cm]{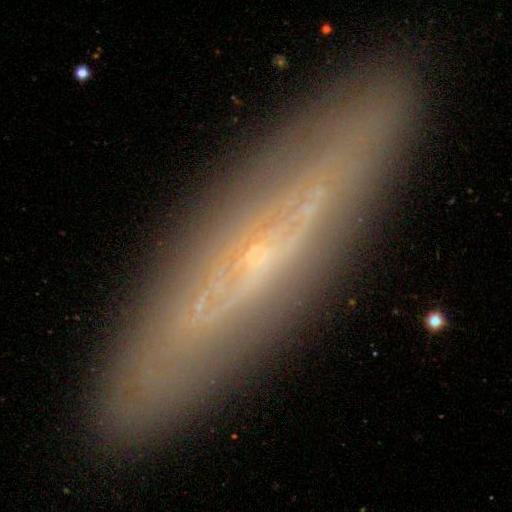} & 
\includegraphics[trim=1cm 1cm 0.5cm 1.5cm, clip=true, height=2.5cm]{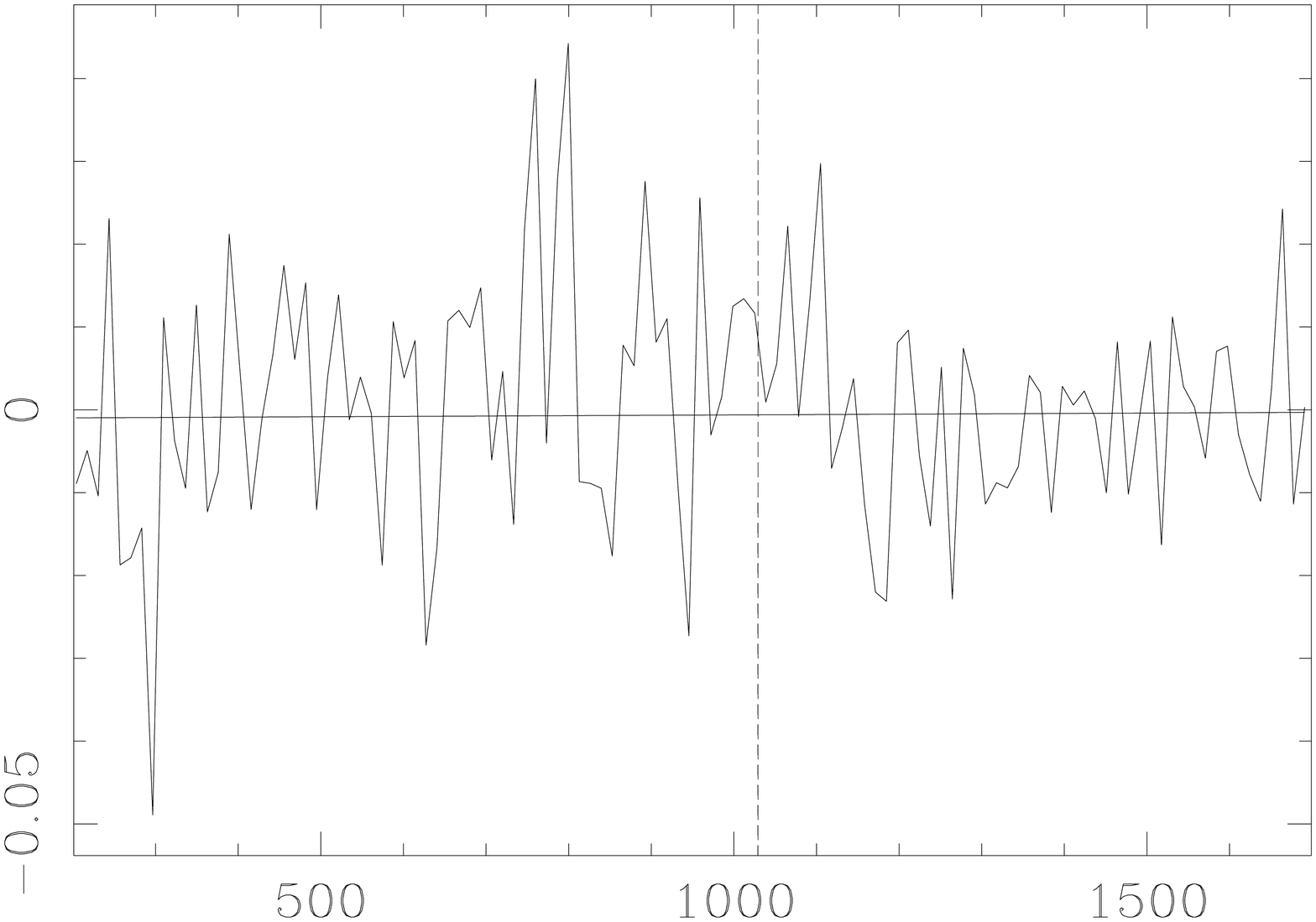}\\
[5pt]
\multicolumn{2}{c}{IC3247 ($3.3' \times 3.3'$) -- $\mathrm{Def_{HI}}=0.073$}
 & &
\multicolumn{2}{c}{IC2957 ($1.6' \times 1.6'$) -- $\mathrm{Def_{HI}}=0.55$} \\
\includegraphics[height=2.5cm]{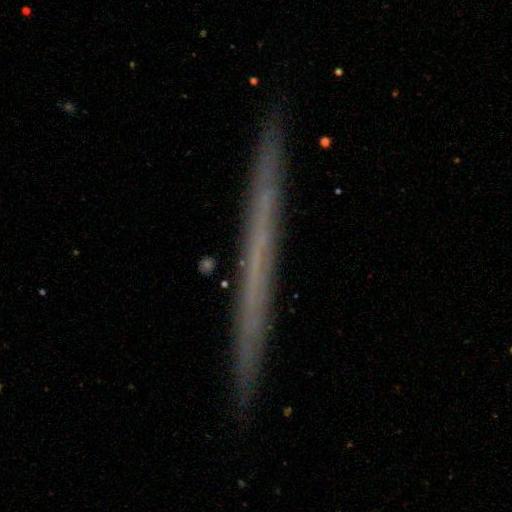} & 
\includegraphics[trim=1cm 1cm 0.5cm 1.5cm, clip=true, height=2.5cm]{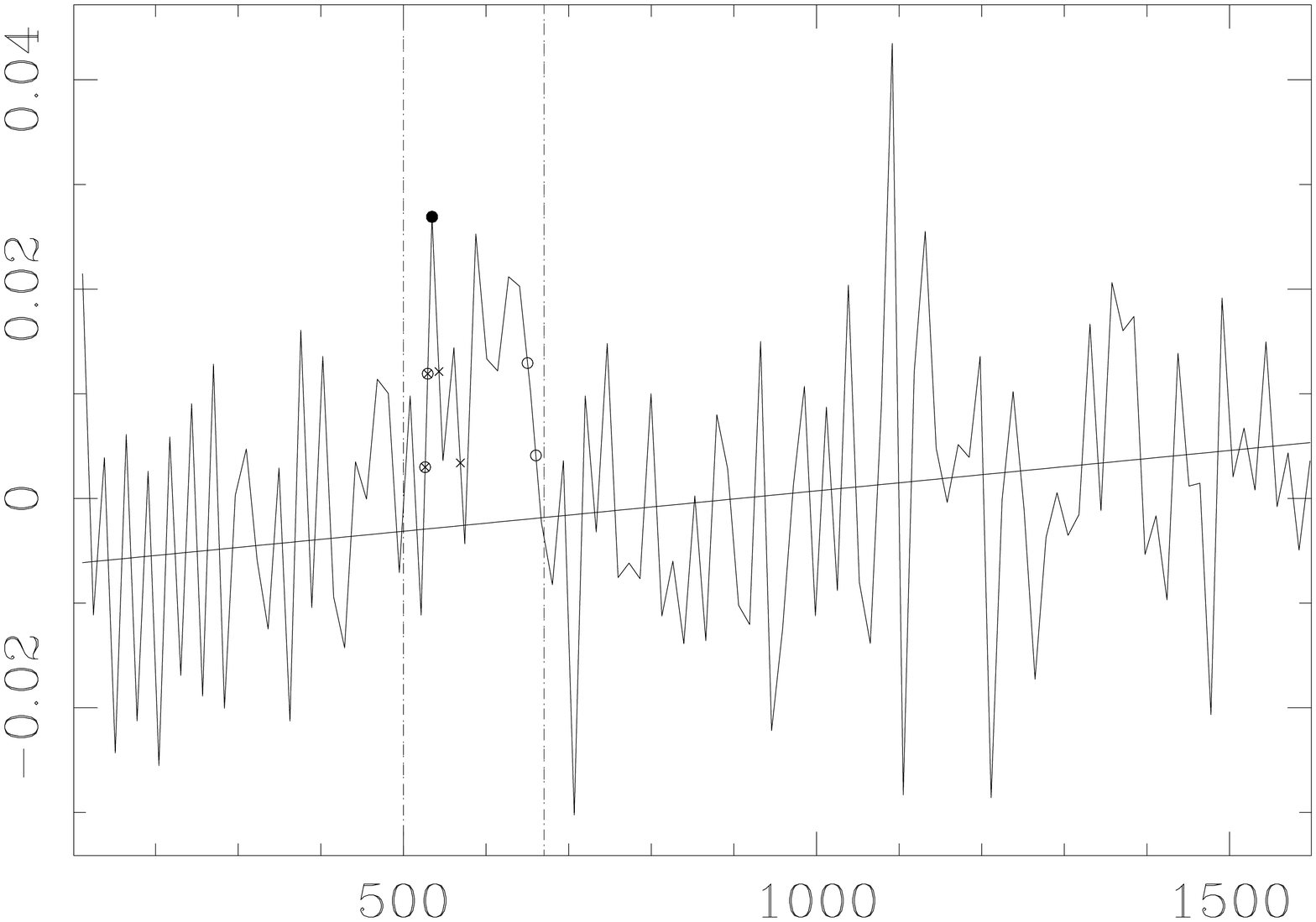} & 
 & 
\includegraphics[height=2.5cm]{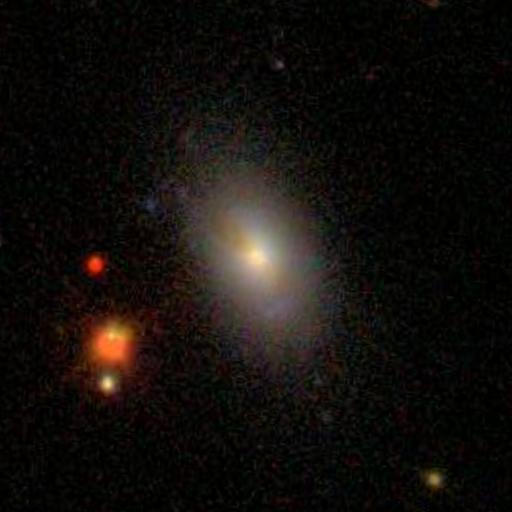} & 
\includegraphics[trim=1cm 1cm 0.5cm 1.5cm, clip=true, height=2.5cm]{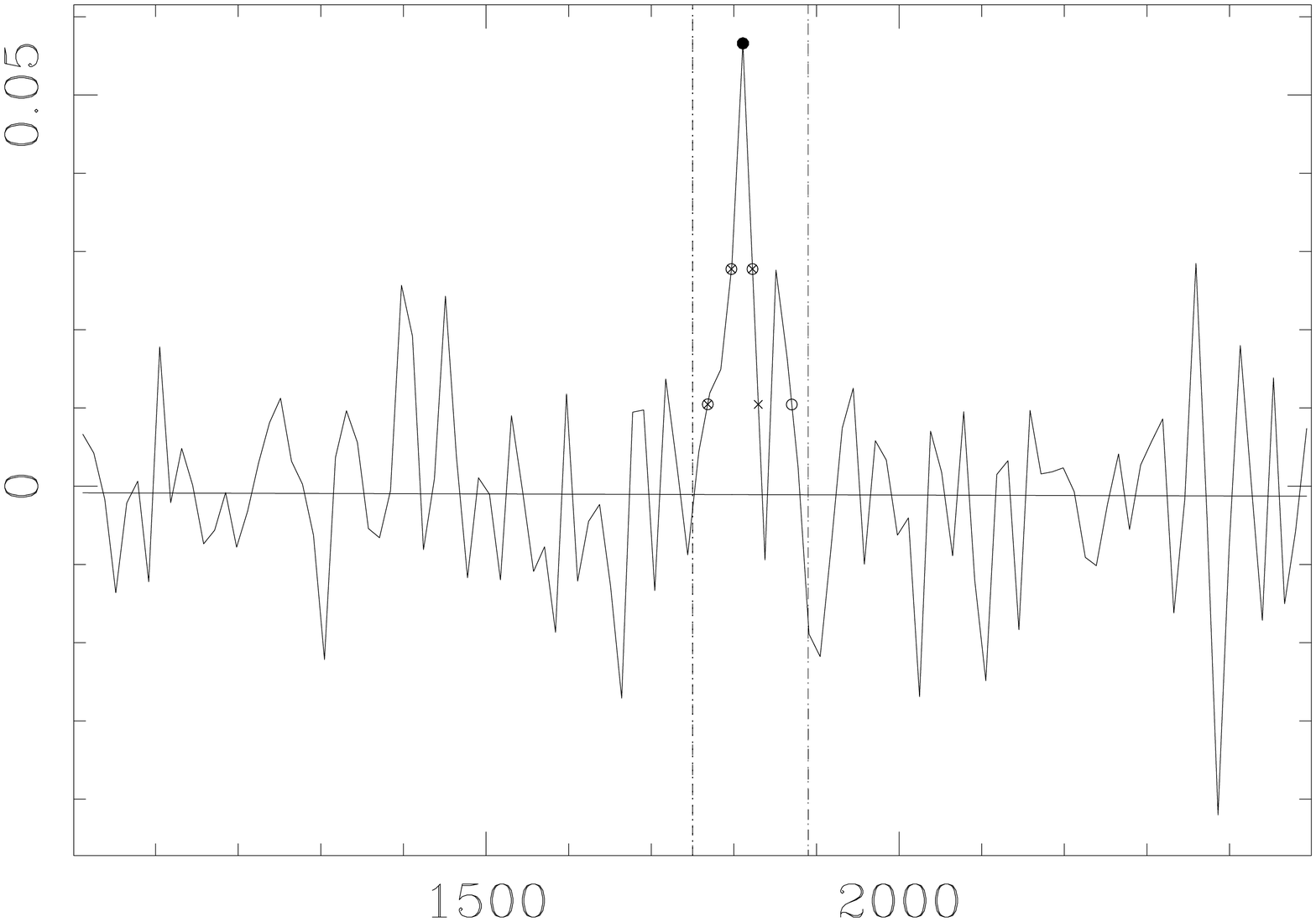}\\
[5pt]
\multicolumn{2}{c}{A1138+35 ($1.6' \times 1.6'$) -- $\mathrm{Def_{HI}}=0.21$}
 & &
\multicolumn{2}{c}{UGC07320 ($1.6' \times 1.6'$) -- $\mathrm{Def_{HI}}=0.07$}\\
\includegraphics[height=2.5cm]{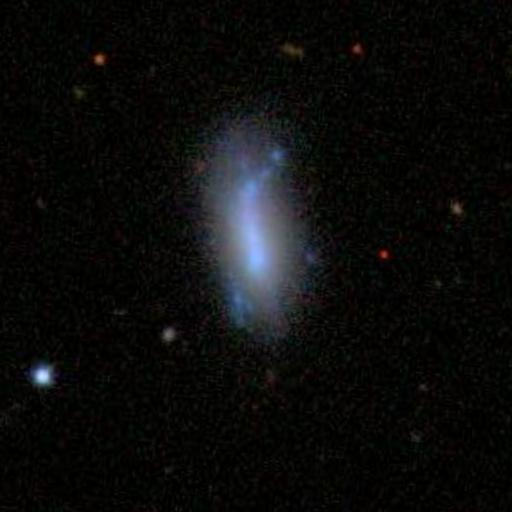} & 
\includegraphics[trim=1cm 1cm 0.5cm 1.5cm, clip=true, height=2.5cm]{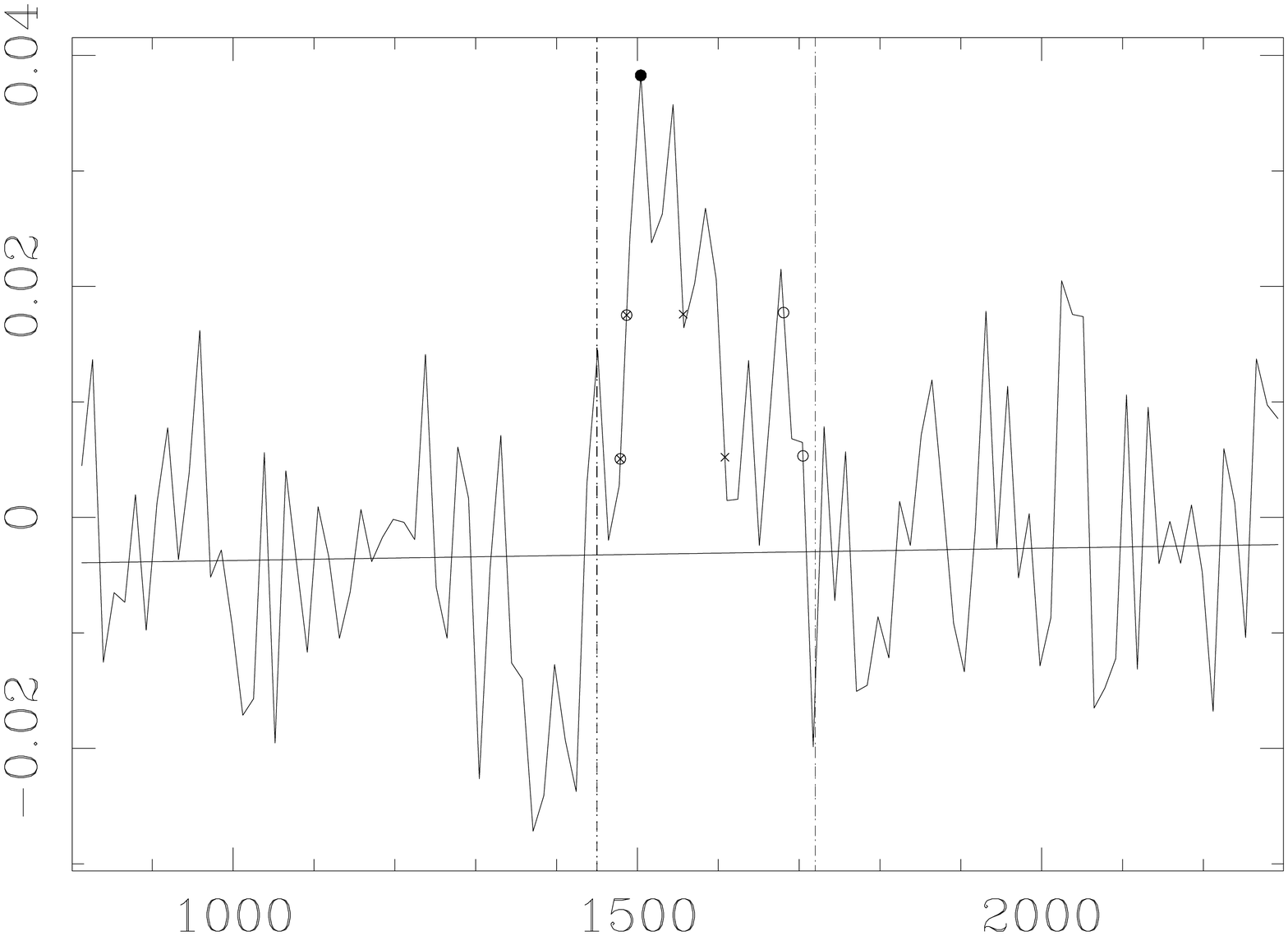} & 
 & 
\includegraphics[height=2.5cm]{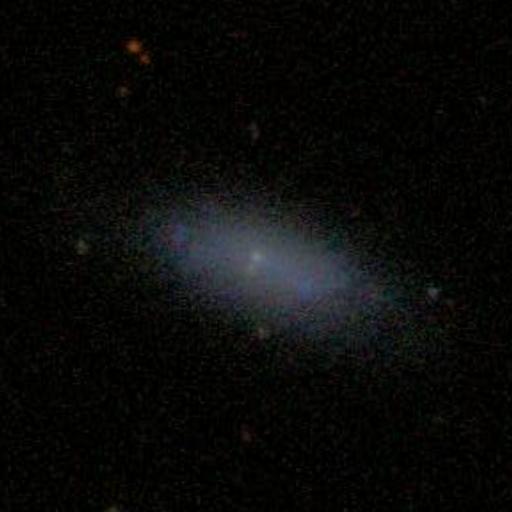} & 
\includegraphics[trim=1cm 1cm 0.5cm 1.5cm, clip=true, height=2.5cm]{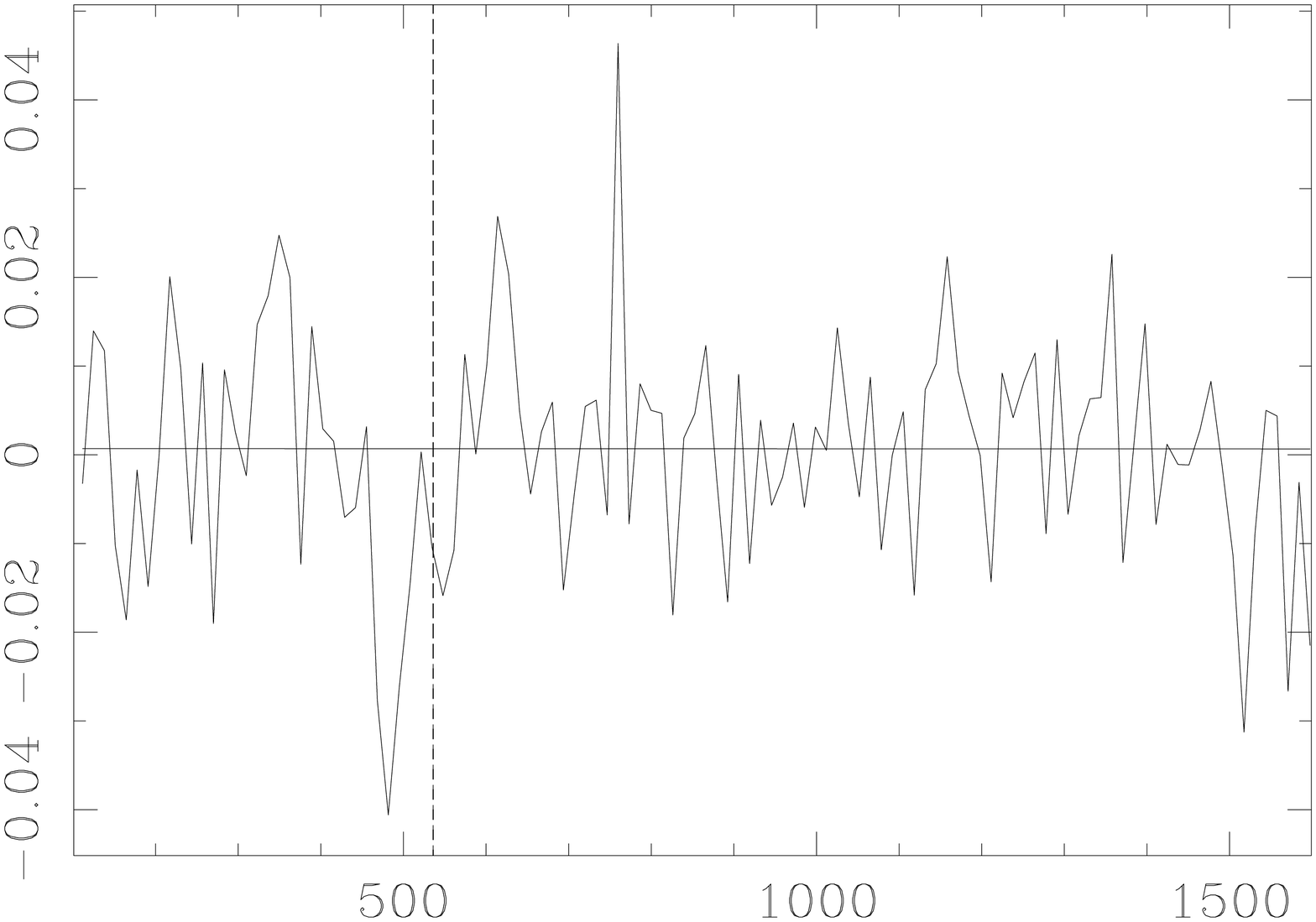}\\
[5pt]
\multicolumn{2}{c}{IC3215 ($3.3' \times 3.3'$) -- $\mathrm{Def_{HI}}=0.04$} & & \\
\includegraphics[height=2.5cm]{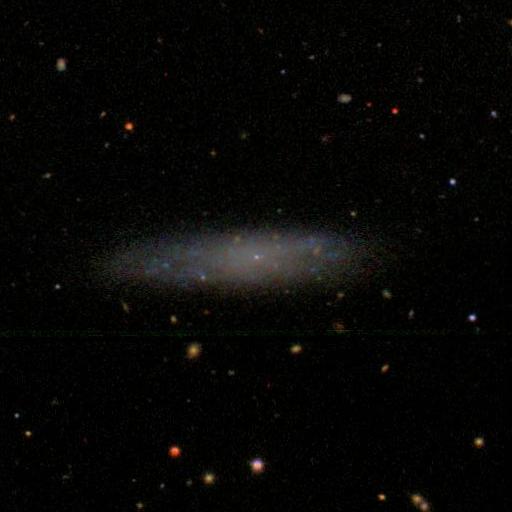} & 
\includegraphics[trim=1cm 1cm 0.5cm 1.5cm, clip=true, height=2.5cm]{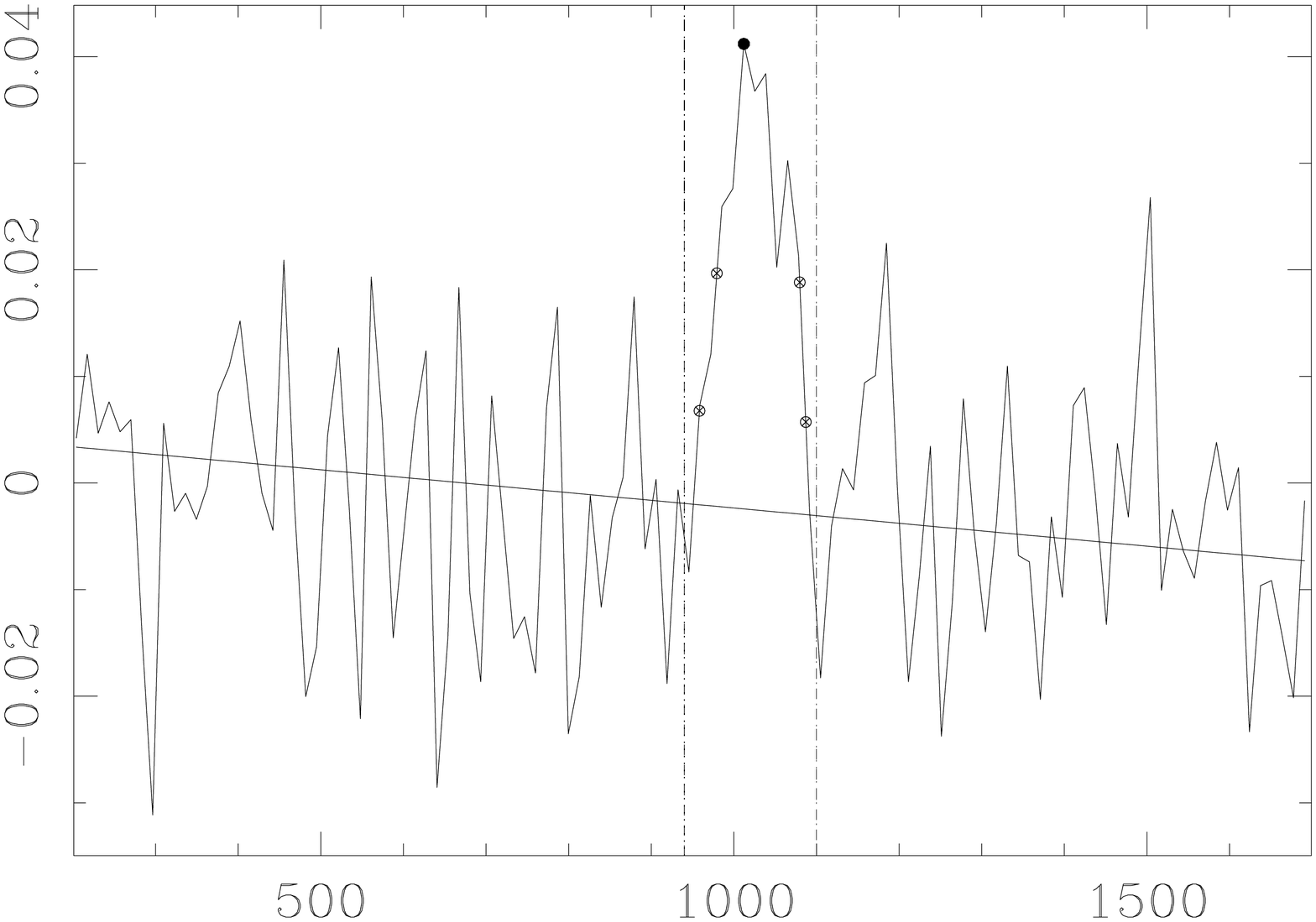} 
& & \\
\end{tabular}
\caption{Multi-colour SDSS images of HIJASS non-detections centred on the optical positions together with the HIJASS spectra (obtained at the optical position). These galaxies are not amongst the 5$\sigma$ detections presented in \citet{wolfinger2013}. The HI deficiency decreases from the left to the right and from the top to the bottom. Galaxy name, image size (in brackets) and HI deficiency is given above each SDSS image and spectrum. The HI spectra have the optical velocities ($cz$ in km~s$^{-1}$) along the abscissas and the flux densities (in Jy) along the ordinates. Galaxy spectra with one dashed vertical line have no obvious HI line emission (the emission should be at the recession velocity of the dashed line; an upper limit on the HI mass is adopted); whereas for galaxies with two dashed vertical lines in the spectra an HI line emission analysis has been conducted similar to \citet{wolfinger2013}.}
\label{fig:HIJASSdef}
\end{figure*}

\begin{table*}
\begin{scriptsize}
\begin{center}
\caption[HI properties of HI deficient galaxies.]{HI properties of HI deficient galaxies as obtained from HIJASS (stars in Figure~\ref{fig:HIJASSdef}). Note that galaxies with $\mathrm{Def_{HI}} < 0.6$ are not considered to be HI deficient.}
\begin{tabular}{lccrcccccc}
\hline\hline\\[-2ex]
Name & RA & Dec. & $v_{\mathrm{LG}}$ & $v_{\mathrm{HI}}$ & $S_{\mathrm{int}}$ & $\mathrm{log(M_{HI, \: pred}})$ & $\mathrm{log(M_{HI}})$ & $\mathrm{DEF_{HI}}$ & group ID \\
(C1) & (C2) & (C3) & (C4) & (C5) & (C6) & (C7) & (C8) & (C9) & (C10)\\
\hline\\[-2ex]
NGC4448  &  12:28:15.46  &  28:37:13.1  &  563  &  -  &  3.55 $^{a}$ &  8.92  &  7.7 $^{b}$ &  1.22  & NGC4278 \\
NGC4314  &  12:22:32.01  &  29:53:43.8  &  879  &  -  &  3.55 $^{a}$ &  9.27  &  8.08 $^{b}$ &  1.19 & NGC4274  \\
UGCA298  &  12:46:55.41  &  26:33:51.4  &  841  &  837  &  0.64  &  8.46  &  7.3  &  1.16  & NGC4725 \\
NGC4080  &  12:04:51.83  &  26:59:33.1  &  622  &  560  &  1.59  &  8.44  &  7.34  &  1.1 & - \\
NGC4245  &  12:17:36.78  &  29:36:28.9  &  709  &  -  &  3.55 $^{a}$ &  8.96  &  7.9 $^{b}$ &  1.06 & NGC4274 \\
NGC4220  &  12:16:11.71  &  47:52:59.7  &  1029  &  -  &  3.55 $^{a}$ &  9.16  &  8.22 $^{b}$ &  0.94  & -\\
IC3247  &  12:23:13.99  &  28:53:37.5  &  499  &  593  &  2.27  &  8.28  &  7.55  &  0.73  & NGC4278\\
IC2957  &  11:45:36.93  &  31:17:58.4  &  1563  &  1819  &  1.7  &  8.95  &  8.4  &  0.55  & -\\
A1138+35  &  11:40:49.04  &  35:12:09.7  &  1773  &  1592  &  3.87  &  8.85  &  8.64  &  0.21 & - \\
UGC07320  &  12:17:28.55  &  44:48:40.4  &  536  &  -  &  3.55 $^{a}$ &  7.72  &  7.65 $^{b}$ &  0.07 & NGC4449 \\
IC3215  &  12:22:10.18  &  26:03:07.7  &  931  &  1023  &  3.67  &  8.27  &  8.23  &  0.04  & -\\
\hline\\[-2ex]
\multicolumn{9}{l}{$^a$ For galaxies without definite HI emission in the HIJASS spectra, an upper limit is assumed.} \\
\multicolumn{9}{l}{$^b$ This is an upper limit on the HI mass for galaxies without definite HI emission in the HIJASS spectra.} \\
\end{tabular}
\label{tab:HI_def}
\end{center}
\end{scriptsize}
\end{table*}

The HI deficient galaxies shown in Figure~\ref{fig:pred_meas} \textit{right} tend to reside in galaxy groups (the majority of data points with positive HI deficiency values are shown in red). There are only a few HI deficient galaxies where HIJASS data is utilized (stars and squares). NGC4278 is the only HI deficient HIJASS detection with $\mathrm{Def_{HI}} > 1$, which is shown in Figure~\ref{fig:HIdefI} (similar to Figure~\ref{fig:HIexcI}). Given that HIJASS J1220+29 is a confused HI detection, the predicted HI mass is calculated as the sum of predicted HI mass of NGC4278 and NGC4286 (circled to the left of NGC4278). The measured HI flux appears to be between the two galaxies and not evenly distributed around NGC4278. Previous WSRT observations of NGC4278 are presented in \citet{raimond1981}. The measured integrated HI flux obtained from the WSRT observations is in good agreement with the HIJASS measurement, suggesting that NGC4286 does not contribute significantly to the measured integrated HI flux of HIJASS J1220+29. \citet{morganti2006} find in deep WSRT observations that NGC4278 shows an extended regular HI disc with two faint tail-like structures offset from the host galaxy.

Given that HIJASS is a shallow HI survey (HI masses in the range of 10$^{7}$--10$^{10.5}$~M$_{\odot}$), HI deficient galaxies are difficult to find in the 5$\sigma$ peak flux HI catalogue as their HI masses still have to be larger than the detection limit. To investigate possible HI deficient galaxies, the predicted HI masses of galaxies within the boundaries of the blind HI survey are plotted versus their LG velocities (see Figure~\ref{fig:HIdefII}). Assuming the catalogue completeness of 99~per~cent at $S\mathrm{_{int}} =$~3.55~Jy~km~s$^{-1}$ as discussed in \citet{wolfinger2013}, the HI completeness limit can be calculated, which is indicated by the black line. The blue points mark galaxies that are not listed in the HIJASS catalogue, which have predicted HI masses below the completeness limit. The red stars mark galaxies, with predicted HI masses larger than the completeness limit, which are HIJASS non-detections and therefore likely to be HI deficient. The red triangle marks NGC4278 (and NGC4286), which is the only HI deficient galaxy found in the 5$\sigma$ peak flux HIJASS catalogue discussed above. Note that the HI excess galaxies (with $300 \leq v_{\mathrm{LG}} \leq 1900$~km~s$^{-1}$) found in the HIJASS data are marked with green diamonds.

The 11 HIJASS non-detections with predicted HI masses above the completeness level are further investigated (red stars in Figure~\ref{fig:HIdefII}). The galaxy parameters are measured from the original HIJASS cube similar to \citet{wolfinger2013} by using the optical position (RA, Dec and $v_{\mathrm{LG}}$) as input parameters. The measured HI properties are given in Table~\ref{tab:HI_def}. The columns are as follows: column~(C1) gives the galaxy name; columns~(C2)-(C4) list the optical position (RA [J2000], Dec. [J2000]) and recession velocity [km~s$^{-1}$]); column~(C5) gives the systemic HI velocity (km~s$^{-1}$ where available); columns~(C6) lists the measured integrated flux or an upper limit on the integrated flux ($S\mathrm{_{int}}$); columns~(C7) and (C8) contain the predicted and measured HI mass; column~(C9) gives the obtained $\mathrm{Def_{HI}}$ and column~(C10) contains the group membership (ID). The HIJASS spectra and multi-colour SDSS images of the galaxies are shown in Figure~\ref{fig:HIJASSdef}. Galaxy spectra with one dashed vertical line have no obvious HI line emission (the emission should be at the recession velocity of the dashed line). For galaxies without HI line emission, an upper limit on the HI mass is adopted, which is the value of the completeness level (black line) at their recession velocity. One can see that the galaxies with the largest HI deficiencies tend to be red spiral galaxies residing preferentially in galaxy groups in the southern part of the Ursa Major region (at declinations between 26 and 30$^{\circ}$), i.e. towards the Virgo cluster.

\section{Discussion}
\label{sec:discussion}
\subsection{Literature comparison}
\label{sec:lit_comp}

\citet{tully1996} identified 79 high-probability `Ursa Major cluster' members, which lie within a projected 7.5$^{\circ}$ circle centred at $ \alpha = 11^h56.9^m, \delta = +49^{\circ}22'$ and with heliocentric velocities between 700 and 1209~km~s$^{-1}$. The large scale `cluster' definition by Tully et al. (1996; hereafter T96) is best compared to this work when using the FoF algorithm with `looser' linking lengths $D_0=0.43$~Mpc and $V_0=120$~km~s$^{-1}$. Of the 79 original T96 group members, 54 galaxies (68~per cent) are within the MESSIER106 group (centred at RA = 12:04:37.62, Dec. = 46:45:33.2 and $v_{LG} = 760$~km~s$^{-1}$  with 185 members in the full and 52 galaxies in the complete sample; see Figure~\ref{fig:C5_litComp} \textit{top}). Note that galaxy groups are labelled by their brightest group galaxy (BGG). Furthermore, four galaxies are part of the NGC3972 group (centred at RA = 11:54:29.37, Dec. = 55:23:09.8 and $v_{LG} = 740$~km~s$^{-1}$ with 5 group members in the complete and full sample) and another four galaxies are part of the NGC3992 group (centred at RA = 11:57:38.07, Dec. = 53:22:30.8 at $v_{LG} = 1267$~km~s$^{-1}$ with 5 group members in the full and 4 in the complete sample, see Figure~\ref{fig:C5_litComp}). Only three of the T96 members reside within the the NGC3998 group (centred at RA = 11:55:06.46, Dec. = 55:36:03.7 at $v_{LG} = 1064$~km~s$^{-1}$), which is a sizeable conglomeration to the North (26 group members in the full and 15 in the complete sample) also shown in Figure~\ref{fig:C5_litComp} (\textit{top}). The remaining 14 T96 Ursa Major `cluster' members are non-group galaxies according to this study. 

Given that the majority of Ursa Major `cluster' members as defined by \citet{tully1996} are members of the `loose' identified MESSIER106, the NGC3972 and the NGC3992 groups, one can consider the Ursa Major `cluster' to consist of these three groups with a total of 195 group members. Therefore this study increases the number of Ursa Major `cluster' members from the 79 members identified in \citet{tully1996} to 195 galaxies, i.e. the number of members increases by approximately 150~per cent. The properties of the dominant and `loose' idetified MESSIER106 group are: velocity dispersion of 205~km~s$^{-1}$, virial mass of $7.96 \times 10^{13}~\mathrm{M}_{\odot}$, crossing time of $0.14~H_0^{-1}$, total luminosity of $1.19 \times 10^{11}~\mathrm{L}_{\odot}$, mass-to-light ratio of 670, a lower total HI mass limit of $7.96 \times 10^{10}~\mathrm{M}_{\odot}$, the fraction of early-type galaxies is 23~per~cent and the maximal radial extent is 1.62~Mpc. However, using the `looser' linking lengths $D_0=0.43$~Mpc and $V_0=120$~km~s$^{-1}$ result in galaxy groups with a radial extents that are overestimated.

Recently Makarov \& Karachentsev (2011; hereafter MK11) presented an all-sky catalogue of 395 nearby groups (with radial velocities $v_{LG} < 3500$~km~s$^{-1}$, excluding the Milky Way zone $\mid b \mid > 15^{\circ}$) using a variant of the FoF technique combined with a hierarchical approach. The authors identified seven bound groups in the Ursa Major region, which are discussed in more detail in \citet{karachentsev2013} and shown in Figure~\ref{fig:C5_litComp} (\textit{bottom}; the seven bound groups are shown in colour, whereas overlapping groups in projection are shown in black). The structures identified in MK11 and \citet{karachentsev2013} (see Figure~\ref{fig:C5_litComp} \textit{bottom}) are best compared to the groups found using the FoF algorithm with `strict' linking lengths $D_0=0.30$~Mpc and $V_0=150$~km~s$^{-1}$ (see Figure~\ref{fig:C5_litComp} \textit{middle}). For clarity, the MK11 groups are colour-coded according to their group membership, whereas the galaxies in the FoF groups are coloured to indicate their recession velocities ($v_{LG} \leq 600$~km~s$^{-1}$ in red; $600 < v_{LG} \leq 900$~km~s$^{-1}$ in orange; $900 < v_{LG} \leq 1200$~km~s$^{-1}$ in magenta and $1200 < v_{LG} \leq 1500$~km~s$^{-1}$ in violet). Galaxies in the groups are shown with circles that are connected to the group centres (line), whereas non-group galaxies are indicated by small boxes. The brightest group galaxies are labelled in the Figures. The best FoF group matches to the MK11 groups are discussed below sorted by decreasing number of MK11 group members ($N_{m,lit}$): 

\begin{figure}
\centering
\includegraphics[trim=1.1cm 6.1cm 1cm 4.5cm, clip=true, width=7.cm]{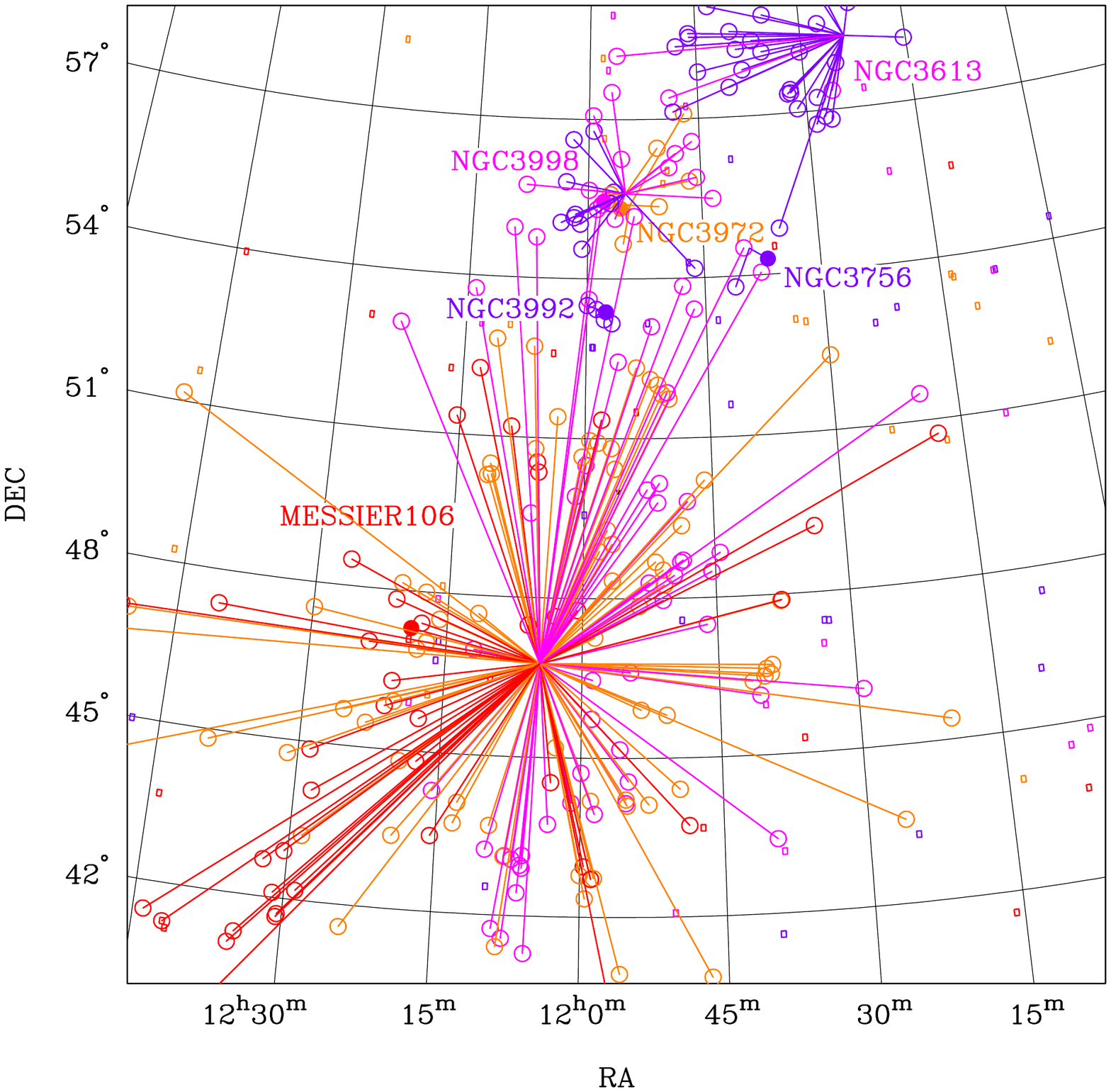}
\includegraphics[trim=1.1cm 6.1cm 1cm 4.5cm, clip=true, width=7.cm]{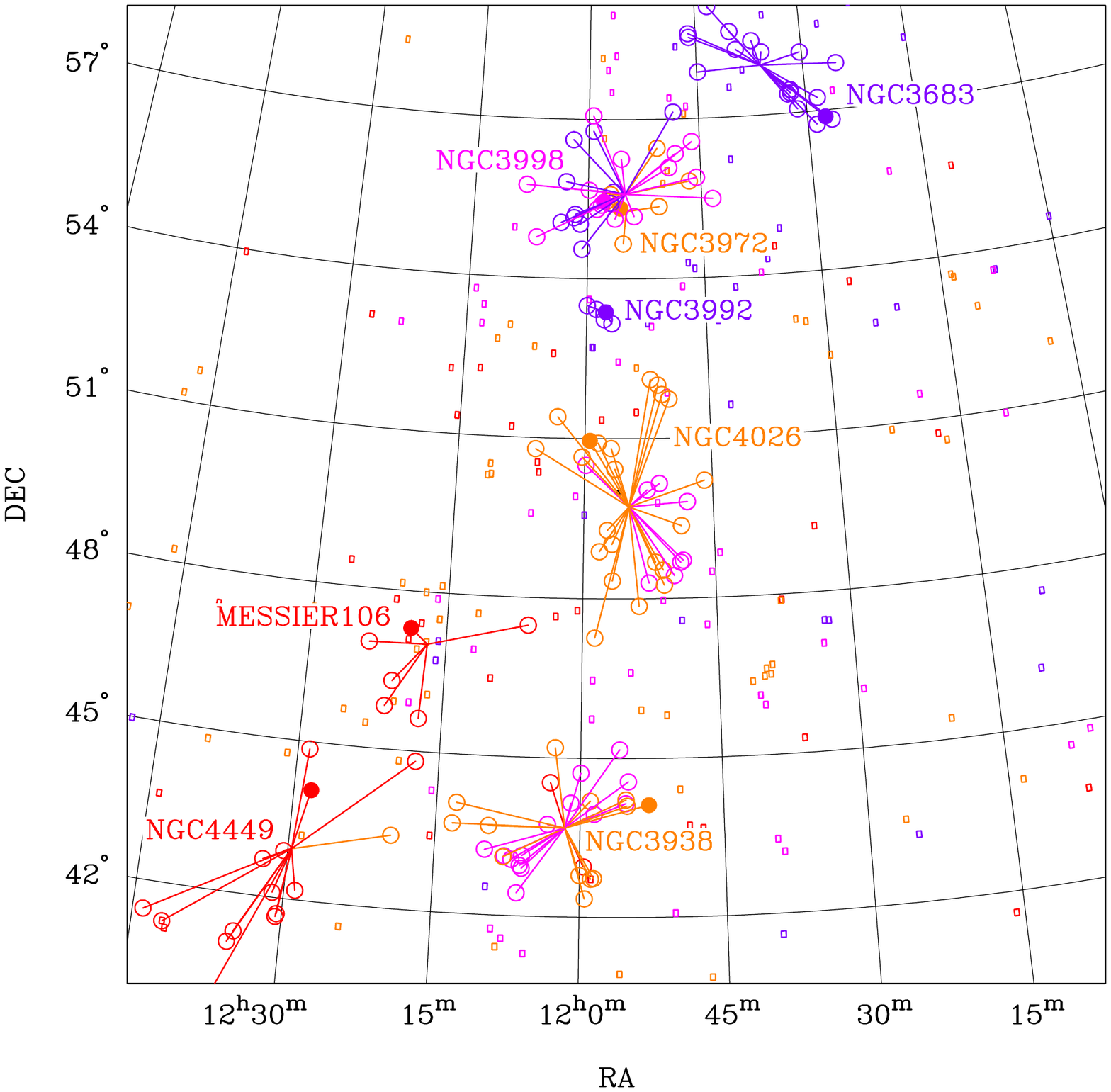}
\includegraphics[trim=1.1cm 4cm 1cm 4.5cm, clip=true, width=7.cm]{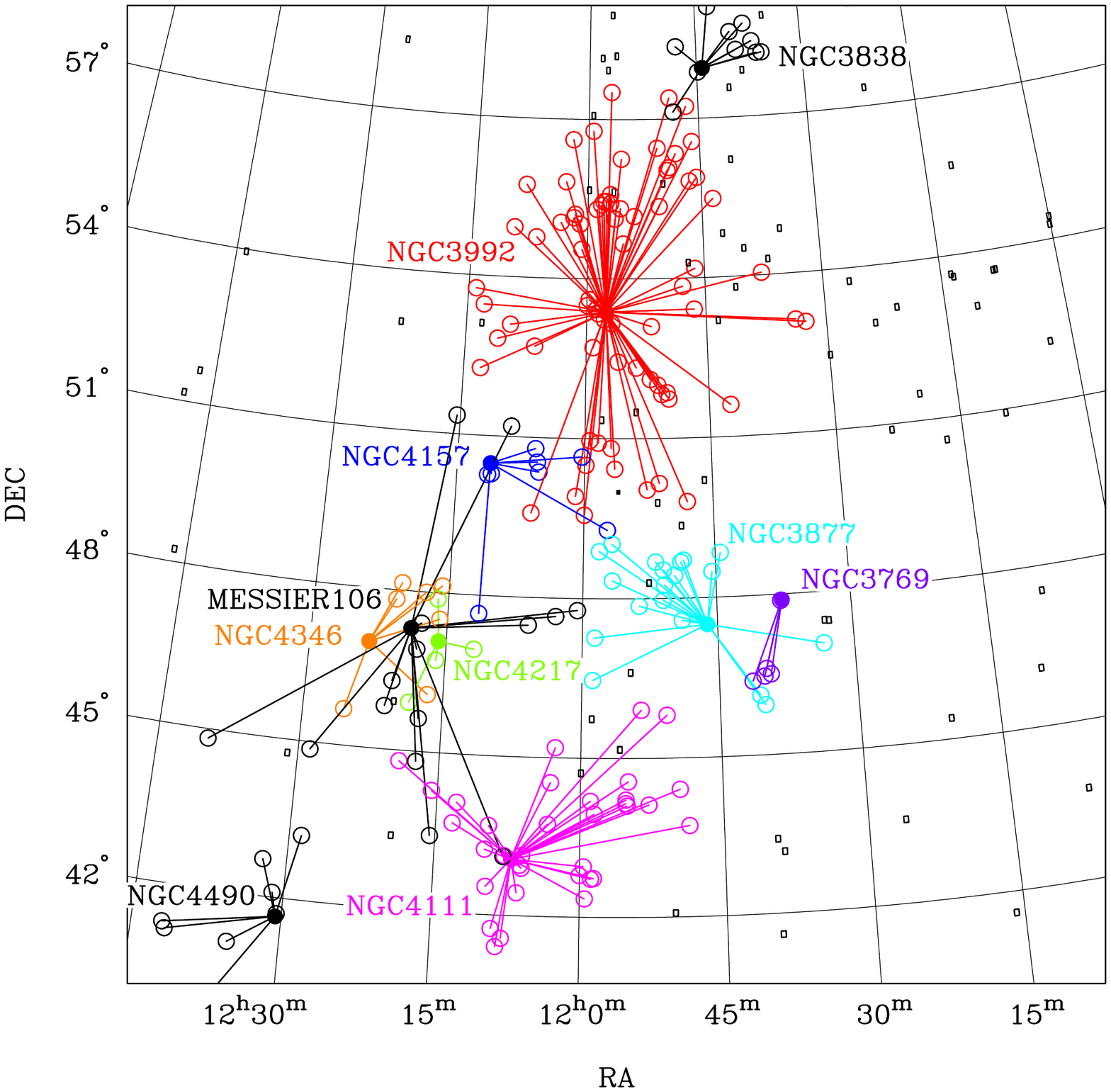}
\caption[Structures identified using the FoF algorithm compared to the Literature.]{Structures identified using the FoF algorithm with linking lengths $D_0=0.43$~Mpc and $V_0=120$~km~s$^{-1}$ (\textit{top}) and $D_0=0.3$~Mpc and $V_0=150$~km~s$^{-1}$ (\textit{middle}). The latter is compared to the groups found in Karachentsev et al. (2012; \textit{bottom}). The galaxies in the FoF groups are colour-coded to indicate the recession velocities ($v_{LG} \leq 600$~km~s$^{-1}$ in red; $600 < v_{LG} \leq 900$~km~s$^{-1}$ in orange; $900 < v_{LG} \leq 1200$~km~s$^{-1}$ in magenta and $1200 < v_{LG} \leq 1500$~km~s$^{-1}$ in violet), whereas the literature groups (\textit{bottom}) are colour-coded according to their group membership for clarity. Group members are shown with circles that are connected to their group centre by a line. The BGG are labelled in the Figures. Non-group galaxies are shown with small boxes.}
\label{fig:C5_litComp}
\end{figure}

(\textit{i}) The dominant structure in MK11 is the NGC3992 group with 74 group members (shown in red - \textit{bottom}). The best match to the NGC3992 group is the NGC3998 group with 29 members (\textit{middle}), of which 25 group members are contained in the MK11 counterpart. Furthermore the galaxies in the smaller NGC3992 and NGC3972 groups {(\textit{middle})} are part of the NGC3992 group identified by \citet{makarov2011}. Therefore, the FoF algorithm used in this study appears to divide the NGC3992 group into its substructures. (\textit{ii}) The NGC4111 group (shown in magenta with 35 group members - \textit{bottom}) is best matched to the NGC3938 group with 29 group members (\textit{middle}) -- 25 galaxies are enclosed in both structures. (\textit{iii}) The MK11 group NGC3877 (21 group members; shown in cyan - \textit{bottom}) is the best match to the NGC4026 group (30 group members - \textit{middle}) -- 11 galaxies are bijective matches. Note that the literature group NGC4157 (\textit{bottom}) forms a substantial part of the NGC4026 group (\textit{middle}), with the centre lying between the two MK11 groups (see \textit{middle}).

(\textit{iv}) The foreground group MESSIER106 group (shown in black - \textit{bottom}) consisting of 18 group members has a smaller counterpart in the FoF groups -- MESSIER106 with only 6 group members (\textit{middle}), of which 5 are included in the literature group. (\textit{v}) Located in the northern part is the NGC3838 group (shown in black - \textit{bottom}), which is best compared to the NGC3683 group with 22 group members (\textit{middle}) -- 9 of the 11 MK11group members can be found in the NGC3683 group. (\textit{vi}) To the South lies the NGC4490 group (8 members, shown in black - \textit{bottom}), which is best matched to the NGC4449 group (15 members - \textit{middle}, 7 galaxies are included in both structures.

By comparing the MK11 groups to the bijective or best matched FoF group (e.g. the NGC3992 to the NGC3998 group, the NGC4111 group to the NGC3938 group, the NGC3877 group to the NGC4026 group, the two MESSIER106 groups etc.) 82 bijective galaxy matches are found out of the 131 galaxies in the six FoF groups and 167 galaxies in the MK11 groups. Approximately half of the group members in the MK11 groups are included in the FoF counterparts. The BGG (and therefore the group name) may differ as Karachentsev et al. (2012) use distance moduli obtained from NED and $K_s$-band luminosities, which are derived from $K_s$-band magnitudes in the 2MASS Survey \citep{jarrett2004}. Note that the few group members of the NGC3992 group (FoF) are included in the NGC3992 group (identified by \citealt{makarov2011}). The conglomeration of the FoF groups (NGC3998 and NGC3992 groups) would result in NGC3992 to be the BGG as in \citealt{makarov2011}.

On average the relative difference in the number of group members between the \citet{makarov2011} groups ($N_{m,lit}$) and the best matched FoF groups (number of galaxies in the full sample; $N_{m,F}$) is 14~per cent given by
\begin{equation} 
\label{eq:lit}
\Delta N_m=\frac{\sum\limits_N \frac{N_{m,F} - N_{m,lit}}{N_{m,lit}} }{N}
\end{equation}
and the average relative difference in the derived velocity dispersions of the groups is 42~per cent (calculated analogously, with $\sigma_{v,F}$ and $\sigma_{v,lit}$ replacing $N_{m,F}$ and $N_{m,lit}$ in equation~\ref{eq:lit}, respectively). This denotes that on average the FoF groups contain 14~per cent more group members and the velocity dispersions are increased by 42~per cent.

\subsection{Will Ursa Major and Virgo merge?}
\label{subsec:UMa_Virgo}

\begin{table}
\begin{scriptsize}
\begin{center}
\caption{High-probability FoF groups to be bound to the Virgo cluster ($P_{bound} > 80$).} 
\begin{tabular}{lcccc}
\hline\hline\\[-2ex] 
                                                & $M_{tot}$                     & $V_r$              & $R_p$ &  $P_{bound}$ \\ 
\multicolumn{1}{c}{Group pair} & ($10^{14} M_{\odot}$) & (km~s$^{-1}$) & (Mpc)   & (per cent) \\
\multicolumn{1}{c}{(D1)}           & (D2)                              & (D3)                 & (D4)     & (D5) \\
[0.5ex] \hline\\[-1.8ex]
NGC4725, Virgo & 8.1 & 69 & 3.4 & 95.7  \\
NGC5033, Virgo & 8.1 & 71 & 6.2 & 94.4  \\
NGC3301, Virgo & 8.0 & 118 & 7.7 & 90.0  \\
NGC4274, Virgo & 8.2 & 173 & 3.8 & 89.8  \\
MESSIER106, Virgo & 8.8 & 215 & 8.1 & 81.5  \\
\hline\\[-2ex]
\end{tabular}
\label{tab:Pbound2}
\end{center}
\end{scriptsize}
\end{table}

\begin{figure*}
\centering
\includegraphics[trim=0.5cm 7cm 1cm 6cm, clip=true, width=13cm]{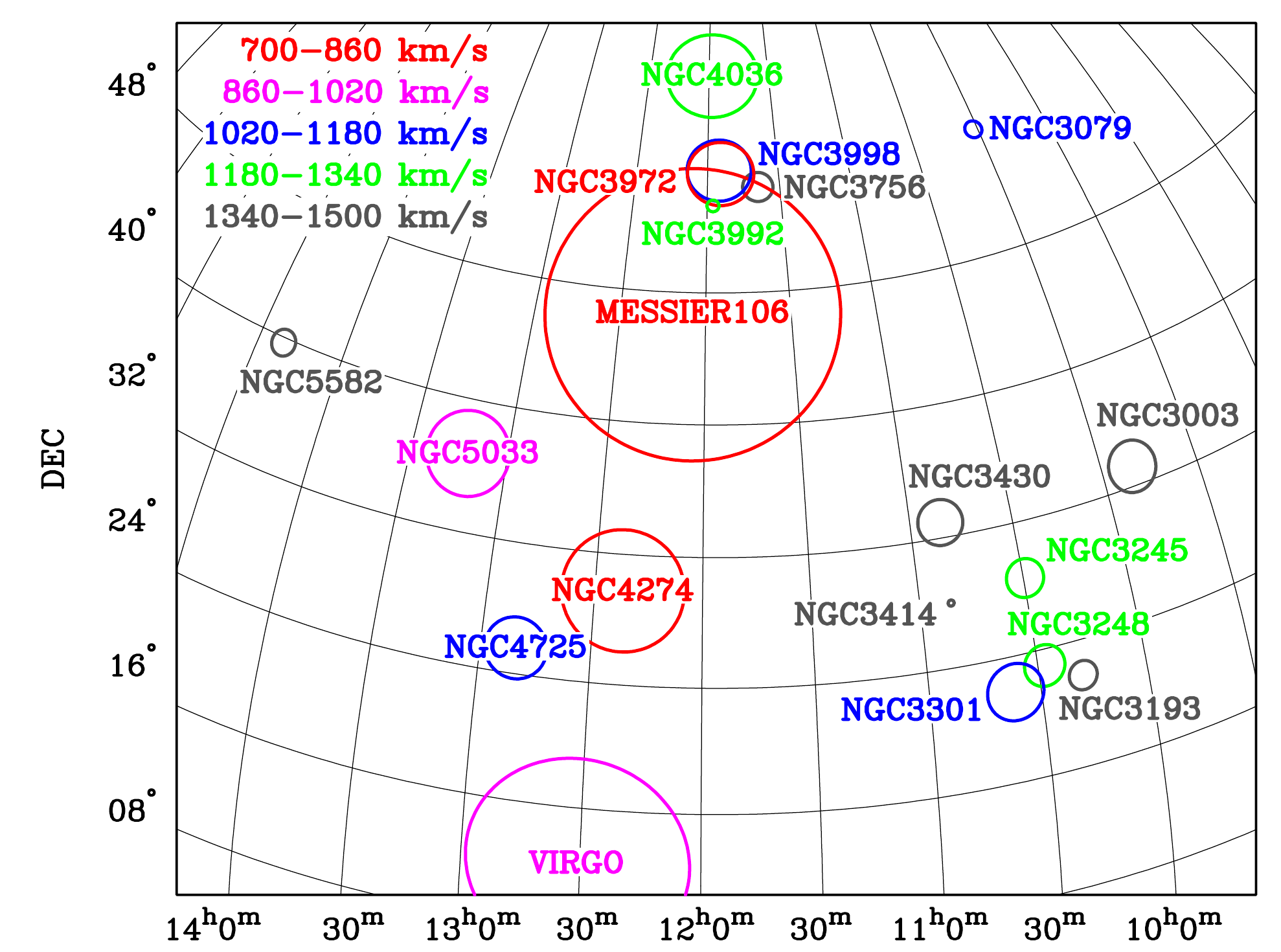} 
\includegraphics[trim=0.5cm 6cm 1cm 6cm, clip=true, width=13cm]{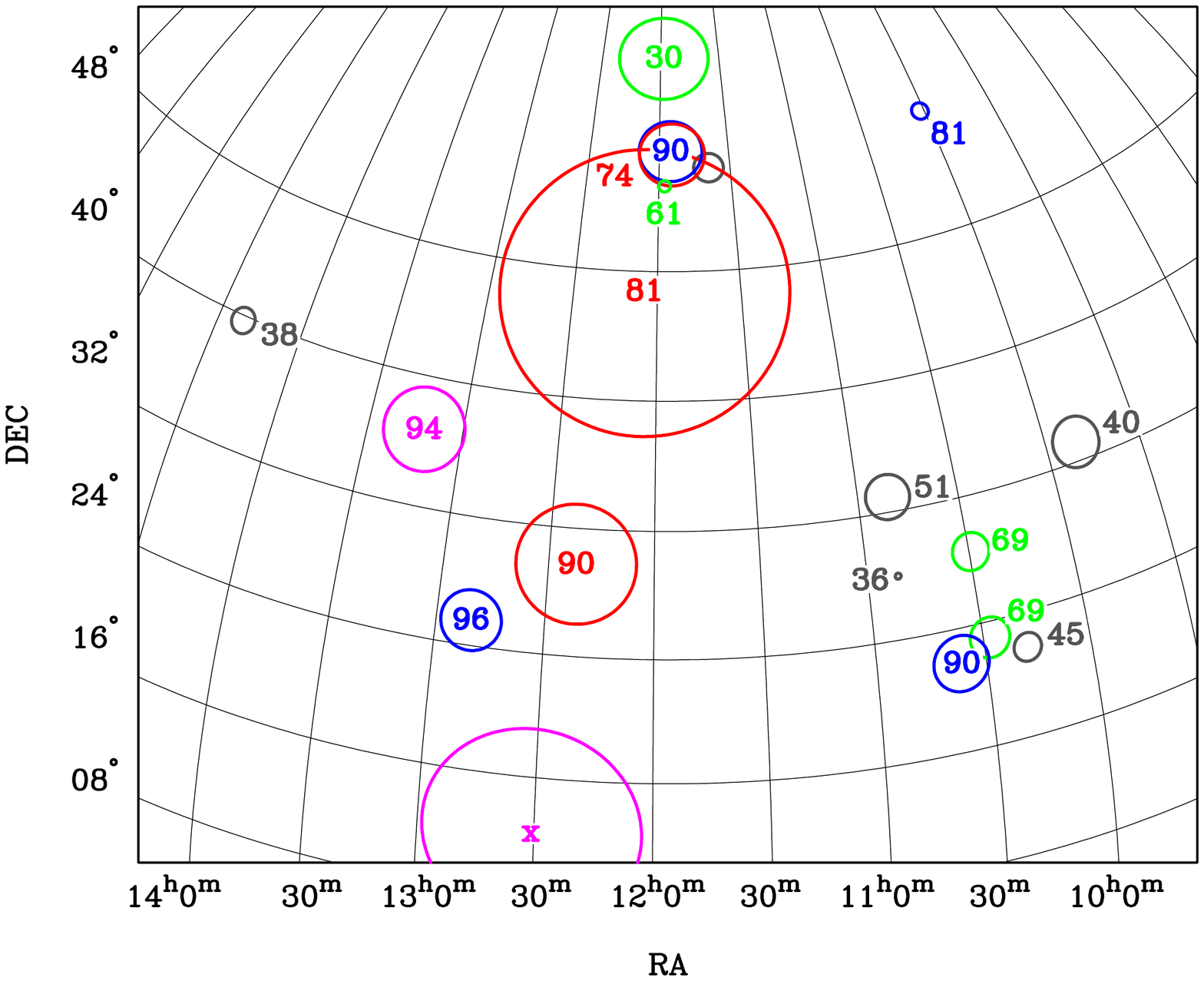} 
\caption{Galaxy groups identified using `looser' linking lengths ($D_0 = 0.43$~Mpc and $V_0 = 120$~km~s$^{-1}$) and their probabilities to be bound to the Virgo cluster. The displayed velocity range is 700 to 1500~km~s$^{-1}$. The groups are labelled in the \textit{top} panel. Probabilities ($> 0$~per cent) of the FoF groups to be bound to the Virgo cluster are given as a numerical value in the \textit{bottom} panel (in per cent). Note that the virial radius $R_v =1.8$~Mpc (Hoffmann et al. 1980) is shown for the Virgo cluster, whereas the maximal radial extent is shown for the FoF groups. The galaxy groups `NGC3998' and `NGC3079' (in the North) are not considered to be bound to the Virgo cluster due to the large projected distance ($R_p > 12$~Mpc) and therefore unfeasible merging times.} 
\label{fig:Virgo}
\end{figure*}

The galaxy groups in the Ursa Major region form an extended filamentary structure in the central region ($11^h < \mathrm{RA} \leq 13^h$) as discussed in Section~\ref{sec:groups}. To the South of the filamentary structure resides the Virgo cluster (centred at $\alpha=187.70$, $\delta=12.34$ and $v_{\mathrm{LG}}=975$~km~s$^{-1}$; \citealt{karachentsev2014}). The analysis of bound structures using the Newtonian binding criterion presented above can be used to determine if the Ursa Major region as a whole is bound to the nearby Virgo cluster. The Virgo cluster has a total mass of $M_T=8.0 \times 10^{14}$~M$_{\odot}$ \citep{karachentsev2014}. 

Assuming that the still forming galaxy groups identified using \textit{strict} linking lengths will eventually merge to form larger structures, the large-scale structures of the Ursa Major region as a whole may be bound to the Virgo cluster. In order to examine if the larger structures are bound to Virgo, the FoF groups identified using \textit{looser} liking lengths $D_0 = 0.43$~Mpc and $V_0 = 120$~km~s$^{-1}$ are used. The probabilities that individual groups are bound to Virgo ($P_{bound} > 80$~per cent) are given in Table ~\ref{tab:Pbound2} with the most likely bound two-body system listed first (similar to Table~\ref{tab:Pbound}). The Ursa Major -- Virgo region is shown in Figure~\ref{fig:Virgo} with the structures labelled in the \textit{top} panel. For illustrative purposes a virial radius of $R_v =1.8$~Mpc \citep{karachentsev2014} is assumed for the Virgo cluster. The two-body systems with probabilities $P _{bound} > 0$~per cent are shown in Figure~\ref{fig:Virgo} (\textit{bottom}; the probabilities are given as a numerical values in per cent). Note that the galaxy groups `NGC3998' and `NGC3079' (in the North) are not considered to be bound to the Virgo cluster due to the large projected distance ($R_p > 12$~Mpc) and therefore unfeasible merging times (not listed in Table~\ref{tab:Pbound2}). The displayed velocity range is 700 to 1500~km~s$^{-1}$ and the structures are at central velocities as noted in the key. There are no groups beyond 1500~km~s$^{-1}$ that are likely to be bound to the Virgo cluster ($P_{bound}$ drops to 0~per~cent). 

The main structure in the Ursa Major region is the MESSIER106 group (constituting of the NGC4026, NGC3938, NGC4449 and MESSIER106 groups when using \textit{strict} linking lengths), which is likely to be bound to the Virgo cluster at the 81~per~cent level. The MESSIER106 group is connected to the Virgo cluster by a filamentary structure including the NGC4274 (constituting of the NGC4274 and NGC4278 groups when using \textit{strict} linking lengths), the NGC4725 and NGC5033 groups, all of which are likely to be bound to the Virgo cluster ($P_{bound} > 90$~per cent). The Virgo cluster appears to be accreting galaxy groups along the filamentary structure including the `Ursa Major cluster' as known in the literature. If Ursa Major and Virgo merge, the total mass of the system will be $M_T=8.8 \times 10^{14}$~M$_{\odot}$. Note that previous studies determined the zone of infall into the Virgo cluster to be 4 times the virial radius \citep{karachentsev2014}, which includes the NGC4725, NGC4274 and NGC5033 groups.

\subsection{The nature of the Ursa Major region}
\label{subsec:supergroup}

\begin{figure*}
\begin{center}
\includegraphics[bb = 0 220 740 750, clip=true, width=18cm]{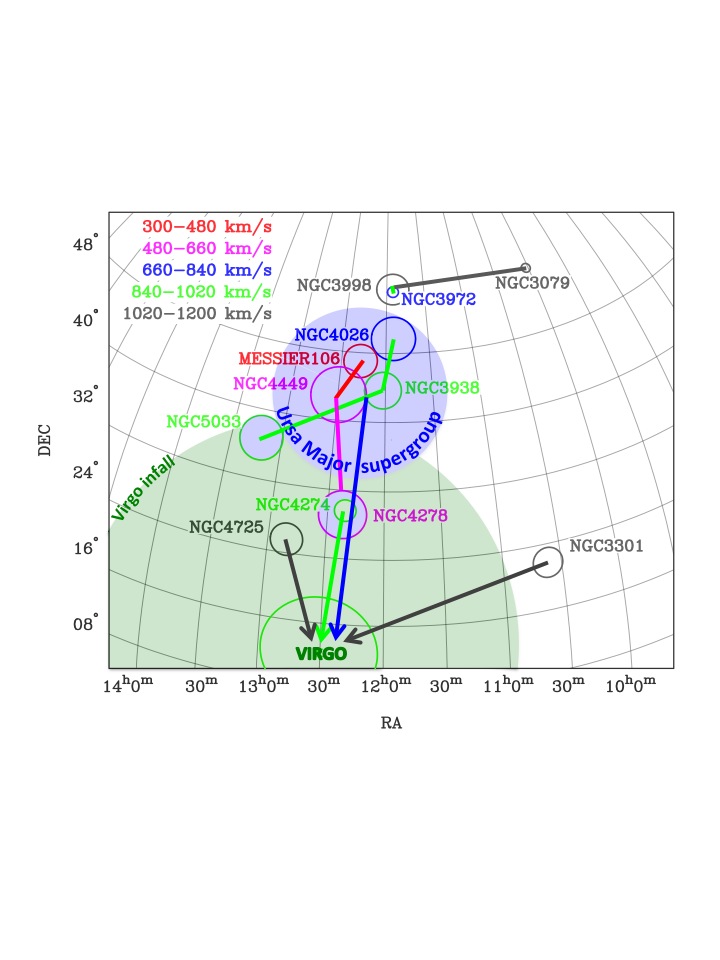}
\caption{Projected overview of the Ursa Major/Virgo region. Note that the virial radius $R_v =1.8$~Mpc (Hoffmann et al. 1980) is shown for the Virgo cluster, whereas the maximal radial extent is shown for the FoF groups. The central group velocities are given in the key. Galaxy groups that are likely to constitute the Ursa Major supergroup are the MESSIER106, NGC4449, NGC4278, NGC4026, NGC3938 and the NGC5033 groups (bound structures are marked by lines -- the red/magenta and blue/green structures are likely bound to one another and therefore constitute to the supergroup, which is highlighted in blue). The zone of infall into the Virgo cluster is highlighted in light green (4 times the virial radius as determined in \citealt{karachentsev2014}).The Virgo cluster is likely to be accreting galaxy groups such as the NGC4274, NGC4725 and NGC3301 groups as well as the Ursa Major supergroup as a whole (marked with arrows towards the Virgo cluster). The NGC3998, NGC3972 and NGC3079 groups are situated in the background and unlikely to be bound to both, the Ursa Major supergroup and the Virgo cluster.}
\label{fig:skymapSupergroup}
\end{center}
\end{figure*}

We identified gravitationally bound structures in the Ursa Major region using a FoF algorithm with \textit{strict} linking lengths ($D_0=0.30$~Mpc and $V_0=150$~km~s$^{-1}$). The properties of the structures emphasize the picture of galaxy groups within the Ursa Major region as opposed to an `Ursa Major cluster' given that (\textit{i}) the number of group members in each structure is small, (\textit{ii}) the fraction of early-type galaxies is low, (\textit{iii}) the velocity dispersions are low and (\textit{iv}) the virial masses are $< 10^{14}~\mathrm{M}_{\odot}$. 

The dynamical analysis of the galaxy groups revealed that many groups are likely bound to one another and in the process of merging. Figure~\ref{fig:skymapSupergroup} shows a projected overview of the region with high-probability two-body bound systems connected by a line. The structure eventually emerging in the foreground (at the centre of mass of the NGC4449, MESSIER106 and NGC4278 groups -- red/magenta) and the structure emerging at intermediate velocity (at the centre of mass of the NGC4026, NGC3938 and NGC5033 groups -- blue/green) are likely bound to one another and therefore constitute to the `Ursa Major supergroup' (highlighted in blue). Using \textit{looser} linking lengths ($D_0=0.43$~Mpc and $V_0=120$~km~s$^{-1}$) results in the NGC4026, NGC3938, NGC4449 and MESSIER106 groups being merged giving an estimate of the properties of the supergroup: a poor cluster with 185 galaxies, a velocity dispersion of 205~km~s$^{-1}$, an early-type fraction of 23~per cent and a virial mass of $7.96 \times 10^{13}$~M$_{\odot}$. 

The zone of infall into the Virgo cluster has previously been determined to be 4 times the virial radius (highlighted in green; \citealt{karachentsev2014}).The Virgo cluster is likely to be accreting galaxy groups such as the NGC4274, NGC4725 and NGC3301 groups as well as the Ursa Major supergroup (marked with arrows towards the Virgo cluster). Note that the NGC4274 group was determined to be likely bound to the NGC3938 and NGC5033 groups (see Section~\ref{sec:dynamics}). However, due to its proximity to the Virgo cluster, the dynamics of the NGC4274 group is likely influenced by the strong sheer caused by the Virgo cluster. The NGC3998, NGC3972 and NGC3079 groups are situated in the background and unlikely to be bound to both, the Ursa Major supergroup and the Virgo cluster.

The evolutionary state of the galaxy groups has also been investigated. The optical galaxy properties as a function of local projected surface density show weak trends, suggesting little transformation occurring in the galaxies of the Ursa Major region yet. The colours, luminosities and morphological types of galaxies residing in the groups are similar to those of galaxies in the field. The little signs of transformation in the optical properties of the galaxies as well as the relatively large offsets of the BGG from the spatial and kinematic group centre suggest that the groups are in an \textit{early evolutionary state}. Note that several galaxy groups may not have had time to virialize yet (groups with $t_c > 0.09 H_0^{-1}$; \citealt{nolthenius1987}). 

Several galaxies with HI excess were found in the Ursa Major region, the majority of which reside in the low density environment. The HI deficient galaxies tend to reside in galaxy groups. We found that galaxies with known disturbed HI content and/or HI tails reside in the central region of the galaxy groups. Given the low velocity dispersion of the groups and the lack of X-ray emission, these disturbances are likely a result of galaxy-galaxy interactions.

\section{Conclusion}

In this paper, galaxy groups are identified whose members are likely to be physically and dynamically associated within the Ursa Major region (within $9^{h}30^{m}\leq\alpha\leq+14^{h}30^{m}$ and $+20^{\circ}\leq\delta\leq+65^{\circ}$). A standard friends-of-friends (FoF) algorithm is extensively tested on mock galaxy lightcones with the same magnitude, velocity and stellar mass cuts as in the real data to determine appropriate linking lengths ($D_0$ and $V_0$). The FoF algorithm is implemented on a complete, velocity-limited ($300 \leq v_{\mathrm{LG}} \leq 3000$~km~s$^{-1}$), absolute-magnitude-limited ($\mathrm{M}_r \leq -15.3$~mag) and stellar-mass-limited ($\mathrm{M_{\ast}} \geq 10^8~\mathrm{M}_{\odot}$) sample of 1209 galaxies in the Ursa Major region as obtained from the Sloan Digital Sky Survey (SDSS DR7; \citealt{abazajian2009}) and the NASA/IPAC Extragalactic Database (NED). Adopting linking lengths $D_0 = 0.3$~Mpc and $V_0 = 150$~km~s$^{-1}$, 41 galaxy groups are identified within the Ursa Major region with 4 or more group members. After defining the luminosity-weighted centroid (RA, Dec. and recession velocity), the maximal radial extent and velocity dispersion, faint galaxies within each cylindrical volume (that are not part of the complete sample) are included as group members and the properties are re-calculated and presented.

The average group properties are $N_{m,F}$=11, $N_{m,C}$=8, $\sigma_v = 90$~km~s$^{-1}$, $r_{max} =360$~kpc, $\mathrm{M}_v = 0.6 \times 10^{13}~\mathrm{M}_{\odot}$, $\mathrm{L}_t = 0.23 \times 10^{11}~\mathrm{L}_{\odot}$, a mass-to-light ratio of $\mathrm{M}_v/\mathrm{L}_t = 480$ and a crossing time of 0.13~$H_0^{-1}$. Note that galaxy groups with $t_c > 0.09 H_0^{-1}$ may not have had time to virialize yet \citep{nolthenius1987}. The obtained properties emphasize the picture of galaxy groups in the Ursa Major region as opposed to an `Ursa Major cluster' given that (\textit{i}) the number of group members is small, (\textit{ii}) the fraction of early-type galaxies in galaxy groups is low, (\textit{iii}) the velocity dispersions are low and (\textit{iv}) the virial masses are $< 10^{14}~\mathrm{M}_{\odot}$. Only 26~per~cent of galaxies in the complete sample reside in groups (with $N_m \geq 4$), i.e. 74~per~cent of galaxies are left ungrouped. In comparison to previous studies (e.g. \citealt{giuricin2000}), the ungrouped fraction is rather high suggesting that many galaxies may reside in pairs and groups with three members.

The galaxy groups in the central Ursa Major region ($11^h < \mathrm{RA} \leq 13^h$) form an extended filamentary structure connected in projection and velocity. The filamentary structure is also connected to the Virgo cluster located to the South of Ursa Major. Behind the filamentary structure is a sparsely populated region (at $1200 < v_{\mathrm{LG}} < 1800$~km~s$^{-1}$). The analysis of the group dynamics using the Newtonian binding criterion, reveals many groups that are likely bound to one another and in the process of merging. The 6 galaxy groups, which are identified to constitute to the Ursa Major supergroup are the NGC4449, MESSIER106, NGC4278, NGC4026, NGC3938 and NGC5033 groups. Eventually, these galaxy groups may form larger structures with properties similar to the galaxy groups identified using \textit{looser} linking lengths ($D_0=0.43$~Mpc and $V_0=120$~km~s$^{-1}$). Assuming the larger structures (groups identified using the \textit{looser} linking lengths), the Ursa Major region as a whole is likely to be bound to the Virgo cluster and will eventually form a larger system with a total mass of $M_T=8.8 \times 10^{14}$~M$_{\odot}$. The Virgo cluster appears to be accreting galaxy groups along the filamentary structure.

The evolutionary state of the galaxy groups can be examined by investigating the location of the brightest group galaxy (BGG) as well as the optical and HI properties as a function of local projected surface density and group/non-group membership: 

The brightest group galaxy (BGG) often shows a relatively large spatial and kinematic offset from the group centre compared to previous studies of BGG in dynamically evolved groups. Therefore, the majority of groups in the Ursa Major region seem to be dynamically immature and in the process of forming. Galaxy groups closest to the Virgo cluster appear more evolved than groups further away, suggesting a change in group properties with distance from the Virgo cluster. However, this may also be a result of the low number of galaxies in the groups closest to the Virgo cluster, where the group centre may be severely biased towards the BGG.

The optical galaxy properties such as colour, morphological type and luminosity as a function of environment are as follows: (\textit{i}) the increase in the fraction of early-type galaxies with local projected surface density is shallow, (\textit{ii}) galaxy colours tend to be redder with increasing density and (\textit{iii}) there is no trend towards brighter galaxies with increasing density.

Several galaxies with HI excess are found in the Ursa Major region, the majority of these reside in low density environments. HI deficient galaxies tend to reside in galaxy groups, preferentially in the southern part of the Ursa Major region, i.e. in the dynamically more evolved groups that are in vicinity of the Virgo cluster. The HI deficiency values appear to be evenly distributed around 0 up to a local projected surface density of $\mathrm{log (} \Sigma_5 \mathrm{)} = 1$ followed by a steep increase towards HI deficient galaxies with increasing density. Note that first signs of galaxy transformation can be seen in several galaxy groups which contain members with disturbed HI content or HI tails. Given the low velocity dispersion of the groups and the lack of X-ray emission, these disturbances are likely due to galaxy-galaxy interactions.

We conclude that the galaxy groups in the Ursa Major region are in an early evolutionary state and the properties of their member galaxies are similar to those in the field. The nature of Ursa Major is likely a supergroup, consisting of several gravitationally bound galaxy groups, which will form a larger structure and eventually also merge with Virgo.

\begin{acknowledgements}
We thank the referee for the careful reading of the article and many useful suggestions, which now implemented have improved the paper. We would also like to thank Max Bernyk for helping to obtain mock catalogues from the Theoretical Astrophysical Observatory (TAO). KW acknowledges financial assistance from a CSIRO top-up scholarship that facilitated collaboration with BK.
\end{acknowledgements}

\bibliographystyle{mn2e}
\bibliography{ref}

\begin{thebibliography}{68}
\expandafter\ifx\csname natexlab\endcsname\relax\def\natexlab#1{#1}\fi

\bibitem[{{Abazajian} {et~al.}(2009){Abazajian}, {Adelman-McCarthy},
  {Ag{\"u}eros}, {Allam}, {Allende Prieto}, {An}, {Anderson}, {Anderson},
  {Annis}, {Bahcall}, \& et~al.}]{abazajian2009}
{Abazajian}, K.~N., {Adelman-McCarthy}, J.~K., {Ag{\"u}eros}, M.~A., {Allam},
  S.~S., {Allende Prieto}, C., {An}, D., {Anderson}, K.~S.~J., {Anderson},
  S.~F., {Annis}, J., {Bahcall}, N.~A., \& et~al. 2009, \apjs, 182, 543

\bibitem[{{Alpaslan} {et~al.}(2015){Alpaslan}, {Driver}, {Robotham},
  {Obreschkow}, {Andrae}, {Cluver}, {Kelvin}, {Lange}, {Owers}, {Taylor},
  {Andrews}, {Bamford}, {Bland-Hawthorn}, {Brough}, {Brown}, {Colless},
  {Davies}, {Eardley}, {Grootes}, {Hopkins}, {Kennedy}, {Liske},
  {Lara-L{\'o}pez}, {L{\'o}pez-S{\'a}nchez}, {Loveday}, {Madore}, {Mahajan},
  {Meyer}, {Moffett}, {Norberg}, {Penny}, {Pimbblet}, {Popescu}, {Seibert}, \&
  {Tuffs}}]{alpaslan2015}
{Alpaslan}, M., {Driver}, S., {Robotham}, A.~S.~G., {Obreschkow}, D., {Andrae},
  E., {Cluver}, M., {Kelvin}, L.~S., {Lange}, R., {Owers}, M., {Taylor}, E.~N.,
  {Andrews}, S.~K., {Bamford}, S., {Bland-Hawthorn}, J., {Brough}, S., {Brown},
  M.~J.~I., {Colless}, M., {Davies}, L.~J.~M., {Eardley}, E., {Grootes}, M.~W.,
  {Hopkins}, A.~M., {Kennedy}, R., {Liske}, J., {Lara-L{\'o}pez}, M.~A.,
  {L{\'o}pez-S{\'a}nchez}, {\'A}.~R., {Loveday}, J., {Madore}, B.~F.,
  {Mahajan}, S., {Meyer}, M., {Moffett}, A., {Norberg}, P., {Penny}, S.,
  {Pimbblet}, K.~A., {Popescu}, C.~C., {Seibert}, M., \& {Tuffs}, R. 2015,
  \mnras, 451, 3249

\bibitem[{{Baldry} {et~al.}(2004){Baldry}, {Balogh}, {Bower}, {Glazebrook}, \&
  {Nichol}}]{baldry2004}
{Baldry}, I.~K., {Balogh}, M.~L., {Bower}, R., {Glazebrook}, K., \& {Nichol},
  R.~C. 2004, in American Institute of Physics Conference Series, Vol. 743, The
  New Cosmology: Conference on Strings and Cosmology, ed. R.~E. {Allen}, D.~V.
  {Nanopoulos}, \& C.~N. {Pope}, 106--119

\bibitem[{{Balogh} {et~al.}(2004){Balogh}, {Baldry}, {Nichol}, {Miller},
  {Bower}, \& {Glazebrook}}]{balogh2004}
{Balogh}, M.~L., {Baldry}, I.~K., {Nichol}, R., {Miller}, C., {Bower}, R., \&
  {Glazebrook}, K. 2004, \apjl, 615, L101

\bibitem[{{Beers} {et~al.}(1990){Beers}, {Flynn}, \& {Gebhardt}}]{beers1990}
{Beers}, T.~C., {Flynn}, K., \& {Gebhardt}, K. 1990, \aj, 100, 32

\bibitem[{{Beers} {et~al.}(1982){Beers}, {Geller}, \& {Huchra}}]{beers1982}
{Beers}, T.~C., {Geller}, M.~J., \& {Huchra}, J.~P. 1982, \apj, 257, 23

\bibitem[{{Bell} {et~al.}(2003){Bell}, {McIntosh}, {Katz}, \&
  {Weinberg}}]{bell2003}
{Bell}, E.~F., {McIntosh}, D.~H., {Katz}, N., \& {Weinberg}, M.~D. 2003, \apjs,
  149, 289

\bibitem[{{Berlind} {et~al.}(2006){Berlind}, {Frieman}, {Weinberg}, {Blanton},
  {Warren}, {Abazajian}, {Scranton}, {Hogg}, {Scoccimarro}, {Bahcall},
  {Brinkmann}, {Gott}, {Kleinman}, {Krzesinski}, {Lee}, {Miller}, {Nitta},
  {Schneider}, {Tucker}, {Zehavi}, \& {SDSS Collaboration}}]{berlind2006}
{Berlind}, A.~A., {Frieman}, J., {Weinberg}, D.~H., {Blanton}, M.~R., {Warren},
  M.~S., {Abazajian}, K., {Scranton}, R., {Hogg}, D.~W., {Scoccimarro}, R.,
  {Bahcall}, N.~A., {Brinkmann}, J., {Gott}, III, J.~R., {Kleinman}, S.~J.,
  {Krzesinski}, J., {Lee}, B.~C., {Miller}, C.~J., {Nitta}, A., {Schneider},
  D.~P., {Tucker}, D.~L., {Zehavi}, I., \& {SDSS Collaboration}. 2006, \apjs,
  167, 1

\bibitem[{{Bernyk} {et~al.}(2014){Bernyk}, {Croton}, {Tonini}, {Hodkinson},
  {Hassan}, {Garel}, {Duffy}, {Mutch}, {Poole}, \& {Hegarty}}]{bernyk2014}
{Bernyk}, M., {Croton}, D.~J., {Tonini}, C., {Hodkinson}, L., {Hassan}, A.~H.,
  {Garel}, T., {Duffy}, A.~R., {Mutch}, S.~J., {Poole}, G.~B., \& {Hegarty}, S.
  2014, ArXiv 1403.5270

\bibitem[{{Binney} \& {Merrifield}(1998)}]{binney1998}
{Binney}, J. \& {Merrifield}, M. 1998, {Galactic Astronomy}

\bibitem[{{Bird} \& {Beers}(1993)}]{bird1993}
{Bird}, C.~M. \& {Beers}, T.~C. 1993, \aj, 105, 1596

\bibitem[{{Blanton} {et~al.}(2005){Blanton}, {Eisenstein}, {Hogg}, {Schlegel},
  \& {Brinkmann}}]{blanton2005}
{Blanton}, M.~R., {Eisenstein}, D., {Hogg}, D.~W., {Schlegel}, D.~J., \&
  {Brinkmann}, J. 2005, \apj, 629, 143

\bibitem[{{Blanton} {et~al.}(2003){Blanton}, {Hogg}, {Bahcall}, {Brinkmann},
  {Britton}, {Connolly}, {Csabai}, {Fukugita}, {Loveday}, {Meiksin}, {Munn},
  {Nichol}, {Okamura}, {Quinn}, {Schneider}, {Shimasaku}, {Strauss}, {Tegmark},
  {Vogeley}, \& {Weinberg}}]{blanton2003}
{Blanton}, M.~R., {Hogg}, D.~W., {Bahcall}, N.~A., {Brinkmann}, J., {Britton},
  M., {Connolly}, A.~J., {Csabai}, I., {Fukugita}, M., {Loveday}, J.,
  {Meiksin}, A., {Munn}, J.~A., {Nichol}, R.~C., {Okamura}, S., {Quinn}, T.,
  {Schneider}, D.~P., {Shimasaku}, K., {Strauss}, M.~A., {Tegmark}, M.,
  {Vogeley}, M.~S., \& {Weinberg}, D.~H. 2003, \apj, 592, 819

\bibitem[{{Brough} {et~al.}(2006{\natexlab{a}}){Brough}, {Forbes}, {Kilborn},
  \& {Couch}}]{brough2006GEMS}
{Brough}, S., {Forbes}, D.~A., {Kilborn}, V.~A., \& {Couch}, W.
  2006{\natexlab{a}}, \mnras, 370, 1223

\bibitem[{{Brough} {et~al.}(2006{\natexlab{b}}){Brough}, {Forbes}, {Kilborn},
  {Couch}, \& {Colless}}]{brough2006}
{Brough}, S., {Forbes}, D.~A., {Kilborn}, V.~A., {Couch}, W., \& {Colless}, M.
  2006{\natexlab{b}}, \mnras, 369, 1351

\bibitem[{{Butcher} \& {Oemler}(1984)}]{butcher1984}
{Butcher}, H. \& {Oemler}, Jr., A. 1984, \apj, 285, 426

\bibitem[{{Conroy} {et~al.}(2009){Conroy}, {Gunn}, \& {White}}]{conroy2009}
{Conroy}, C., {Gunn}, J.~E., \& {White}, M. 2009, \apj, 699, 486

\bibitem[{{Cortese} {et~al.}(2004){Cortese}, {Gavazzi}, {Boselli},
  {Iglesias-Paramo}, \& {Carrasco}}]{cortese2004}
{Cortese}, L., {Gavazzi}, G., {Boselli}, A., {Iglesias-Paramo}, J., \&
  {Carrasco}, L. 2004, \aap, 425, 429

\bibitem[{{Croton} {et~al.}(2006){Croton}, {Springel}, {White}, {De Lucia},
  {Frenk}, {Gao}, {Jenkins}, {Kauffmann}, {Navarro}, \& {Yoshida}}]{croton2006}
{Croton}, D.~J., {Springel}, V., {White}, S.~D.~M., {De Lucia}, G., {Frenk},
  C.~S., {Gao}, L., {Jenkins}, A., {Kauffmann}, G., {Navarro}, J.~F., \&
  {Yoshida}, N. 2006, \mnras, 365, 11

\bibitem[{{de Vaucouleurs} {et~al.}(1991){de Vaucouleurs}, {de Vaucouleurs},
  {Corwin}, {Buta}, {Paturel}, \& {Fouqu{\'e}}}]{devaucouleurs1991}
{de Vaucouleurs}, G., {de Vaucouleurs}, A., {Corwin}, Jr., H.~G., {Buta},
  R.~J., {Paturel}, G., \& {Fouqu{\'e}}, P. 1991, {Third Reference Catalogue of
  Bright Galaxies. Volume I: Explanations and references. Volume II: Data for
  galaxies between 0$^{h}$ and 12$^{h}$. Volume III: Data for galaxies between
  12$^{h}$ and 24$^{h}$.}

\bibitem[{{D{\'e}nes} {et~al.}(2014){D{\'e}nes}, {Kilborn}, \&
  {Koribalski}}]{denes2014}
{D{\'e}nes}, H., {Kilborn}, V.~A., \& {Koribalski}, B.~S. 2014, \mnras, 444,
  667

\bibitem[{{Dressler}(1980)}]{dressler1980}
{Dressler}, A. 1980, \apj, 236, 351

\bibitem[{{Driver} {et~al.}(2008){Driver}, {Popescu}, {Tuffs}, {Graham},
  {Liske}, \& {Baldry}}]{driver2008}
{Driver}, S.~P., {Popescu}, C.~C., {Tuffs}, R.~J., {Graham}, A.~W., {Liske},
  J., \& {Baldry}, I. 2008, \apjl, 678, L101

\bibitem[{{Ellison} {et~al.}(2009){Ellison}, {Simard}, {Cowan}, {Baldry},
  {Patton}, \& {McConnachie}}]{ellison2009}
{Ellison}, S.~L., {Simard}, L., {Cowan}, N.~B., {Baldry}, I.~K., {Patton},
  D.~R., \& {McConnachie}, A.~W. 2009, \mnras, 396, 1257

\bibitem[{{Gavazzi} {et~al.}(2009){Gavazzi}, {Adami}, {Durret}, {Cuillandre},
  {Ilbert}, {Mazure}, {Pell{\'o}}, \& {Ulmer}}]{gavazzi2009}
{Gavazzi}, R., {Adami}, C., {Durret}, F., {Cuillandre}, J.-C., {Ilbert}, O.,
  {Mazure}, A., {Pell{\'o}}, R., \& {Ulmer}, M.~P. 2009, \aap, 498, L33

\bibitem[{{Girardi} {et~al.}(2005){Girardi}, {Demarco}, {Rosati}, \&
  {Borgani}}]{girardi2005}
{Girardi}, M., {Demarco}, R., {Rosati}, P., \& {Borgani}, S. 2005, \aap, 442,
  29

\bibitem[{{Giuricin} {et~al.}(2000){Giuricin}, {Marinoni}, {Ceriani}, \&
  {Pisani}}]{giuricin2000}
{Giuricin}, G., {Marinoni}, C., {Ceriani}, L., \& {Pisani}, A. 2000, \apj, 543,
  178

\bibitem[{{G{\'o}mez} {et~al.}(2003){G{\'o}mez}, {Nichol}, {Miller}, {Balogh},
  {Goto}, {Zabludoff}, {Romer}, {Bernardi}, {Sheth}, {Hopkins}, {Castander},
  {Connolly}, {Schneider}, {Brinkmann}, {Lamb}, {SubbaRao}, \&
  {York}}]{gomez2003}
{G{\'o}mez}, P.~L., {Nichol}, R.~C., {Miller}, C.~J., {Balogh}, M.~L., {Goto},
  T., {Zabludoff}, A.~I., {Romer}, A.~K., {Bernardi}, M., {Sheth}, R.,
  {Hopkins}, A.~M., {Castander}, F.~J., {Connolly}, A.~J., {Schneider}, D.~P.,
  {Brinkmann}, J., {Lamb}, D.~Q., {SubbaRao}, M., \& {York}, D.~G. 2003, \apj,
  584, 210

\bibitem[{{Gonzalez} {et~al.}(2005){Gonzalez}, {Tran}, {Conbere}, \&
  {Zaritsky}}]{gonzalez2005}
{Gonzalez}, A.~H., {Tran}, K.-V.~H., {Conbere}, M.~N., \& {Zaritsky}, D. 2005,
  \apjl, 624, L73

\bibitem[{{Haynes} \& {Giovanelli}(1983)}]{haynes1983}
{Haynes}, M.~P. \& {Giovanelli}, R. 1983, \apj, 275, 472

\bibitem[{{Haynes} {et~al.}(2011){Haynes}, {Giovanelli}, {Martin}, {Hess},
  {Saintonge}, {Adams}, {Hallenbeck}, {Hoffman}, {Huang}, {Kent}, {Koopmann},
  {Papastergis}, {Stierwalt}, {Balonek}, {Craig}, {Higdon}, {Kornreich},
  {Miller}, {O'Donoghue}, {Olowin}, {Rosenberg}, {Spekkens}, {Troischt}, \&
  {Wilcots}}]{haynes2011}
{Haynes}, M.~P., {Giovanelli}, R., {Martin}, A.~M., {Hess}, K.~M., {Saintonge},
  A., {Adams}, E.~A.~K., {Hallenbeck}, G., {Hoffman}, G.~L., {Huang}, S.,
  {Kent}, B.~R., {Koopmann}, R.~A., {Papastergis}, E., {Stierwalt}, S.,
  {Balonek}, T.~J., {Craig}, D.~W., {Higdon}, S.~J.~U., {Kornreich}, D.~A.,
  {Miller}, J.~R., {O'Donoghue}, A.~A., {Olowin}, R.~P., {Rosenberg}, J.~L.,
  {Spekkens}, K., {Troischt}, P., \& {Wilcots}, E.~M. 2011, \aj, 142, 170

\bibitem[{{Heisler} {et~al.}(1985){Heisler}, {Tremaine}, \&
  {Bahcall}}]{heisler1985}
{Heisler}, J., {Tremaine}, S., \& {Bahcall}, J.~N. 1985, \apj, 298, 8

\bibitem[{{Huchra} \& {Geller}(1982)}]{huchra1982}
{Huchra}, J.~P. \& {Geller}, M.~J. 1982, \apj, 257, 423

\bibitem[{{Jarrett}(2004)}]{jarrett2004}
{Jarrett}, T. 2004, \pasa, 21, 396

\bibitem[{{Karachentsev} {et~al.}(2013){Karachentsev}, {Nasonova}, \&
  {Courtois}}]{karachentsev2013}
{Karachentsev}, I.~D., {Nasonova}, O.~G., \& {Courtois}, H.~M. 2013, \mnras,
  429, 2264

\bibitem[{{Karachentsev} {et~al.}(2014){Karachentsev}, {Tully}, {Wu}, {Shaya},
  \& {Dolphin}}]{karachentsev2014}
{Karachentsev}, I.~D., {Tully}, R.~B., {Wu}, P.-F., {Shaya}, E.~J., \&
  {Dolphin}, A.~E. 2014, \apj, 782, 4

\bibitem[{{Kauffmann} {et~al.}(2004){Kauffmann}, {White}, {Heckman},
  {M{\'e}nard}, {Brinchmann}, {Charlot}, {Tremonti}, \&
  {Brinkmann}}]{kauffmann2004}
{Kauffmann}, G., {White}, S.~D.~M., {Heckman}, T.~M., {M{\'e}nard}, B.,
  {Brinchmann}, J., {Charlot}, S., {Tremonti}, C., \& {Brinkmann}, J. 2004,
  \mnras, 353, 713

\bibitem[{{Kilborn} {et~al.}(2009){Kilborn}, {Forbes}, {Barnes}, {Koribalski},
  {Brough}, \& {Kern}}]{kilborn2009}
{Kilborn}, V.~A., {Forbes}, D.~A., {Barnes}, D.~G., {Koribalski}, B.~S.,
  {Brough}, S., \& {Kern}, K. 2009, \mnras, 400, 1962

\bibitem[{{Kova{\v c}} {et~al.}(2009){Kova{\v c}}, {Oosterloo}, \& {van der
  Hulst}}]{kovac2009}
{Kova{\v c}}, K., {Oosterloo}, T.~A., \& {van der Hulst}, J.~M. 2009, \mnras,
  400, 743

\bibitem[{{Lang} {et~al.}(2003){Lang}, {Boyce}, {Kilborn}, {Minchin}, {Disney},
  {Jordan}, {Grossi}, {Garcia}, {Freeman}, {Phillipps}, \& {Wright}}]{lang2003}
{Lang}, R.~H., {Boyce}, P.~J., {Kilborn}, V.~A., {Minchin}, R.~F., {Disney},
  M.~J., {Jordan}, C.~A., {Grossi}, M., {Garcia}, D.~A., {Freeman}, K.~C.,
  {Phillipps}, S., \& {Wright}, A.~E. 2003, \mnras, 342, 738

\bibitem[{{Lewis} {et~al.}(2002){Lewis}, {Balogh}, {De Propris}, {Couch},
  {Bower}, {Offer}, {Bland-Hawthorn}, {Baldry}, {Baugh}, {Bridges}, {Cannon},
  {Cole}, {Colless}, {Collins}, {Cross}, {Dalton}, {Driver}, {Efstathiou},
  {Ellis}, {Frenk}, {Glazebrook}, {Hawkins}, {Jackson}, {Lahav}, {Lumsden},
  {Maddox}, {Madgwick}, {Norberg}, {Peacock}, {Percival}, {Peterson},
  {Sutherland}, \& {Taylor}}]{lewis2002}
{Lewis}, I., {Balogh}, M., {De Propris}, R., {Couch}, W., {Bower}, R., {Offer},
  A., {Bland-Hawthorn}, J., {Baldry}, I.~K., {Baugh}, C., {Bridges}, T.,
  {Cannon}, R., {Cole}, S., {Colless}, M., {Collins}, C., {Cross}, N.,
  {Dalton}, G., {Driver}, S.~P., {Efstathiou}, G., {Ellis}, R.~S., {Frenk},
  C.~S., {Glazebrook}, K., {Hawkins}, E., {Jackson}, C., {Lahav}, O.,
  {Lumsden}, S., {Maddox}, S., {Madgwick}, D., {Norberg}, P., {Peacock}, J.~A.,
  {Percival}, W., {Peterson}, B.~A., {Sutherland}, W., \& {Taylor}, K. 2002,
  \mnras, 334, 673

\bibitem[{{Lin} \& {Mohr}(2004)}]{lin2004}
{Lin}, Y.-T. \& {Mohr}, J.~J. 2004, \apj, 617, 879

\bibitem[{{Makarov} \& {Karachentsev}(2011)}]{makarov2011}
{Makarov}, D. \& {Karachentsev}, I. 2011, \mnras, 412, 2498

\bibitem[{{Morganti} {et~al.}(2006){Morganti}, {de Zeeuw}, {Oosterloo},
  {McDermid}, {Krajnovi{\'c}}, {Cappellari}, {Kenn}, {Weijmans}, \&
  {Sarzi}}]{morganti2006}
{Morganti}, R., {de Zeeuw}, P.~T., {Oosterloo}, T.~A., {McDermid}, R.~M.,
  {Krajnovi{\'c}}, D., {Cappellari}, M., {Kenn}, F., {Weijmans}, A., \&
  {Sarzi}, M. 2006, \mnras, 371, 157

\bibitem[{{Mulchaey} \& {Zabludoff}(1998)}]{mulchaey1998}
{Mulchaey}, J.~S. \& {Zabludoff}, A.~I. 1998, \apj, 496, 73

\bibitem[{{Nolthenius} \& {White}(1987)}]{nolthenius1987}
{Nolthenius}, R. \& {White}, S.~D.~M. 1987, \mnras, 225, 505

\bibitem[{{Oegerle} \& {Hill}(2001)}]{oegerle2001}
{Oegerle}, W.~R. \& {Hill}, J.~M. 2001, \aj, 122, 2858

\bibitem[{{Osmond} \& {Ponman}(2004)}]{osmond2004}
{Osmond}, J.~P.~F. \& {Ponman}, T.~J. 2004, \mnras, 350, 1511

\bibitem[{{Pisano} {et~al.}(2011){Pisano}, {Barnes}, {Gibson},
  {Staveley-Smith}, {Freeman}, \& {Kilborn}}]{pisano2011}
{Pisano}, D.~J., {Barnes}, D.~G., {Gibson}, B.~K., {Staveley-Smith}, L.,
  {Freeman}, K.~C., \& {Kilborn}, V.~A. 2011, {Measuring the Halo Mass Function
  in Loose Groups}, ed. I.~{Ferreras} \& A.~{Pasquali}, 47

\bibitem[{{Raimond} {et~al.}(1981){Raimond}, {Faber}, {Gallagher}, \&
  {Knapp}}]{raimond1981}
{Raimond}, E., {Faber}, S.~M., {Gallagher}, III, J.~S., \& {Knapp}, G.~R. 1981,
  \apj, 246, 708

\bibitem[{{Ramella} {et~al.}(1995){Ramella}, {Geller}, {Huchra}, \&
  {Thorstensen}}]{ramella1995}
{Ramella}, M., {Geller}, M.~J., {Huchra}, J.~P., \& {Thorstensen}, J.~R. 1995,
  \aj, 109, 1458

\bibitem[{{Rasmussen} {et~al.}(2006){Rasmussen}, {Ponman}, \&
  {Mulchaey}}]{rasmussen2006}
{Rasmussen}, J., {Ponman}, T.~J., \& {Mulchaey}, J.~S. 2006, \mnras, 370, 453

\bibitem[{{Robotham} {et~al.}(2011){Robotham}, {Norberg}, {Driver}, {Baldry},
  {Bamford}, {Hopkins}, {Liske}, {Loveday}, {Merson}, {Peacock}, {Brough},
  {Cameron}, {Conselice}, {Croom}, {Frenk}, {Gunawardhana}, {Hill}, {Jones},
  {Kelvin}, {Kuijken}, {Nichol}, {Parkinson}, {Pimbblet}, {Phillipps},
  {Popescu}, {Prescott}, {Sharp}, {Sutherland}, {Taylor}, {Thomas}, {Tuffs},
  {van Kampen}, \& {Wijesinghe}}]{robotham2011}
{Robotham}, A.~S.~G., {Norberg}, P., {Driver}, S.~P., {Baldry}, I.~K.,
  {Bamford}, S.~P., {Hopkins}, A.~M., {Liske}, J., {Loveday}, J., {Merson}, A.,
  {Peacock}, J.~A., {Brough}, S., {Cameron}, E., {Conselice}, C.~J., {Croom},
  S.~M., {Frenk}, C.~S., {Gunawardhana}, M., {Hill}, D.~T., {Jones}, D.~H.,
  {Kelvin}, L.~S., {Kuijken}, K., {Nichol}, R.~C., {Parkinson}, H.~R.,
  {Pimbblet}, K.~A., {Phillipps}, S., {Popescu}, C.~C., {Prescott}, M.,
  {Sharp}, R.~G., {Sutherland}, W.~J., {Taylor}, E.~N., {Thomas}, D., {Tuffs},
  R.~J., {van Kampen}, E., \& {Wijesinghe}, D. 2011, \mnras, 416, 2640

\bibitem[{{Schlegel} {et~al.}(1998){Schlegel}, {Finkbeiner}, \&
  {Davis}}]{schlegel1998}
{Schlegel}, D.~J., {Finkbeiner}, D.~P., \& {Davis}, M. 1998, \apj, 500, 525

\bibitem[{{Shimasaku} {et~al.}(2001){Shimasaku}, {Fukugita}, {Doi}, {Hamabe},
  {Ichikawa}, {Okamura}, {Sekiguchi}, {Yasuda}, {Brinkmann}, {Csabai},
  {Ichikawa}, {Ivezi{\'c}}, {Kunszt}, {Schneider}, {Szokoly}, {Watanabe}, \&
  {York}}]{shimasaku2001}
{Shimasaku}, K., {Fukugita}, M., {Doi}, M., {Hamabe}, M., {Ichikawa}, T.,
  {Okamura}, S., {Sekiguchi}, M., {Yasuda}, N., {Brinkmann}, J., {Csabai}, I.,
  {Ichikawa}, S.-I., {Ivezi{\'c}}, Z., {Kunszt}, P.~Z., {Schneider}, D.~P.,
  {Szokoly}, G.~P., {Watanabe}, M., \& {York}, D.~G. 2001, \aj, 122, 1238

\bibitem[{{Springel} {et~al.}(2005){Springel}, {White}, {Jenkins}, {Frenk},
  {Yoshida}, {Gao}, {Navarro}, {Thacker}, {Croton}, {Helly}, {Peacock}, {Cole},
  {Thomas}, {Couchman}, {Evrard}, {Colberg}, \& {Pearce}}]{springel2005}
{Springel}, V., {White}, S.~D.~M., {Jenkins}, A., {Frenk}, C.~S., {Yoshida},
  N., {Gao}, L., {Navarro}, J., {Thacker}, R., {Croton}, D., {Helly}, J.,
  {Peacock}, J.~A., {Cole}, S., {Thomas}, P., {Couchman}, H., {Evrard}, A.,
  {Colberg}, J., \& {Pearce}, F. 2005, \nat, 435, 629

\bibitem[{{Springob} {et~al.}(2005){Springob}, {Haynes}, {Giovanelli}, \&
  {Kent}}]{springob2005}
{Springob}, C.~M., {Haynes}, M.~P., {Giovanelli}, R., \& {Kent}, B.~R. 2005,
  \apjs, 160, 149

\bibitem[{{Strateva} {et~al.}(2001){Strateva}, {Ivezi{\'c}}, {Knapp},
  {Narayanan}, {Strauss}, {Gunn}, {Lupton}, {Schlegel}, {Bahcall}, {Brinkmann},
  {Brunner}, {Budav{\'a}ri}, {Csabai}, {Castander}, {Doi}, {Fukugita}, {Gy{\H
  o}ry}, {Hamabe}, {Hennessy}, {Ichikawa}, {Kunszt}, {Lamb}, {McKay},
  {Okamura}, {Racusin}, {Sekiguchi}, {Schneider}, {Shimasaku}, \&
  {York}}]{strateva2001}
{Strateva}, I., {Ivezi{\'c}}, {\v Z}., {Knapp}, G.~R., {Narayanan}, V.~K.,
  {Strauss}, M.~A., {Gunn}, J.~E., {Lupton}, R.~H., {Schlegel}, D., {Bahcall},
  N.~A., {Brinkmann}, J., {Brunner}, R.~J., {Budav{\'a}ri}, T., {Csabai}, I.,
  {Castander}, F.~J., {Doi}, M., {Fukugita}, M., {Gy{\H o}ry}, Z., {Hamabe},
  M., {Hennessy}, G., {Ichikawa}, T., {Kunszt}, P.~Z., {Lamb}, D.~Q., {McKay},
  T.~A., {Okamura}, S., {Racusin}, J., {Sekiguchi}, M., {Schneider}, D.~P.,
  {Shimasaku}, K., \& {York}, D. 2001, \aj, 122, 1861

\bibitem[{{Strauss} {et~al.}(2002){Strauss}, {Weinberg}, {Lupton}, {Narayanan},
  {Annis}, {Bernardi}, {Blanton}, {Burles}, {Connolly}, {Dalcanton}, {Doi},
  {Eisenstein}, {Frieman}, {Fukugita}, {Gunn}, {Ivezi{\'c}}, {Kent}, {Kim},
  {Knapp}, {Kron}, {Munn}, {Newberg}, {Nichol}, {Okamura}, {Quinn}, {Richmond},
  {Schlegel}, {Shimasaku}, {SubbaRao}, {Szalay}, {Vanden Berk}, {Vogeley},
  {Yanny}, {Yasuda}, {York}, \& {Zehavi}}]{strauss2002}
{Strauss}, M.~A., {Weinberg}, D.~H., {Lupton}, R.~H., {Narayanan}, V.~K.,
  {Annis}, J., {Bernardi}, M., {Blanton}, M., {Burles}, S., {Connolly}, A.~J.,
  {Dalcanton}, J., {Doi}, M., {Eisenstein}, D., {Frieman}, J.~A., {Fukugita},
  M., {Gunn}, J.~E., {Ivezi{\'c}}, {\v Z}., {Kent}, S., {Kim}, R.~S.~J.,
  {Knapp}, G.~R., {Kron}, R.~G., {Munn}, J.~A., {Newberg}, H.~J., {Nichol},
  R.~C., {Okamura}, S., {Quinn}, T.~R., {Richmond}, M.~W., {Schlegel}, D.~J.,
  {Shimasaku}, K., {SubbaRao}, M., {Szalay}, A.~S., {Vanden Berk}, D.,
  {Vogeley}, M.~S., {Yanny}, B., {Yasuda}, N., {York}, D.~G., \& {Zehavi}, I.
  2002, \aj, 124, 1810

\bibitem[{{Tran} {et~al.}(2009){Tran}, {Saintonge}, {Moustakas}, {Bai},
  {Gonzalez}, {Holden}, {Zaritsky}, \& {Kautsch}}]{tran2009}
{Tran}, K.-V.~H., {Saintonge}, A., {Moustakas}, J., {Bai}, L., {Gonzalez},
  A.~H., {Holden}, B.~P., {Zaritsky}, D., \& {Kautsch}, S.~J. 2009, \apj, 705,
  809

\bibitem[{{Tully}(1987)}]{tully1987}
{Tully}, R.~B. 1987, \apj, 321, 280

\bibitem[{{Tully} {et~al.}(2009){Tully}, {Rizzi}, {Shaya}, {Courtois},
  {Makarov}, \& {Jacobs}}]{tully2009}
{Tully}, R.~B., {Rizzi}, L., {Shaya}, E.~J., {Courtois}, H.~M., {Makarov},
  D.~I., \& {Jacobs}, B.~A. 2009, \aj, 138, 323

\bibitem[{{Tully} {et~al.}(2008){Tully}, {Shaya}, {Karachentsev}, {Courtois},
  {Kocevski}, {Rizzi}, \& {Peel}}]{tully2008}
{Tully}, R.~B., {Shaya}, E.~J., {Karachentsev}, I.~D., {Courtois}, H.~M.,
  {Kocevski}, D.~D., {Rizzi}, L., \& {Peel}, A. 2008, \apj, 686, 1523

\bibitem[{{Tully} {et~al.}(1996){Tully}, {Verheijen}, {Pierce}, {Huang}, \&
  {Wainscoat}}]{tully1996}
{Tully}, R.~B., {Verheijen}, M.~A.~W., {Pierce}, M.~J., {Huang}, J.-S., \&
  {Wainscoat}, R.~J. 1996, \aj, 112, 2471

\bibitem[{{Verheijen} \& {Sancisi}(2001)}]{verheijen2001}
{Verheijen}, M.~A.~W. \& {Sancisi}, R. 2001, \aap, 370, 765

\bibitem[{{Wolfinger} {et~al.}(2013){Wolfinger}, {Kilborn}, {Koribalski},
  {Minchin}, {Boyce}, {Disney}, {Lang}, \& {Jordan}}]{wolfinger2013}
{Wolfinger}, K., {Kilborn}, V.~A., {Koribalski}, B.~S., {Minchin}, R.~F.,
  {Boyce}, P.~J., {Disney}, M.~J., {Lang}, R.~H., \& {Jordan}, C.~A. 2013,
  \mnras, 428, 1790

\bibitem[{{Yahil} {et~al.}(1977){Yahil}, {Tammann}, \& {Sandage}}]{yahil1977}
{Yahil}, A., {Tammann}, G.~A., \& {Sandage}, A. 1977, \apj, 217, 903

\bibitem[{{Zabludoff} {et~al.}(1990){Zabludoff}, {Huchra}, \&
  {Geller}}]{zabludoff1990}
{Zabludoff}, A.~I., {Huchra}, J.~P., \& {Geller}, M.~J. 1990, \apjs, 74, 1

\end{thebibliography}

\onecolumn
\section*{Appendix}

\subsection*{The main structures in the southern part of the Ursa Major region}
\label{subsec:south}

\begin{figure*}
\begin{center}
\begin{tabular}{p{4.3cm} p{4.7cm} p{5cm}}
\mbox{\includegraphics[trim=3cm 7cm 1cm 4cm, clip=true, height=4.6cm]{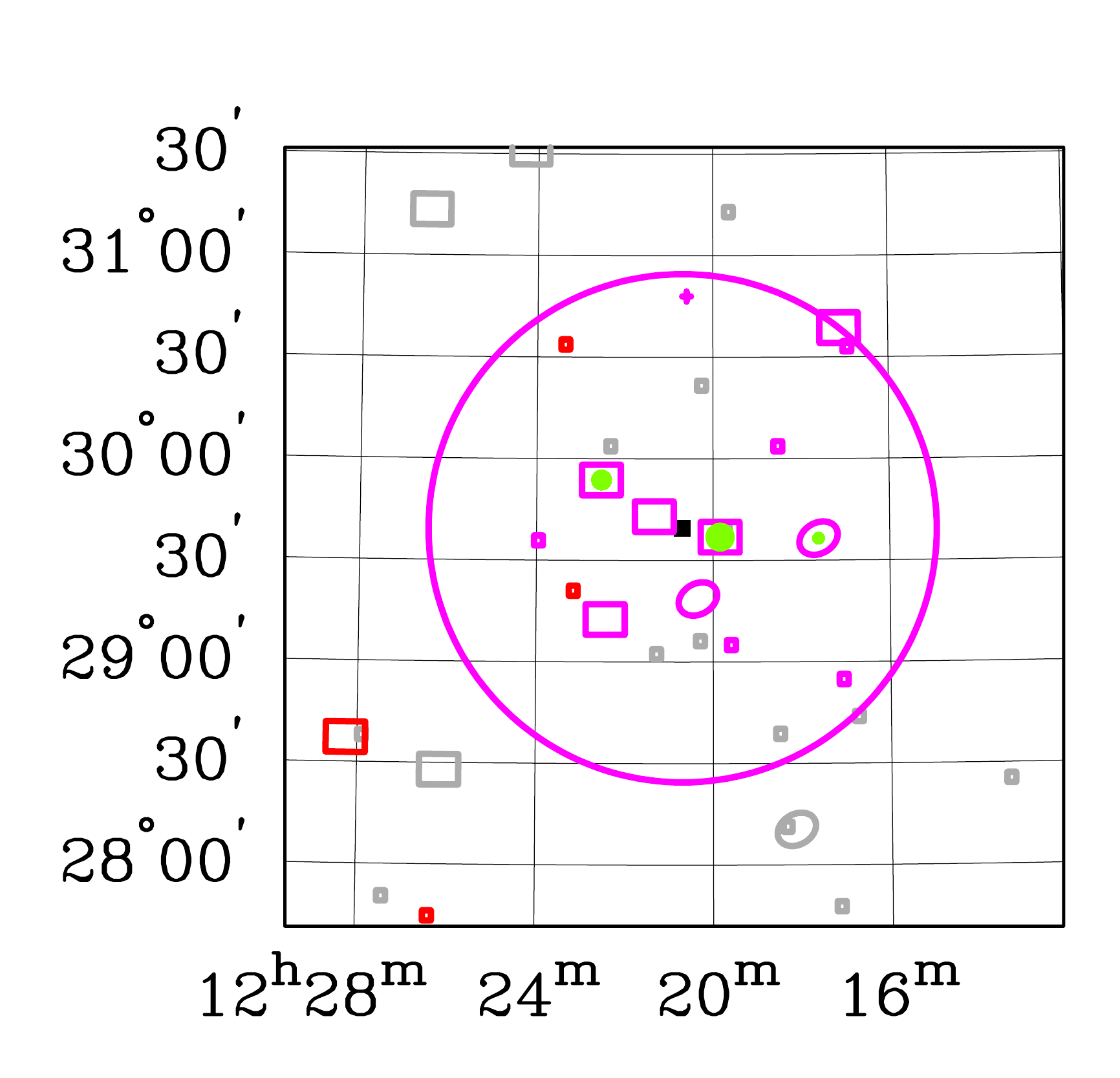}} & 
\mbox{\includegraphics[trim=0cm 0cm 0.5cm 0.5cm, clip=true, height=4.5cm]{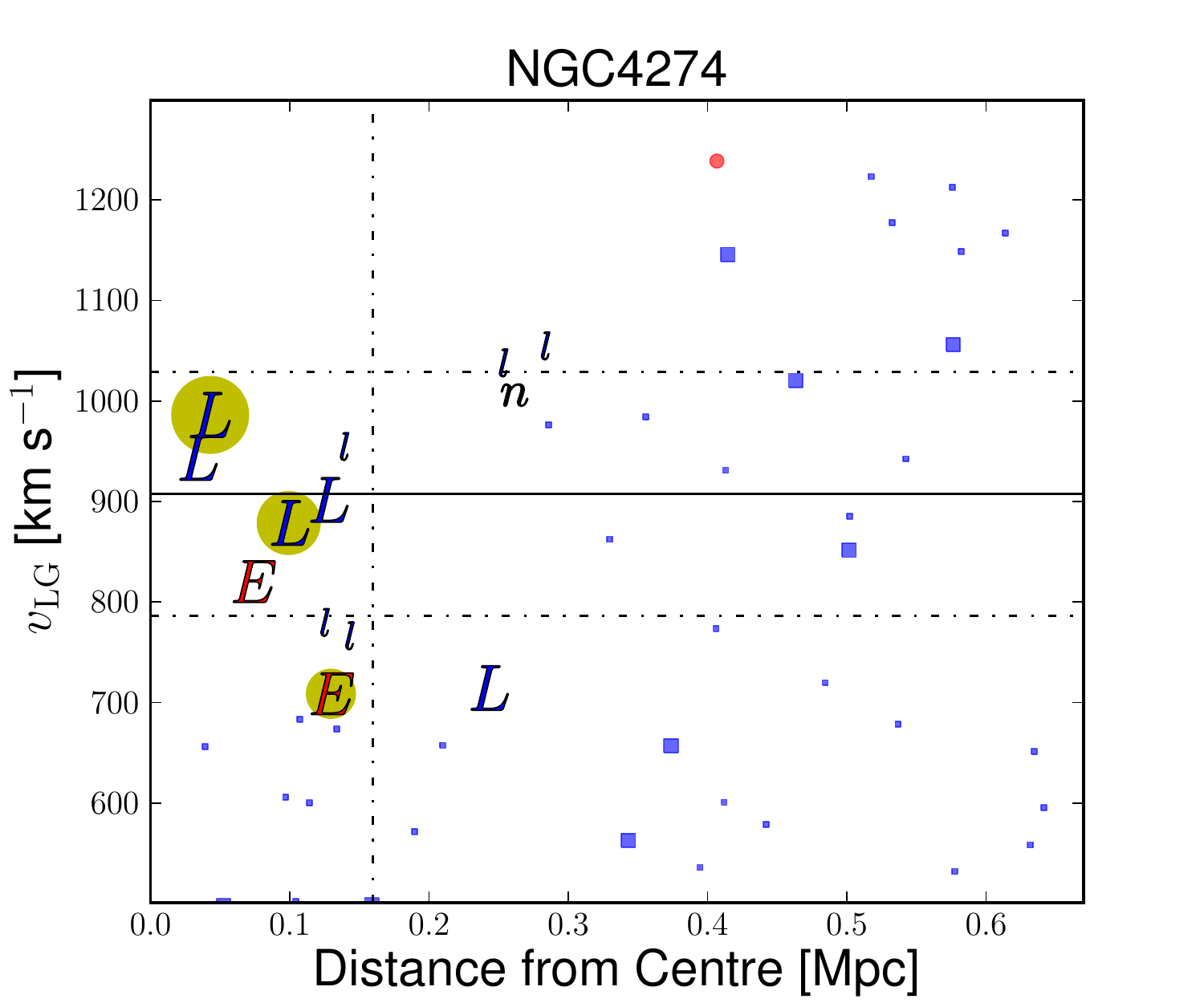}} &
\mbox{\includegraphics[trim=0.5cm 0cm 0.5cm 0.5cm, clip=true, height=4.5cm]{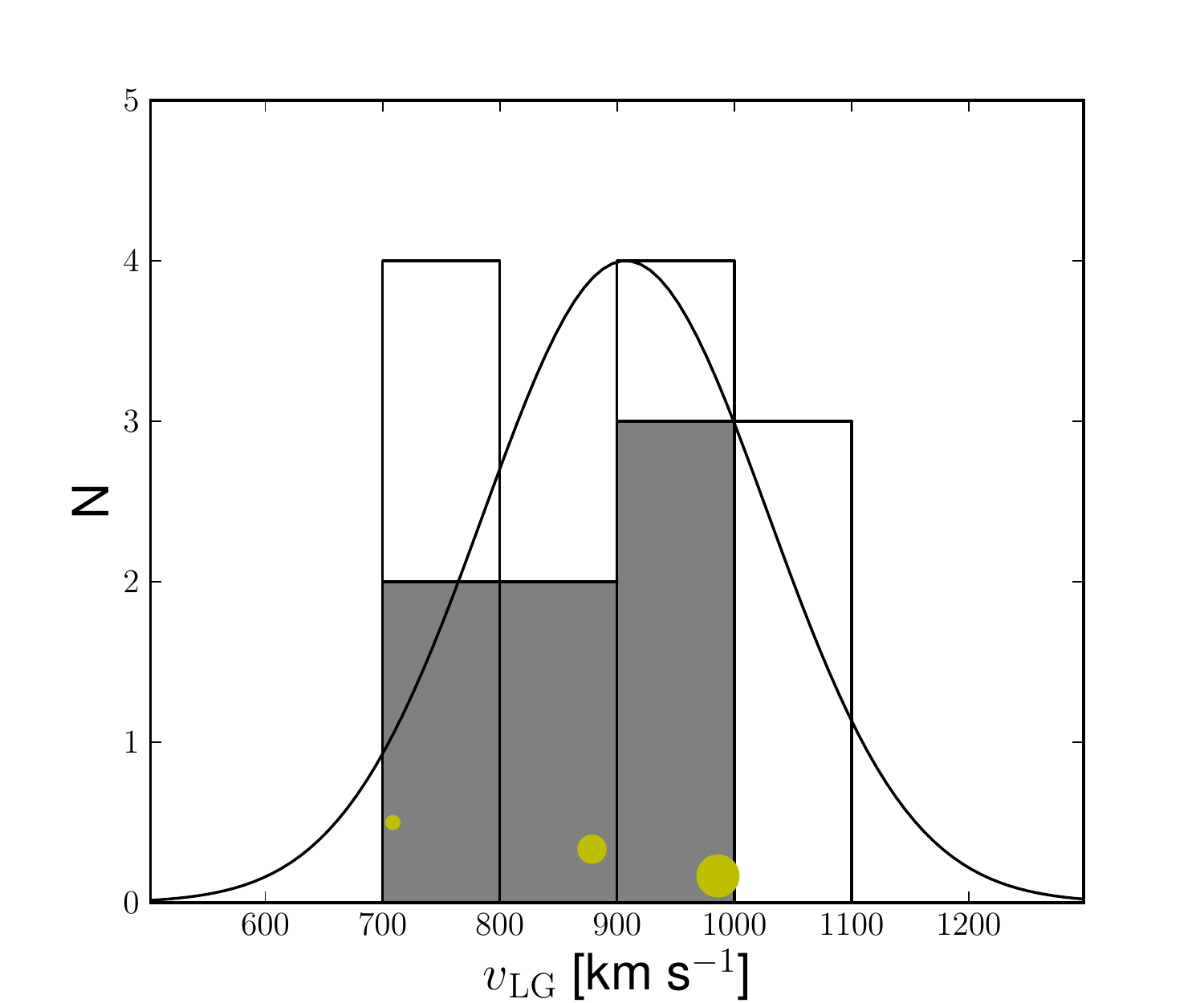}} \\
[6pt]
\mbox{\includegraphics[trim=3cm 6.5cm 1cm 4cm, clip=true, height=4.6cm]{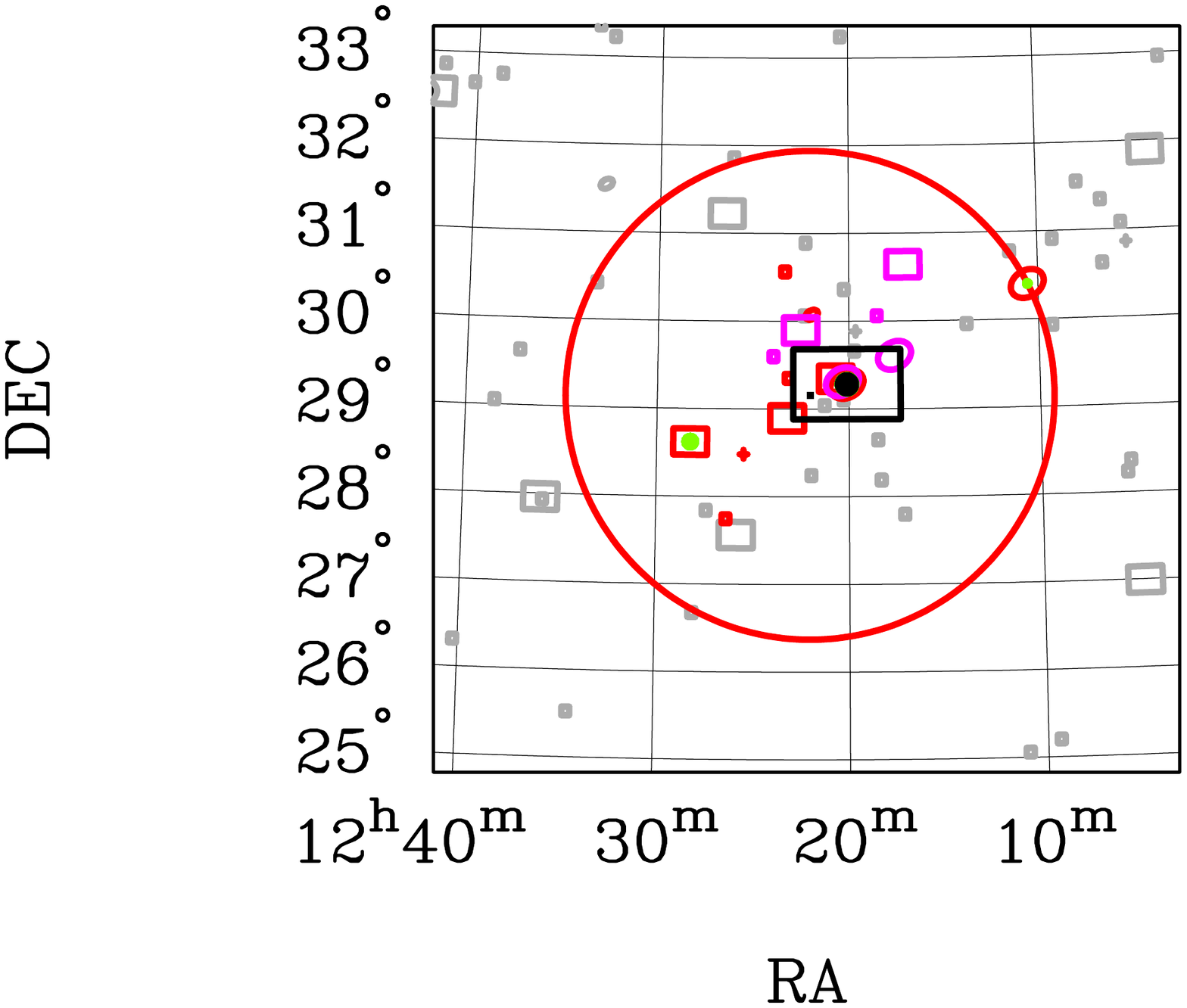}} & 
\mbox{\includegraphics[trim=0cm 0cm 0.5cm 0.5cm, clip=true, height=4.5cm]{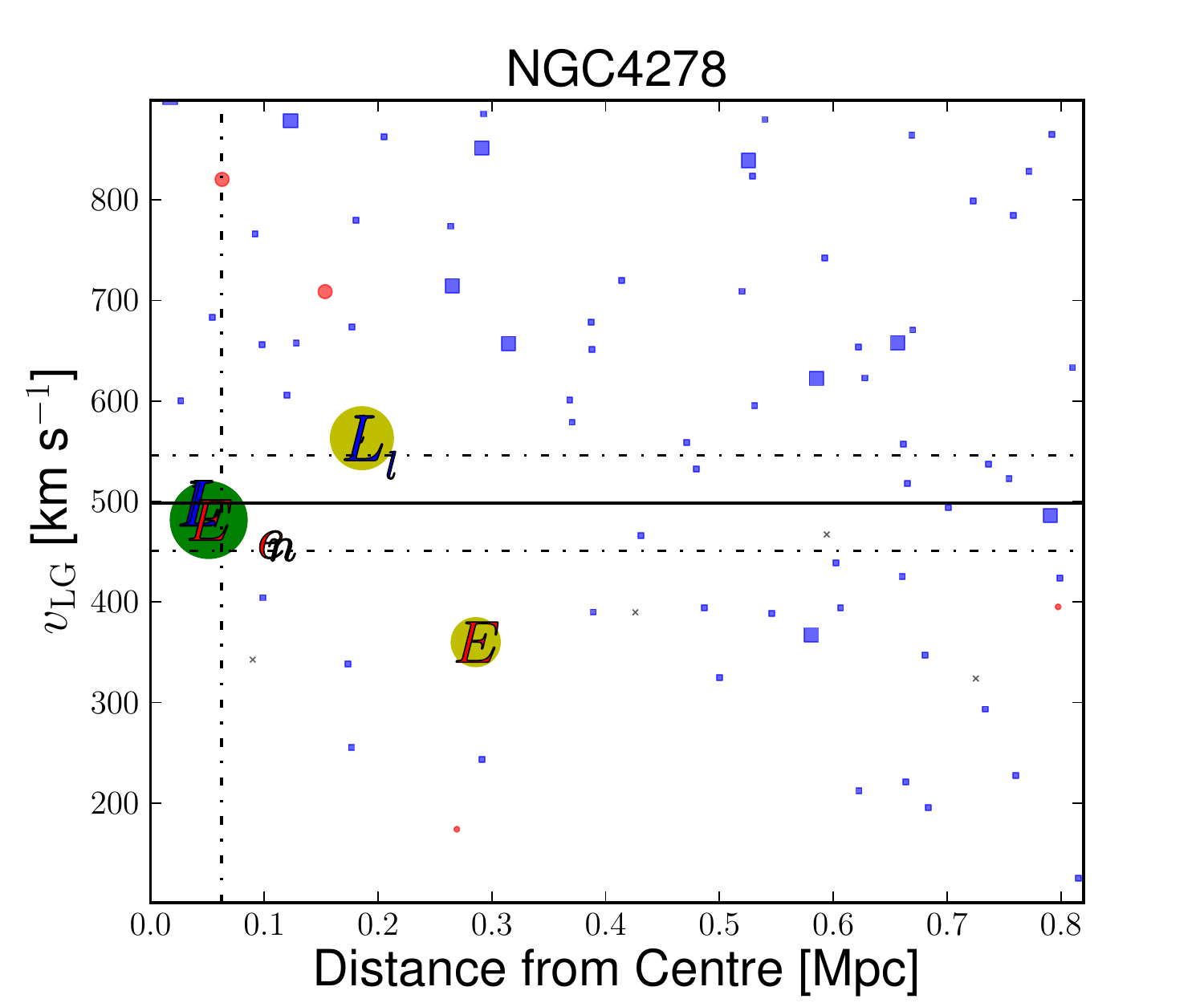}} &
\mbox{\includegraphics[trim=0.5cm 0cm 0.5cm 0.5cm, clip=true, height=4.5cm]{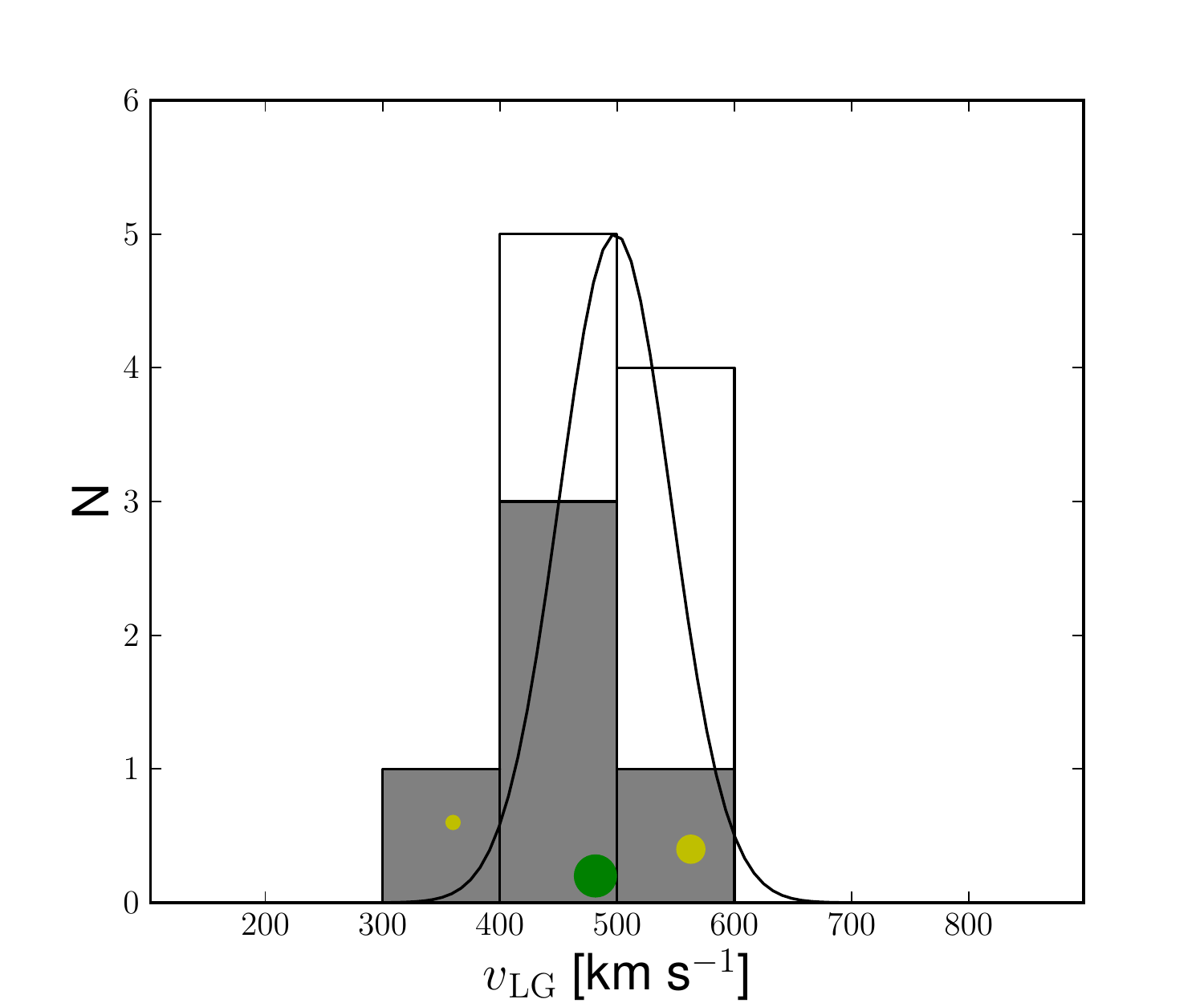}} \\
[6pt]
\mbox{\includegraphics[trim=3cm 6.5cm 1cm 4cm, clip=true, height=4.6cm]{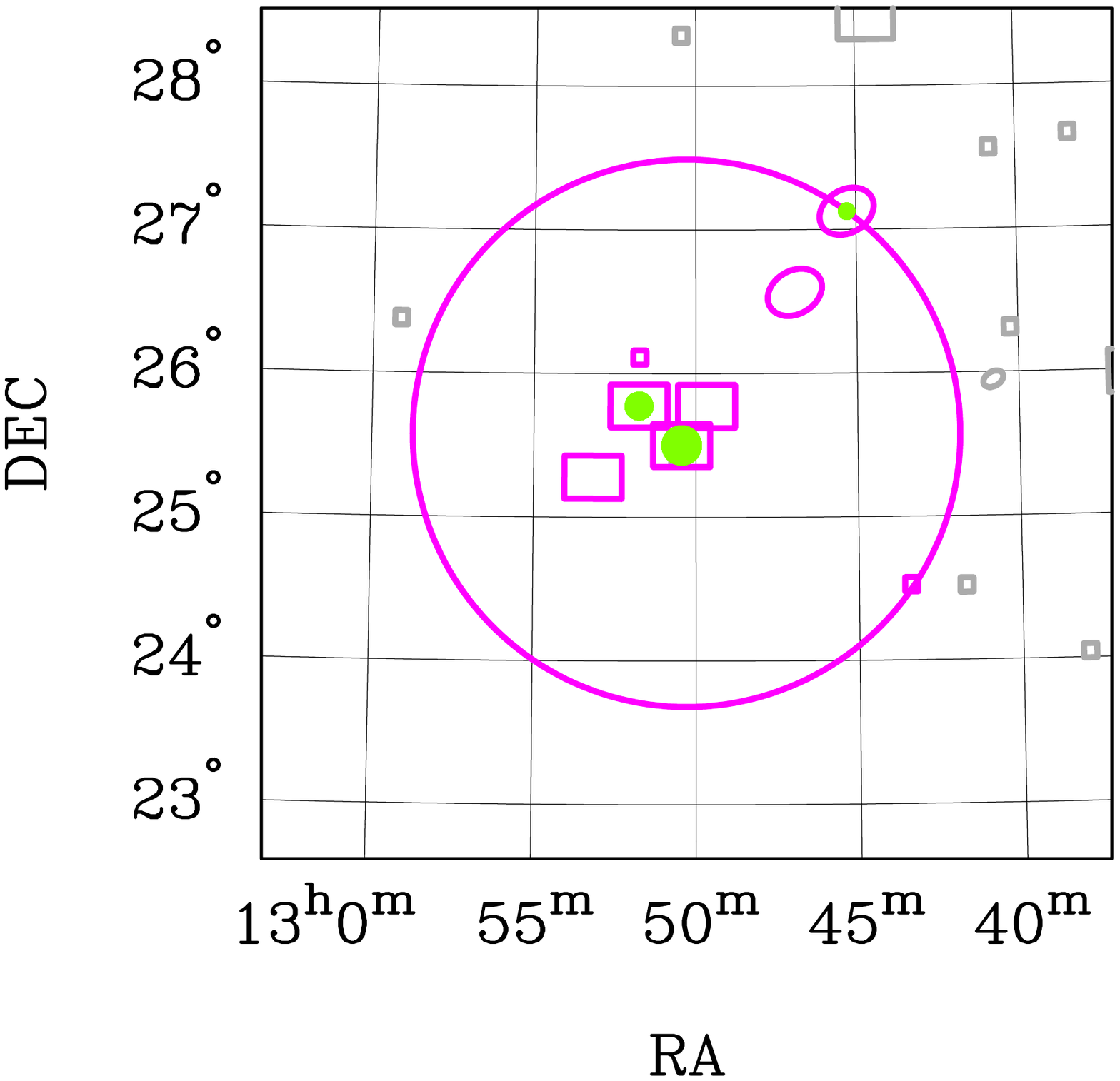}} & 
\mbox{\includegraphics[trim=0cm 0cm 0.5cm 0.5cm, clip=true, height=4.5cm]{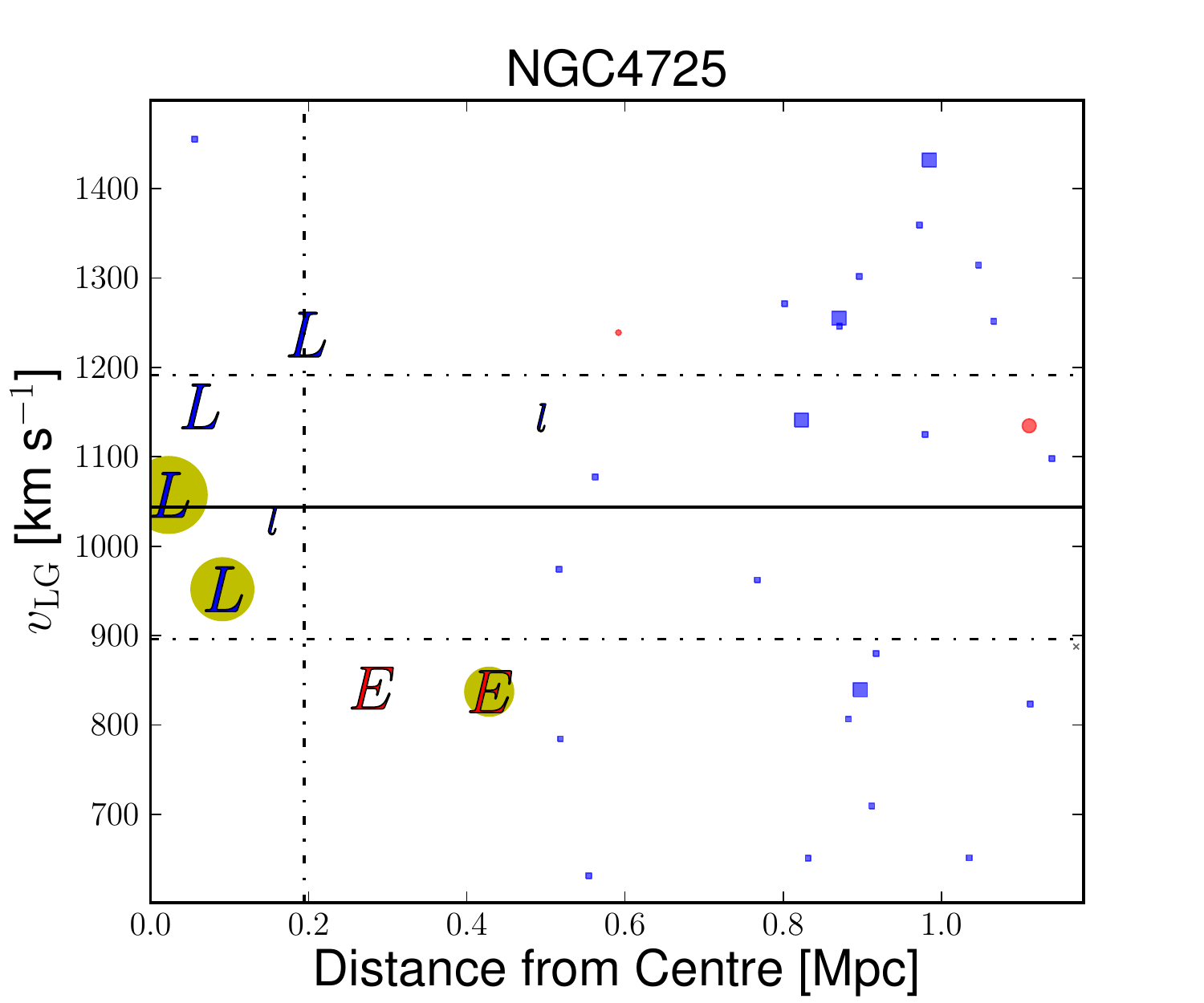}} &
\mbox{\includegraphics[trim=0.5cm 0cm 0.5cm 0.5cm, clip=true, height=4.5cm]{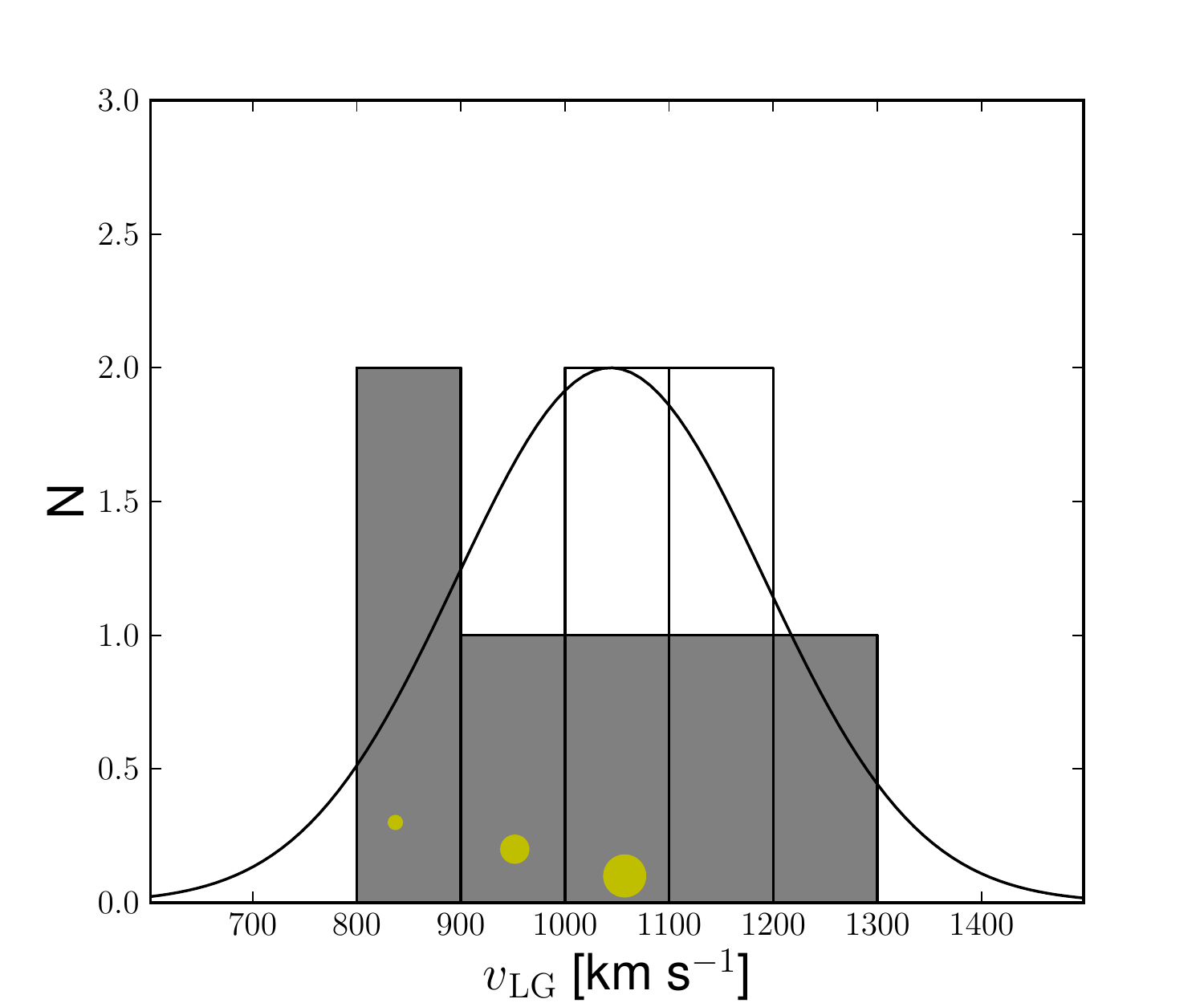}} \\
\end{tabular} 
\caption{The main structures in the southern part of the Ursa Major region (found using linking lengths $D_0=0.3$~Mpc and $V_0=150$~km~s$^{-1}$). Similar to Figure~\ref{fig:middle1}. The NGC4278 galaxy, which shows an HI cloud offset from the host galaxy \citep{morganti2006} is highlighted with a black dot (skymap) and a green circle (distance-velocity diagram and velocity histogram).}
\label{fig:south1}
\end{center}
\end{figure*}

The dominant structures in the southern part of the central Ursa Major region ($11^{h}\leq\alpha\leq+13^{h}$ and $20^{\circ}\leq\delta\leq+35^{\circ}$) are the NGC4274, the NGC4278 and NGC4725 groups (Figure~\ref{fig:south1}). All three groups have central velocities below 1100~km~s$^{-1}$ (see Table~\ref{tab:group_prop} for more details on the group properties).The NGC4278 and NGC4274 groups are located at a similar right ascension and declination, but with an offset in velocity -- the central group velocities are 498~km~s$^{-1}$ and 907~km~s$^{-1}$, respectively.

The NGC4274 group is the richest group in the southern part of the Ursa Major region with 13 group members (7 in the complete sample). NGC4274 (BGG) is located 41~kpc and 79~km~s$^{-1}$ from the luminosity-weighted group centroid. The group is dominated by late-type galaxies -- the early-type fraction is 29~per cent. NGC4283 is listed in the Extragalactic Distance Database (EDD; \citealt{tully2009}) at a quality distance of 15.7~Mpc. Assuming Hubble flow, NGC4283 resides at a distance of 11.2~Mpc and the group centre is at 12.4~Mpc, which is closer than the redshift-independent distance measurement. The group's velocity dispersion is 121~km~s$^{-1}$.

The BGG in the NGC4278 group lies close to the spatial and kinematic centre of the group, with an offset of 52~kpc and 17~km~s$^{-1}$ respectively. The NGC4278 group is located at similar position on the sky as the previously discussed NGC4274 group, but located in the foreground at a central velocity of 498~km~s$^{-1}$, which corresponds to a distance of 6.8~Mpc (Hubble flow). Two group members are listed in EDD \citep{tully2009} with quality distance measurements -- the bulge dominated galaxies NGC4278 (BGG) at 16.1~Mpc and NGC4150 ($3^{rd}$ BGG) at 13.7~Mpc. The redshift-independent distance measurements place the NGC4278 group at similar distance to the NGC4274 group. Large peculiar velocities may explain the inconsistency in quality and redshift distance measurements. The NGC4278 group may in fact be located behind the NGC4274 group and infalling, which would lead to blue-shifted redshift measurements, i.e. the distance obtained assuming Hubble flow is closer than the real distance. The group has an early-type fraction of 40~per cent and a low velocity dispersion of only 47~km~s$^{-1}$. Note that the BGG shows an HI tail (\citealt{morganti2006}; the galaxy and its surroundings are marked with a black box in Figure~\ref{fig:south1}; NGC4278 is highlighted with a green circle in the distance-velocity diagram and in the velocity histogram).

The group members of the NGC4725 group are distinct from other nearby galaxies as evidenced by a clear gap between group members and non-group galaxies in the distance-velocity diagram (for the complete sample, i.e. capital letters and large symbols). NGC4725 (BGG) is at the spatial and kinematic group centre with an offset of 23~kpc and 14~km~s$^{-1}$ respectively. The redshift-independent distance measurement for the BGG is 12.4~Mpc (EDD; \citealt{tully2009}), which is slightly less than the distance obtained assuming Hubble flow (14.3~Mpc). The group has a velocity dispersion of 148~km~s$^{-1}$ and an early-type fraction of 33~per cent.

\subsection*{The main structures in the northern part of the Ursa Major region}
\label{subsec:north}

\begin{figure*}
\begin{center}
\begin{tabular}{p{4.3cm} p{4.7cm} p{5cm}}
\mbox{\includegraphics[trim=3cm 6.5cm 1cm 4cm, clip=true, height=4.6cm]{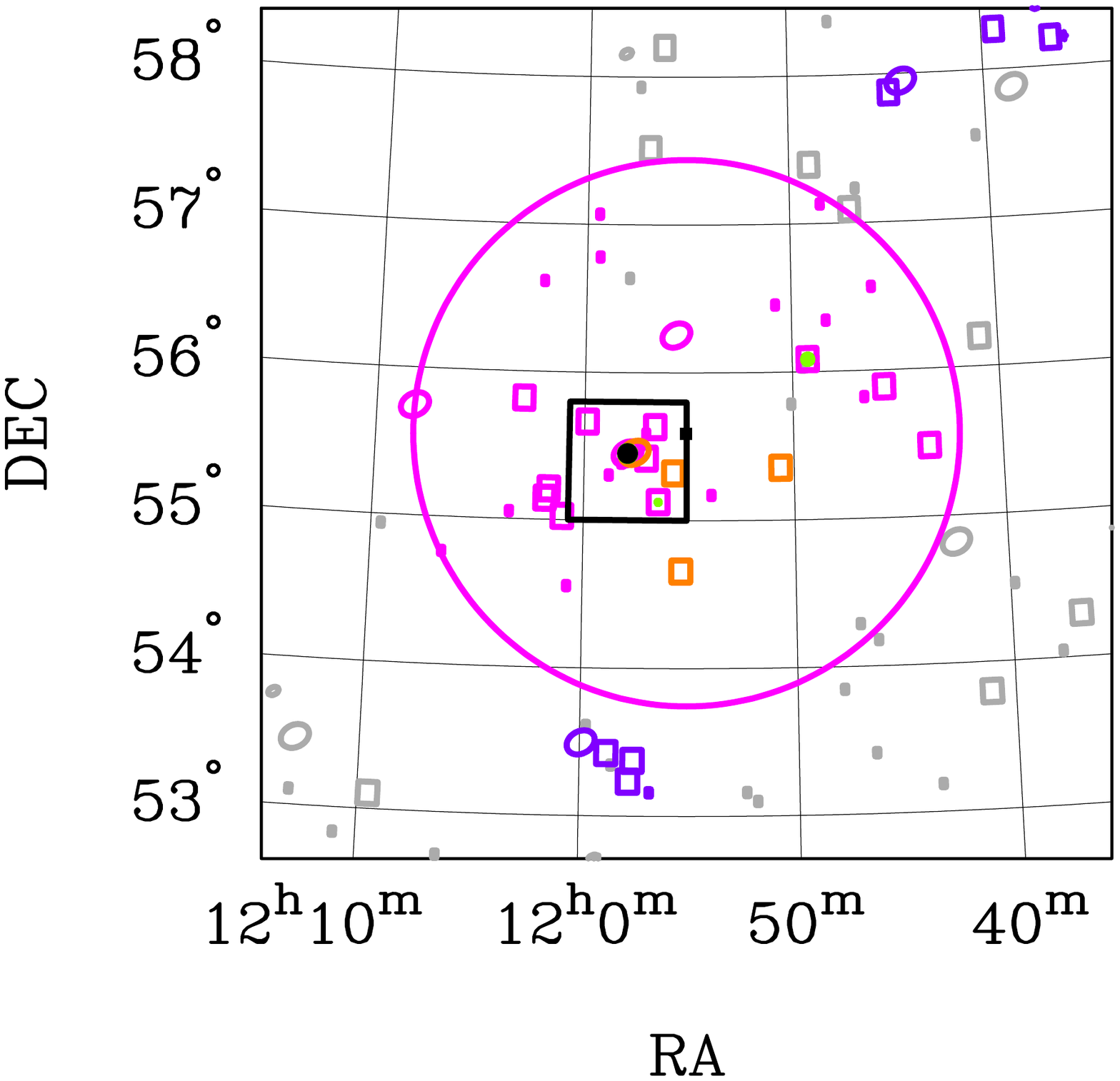}} & 
\mbox{\includegraphics[trim=0cm 0cm 0.5cm 0.5cm, clip=true, height=4.5cm]{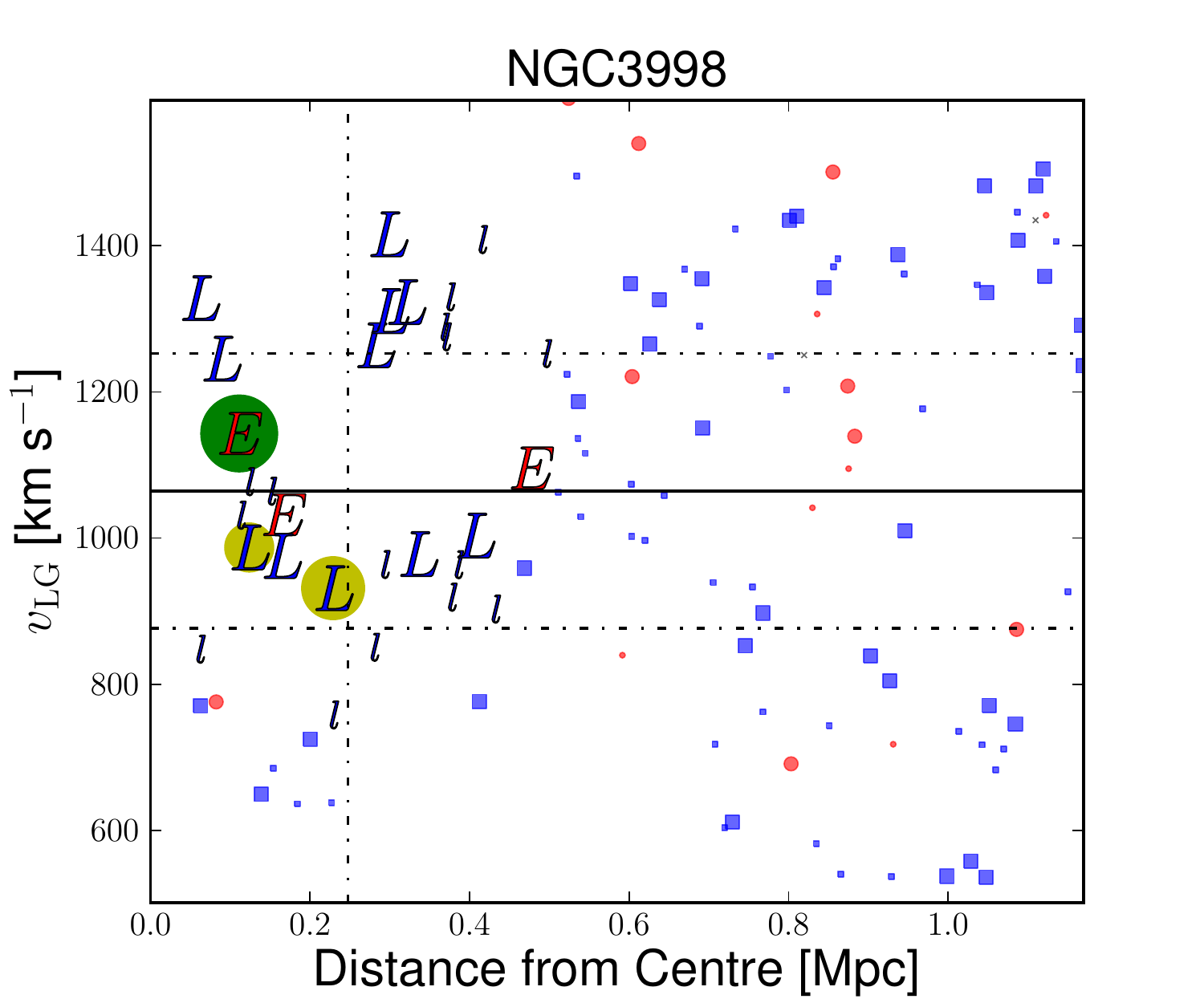}} &
\mbox{\includegraphics[trim=0.5cm 0cm 0.5cm 0.5cm, clip=true, height=4.5cm]{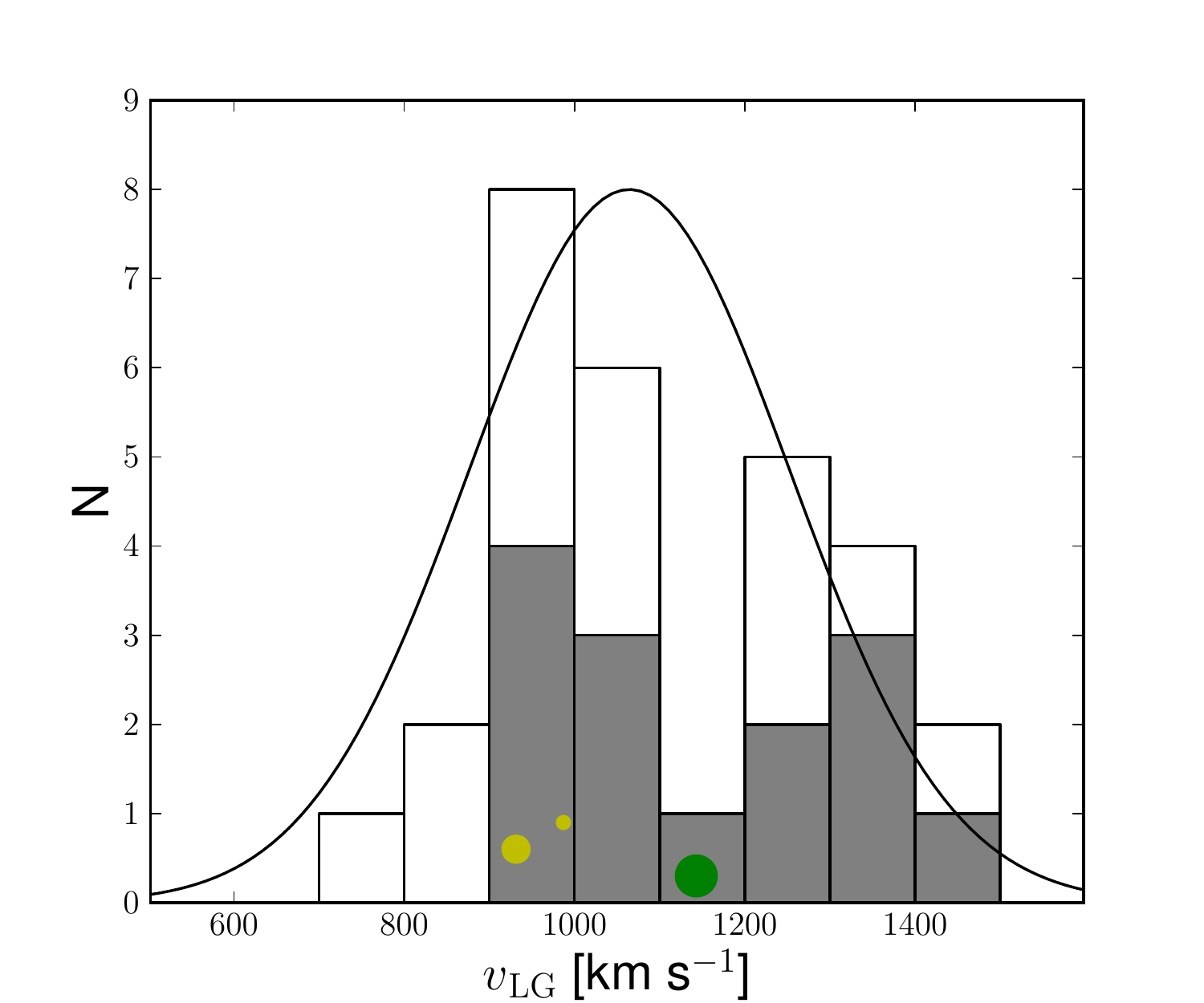}} \\
[12pt]
\mbox{\includegraphics[trim=3cm 6.5cm 1cm 4cm, clip=true, height=4.6cm]{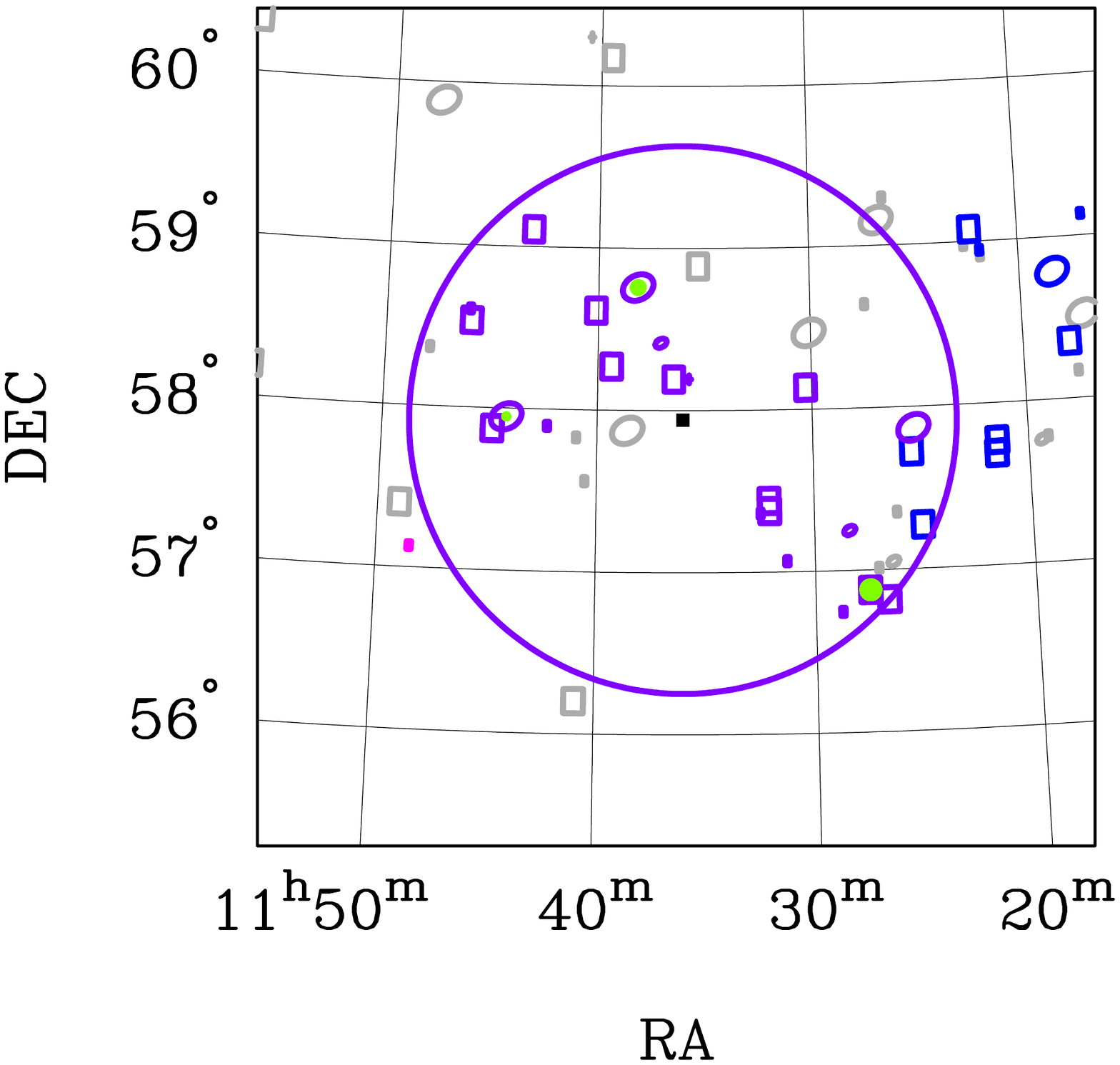}} & 
\mbox{\includegraphics[trim=0cm 0cm 0.5cm 0.5cm, clip=true, height=4.5cm]{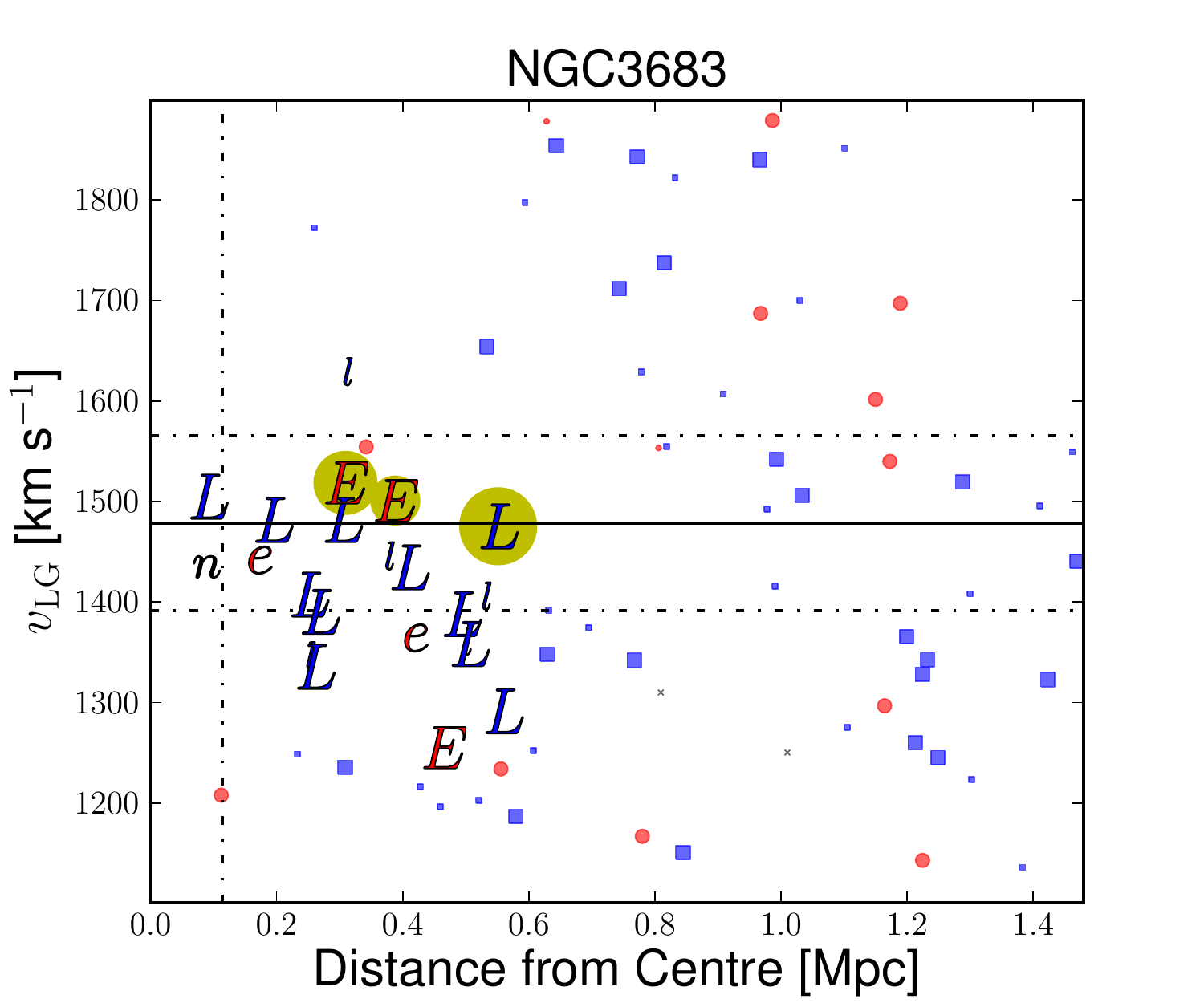}} &
\mbox{\includegraphics[trim=0.5cm 0cm 0.5cm 0.5cm, clip=true, height=4.5cm]{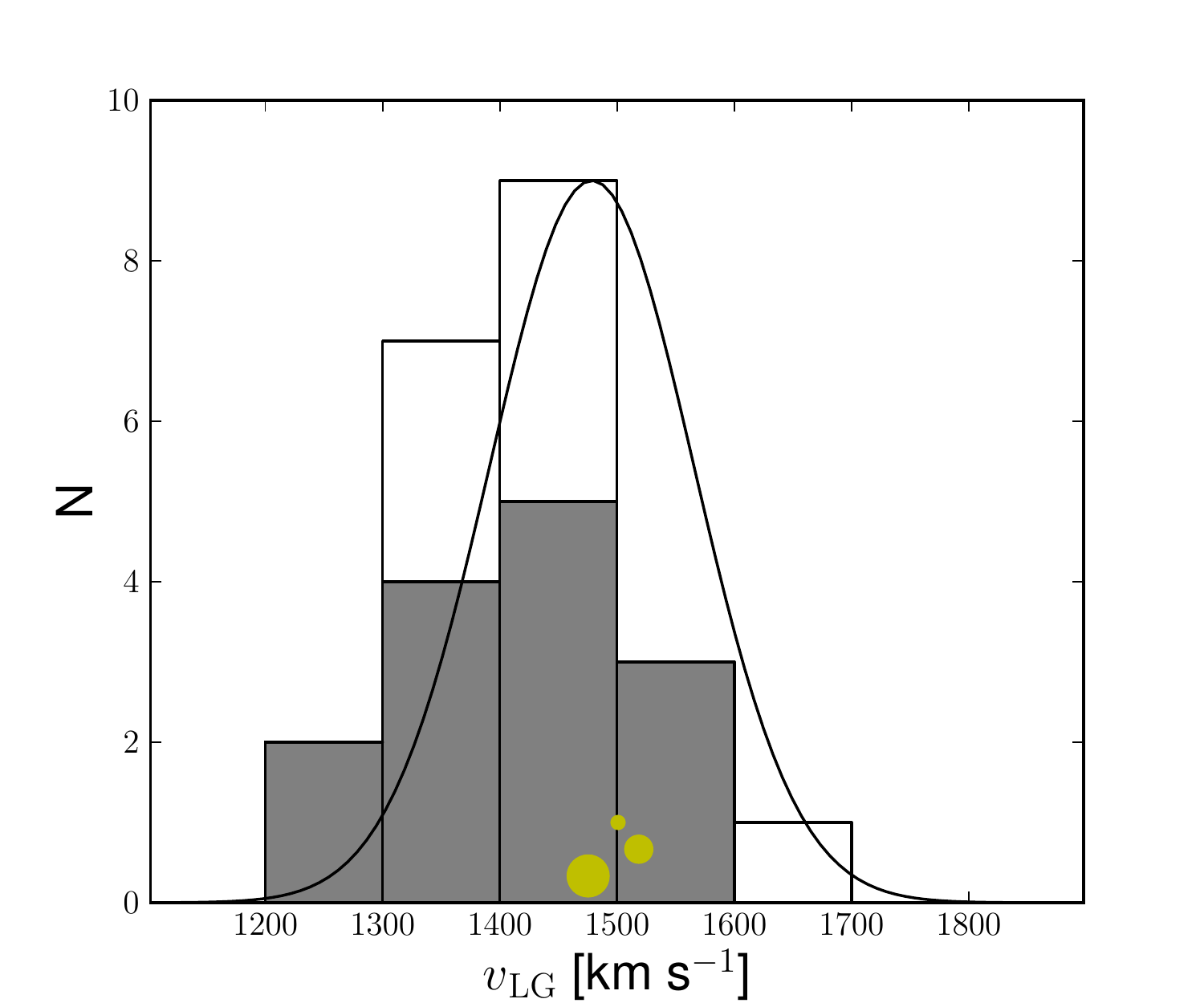}} \\
[12pt]
\mbox{\includegraphics[trim=4cm 7.5cm 1cm 3.5cm, clip=true, height=4.7cm]{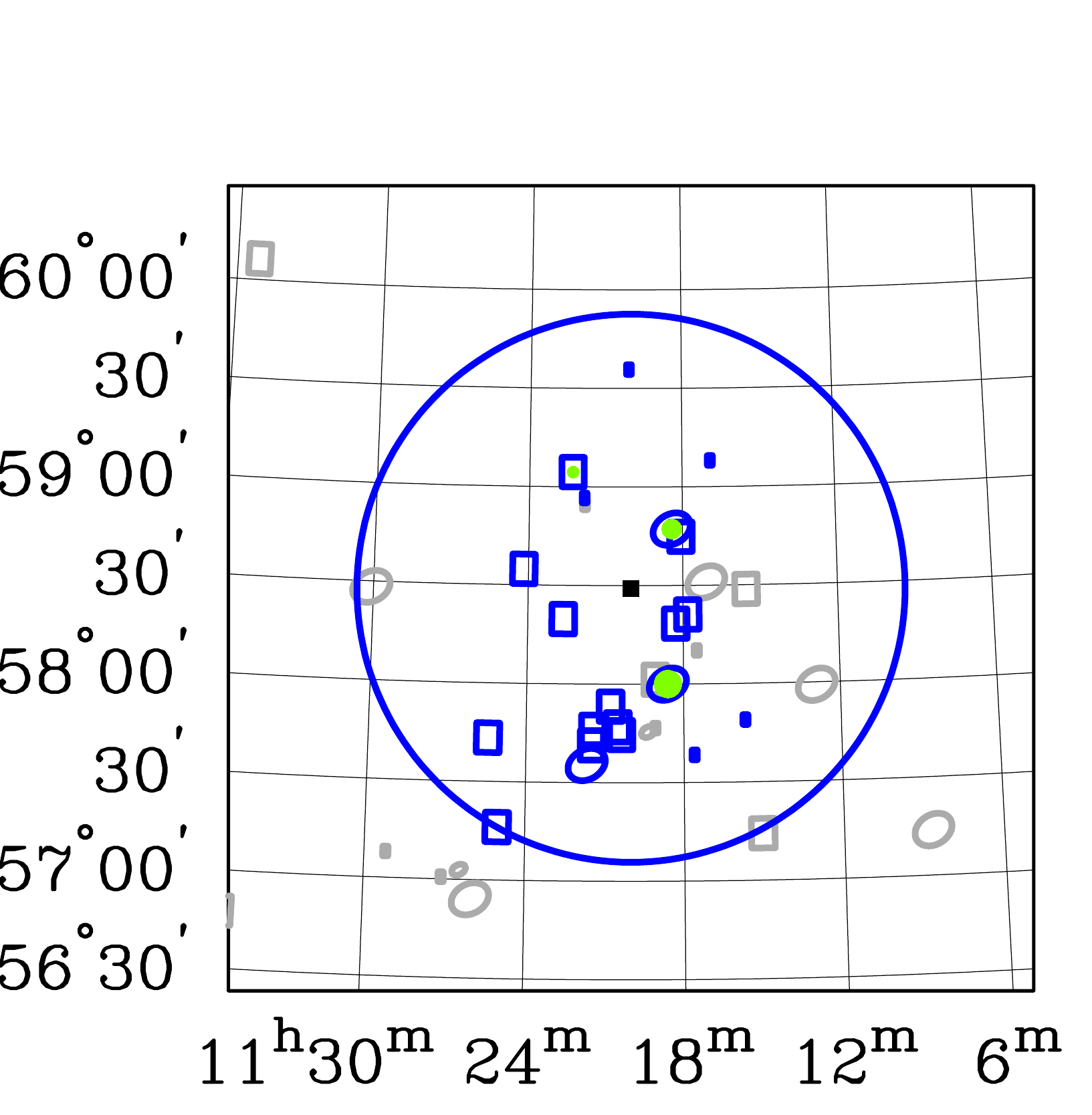}} & 
\mbox{\includegraphics[trim=0cm 0cm 0.5cm 0.5cm, clip=true, height=4.5cm]{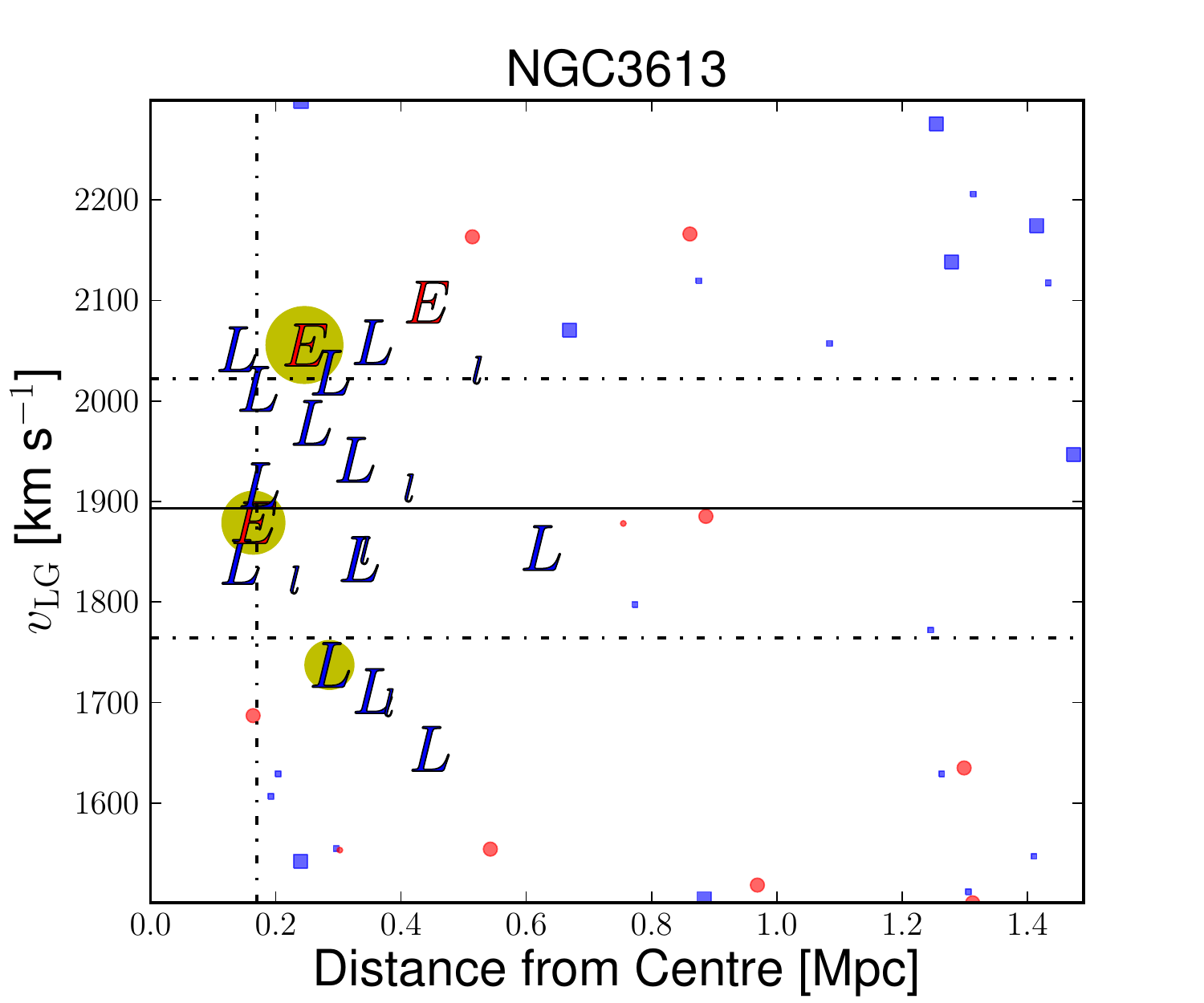}} &
\mbox{\includegraphics[trim=0.5cm 0cm 0.5cm 0.5cm, clip=true, height=4.5cm]{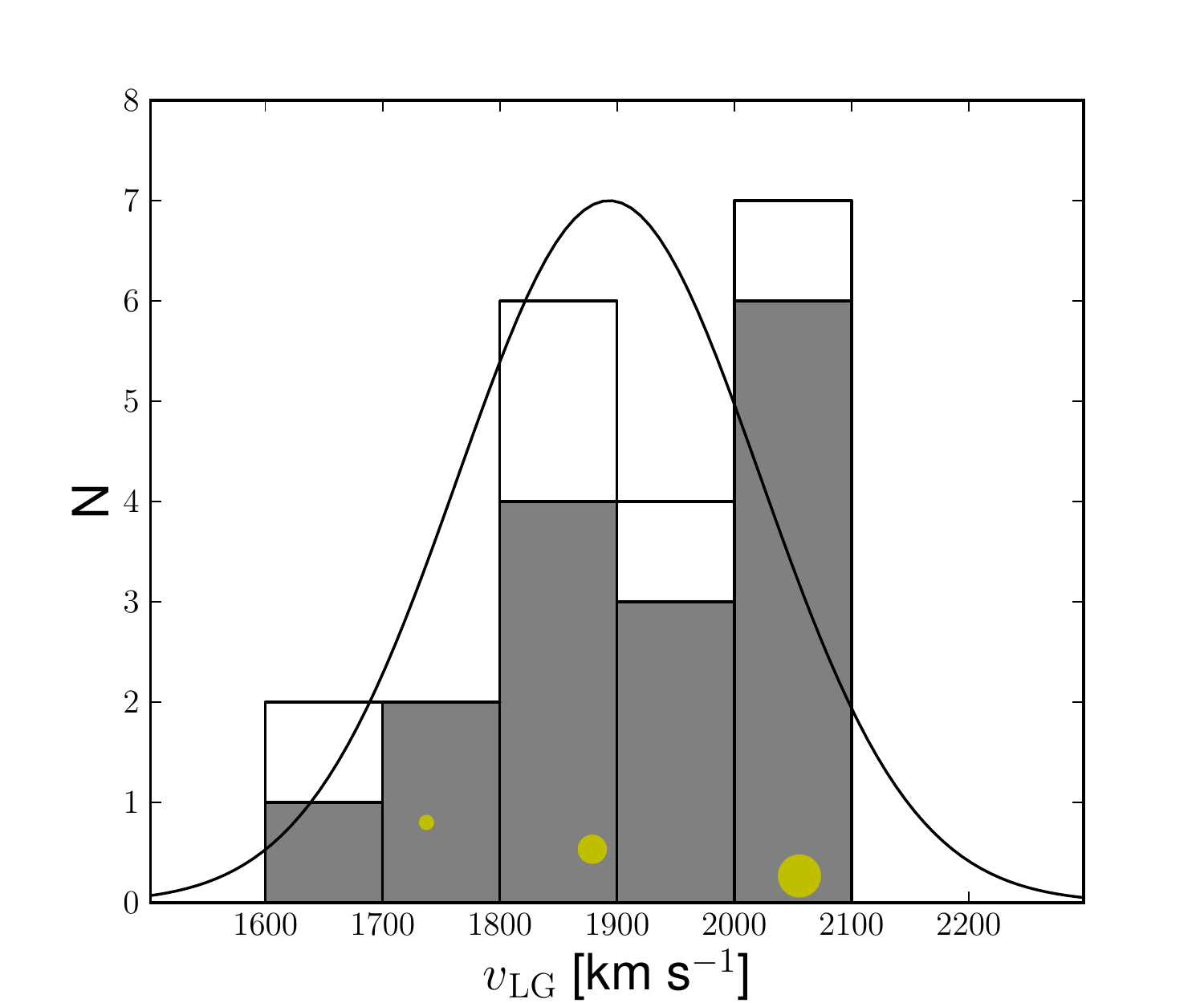}} \\
[12pt]
\end{tabular} 
\caption{The main structures in the northern part of the Ursa Major region (using linking lengths $D_0=0.3$~Mpc and $V_0=150$~km~s$^{-1}$). Similar to Figure~\ref{fig:middle1}. The NGC3998 galaxy, which shows a disturbed HI content (Verheijen \& Zwaan 2001) is highlighted with a black dot (skymap) and a green circle (distance-velocity diagram and velocity histogram).}
\label{fig:north1}
\end{center}
\end{figure*}

The dominant structures in the northern part of the Ursa Major region ($11^{h}\leq\alpha\leq+13^{h}$ and $50^{\circ}\leq\delta\leq+65^{\circ}$) are the NGC3998, NGC3683 and NGC3613 groups. The groups form a filamentary structure connected in projection, with increasing central velocities of 1065~km~s$^{-1}$, 1478~km~s$^{-1}$ and 1893~km~s$^{-1}$, respectively (see Table~\ref{tab:group_prop} for more details on the group properties). The northern structures are linked to the central groups at low recession velocities via the NGC3998 and the much smaller NGC3992 group (at $v_{LG} = 1065$~km~s$^{-1}$ and $v_{LG} = 1267$~km~s$^{-1}$, respectively; see Figure~\ref{fig:overview}).

The richest group in the northern part of the Ursa Major region is the NGC3998 group with 29 group members (14 in the complete sample). The group has a velocity dispersion of 188~km~s$^{-1}$ and an early-type fraction of 21~per cent. NGC3998 (BGG) is located at 107~kpc and 79~km~s$^{-1}$ off the group centre and is one of the three brightest lenticulars in the Ursa Major region (together with NGC4026 and NGC3938; \citealt{verheijen2001}). All of them show signs of transformation occurring to the galaxies, such as tidal tails and disturbed HI content (see \citealt{verheijen2001}; marked with a black box in Figures~\ref{fig:middle1} and \ref{fig:north1}). Two group members have quality distance measurements listed in EDD \citep{tully2009} -- the BGG at 14.1~Mpc and NGC3982 ($3^{rd}$ BGG) at 20.5~Mpc, which implies a structure of more than 6~Mpc across. Assuming Hubble flow, the group is located at 14.6~Mpc, which is in good agreement with the redshift-independent distance measurement of NGC3998. The velocity distribution of the NGC3998 group shows two peaks. The inconsistency in quality distances and the non-Gaussian velocity distribution indicate that the group is unrelaxed. Note that the much smaller NGC3972 group (4 group members in complete and full sample) is located at similar position on the sky (see Figure~\ref{fig:overview}) as the NGC3998 group, but at lower recession velocity (738~km~s$^{-1}$ for the NGC3972 group as opposed to 1065~km~s$^{-1}$ for the NGC3998 group).

The BGG in the NGC3683 group is located at the kinematic centre (the velocity offset is only 3~km~s$^{-1}$). However, the projected spatial offset is 552~kpc, i.e. the BGG resides in the outskirts of the group at nearly maximal radial extent ($r_{max}$). No redshift-independent distance measurements are available for any of the 22 group members (14 in the complete sample). Assuming Hubble flow, the group resides at a distance of 20.3~Mpc. The velocity dispersion of the group is 87~km~s$^{-1}$ and the early-type fraction is 21~per cent.

The NGC3613 group resides in the background of the Ursa Major region at 25.9~Mpc (Hubble flow). The group comprises 21 group members (16 in the complete sample) with a velocity dispersion of 129~km~s$^{-1}$ and an early-type fraction of 19~per cent. Two quality distances are listed in EDD \citep{tully2009}, which are inconsistent -- NGC3613 (BGG) is at 29.1~Mpc, whereas NGC3610 ($2^{nd}$ BGG) is located at 21.4~Mpc suggesting a structure of approximately 8~Mpc across. The spatial and kinematic offsets of the BGG from the group centroid are 236~kpc and 162~km~s$^{-1}$.

\end{document}